\documentclass[twoside]{book}

\usepackage{quotchap}

\usepackage{minitoc}

\usepackage{fancyhdr}
\usepackage{amssymb}
\usepackage{amsmath}
\usepackage{bm}
\usepackage{graphicx}
\usepackage{fancybox}
\usepackage{mathdots}
\usepackage{rotating}
\usepackage{makeidx}
\usepackage{color}
\usepackage{framed}
\usepackage{wasysym}
\usepackage{gravity}

\usepackage{graphicx,epsf,epsfig,amssymb} 
\usepackage{cite} 
\usepackage[
  colorlinks=true,
  linkcolor=blue,
  urlcolor=magenta,
  filecolor=blue,
  citecolor=red,
  pdfstartview=FitV,
  pdftitle={},
  pdfauthor={},
  pdfsubject={},
  pdfkeywords={},
  pdfpagemode={},
  bookmarksopen=true
]{hyperref} 

\usepackage{float}

\usepackage{geometry}
\usepackage{nameref}
\usepackage{xparse}

\usepackage{etoolbox} 

\makeatletter
\NewDocumentCommand{\codefile}{O{} m}{%
\item \texttt{#2}%
  \phantomsection%
  \def\@currentlabelname{#2}%
  \label{#1}%
}
\makeatother

\newcommand{\coderef}[1]{%
  \begingroup
  \hypersetup{hidelinks}%
  \textcolor{red}{\hyperref[#1]{\texttt{\nameref{#1}}}}%
  \endgroup
}
\definecolor{ss}{rgb}{0.54, 0.17, 0.89} 

\newcounter{mnotecount}[section]

\usepackage{epigraph} 
\usepackage{lipsum}

\usepackage[T1]{fontenc} 
\usepackage{lmodern} 

\usepackage{textcomp, gensymb}
\usepackage{bm} 

\usepackage{mathrsfs} 

\def\h{\hfill\break\noindent}

\def\be{
\begin{equation}}
  \def\ee{
\end{equation}}
\def\beq{
\begin{eqnarray}}
  \def\eeq{
\end{eqnarray}}

\newcommand{\beqn}{
\begin{eqnarray*}}
  \newcommand{\eeqn}{
\end{eqnarray*}}

\def\op{ \ $ }
\def\cl{$ \ }

\def\l2{\op \bf L^2\cl}

\def\l{{\cal L}}

\renewcommand{\vec}[1]{\boldsymbol{#1}}

\makeatletter

\let\oldchaptermark\chaptermark
\renewcommand{\chaptermark}[1]{\oldchaptermark{\@chaptermark}}
\let\old@chapter\@chapter
\def\@chapter[#1]#2{%
  \def\@chaptermark{#1}
  \old@chapter[#2]{#2}%
}

\let\oldsectionmark\sectionmark
\renewcommand{\sectionmark}[1]{\oldsectionmark{\@sectionmark}}
\let\oldsubsectionmark\subsectionmark
\renewcommand{\subsectionmark}[1]{\oldsubsectionmark{\@subsectionmark}}
\let\old@sect\@sect
\def\@sect#1#2#3#4#5#6[#7]#8{%
  \@namedef{@#1mark}{#7}
  \old@sect{#1}{#2}{#3}{#4}{#5}{#6}[#8]{#8}%
}

\usepackage{lmodern} \normalfont 
\DeclareFontShape{T1}{lmr}{bx}{sc} { <-> ssub * cmr/bx/sc }{}

\renewcommand
\chapter{\clearpage
  \thispagestyle{plain}%
  \global\@topnum\z@
  \@afterindentfalse
\secdef\@chapter\@schapter}
\makeatother
\makeindex
\begin{document}

\dominitoc
\setcounter{minitocdepth}{1}

\pagenumbering{roman}
\thispagestyle{empty}

\begin{center}

  \Huge \sc The Physics of Black Holes and Their Environments: \\[0.2cm]
  \huge \sc Consequences for Gravitational Wave Science \\[1.2in]

  \Large {\scshape \textsl{Lecture notes from}} \\[0.1cm]
  \huge \sc Beyond the Horizon: Testing the Black Hole Paradigm \\[0.2cm]
  \Large {\scshape \textsl{ @ ICTS–TIFR, Bengaluru (2025)}}\\[0.6in]
  \huge \sc Vitor Cardoso\textsuperscript{1,2,$\dagger$} \\[0.6in]
  \begin{tabular}{c}
    \huge \sc Shauvik Biswas\textsuperscript{3,4$\ddagger$} \quad Subhodeep Sarkar\textsuperscript{5\textsection}
  \end{tabular} \\[0.9in]

  \LARGE \sc (Dated: 19 November 2025)\\[1in]


  \begin{minipage}{0.9\textwidth}
    \hrule height 0.5pt
    \vspace{0.3em}
    \small
    \raggedright
    \textsuperscript{1} Center of Gravity, Niels Bohr Institute, Blegdamsvej 17, 2100 Copenhagen, Denmark\\
    \textsuperscript{2} CENTRA, Departamento de Física, Instituto Superior Técnico – IST, Universidade de Lisboa – UL, Avenida Rovisco Pais 1, 1049 Lisboa, Portugal\\

    \textsuperscript{3} Indian Institute of Technology Gandhinagar, Gujarat 382055, India\\
    \textsuperscript{4} School of Physical Sciences, Indian Association for the Cultivation of Science, Kolkata 700032, India\\
    \textsuperscript{5} Centre for Strings, Gravitation and Cosmology, Department of Physics, Indian Institute of Technology Madras, Chennai 600036, India\\[0.3em]
    \centering
    \textsuperscript{$\dagger$} {\tt vitor.cardoso@nbi.ku.dk} \quad \\[0.2em]
    \textsuperscript{$\ddagger$} {\tt shauvikbiswas2014@gmail.com} ~~ \textsuperscript{\textsection} {\tt subhodeep.sarkar1@gmail.com}
    \vspace{0.3em}
    \hrule height 0.5pt
  \end{minipage}\\[0.8in]


\end{center}


\vspace*{1.5in}

\begin{center}
    {\Huge \sc Abstract}
\end{center}

\vspace{0.5in}

\begin{center}
    \begin{minipage}{0.85\textwidth}
        \normalsize

        Ten short years ago, we had the rare privilege of witnessing the onset of a renaissance in science: humanity finally succeeded in its arduous quest to directly detect gravitational waves. This breakthrough did not occur in a vacuum: it was the natural culmination of decades of research dedicated towards understanding the nature of gravitation based on Einstein’s General Theory of Relativity. It is a story of false starts, perseverance, and remarkable insights, propelled as much by technological progress as by human curiosity.
        We now proudly live in the new golden age of gravitational physics. The detection of gravitational wave signals from the merger of binary black holes and neutron stars are becoming routine. Coupled with our theoretical understanding of phenomena in the strong gravity regime, black hole physics has become a precision science.

        We present here a set of lecture notes that aim to help the reader make sense of this paradigm shift in how we study the cosmos. They are not, however, a historical overview of the field. Rather, the purpose of these notes is to help the reader \emph{understand} the language and framework of this rapidly evolving subject, and to develop the ability to interpret, think, and discuss ideas that lie at the confluence of gravitational wave astronomy and black hole physics.
        It is our hope that these notes will prepare students and colleagues for the \emph{next} revolution when gravitational wave events become commonplace and we begin to observe unexpected features in the signal, indicating either surprising astrophysical scenarios or a strong need to modify the theoretical description of gravitational interactions.
        We provide first principles analysis of black hole and gravitational wave physics, and sometimes a very personal interpretation of results. We share with the readers a number of notebooks that will allow them to reproduce some of the most important results in the field, and could even help in carrying out state-of-the-art research. We also include a few original results that we think are helpful in understanding the broader picture. We hope that the reader enjoys reading these lectures as much as we have enjoyed preparing them.

    \end{minipage}
\end{center}

\newpage

\vfill

\begin{center}
    \vspace*{\fill}
    \emph{We dedicate this manuscript to the enduring legacy and memory of Jayant Vishnu Narlikar and Naresh Dadhich, in deep appreciation for their scholarship and leadership, which shall continue to be a guiding light and a profound source of inspiration for scholars across time and space.}
    \vspace*{\fill}
\end{center}

\vfill
\tableofcontents
\pagenumbering{arabic}

\chapter{Part I}
\label{ch:part1}
\allowdisplaybreaks
\minitoc
\section{Introduction}
The study of strong field gravity has recently undergone a revolution. From a field mostly disconnected from observations, it is now one of the most vibrant areas of research, tying gravitational-wave and electromagnetic observations to tests of General Relativity and some of its most astounding predictions. The advent of very large baseline interferometry, and of gravitational wave astronomy that can monitor the coalescence of compact binaries across millions of light-years, provides access to the most mysterious macroscopic objects: black holes.

Black holes display extreme properties: regions where light can be trapped, regions where any form of matter is forced to co-rotate with the spacetime, and regions of no return. Perhaps more importantly, the interior of black holes contains singularities. Our theoretical description of black hole interiors is, at best, incomplete. This failure of General Relativity to describe black hole interiors is perhaps the most profound gap in our understanding of fundamental interactions. Thus, we want (we need!) to test General Relativity's predictions in the strong-field regime. This program is challenging and involves testing the general relativistic predictions for the motion of two black holes or for the relaxation of black holes as they vibrate. The cosmos is full of matter, which is also an opportunity to test the environments where black holes evolve. It is also a necessity: to ever claim new physics, one must have the systematic effects caused by black hole environments well under control.

These lectures\footnote{This is a collection of notes based on a set of lectures delivered at ICTS-TIFR, Bengaluru in {April} 2025 by V.C. with S.B. and S.S. acting as his teaching assistants. The lectures on which this material is based are available at \url{https://www.icts.res.in/program/beyondhorizon/talks}. The written notes sometimes go substantially beyond the scope of the recorded lectures.} are an informal introduction to the physics of compact objects and black holes, and are intended to help start a discussion on how to test a number of foundational issues:
\begin{itemize}
    \item Do black holes exist?
    \item Are they the solutions predicted by General Relativity?
    \item Do black holes carry charge?
    \item Is cosmic censorship preserved?
    \item What is the maximum possible luminosity from any cosmological event?
    \item Are gravitons massless?
    \item How did these black holes form? How many are there and how do they grow over time?
    \item Can gravitational waves from black holes help us understand the nature of dark matter or tell us about the environment in which they live?
\end{itemize}

We also provide a number of (solved) exercises and \textsc{Mathematica}\textsuperscript{\textregistered} notebooks (listed in \textbf{Appendix \ref{appendix_codes}}), which, in our experience, help in cementing knowledge, and in understanding the state of the art (since many of the current calculations are done within a similar framework). Some of the results discussed here have not appeared previously in the literature, and some of the exercises are variants of well-known results. We hope that you enjoy reading these lectures as much as we have enjoyed preparing them.

\section{Setting up the stage}
In these lecture notes, we will always have General Relativity in mind, and when we depart from it, we will always assume that gravity is still described by a metric theory. Then, the fundamental object is the spacetime metric,

\be
ds^2=g_{ab}dx^adx^b\,. \nonumber
\ee
We will use the Einstein equations at all times,
\be
G_{\mu\nu}\equiv R_{\mu\nu}-\frac{1}{2} g_{\mu\nu}R=8\pi \frac{G}{c^4} T_{\mu\nu} \,,
\ee
but moving forward, we will use geometrized units with $G=c=1$.
Note that mass, time, and length have the same dimensions in these units,
\be
[M]=[T]=[L]\,.
\ee
Thus, we can calculate mass in kilograms, kilometers or even seconds. For example, consider the mass of our Sun, $M_{\odot}$ in these geometrized units:
\begin{align}
    M_{\odot}              & = 1.99\times 10^{30}\,{\rm kg}\,, \\
    \frac{G}{c^2}M_{\odot} & = 1.47\,{\rm km}\,,               \\
    \frac{G}{c^3}M_{\odot} & = 4.92\times 10^{-6}\,{\rm s}\,.
\end{align}

It is a classical result, worked out in the notebook \coderef{code:Field-Equations.nb},
that the unique {\bf vacuum}, spherically symmetric, and asymptotically flat solution is the Schwarzschild solution,
\be
ds^2=-\left(1-\frac{2M}{r}\right)dt^2+\frac{dr^2}{\left(1-\frac{2M}{r}\right)}+r^2d\Omega^2\,.\label{eq:Schwarzschild_geometry}
\ee
Here, $M$ is the gravitational mass of the black hole as measured by observers far away (also see Eq.~\eqref{eq:geodesics_mass} below). The spacetime has singularities at $r=0$ and at $r=2M$. To understand the nature of these singularities, let us calculate curvature scalars, something that all observers equipped with an arbitrary coordinate chart should agree upon.

Since this is a vacuum spacetime, the Ricci tensor---and therefore the Ricci scalar---vanishes by construction. Then the next higher-order scalar built out of the geometry is the Kretschmann scalar (see notebook \coderef{code:Field-Equations.nb}),
\be
K=R_{\mu\nu\alpha\beta}R^{\mu\nu\alpha\beta}=\frac{48M^2}{r^6}\,.
\ee
So from this standpoint, $r=0$ is indeed a spacetime singularity, since it leads to a diverging curvature at the origin of the coordinate system. An observer falling into a black hole {\it will} hit the singularity in finite proper time~\cite{Lewis:2007gc,Toporensky:2019ueu}.

Note that there is nothing special about the $r=2M$ surface from the point of view of curvature. Let us change from $(t,r)$ to $(T,X)$ coordinates, such that,
\beq
X^2-T^2&=&\left(\frac{r}{2M}-1\right)e^{r/2M}\,,\\
\frac{T}{X}&=&\tanh\left(\frac{t}{4M}\right)\,.
\eeq
We then find put Eq.~\ref{eq:Schwarzschild_geometry} in the following form,
\be
ds^2=\frac{32M^3}{r}e^{-r/2M}\left(dX^2-dT^2\right)+r^2d\Omega^2\,.
\ee
Even though $r=0$ is still singular, the $r=2M$
surface is now perfectly regular. An observer falling onto a black hole crosses the $r=2M$ surface in a finite proper time and an infinite Schwarzschild coordinate time (see the discussion on the radial geodesics following Eq.~\eqref{eqn:dt/dtau}). The surface $r=2M$ is the event horizon of the black hole.

The existence and formation of a singularity is inevitable within Einstein's theory. Indeed, gravitational collapse of ``reasonable'' matter to form a trapped surface results in a ``singularity'', where at least one of the following holds~\cite{Penrose:1964wq,Penrose:1969pc}:
\vskip 2mm
\begin{enumerate}
    \item [a.] Negative local energy occurs.
    \item  [b.] The Einstein equations are violated.
    \item [c.] The spacetime manifold is incomplete.
    \item [d.] The concept of spacetime loses its meaning at very high curvatures---possibly because of quantum effects.
\end{enumerate}

The existence of spacetime singularities inside black holes implies that the ultimate description of black hole interiors cannot be achieved within the theory from which they were born.

\section{Uniqueness}

The black holes out there are obviously not isolated, and we will look into the consequences later on. But for now, let us see whether we can find other black hole solutions in the presence of a very specific and simple type of matter: a minimally coupled scalar field $\Phi$, with some possible self-interactions described by a potential $V(\Phi)$ and an action,
\be
S=\int d^4x \sqrt{-g}\left(\frac{R}{k}-\frac{1}{2}g^{\mu\nu}\partial_\mu\Phi\partial_\nu\Phi-V(\Phi)\right)\,.\label{eq_chapter1_scalar}
\ee
Let us look for spherically symmetric, static solutions, which can always be written as,
\be
ds^2=-fdt^2+\frac{dr^2}{g}+r^2d\Omega^2\,.
\ee
A black hole solution corresponds to having a zero of $g$.

The equation of motion for the scalar field (we won't need the full set of Einstein equations), is simply
\be
\frac{1}{r^2}\sqrt{\frac{g}{f}}\,\frac{d}{dr}\left(r^2\sqrt{fg}\,\Phi'\right)=\frac{dV}{d\Phi}\,.
\ee

Now, assume that there is a value $\Phi_0$ that minimizes $V$ (for example, $\Phi_0=0$ for $V=\Phi^2$). We rewrite the above equation as,
\be
(\Phi-\Phi_0)\frac{d}{dr}\left(r^2\sqrt{fg}\,\Phi'\right)=(\Phi-\Phi_0)r^2\sqrt{\frac{f}{g}}\,\frac{dV}{d\Phi}\,,
\ee
and integrate from the horizon (at $r=r_+$, that is, the zero of $g$) to infinity. We find
\begin{align}
    \begin{split}
        \int_{r_+}^{\infty}\frac{d}{dr}\left(r^2\sqrt{fg}\,(\Phi-\Phi_0)\Phi'\right)
        ={} & \int_{r_+}^{\infty}r^2\sqrt{\frac{f}{g}}\Big(g\,\left(\Phi'\right)^2 {}  +(\Phi-\Phi_0)\frac{dV}{d\Phi}\Big)\,.
    \end{split}
\end{align}
Now, $g(r_+)=0$ by definition of horizon. Thus, both boundary terms on the left vanish. The right-hand side is positive-definite if there is a single minimum. In such a case, the only way to satisfy the above identity is by having $\Phi=0$. We call this a no-hair property. These results are part of the uniqueness results in General Relativity. It turns out that the Kerr spacetime is the only asymptotically flat, axisymmetric, stationary, and vacuum solution of the field equations~\cite{Bekenstein:1996pn,Carter:1997im,Chrusciel:2012jk,Robinson:1975bv}. It is parametrized only by its mass and spin. \emph{Black holes are indeed simple!} We will see later on just how this final vacuum state is approached dynamically when we start from general initial conditions.

However, note the assumptions in the theorem: that the Universe is vacuum and that spacetime is stationary. Both are, obviously, violated in the Universe---if only, because humankind exists and is dynamic. To what extent, then, are the uniqueness results useful?

Once we give up vacuum, there {\it are} black hole solutions with ``hair’’ even under the stationarity approach. Such black holes have been built with fluids---modeling galactic halos or other structures~\cite{Cardoso:2021wlq,Figueiredo:2023gas,Speeney:2024mas}, or with massive scalar or vector fields~\cite{Herdeiro:2015waa,Sotiriou:2015pka}. The physical reason why these objects are possible is that the fluids are totally anisotropic, and hence behave as matter sustained by the centrifugal force. On the other hand, solutions with scalars or vectors are possible because, even though the geometry is stationary, the fundamental fields are time-dependent, and supported through superradiance (more on this effect in Section~\ref{sec:superradiance}). For a black hole supporting structures in its exterior that are made up of scalar or vector fields, see e.g.~\cite{Herdeiro:2015waa,Yazadjiev:2025ezx}.

It is also important to keep in mind that most structures in the Universe are not stationary. The Milky Way is a complex, evolving structure. The geometry describing the black hole at the center of our galaxy is therefore neither stationary nor vacuum. It might be a ``good’’ approximation on some length or timescales to consider it vacuum and stationary, but it remains to be quantified what error one incurs by doing so.

\section{Motion}
To study black hole spacetimes and strong-field gravity, one needs to understand how matter moves and the important features of such motion. Consider the motion of test particles---idealizations of objects moving freely in a gravitational field. In General Relativity, the motion is encoded in the field equations. For point particles, it is embedded in the Lagrangian density~\cite{Chandrasekhar:1985kt}

\begin{equation}
    {\cal L}=\frac{1}{2}g_{ab}\dot{x}^a\dot{x}^b \label{eqn:lagrangian_density}
\end{equation}
where
\be
g_{ab}\dot{x}^a\dot{x}^b=\delta_1\equiv0,\,-1\,\quad {\rm for \,\,null\,\,and \,\,timelike\,\, particles\,\, respectively}\label{geodesic_normalization}
\ee
The Euler-Lagrange equations are equivalent to the geodesic equations,
\be
\ddot{x}^\alpha+\Gamma^\alpha_{\beta\gamma}\dot{x}^\beta\dot{x}^\gamma=0\nonumber\,,
\ee
and follow from the vanishing of the divergence of stress-tensor (see Sec. 19.1 of Ref.
~\cite{Poisson:2011nh}).
\be
T^{\mu\nu}=m_0\int d\tau v^\mu v^\nu \frac{\delta(x^\mu-r_p^\mu(\tau))}{\sqrt{-g}}\label{eqn:particle_stress_tensor}\,,
\ee

For the Schwarzschild spacetime ($f=g=1-2M/r$), the Lagrangian does not depend on the $(t,\phi)$ coordinates, hence we have two conserved quantities: the specific (per unit rest mass) energy $E$ and angular momentum $L$,
\beq
E&=&f\dot{t}\,\label{eqn:tdot_geodesic},\\
L&=&r^2\dot{\phi} \label{eqn:phidot_geodesic} \,,
\eeq
and the normalization
\be
\dot{r}^2=E^2-f\frac{L^2}{r^2}+f\delta_1\equiv V\,.\label{eq_Eff_potential_geodesics}
\ee

Let's look at motion far away, where spacetime is asymptotically flat. We get
\be
2\dot{r}\ddot{r}=-\dot{f}\left(\frac{L^2}{r^2}-\delta_1\right)+f\frac{2L^2\dot{r}}{r^3}\,.
\ee
Now, $\dot{f}=2M\dot{r}/r^2$, therefore
\be
2\ddot{r}=-\frac{2M}{r^2}\left(\frac{L^2}{r^2}-\delta_1\right)-f\frac{2L^2}{r^3} \sim -\frac{2M}{r^2}\,.\label{eq:geodesics_mass}
\ee
We can compare this to the Newtonian expression to conclude that $M$ is indeed the gravitational mass.

\subsection{Circular orbits}
%
\begin{figure}
    \centering
    \includegraphics[width=1\textwidth]{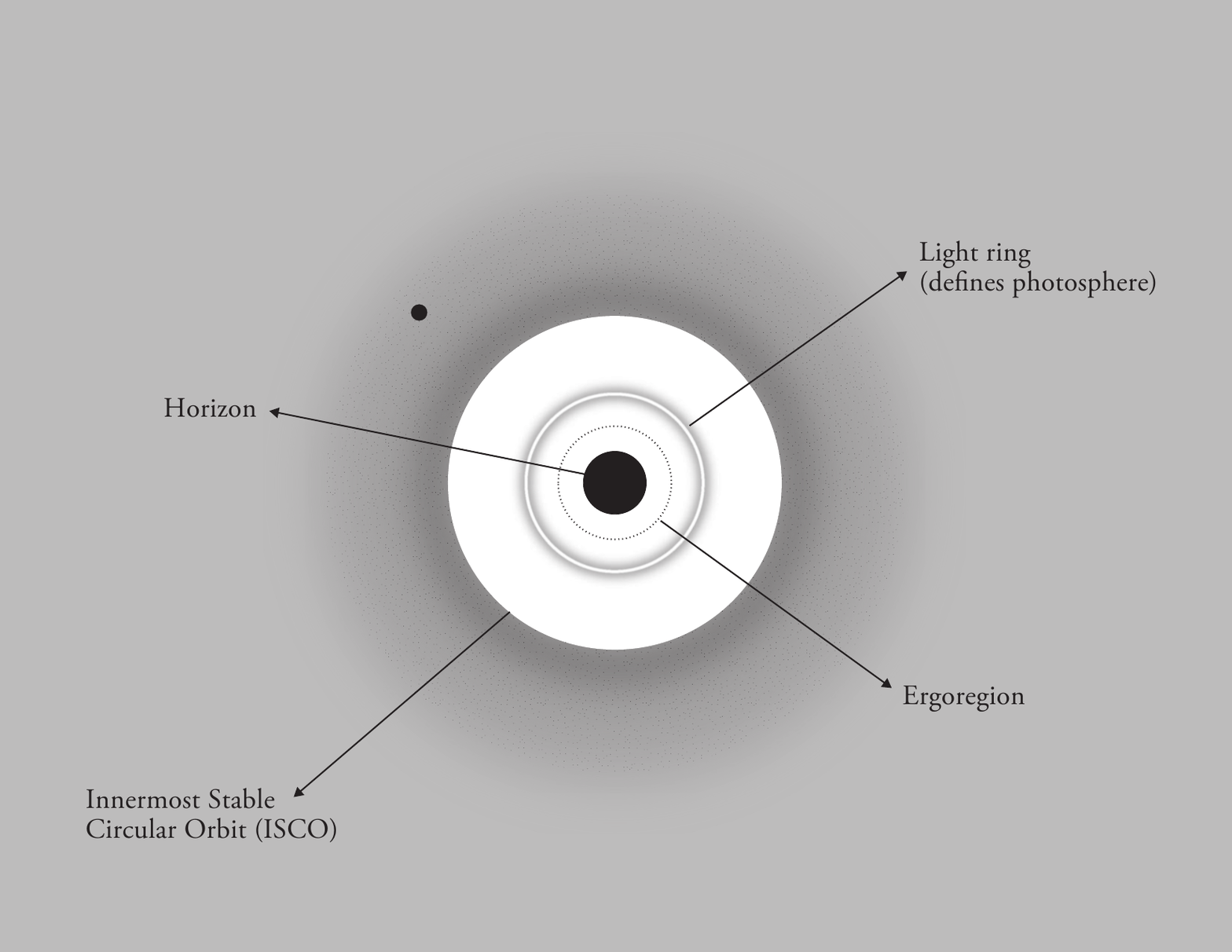}
    \caption{ Equatorial slice of a black hole spacetime, depicting interesting aspects of motion of point particles. Far away from the black hole, the gravitational interaction is well described by Newton's law and stable close circular motion is possible. As we move inwards, General Relativistic effects become important and within the Innermost Stable Circular Orbit (at $6M$ for non-spinning geometries) circular stable motion is no longer possible. Moving closer to the horizon, one finds the photon sphere, or light ring, where light or high frequency gravitational waves can move on a closed our bounded motion. This region effectively traps high frequency radiation and is responsible for the late time relaxation of black holes: the sound of black holes is produced close to the light ring. For spinning black holes, an ergoregion exists close to the horizon, forcing all matter to co-rotate with the hole and facilitating energy extraction from vacuum. Finally, the one-way horizon causally disconnects from outside observers the physics of the interior. Image by Ana Carvalho.
    }
    \label{fig:equatorial_slice}
\end{figure}
In Newtonian dynamics, circular motion for matter is both possible and stable. However, in General Relativity, the situation is slightly more intricate.
Using the effective potential \eqref{eq_Eff_potential_geodesics}, the conditions for circular motion are $V=0$ (i.e., that $\dot{r}=0$) and $V'=0$ (i.e., that $\ddot{r}=0$ such that the particle remains on a circular orbit).
Let us first consider null matter ({\bf A}), followed by a discussion on massive particles ({\bf B}):

\vspace{1em}

\noindent $\blacksquare$ {\bf A.} For $\delta_1=0$, describing null particles,
\be
V'=0 \Rightarrow \left(\frac{f}{r^2}\right)'=0\Rightarrow r=3M\,. \label{eqn:photon_sphere_eq}
\ee
In other words, in General Relativity, circular motion for null particles (which describe the propagation of high frequency photons and gravitons) in vacuum {\it is} possible. It exists at a single location, defining the photon sphere (in the case of spherical symmetry). The projection onto the equatorial plane is also frequently referred to as the light ring.

Although a circular orbit is possible, the motion is unstable. Indeed, let us consider a null particle be slightly displaced from the light ring, via $r=3M+\delta$, and find
\be
\dot{r}^2=V \Rightarrow \dot{\delta}^2=\frac{1}{2}(r-3M)^2V''(3M)=\frac{1}{2}\delta^2V''(3M)\,.
\ee
The solution to this relation is simply
\be
\delta\sim e^{\lambda t}\,,\quad \lambda=\sqrt{\frac{V''\,f^2}{2E^2}}=\frac{1}{3\sqrt{3}M}\,.\label{null_geoedesic_decay}
\ee
Thus, any small fluctuation to a high frequency photon or graviton in the light ring will cause it to disperse or to fall onto the black hole, on scales of
\be
\frac{1}{\lambda}=3\sqrt{3}M=1.5\,{\rm ms}\,\frac{M}{30 M_{\odot}}\,.
\ee
Note also that the frequency of particles on the light ring, as perceived by distant observers, is also $\Omega_{\rm LR}=\dot{\phi}/\dot{t}=({3\sqrt{3}M})^{-1}$.

We think of the light ring, or the photon sphere, as the location where the characteristic modes (called quasinormal modes or ringdown modes) of black holes are localized. The trapping property of light rings causes such particles to remain there on timescales of $3\sqrt{3}M$ or smaller. Thus, in a dynamical process where black holes are excited with gravitational radiation or light, the waves trapped in the light ring dominate the late time response, since they remain trapped longer: at late times, the signal is an exponentially damped sinusoid~\cite{Cardoso:2008bp,Cardoso:2016rao,Cardoso:2019rvt}. More later on this in Section~\ref{sec:response}.
\vspace{1em}

\noindent $\blacksquare$ {\bf B.} For $\delta_1=-1$,
\beq
V'&=&0 \Rightarrow L^2=\frac{ME^2r}{f}\,,\\
V&=&0\Rightarrow E^2=\frac{rf^2}{r-3M}\,.
\eeq
We can immediately conclude that circular motion is possible only for $r \geq 3M$, a unique feature of General Relativity. In addition, one can easily verify that circular motion is stable {\it only} for $r \geq 6M$. This is another interesting consequence of General Relativity: matter, like plasma or dust, is expected to orbit black holes, but only at radii $r>6M$. When it gets closer to this radius, circular motion becomes unstable and matter is expected to fall into the black hole. The circular orbit at radius $r=6M$ is, therefore, usually referred to as the Innermost Stable Circular Orbit (ISCO). This also has curious implications for the relaxation of black holes: even in the presence of matter, the light ring is expected to be relatively devoid of matter.

A cartoonish depiction of the possible structure of a black hole and motion of matter is shown in Fig.~\ref{fig:equatorial_slice}.
\subsection{Radial motion}
Radial geodesics are described by $L=0$. So we have the following expression from Eq.~\eqref{eq_Eff_potential_geodesics},
\be
\dot{r}^2=E^2+f\delta_1\,.\label{geodesics_timelike}
\ee
This means that a particle falling in from $r_0$ to $r=2M$ takes a proper time given by
\be
\Delta \tau=\int_{2M}^{r_0}\frac{dr}{\sqrt{E^2-f}}\,.
\ee
For instance, if we consider a particle falling from rest, $E=1$ then
\be
\Delta \tau=\int_{2M}^{r_0}dr\frac{\sqrt{r}}{\sqrt{2M}}\Leftrightarrow \Delta \tau/M=\frac{\sqrt{2}(r_0/M)^{3/2}-4}{3}\,.
\ee
Therefore, a particle takes a finite proper time to reach the $r=2M$ surface. But on the other hand, from Eq.~\eqref{eqn:tdot_geodesic} the change in $\tau$ measured with respect to time $t$ (time ``at infinity'') is given by,
\be
\frac{dt}{d\tau}=E/f\,, \label{eqn:dt/dtau}
\ee
and therefore the time, as measured by a stationary observer at infinity, diverges logarithmically. This is one reason why objects with surface $R<2M$ were also called frozen stars. In other words, the horizon of a black hole can never be probed by outside observers in a finite amount of time. There is no experiment that can ever tell us that a black hole exists for a fact. The singularity inside the black hole causally disconnected from external observers.

Let us now imagine an infalling test particle which is a radio transmitter, emitting some standard frequency $\omega_e$. The emitter follows some time-like radial geodesic and at event $e$ emits a photon of frequency $\omega_e=-(v_\mu k^{\mu})_e$, where $k$ is the photon's momentum and $v$ the emitter's four-velocity. The photon is observed at event $o$, where it intersects the world-line of the observer, such that it is observed with frequency $\omega_o=-(u_\mu k^\mu)_o$. We assume the observer to be at rest at infinity.

We know the null geodesics in full generality. The momentum of the photon is written as
\be
k^\mu=E_f\left(1/f,1,0,0\right)\,,\quad k_{\mu}=E_f\left(-1,1/f,0,0\right)\,.
\ee
The emitter's four-velocity is ($\gamma$ is the emitter's Lorentz factor at infinity)
\be
v_e^\mu=\gamma\left(1/f,\pm\sqrt{\gamma^2-f},0,0\right)\,,\quad v_{e\,\mu}=\gamma\left(-1,\pm\sqrt{\gamma^2-f}/f,0,0\right)\,,
\ee
where the plus sign is for outwards directed test particles, the minus sign is for infalling particles.
We also have by definition of the observer that
\be
v_o^\mu=\left(1,0,0,0\right)\,,\quad v_{o\,\,{\mu}}=\left(-1,0,0,0\right)\,.
\ee
We thus find that
\be
\frac{\omega_e}{\omega_o}=\frac{\gamma \mp \sqrt{\gamma^2-f}}{f}
\ee
If the signal is emitted close to the horizon at $r=2M$, we have
\be
\frac{\omega_e}{\omega_o}\sim \frac{1}{2\gamma} \,,\quad \textrm{for outward directed transmitters}
\ee
\be
\frac{\omega_e}{\omega_o}\sim \frac{2\gamma}{1-2M/r} \,,\quad \textrm{for infalling transmitters} \label{redshiftinfall}
\ee
It is not surprising that infalling or outgoing give rise to different redshifts. Perhaps it is not appreciated however that the redshift
of a signal emitted close to the horizon is infinite for infalling transmitters, but is finite for outgoing.

So far, we have computed the redshift for a fixed observer at a fixed proper time. Let's see how the redshift evolves in time.
We know that for the infalling transmitter
\be
\frac{dr}{dt}=-\frac{f\sqrt{\gamma^2-f}}{\gamma}\sim -\frac{(r-2M)}{2M}\,.
\ee
Thus, close to the horizon, $t\sim -2M\log(r-2M)$. Now, this $r$ is {\it not} the place where the observer actually sees the light. Light requires some time $t_{\rm light}$ to traverse the distance to the observer. This time is easily computed from null geodesics as
\be \label{eqn:tlight}
\frac{dt_{\rm light}}{dr}=\frac{2M}{r-2M} \,.
\ee
Once again, close to the horizon $t_{\rm light} \sim 2M \log(r-2M)$. Finally, the time it takes for the observer to see the light is
\be
t_{\rm total}\approx -4M\log{(r-2M)}\,.
\ee
If we substitute this in equation (\ref{redshiftinfall}) we find therefore
\be
\frac{\omega_e}{\omega_o}\sim e^{t_{\rm total}/(4M)}\,.
\ee
Therefore, although the redshift takes mathematically an infinite observer time to diverge, it diverges exponentially with an e-folding time of $4M$. Thus, there is a sharp transition to very large redshifts after roughly $4M$. For a solar mass black hole, $4M_{\odot}=20\times 10^{-6}\,\textrm{sec}$.

Note that these are only radially directed photons. As we argued, photons that are trapped close to the light ring take a longer timescale to dissipate. Hence, when a generic source falls into the black hole, emitting in all directions, observers will see both a redshift effect and photons from a relaxing light ring~\cite{1965SvA.....8..868P,1968ApJ...151..659A,Cardoso:2021sip}.

\subsection{Scattering and cross-section for absorption}
Let us now consider null particles with some finite impact parameter at infinity $b$, where we define $b\equiv L/E$, then Eq.~\eqref{eq_Eff_potential_geodesics} can be recast as,
\be
\dot{r}^2/E^2=1-\frac{fb^2}{r^2}=1-\left(\frac{1}{r^2}-\frac{2M}{r^3}\right)b^2\,.\label{eq_Eff_potential_null_geodesics}
\ee
The large $r$ behavior shows there is a turning point at large $b$. Indeed, for any large $b$ there is a turning point at $r\sim b$: photons shot at a black hole with enough angular momentum from far away, start probing smaller and smaller $r$ and at $r\sim b$ are deflected back.

Let us define $y=r/(2M),\,b=M\hat{b}$, then the previous equation gives us, for $E=1$ (which corresponds to re-parametrization of the affine parameter),
\beq
\dot{r}^2&=&1-\frac{1}{y^2}\left(1-\frac{1}{y}\right)\frac{\hat{b}^2}{4}\nonumber\\
&=&\left(1-\frac{\hat{b}}{2}\frac{1}{y}\sqrt{1-\frac{1}{y}}\right)\left(1+\frac{\hat{b}}{2}\frac{1}{y}\sqrt{1-\frac{1}{y}}\right)\nonumber
\eeq
\begin{figure}
    \centering
    \includegraphics[width=1\textwidth]{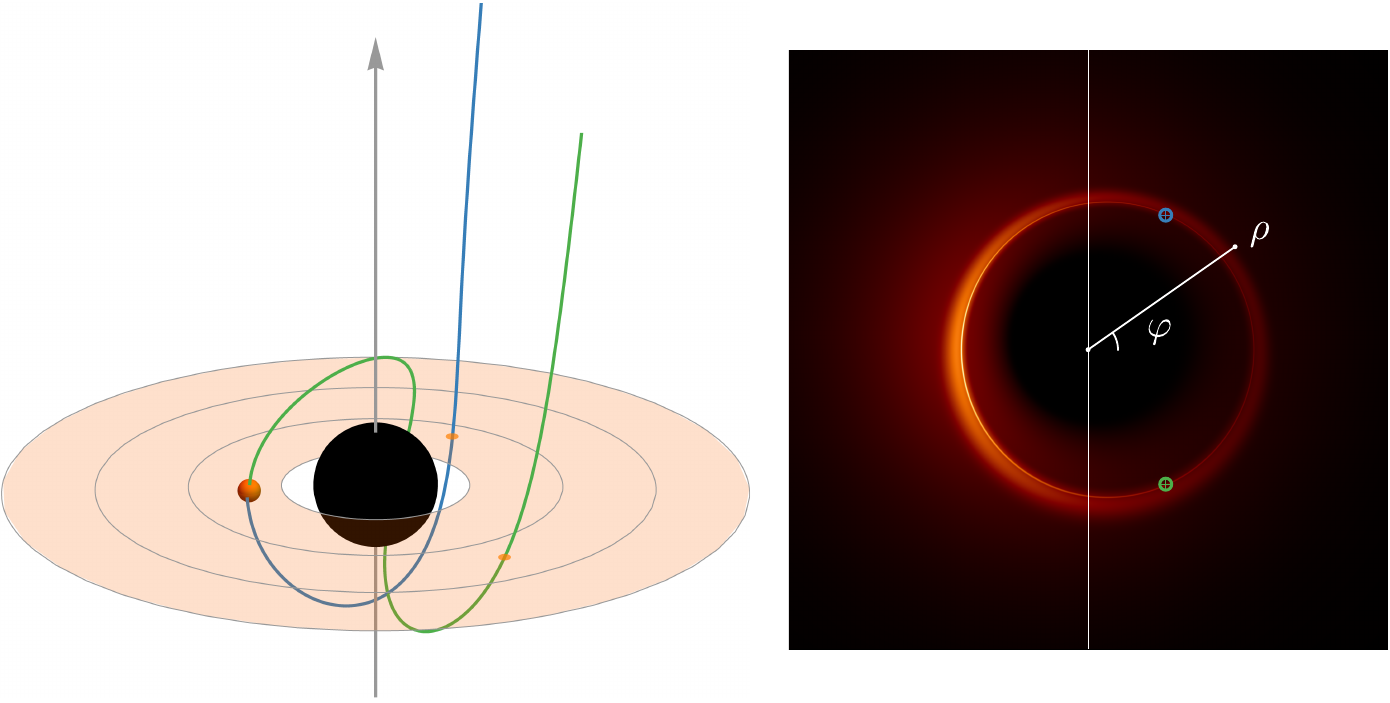}
    \caption{Possible example of a black hole illuminated by an accretion disk.
            {\bf Left:} Null geodesics emanating from a ``flare'' (orange sphere) in an optically thin equatorial emission disk around a Kerr BH of spin $a=0.94M$. The geodesics were chosen such that they are near critical and suffer extreme lensing. Frame dragging is seen for both geodesics, most specially for the green curve.
            {\bf Right:} image of the disk as would be seen by an infinite-resolution far observer at an inclination $\theta_o=17^\circ$. Strongly lensed light rays, which undergo multiple half-orbits, appear on the observer screen close to the ``critical curve'', displaying enhanced brightness, and compose the photon ring. Correlated images of the same spacetime event--the flare for example--appear at different angles and times along the ring (blue and green dots on the right image). Frame-dragging is apparent also here. From Ref.~\cite{Hadar:2020fda}.
    }
    \label{fig:photon_shell_ring}
\end{figure}
The second term is positive. We need to look for roots of $y^3-\hat{b}^2y/4+\hat{b}^2/4$. Its discriminant is $(\hat{b}^2/4)^2(\hat{b}^2-27)$.
Thus, light from infinity with impact parameter
\be
b<3\sqrt{3}M\,,\label{critical_b_light}
\ee
falls into the black hole, which therefore gives us the cross-section
\be
\sigma_{\rm abs}^{\rm light}=27\pi M^2\,.\label{cross_section_light}
\ee
Light shot in all directions will behave in a variety of ways. For example, outgoing photons with large impact parameter will reach infinity, but those moving inwards head-on will inevitably fall into the horizon. It follows that certain photons on critical trajectories will circle the black hole a number of times---asymptoting to the light ring---before scattering or being absorbed. In other words, black holes cast shadows and some of its most salient features are defined by the light ring properties~\cite{Ayzenberg:2023hfw,Johnson:2023ynn}, see e.g. Fig.~\ref{fig:photon_shell_ring}.

For low-energy particles with $E=1$, we find from Eq.~\eqref{geodesics_timelike} with $\delta=-1$
\beq
\dot{r}^2&=&\frac{2M^3}{r^3}\left(\frac{r}{M}-\frac{L^2/M^2+L/M\sqrt{L^2/M^2-16}}{4}\right)\nonumber\\
&\times &\left(\frac{r}{M}-\frac{L^2/M^2-L/M\sqrt{L^2/M^2-16}}{4}\right)\,.
\eeq
Thus, there is a turning point if and only if $L^2>16M^2$. The turning point lies at $r>2M$. The critical angular momentum for absorption is therefore $L_{\rm crit}=4M$ which translates into a critical impact parameter $b=4M/v$ (note that for timelike particles, $b=L/v$).

The cross-section for absorption is then
\be
\sigma_{\rm abs}^{\rm low-energy}=\frac{16 \pi M^2}{v^2}\,.\label{cross_section_lowE}
\ee

The apparent divergence of the cross-section when $v\to 0$ only states the obvious: that particles at rest with respect to the black hole will be accreted eventually, wherever they are.
Impact parameters for spinning black holes at different energies are also discussed in Ref.~\cite{Berti:2010ce}.

There is nothing surprising in the numbers above: black holes absorb matter that comes close to them. The exact coefficient and dependence of the absorption cross-section on the relative velocity will determine the impact of accretion on moving black holes, as we will discuss in the exercises at the end of this chapter.
\section{Adding rotation}
%
\begin{figure}
    \centering
    \includegraphics[width=0.8\textwidth]{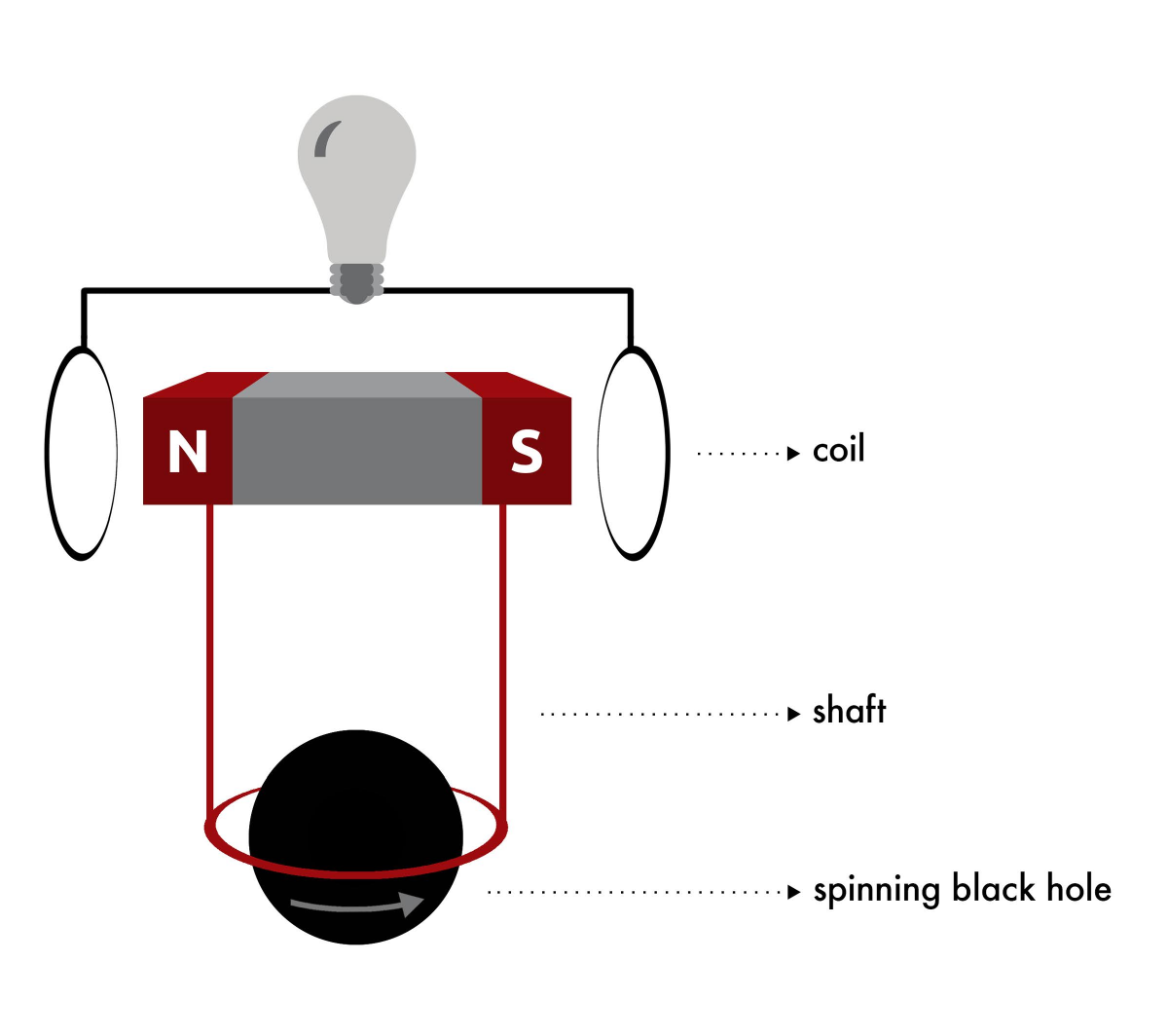}
    \caption{A primitive version of a device to extract energy from a spinning black hole. A metal ring of radius small enough that it fits within the ergoregion of a spinning black hole is lowered onto it, where it is forced to co-rotate with spacetime. Metal shafts attached to the ring cause magnets to spin and Faraday's induction law then tell us that electricity flows in an external circuit, extracting energy from the spinning black hole. Note that such a device would work if planted on any spinning object, like a rotating star or planet. The exquisite aspect of black holes, is that energy is extracted from {\it vacuum}. From Ref.~\cite{Brito:2015oca}.
    }
    \label{fig:energy extraction1}
\end{figure}
Finding the generalization of Schwarzschild solution to describe a spacetime with angular momentum is far from trivial. It turns out that it exists, and is the most general vacuum, regular solution of Einstein equations, which is stationary, axisymmetric and asymptotically flat \cite{Carter:1971zc}. It is known as the Kerr family of geometries, described by
\beq
ds^2&=&-\left(1-\frac{2Mr}{\Sigma}\right)dt^2-\frac{4Mar\sin^2\theta}{\Sigma} dtd\phi\nonumber\\
&+&\frac{\Sigma}{\Delta} dr^2+\Sigma d\theta^2+\left(r^2+a^2+2Ma^2r\sin^2\theta/\Sigma\right)\sin^2\theta d\phi^2\,,\label{kerr_geometry}\\
\Delta &=&r^2-2Mr+a^2\,,\quad \Sigma=r^2+a^2\cos^2\theta\,,
\eeq
and with gravitational mass $M$ and angular momentum $J=aM$~\cite{Papapetrou:1948jw,Hartle:1967he,Hartle:1968si}.
One can (and should) repeat the previous analysis to find curvature and coordinate singularities. It is straightforward to check that there is an outer horizon at the root of $\Delta$, i.e., at $r_+=M+\sqrt{M^2-a^2}$. For $a>M$ there is no horizon and one is instead left with an exposed singularity. In other words, there's a maximum limit on spin parameter:
\be
\frac{Jc}{GM^2}\equiv \frac{a}{M}\leq 1 \,,\label{Kerr_bound}
\ee
where we momentarily re-instated Newton's constant and the speed of light in the l.h.s.

It is not difficult to estimate the angular momentum per unit mass of the Sun, to find $J/M_\odot^2\sim 1$. For the Earth, $J=I\Omega \sim M_{\rm Earth}R_{\rm Earth}^2\Omega_{\rm Earth}$ under a constant rotation approximation. For a period of one day one gets $J/M_{\rm Earth}^2\sim 10^3$. Indeed, any child can produce a spinning top with $J/M^2\gtrsim 10^{16}$. From this point of view\footnote{They don't, in fact they can spin close to the speed of light near extremality.}, black holes spin slowly, and this fact provides possible thought experiments to destroy black holes, one of which we will soon test.

There are many interesting aspects of Kerr black holes, we will only consider one more. Take a timelike observer with a powerful spaceship and engines on, such that it is at rest relative to distant observers, with four-velocity
\be
v^\mu=(v^t,0,0,0)\,,\qquad v_\mu=g_{tt}v^t\,,\qquad v^2=(v^t)^2g_{tt} \,.
\ee
But condition \eqref{geodesic_normalization} imposes that $v^2<0$. Thus, in regions where $g_{tt}>0$ no timelike particle can remain at rest. Such regions are called ergoregions, and for the Kerr geometry it is easy to see that it comprises the exterior of the horizon with $r<2M$ along the equator for example. Ergoregions provide a very natural source of energy~\cite{Penrose:1964wq,Brito:2015oca}, as depicted in Fig.~\ref{fig:energy extraction1}.
\subsection{The Penrose process and the irreducible mass}
%
\begin{figure}[!ht]
    \centering
    \includegraphics[width=0.7\textwidth]{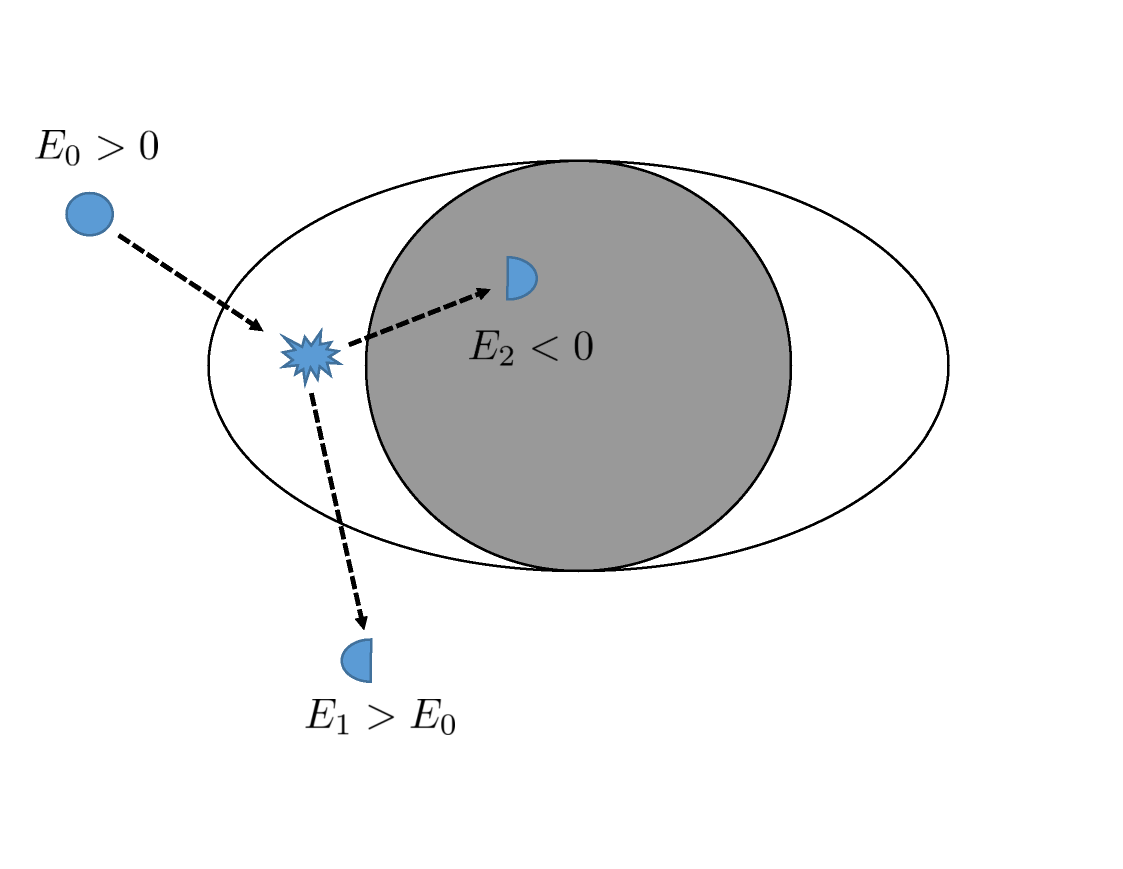}
    \caption{Extracting energy from a black hole via the Penrose process. A particle falling from rest at infinity disintegrates into two within the ergoregion of a Kerr black hole. From Ref.~\cite{Brito:2015oca}.}
    \label{fig:energy extraction}
\end{figure}
Due to its historical relevance, let's take a closer look at a particular energy extraction mechanism, the so-called Penrose process~\cite{Penrose:1964wq,Brito:2015oca}. The process is depicted in Fig.~\ref{fig:energy extraction}.
Here, one considers a particle ``$0$'' at rest at infinity ($E^{(0)}=1$), with energy $E^{(0)}\mu_0=\mu_0$, falling along a geodesic in the $\theta=\frac{\pi}{2}$ plane. Inside the ergoregion, the particle gets fragmented (at radial location $r_{0}$) into two identical particles of rest mass $\mu_{\text{fin}}$ and specific energy $E^{(i)}$($i=1,2)$ respectively. Therefore, $r_{0}$ is the turning point for both of the particles.

Particle ``1'' reaches infinity, where its energy is measured. Particle ``2'' is irretrievably lost to the black hole. It is straightforward to see that the radial geodesic equation is
\beq
r^2\dot{r}^2&=&r^2E^2+\frac{2M}{r}(a E-L)^2+(a^2 E^2-L^2)+\delta_1 \Delta\ ,\label{eqn:radial_geodesic}
\eeq
{where $\delta_{1}= 0,-1$ for null and timelike geodesics respectively. Now, imposing the condition that $\dot{r}=0$ at the turning point $r_0$, we get the following},
\begin{align}
    L^{(0)} & =\frac{-2aM+\sqrt{2Mr_0\Delta}}{r_0-2M}\,\quad (E^{(0)}=1)\,,                                        \\
    L^{(i)} & =\frac{-2aME^{(i)}\pm\sqrt{r_0\Delta\left(r_0((E^{(i)})^2+\delta_1)-2M\delta_{1}\right)}}{r_0-2M}\,.
\end{align}
Here, the plus sign corresponds to $i=1$ and the minus sign corresponds to $i=2$.
Conservation of energy and angular momentum yields,
\be
\mu_0=\mu_{\rm fin}E^{(1)}+\mu_{\rm fin}E^{(2)}\,,\quad \mu_0 L^{(0)}=\mu_{\rm fin}L^{(1)}+\mu_{\rm fin}L^{(2)}\,,
\ee
and one finds for $\delta_1=-1$
\be
\mu_{\rm fin}E^{(i)}=\frac{\mu_{0}}{2}\left(1\pm\sqrt{\frac{2M(1-4\mu_{\rm fin}^2/\mu_0^2)}{r_0}}\right)\,,\label{null_penrose1}
\ee
and for $\delta_{1}=0$, we have
\begin{align}
    \mu_{\text{fin}}E^{(i)}=\frac{1}{2}\mu_{\text{0}}\left[1\pm\sqrt{\frac{2M}{r_{0}}}\right]~.  \label{null_penrose2}
\end{align}
Note that \eqref{null_penrose2} is, as it should be, the limit of \eqref{null_penrose1} when $\mu_{\rm fin}\to 0$.

This expression shows that there is energy gain (i.e. the energy $\mu_{\rm fin}E^{(i)}$ of particle 1 is larger than the energy $\mu_0$ of particle 0) if some conditions are met, in particular for $r_0<2M(1-4\mu_{\rm fin}^2/\mu_0^2)<2M$ (note: $0<1-4\mu_{\rm fin}^2/\mu_0^2<1$). If the end state are two photons, for example, then $\mu_{\rm fin}=0$ and the energy gain is $\frac{1+\sqrt{2}}{2}$ for an extremal black hole.

The energy gain is maximum at the horizon, so let's see how much energy one can extract. For particle 2, going down the horizon,
\be
L^{(2)}=\frac{-2aME^{(2)}}{r_+-2M}\Rightarrow E^{(2)}=\frac{a}{2Mr_+}L^{(2)}=\frac{a}{r_+^2+a^2}L^{(2)}\,.
\ee
Thus, when this particle is absorbed, the black hole parameters change as
\be
dM=\frac{J/M}{(M+\sqrt{M^2-J^2/M^2})^2+J^2/M^2}dJ\,.
\ee
Define $J=j M^2$ and find
\beq
&&dM=\frac{j}{(1+\sqrt{1-j^2})^2+j^2}(Mdj+2jdM)\nonumber\\
&&\Rightarrow dM\left(1-\frac{2j^2}{(1+\sqrt{1-j^2})^2+j^2}\right)=\frac{jMdj}{(1+\sqrt{1-j^2})^2+j^2}\nonumber\\
&&\Rightarrow \frac{dM}{M}=\frac{jdj}{2-2j^2+2\sqrt{1-j^2}}\,. \label{eq:dM/M}
\eeq
The last equation can be integrated to
\be
\int_{M_{\rm irr}}^{M}\frac{dM}{M}=\int_0^{j}\frac{jdj}{2-2j^2+2\sqrt{1-j^2}}\,,
\ee
which results in
\be
M^2=\frac{2M_{\rm irr}^2}{1+\sqrt{1-j^2}}\,\quad{\rm or} \,\quad M^2=M_{\rm irr}^2+\frac{J^2}{4M_{\rm irr}^2}\,.
\ee
This is Christodoulou's irreducible mass~\cite{Christodoulou:1970wf}. It is a fundamental quantity that tells us how much energy is available to extract from a black hole. For example, the irreducible mass of a non-spinning black hole, from the above, is $M_{\rm irr}=M$; a nonspinning black hole has no energy to give. However, an extremal black hole with $J=M^2$ now has $M_{\rm irr}=M/\sqrt{2}$, and one can in principle extract $M-M_{\rm irr}$ using this process.

The fundamental ingredient in the above calculation is the existence of negative-energy states, which are provided by an ergoregion~\cite{Vicente:2018mxl,Brito:2015oca}.

Finally, as the extraction of energy proceeds, the black hole evolves. In particular, the area of the horizon
\be
A=\int d\theta d\phi \sqrt{g^{(2)}}=4\pi (r_+^2+a^2)=8 \pi M^2\left(\left(1+\sqrt{1-j^2}\right)^2+j^2\right)\,,
\ee
where $g^{(2)}$ is the $2D$ metric on the surface $t=$ constant and $r=r_+$, and $a=jM$.
Therefore,
\be
\frac{dA}{16\pi M r_+}=\frac{dM}{M}-\frac{jdj}{2-2j^2+2\sqrt{1-j^2}}=0\,,
\ee
where we have used \eqref{eq:dM/M} in writing the last equality. So we see that in this version of the Penrose process, the area does not change and the entropy does not increase.

\subsection{Testing the Cosmic Censor}
%
\begin{figure}[!ht]
    \centering
    \includegraphics[width=0.8\textwidth]{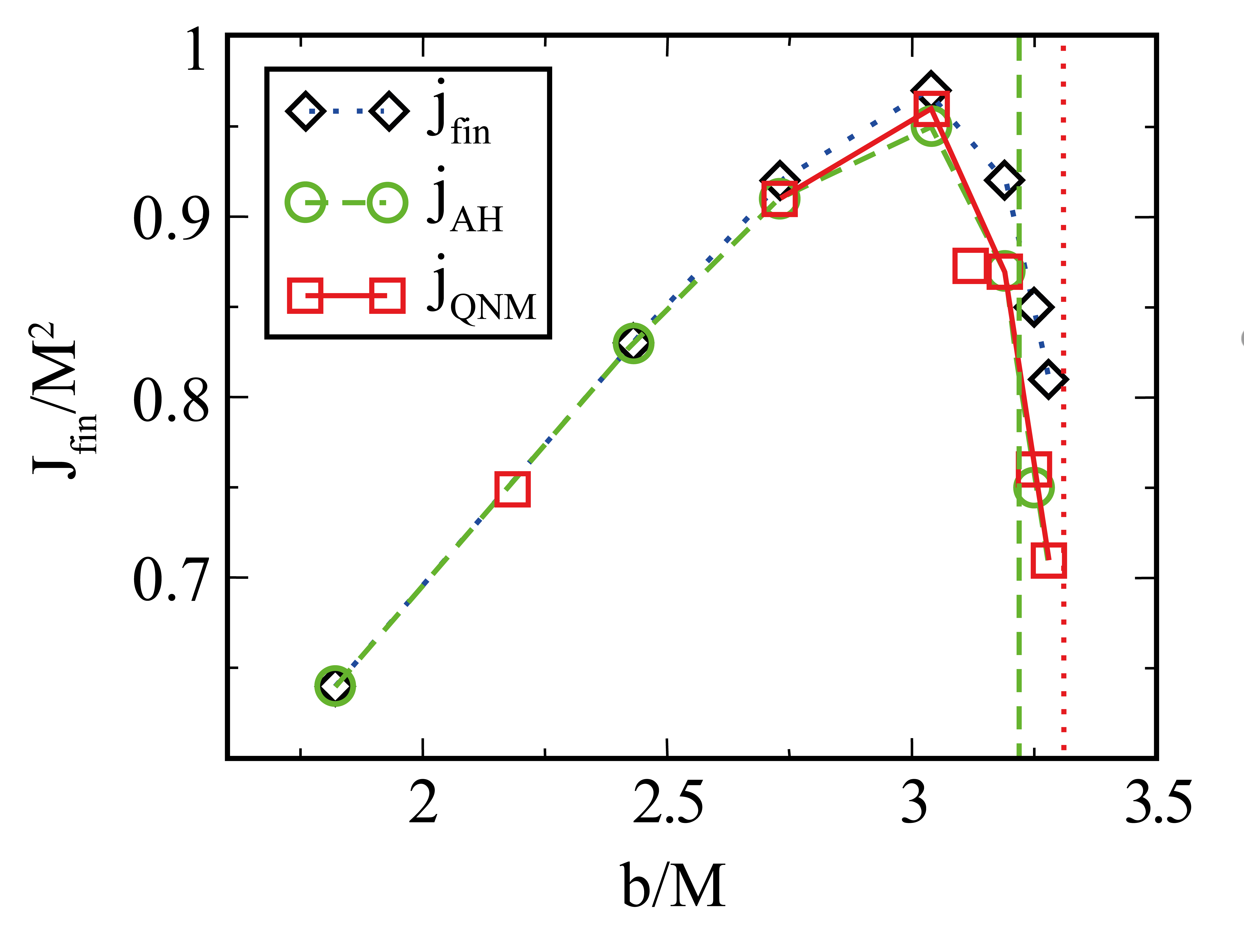}
    \caption{
        {\small High energy collision of two equal-mass black holes, colliding at $v=0.75$ in the center of mass. The figure shows the spin of the final black hole (different lines are different estimates for its spin) as the impact parameter is varied. For fine-tuned collisions the final black hole is near-extremal. A horizon is always present. Adapted from Ref.~\cite{Sperhake:2009jz}.}
    }
    \label{fig:Imagem_collision_1}
\end{figure}
We now have a framework to at least discuss energy extraction from black holes. We can use a similar framework to think about violating the Kerr bound~\eqref{Kerr_bound}. If this happens, then the horizon would be destroyed and one would be left with a ``singularity'' visible to external observers, a singularity for which the theory lacks a description. This is the subject of the weak Cosmic Censorship, that basically asserts that such a possibility does not happen~\cite{Wald:1974hkz}. The conjecture is not proven but supported by a number of partial results, one of which is the following.

Let's try to violate bound \eqref{Kerr_bound}, by throwing in pointlike particles that to a first approximation travel on geodesics. If we start from a non-spinning black hole, it is easy to spin it up. We will assume that we managed to make it extremal (for the detailed evolution of black hole mass and spin as the process develops, see exercise at the end of this Chapter). We now will throw a particle with angular momentum fine tuned such that it will spin the black hole up. The equation of motion of a particle in the $\theta=\frac{\pi}{2}$ plane can be written as
\beq
r^2\dot{r}^2&=&r^2E^2+\frac{2M}{r}(a E-L)^2+(a^2E^2-L^2) - \Delta\,,\\
&=&\frac{2Mr^2-L^2r+2M(L-M)^2}{r}\,,\qquad {\rm for\,\,} E=1, a=M \,,\\
&=&2M\frac{(r-r_1)(r-r_2)}{r}\,,
\eeq
with
\be
r_{1,2}=\frac{L^2\pm (L-2M)\sqrt{L^2+4LM-4M^2}}{4M}\,.
\ee

If $L>2M$ there is always a turning point outside the horizon (for extremal black holes, the horizon is located at $r_+=M$). Therefore,
$L_{\rm crit}=2M$ is the critical threshold separating scattering from absorption. Let's tune the particle to have the maximum possible angular momentum while still being absorbed, that is, {$L=\delta J/ \delta M = 2 M$}. Then, the rotation of the black hole changes according to
\be
j\equiv \frac{J}{M^2}\Rightarrow \delta j=\frac{\delta J}{M^2}-2\frac{\delta M}{M^3}J=0 \,,
\ee
{where we have used $\delta J = 2 M \delta M \Rightarrow J = M^2 $ in writing down the last equality. So we can conclude that an extremal black hole stays extremal.} Particles with too much angular momentum are simply not absorbed. The field equations have a built-in protection mechanism against the destruction of horizons. We note that the same result holds going to other types of matter or higher order in perturbation theory~\cite{Natario:2016bay,Sorce:2017dst}. Indeed, full nonlinear simulations of colliding black holes tuned to have large angular momentum, always result in a common horizon, and a black hole as final end-state, in four-dimensional asymptotically flat spacetimes~\cite{Sperhake:2009jz}. Typical results for the spin of the final black hole as a function of the impact parameter of the collision are shown in Fig.~\ref{fig:Imagem_collision_1}. The final spin is large, but always below the Kerr bound. Violations of cosmic censorship (horizon destruction, signaled by large curvatures in the spacetime) have been reported in higher dimensions~\cite{Okawa:2011fv,Andrade:2020dgc,Lehner:2010pn,Figueras:2017zwa}, indicating that if a protection mechanism in built into the partial differential equations, it's intrinsically tied to the spacetime dimensionality.

\section{Black hole formation: the hoop conjecture}
\begin{figure}[!ht]
    \centering
    \includegraphics[width=0.5\textwidth]{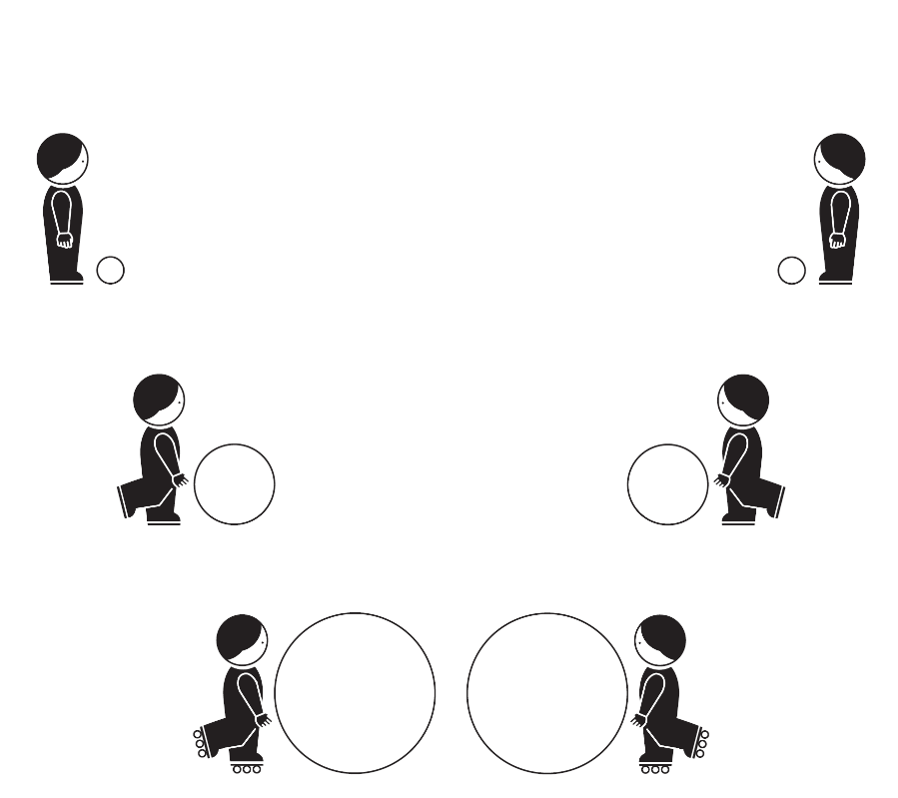}
    \caption{
        {\small The Hoop used to test black hole formation. A loop of radius $GM/c^2$ is used to test whether an object of mass $M$ is a black hole or not. If the object is free to pass inside the hoop in all directions, then it is a black hole. The hoop size is greatly enlarged in the cartoon, and scales with the Lorentz boost of the object.}
    }
    \label{fig:hoop}
\end{figure}
It is hard, or impossible, to destroy black holes. But how do they form in the first place? The canonical way to form a black hole in astrophysics is via the gravitational collapse of an old massive star. There is one very general criteria that should describe black hole formation, and which is useful in a variety of circumstances.
This is the Hoop Conjecture~\cite{1972mwm..book..231T,Choptuik:2009ww} and states that ``An imploding object forms a black hole when a circular hoop with circumference $2\pi$ times the Schwarzschild radius of the object can be made that encloses the object in all directions.''

We can use this to estimate when an environment of constant density should collapse to a black hole. Take a sphere of radius $R$, which encloses a mass $M=\sim \rho R^3$. Then the hoop will enclose this mass if $\rho R^2\sim 1$: any dilute environment which is sufficiently extended must collapse to a black hole.

The hoop conjecture can also inform us on ways to produce a black hole. A cartoon is shown in Fig.~{fig:hoop}. The semi-classical radius of the electron, for example, is $R_e\sim 10^{-15} {\rm m}$, but its Schwarzschild radius is $10^{-57} {\rm m}$. The electron is not a black hole, which is satisfying to know. However, all forms of energy gravitate. If we boost two electrons at rest to Lorentz factors $\gamma \gtrsim 10^{42}$, then the Hoop radius will be larger than the electron size: black holes will be created. Thus, (very) high energy processes always end up in black hole production.

\section{Exercises}

\noindent{\bf $\blacksquare$\ Q.1.a} Using the notebook \coderef{code:Field-Equations.nb}, show the following,
\begin{itemize}
    \item []\textbf{\textbullet} Starting with the general spherically symmetric ansatz,
        \begin{equation}
            ds^2 = -f(t,r)dt^2+g(t,r)^{-1} dr^2 + r^2 d \theta^2 + r^2 \sin \theta^2 d \phi^2 \nonumber
        \end{equation}
        obtain the Einstein field equations. Solve the field equations to obtain the Schwarzschild solution. Calculate the Kretschmann scalar and verify that you get $48M^2/r^6$.
    \item [] \textbf{\textbullet} Write down the Kerr metric in Boyer–Lindquist coordinates and show that it satisfies the vacuum Einstein Field Equations.
\end{itemize}

\vspace{0.25cm}

\noindent{\bf $\blacksquare$\ Q.1.b} We want to spin up a non-rotating black hole, by letting particles from infinity fall with some angular momentum. We want to be as efficient as possible.
Calculate the amount of mass that you need to throw in to make the black hole extremal.

\vspace{0.25cm}

\noindent \textbf{ {$\square$}\ Solution:} We want to be as efficient as possible, so we throw particles with the maximum possible angular momentum but at the same time, we must ensure that they are still absorbed by the black hole.

We start by noting that we can rewrite the radial geodesic equation for timelike particles, i.e., \eqref{eqn:radial_geodesic} with $\delta_1=-1$ as,
\begin{equation}
    \dfrac{E^2-1}{2} = \dfrac{\dot{r}^2}{2} + V_\mathrm{eff}, \label{eqn:kerr_radial_geodesic_Veff}
\end{equation}
where the effective potential $V_\mathrm{eff}$,
\begin{equation}
    V_\mathrm{eff} = -\dfrac{M}{r} + \dfrac{L^2-a^2(E^2-1)}{2r^2}- \dfrac{M(L-a E)^2}{r^3}
\end{equation}
Let us now write the spin (or the angular momentum per unit mass) of the black hole as $a=j M$, and we write the angular momentum per unit rest mass of the particle as $L=M h$. We also set the energy per unit rest mass of the particle $E$ to unity.

Then, the effective potential reduces to
\be
2 r^2V=M\left(-2r+h^2M-\frac{2M^2(h-j)^2}{r}\right)\,.
\ee
If we now solve for the turning points, that is, put $\dot{r}=0$, we have to solve for $V(r_0)=0$  and it's a simple quadratic equation. We can therefore readily write
\be
4r_0/M=h^2\pm\sqrt{h^4-16h^2+32hj-16j^2}\,,
\ee
with critical values $h=2(1 \pm \sqrt{1-j})$ (which you may obtain by setting the discriminant to zero).

We can now evolve the black hole: if one throws a particle of
energy $\delta M$ and angular momentum $ \delta J = M h \delta M$ into a black hole of mass $M$ and angular momentum $J = M^2j$, then we can calculate the change in the spin of the black hole as
\be
dj=d\left(\frac{J}{M^2}\right)=\frac{dJ}{M^2}-\frac{2JdM}{M^3}=\frac{hdM}{M}-\frac{2jdM}{M}=\frac{(h-2j)dM}{M}\nonumber
\ee
Using the critical value of the angular momentum $h=2 + 2\sqrt{1-j}$ to ensure that we are being as efficient as possible, we write the above equation as,
\begin{equation}
    \dfrac{d j}{2+2\sqrt{1-j}-2j} =\dfrac{d M}{M}.
\end{equation}
Recall that we are spinning up a non-rotating black hole, so we integrate the above equation with the initial condition $j(M_0)=0$, $M_0$ being the initial mass of the black hole, and find
\beq
j=4M_0\frac{M-M_0}{M^2}\,,
\eeq
To make the black hole extremal, we want $j=1$, and so we deduce that the final mass of the black hole should be $M=2M_0$. In other, we started with a non-spinning black hole and threw a flux of point-like particles (in an adiabatic manner), each with a critical impact parameter (which depends on black hole mass at each instant). After a total mass $M_0$ was thrown, the black hole reached extremality.

\vspace{0.5cm}
\noindent{\bf $\blacksquare$\ Q.2.} Let us now get down to some serious business: we wish to rigorously study how a black hole evolves as it swallows matter from its surroundings. Specifically, we shall consider accretion from a disk of gas orbiting a Kerr black hole. Following Bardeen's pioneering calculation \cite{Bardeen:1970zz}, consider a disk in the equatorial plane of the black hole. Assuming that the gas is being directly dumped across the horizon from the innermost stable circular orbit (ISCO), argue that
\begin{align}
    E = & \left(1 - \dfrac{2}{3z}\right)^{1/2},              \\
    L = & \dfrac{2M}{3^{3/2}}\left(1 + 2(3z-2)^{1/2}\right), \\
    j = & \dfrac{1}{3}z^{/2}\left( 4 -(3z-2)^{1/2}\right),
\end{align}
where $z \equiv r_\mathrm{ISCO}/M$. Use the last equation to show how the ISCO radius $z$ changes with $j$. Proceed to relate the change in the ISCO radius to the accreted mass, viz.,
\begin{equation}
    \dfrac{z}{z_0} =\left(\frac{M_0}{M}\right)^2,
\end{equation}
where $M_0$ and $z_0$ are the initial values corresponding to those of a Schwarzschild black hole.
Finally express the final mass of the black hole $M$ in terms of the rest mass accreted $\mu_0$.

\vspace{0.25cm}

\noindent\textbf{{$\square$}\ Solution:} We start from the radial geodesic equation for a timelike particle in the equatorial plane of the Kerr black hole. You should refer to the notebook \coderef{code:Kerr-Accretion.nb} to see how you can obtain the radial geodesic equation (Eqs.~\eqref{eqn:radial_geodesic} and \eqref{eqn:kerr_radial_geodesic_Veff}) from the Lagrangian density given by Eq.~{\eqref{eqn:lagrangian_density}}. Our goal is to solve \eqref{eqn:kerr_radial_geodesic_Veff} and express $E$, $L$ and $j$ in terms of $r_\mathrm{ISCO}$. We start by defining the potential,
\begin{equation}
    V(r) = -2r^2\left(V_\mathrm{eff} - \dfrac{E^2-1}{2}\right).
\end{equation}
We can get three algebraic equations for the three unknowns by noting that the conditions for a particle to remain marginally in a circular orbit are $$V(z)=V'(z)=V''(z)=0\,.$$

Note that the three aforementioned conditions hold at the ISCO radius $z$, and they give us the following equations:
\begin{align}
    a^2 (E^2-1 ) + \frac{2 \left( L-a E  \right)^2}{z} + 2 M^2 z + ( E^2-1) M^2 z^2 -L^2 & = 0, \\
    M^2 + (E^2-1) M^2 z - \frac{ \left( L-a E  \right)^2}{z^2}                           & = 0, \\
    \frac{2 \left( L -a E  \right)^2}{z} + ( E^2-1) M^2 z^2                              & = 0.
\end{align}

We can solve these equations in  \textsc{Mathematica}\textsuperscript{\textregistered} but we encourage you to try solving them the old-fashioned way as well ({see \cite{bolin_angular_2015} for explicit details}). In any case, these three equations will have eight solutions. Neglecting the set of solutions having $E<0$, we are still left with four solutions. We can easily discard two solutions by noting that in the Schwarzschild limit, $z_\mathrm{ISCO}=6$, they give $|j|\neq0$ which is unphysical since we expect $j=0$ for a Schwarzschild black hole. Finally, from the remaining two solutions, we reject the one which gives $a=-M$ for $z=1$. We therefore get the following three solutions,
\begin{align}
    E = & \left(1 - \dfrac{2}{3z}\right)^{1/2},\label{eqn:E(z)}                  \\
    j = & \pm \dfrac{1}{3}z^{1/2}\left( 4 -(3z-2)^{1/2}\right), \label{eqn:j(z)} \\
    L = & \pm \dfrac{2M}{3^{3/2}}\left(1 + 2(3z-2)^{1/2}\right),                 \\
\end{align}
The details have been worked out in the notebook \coderef{code:Kerr-Accretion.nb}. We have therefore successfully expressed $E$, $L$ and $j$,in terms of the ISCO radius.

\begin{figure}[h!]
    \centering
    \includegraphics[width=0.75\textwidth]{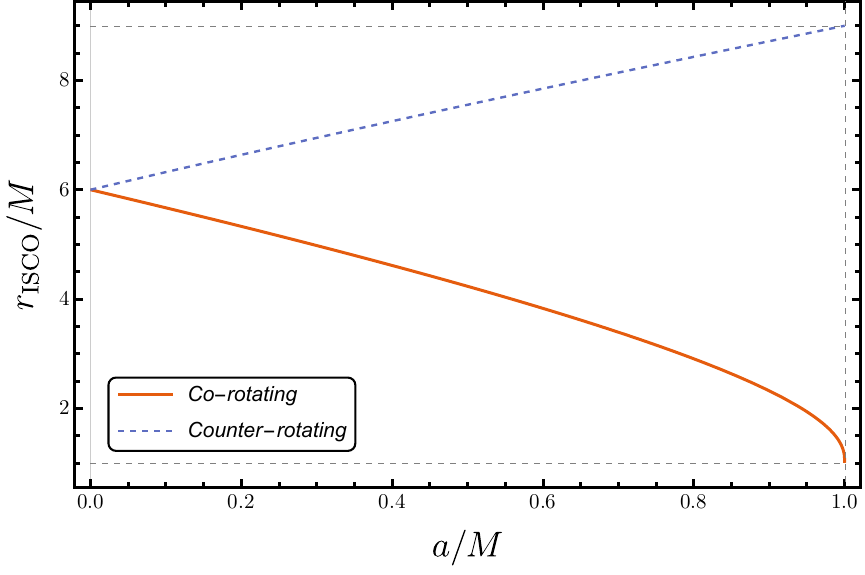}
    \caption{%
    Variation of the innermost stable circular orbit (ISCO) radius $z=r_{\mathrm{ISCO}}/M$ as a function of the dimensionless spin parameter $a/M$ for a Kerr black hole.
    The blue solid curve corresponds to co-rotating orbits (aligned with the black hole spin), while the orange dashed curve represents counter-rotating orbits (anti-aligned with the spin).
    In the non-rotating limit $a = 0$, the ISCO radius for both cases converges to $6M$, the Schwarzschild value.
    As the spin increases toward the extremal Kerr limit $a = M$, the ISCO radius for co-rotating orbits decreases to $r_{\mathrm{ISCO}} = M $, allowing stable orbits closer to the event horizon.
    In contrast, for counter-rotating orbits, the ISCO radius increases with spin, moving away from the black hole and approaching $r_{\mathrm{ISCO}} = 9M$ as $a \to M$.
    }
    \label{fig:ISCO_kerr}
\end{figure}

Now, as the black hole accretes matter, the BH parameters $M$ and $J$ must change. Consequently, the ISCO radius will also change. We can use the second equation of the above set to solve for $z(j)$. We can then plot the resultant expression by varying $j$ to see how the ISCO radius changes for particles which are co-rotating and counter-rotating with respect to the black hole. The result is shown in Fig.~\ref{fig:ISCO_kerr}. Refer to the notebook \coderef{code:Kerr-Accretion.nb} to see how the plot has been generated.

To see how the ISCO radius $z$  changes as a function of the mass accreted, we note that if the black hole swallows matter of rest mass $\delta \mu_0$, the change in the mass $\delta M$ and the angular momentum $\delta J$ is given by
\begin{equation}
    \delta M = E \delta \mu_0,~~ \delta J = L \delta \mu_0. \label{eqn:evolveMJ}
\end{equation}
Recall, the spin of the BH evolves as,
\be
dj=d\left(\frac{J}{M^2}\right)=\frac{dJ}{M^2}-\frac{2JdM}{M^3} = \left(\dfrac{L}{E M^2}-\dfrac{2j}{M} \right) d M,
\ee
where we have used \eqref{eqn:evolveMJ} to write the last equality.

Using the chain rule, we can rewrite the above equation and get
\begin{equation}
    M\dfrac{dz}{dM}\dfrac{dj}{dz} = \dfrac{L}{EM}-2j.
\end{equation}
Using the results we have just derived, we get a separable differential equation
\begin{equation}
    \left(2z + M \dfrac{dz}{d M}\right) = 0
\end{equation}
The above differential equation has the solution,
\begin{equation}
    z = \dfrac{C}{M^2}.
\end{equation}
If we choose the initial conditions that $z=z_0$  when $M_0$, we get,
\begin{equation}
    \dfrac{z}{z_0} =\left(\dfrac{M_0}{M}\right)^2, \label{eqn:z/z0}
\end{equation}
Note that for the Schwarzschild black hole, $z_0=6$. We can now express the behavior of $j$ as a function of $M$. Plugging \eqref{eqn:z/z0} into \eqref{eqn:j(z)} with $z_0=6$, we get,
\begin{equation}
    j(M)=\left(\dfrac{2}{3}\right)^{1/2}\dfrac{M_1}{M}\left(4-\left(\frac{M_1}{M}\right)^2-2\right)\,.
\end{equation}

We now want to relate the actual rest mass accreted $\mu_0$ to the initial mass $M_0$ and the final black hole mass $M$. So we use the first equation of \eqref{eqn:evolveMJ} to write the following differential equation,
\begin{equation}
    \dfrac{d M}{d \mu_0} = E(z).
\end{equation}
Plugging \eqref{eqn:z/z0} and \eqref{eqn:E(z)} into the above equation, we get the following equation,
\begin{equation}
    \dfrac{dM}{\sqrt{1-M^2/9M_0^2}} = d \mu_0
\end{equation}
Substituting $M=3M_0 \sin \vartheta$ into the above integrand, we can easily obtain,
\begin{equation}
    \mu_0 = 3 M_0 \arcsin{\dfrac{M}{3 M_0}} + C.
\end{equation}
We fix $C$ by noting that $M=M_0$ when $\mu_0=0$, and obtain,
\begin{equation}
    \mu_0 = 3 M_0\left( \arcsin \dfrac{M}{3 M_0} - \arcsin\dfrac{1}{3} \right).
\end{equation}
Finally, we can invert the above equation to express the final mass of the black hole $M$ in terms of the rest mass accreted $\mu_0$,
\begin{equation}
    \dfrac{M}{M_0}= 3 \sin \left(\dfrac{\mu_0}{3 M_0} +\arcsin{\dfrac{1}{3}}\right)\,.
\end{equation}
We can now use the elementary trigonometric identities $\sin(A+B)=\sin A \cos B + \cos A \sin B$ and $\cos^2 (\arcsin \vartheta) + \sin^2(\arcsin \vartheta) = 1$ to recast the above equation as
\begin{equation}
    \dfrac{M}{M_0} = 2 \sqrt{2} \sin \left( \dfrac{\mu_0}{2 M_0} \right)+ \cos\left( \dfrac{\mu_0}{3 M_0}\right)\,.
\end{equation}
We therefore recover Bardeen's equations~\cite{Bardeen:1970zz,Bardeen:1972fi}. Here we have provided the details missing in the original publications (helpful details can be found in Ref.~\cite{bolin_angular_2015}, which we used as a guide).

Note also that Bardeen's result~\cite{Bardeen:1970zz} is a milestone because it began the conversation about the inadequacy of the Schwarzschild metric to describe \emph{real} black holes in our Universe. In a scenario where black holes grow by accretion,  most astrophysical black holes would be spinning very fast~\cite{Thorne:1974ve}.


\chapter{Part II}
\label{ch:part2}
\allowdisplaybreaks
\minitoc
\section{Dynamics of black holes: the perturbative framework}
So far, we have looked at the motion of matter close to black holes. The geometry itself was held fixed, and under that assumption matter follows geodesics. The effect of backreaction on the evolution of the black hole itself can be calculated perturbatively. This framework is suitable for high frequency radiation~\cite{Seitz:1994xf,Perlick:2004tq,Dolan:2018ydp} which follows null geodesics. But it is ill-suited to capture phenomena which is intrinsically wavelike: for example, we may want to study gravitational radiation emitted when two black holes collide or the evolution of some prescribed initial conditions for the gravitational field. We now turn to this problem.

To simplify the discussion we start with a simple theory of a minimally coupled scalar field $\Phi$, described by the action
\be
S=\int d^4x \sqrt{-g}\left(\frac{R}{k}-\frac{1}{2}g^{\mu\nu}\partial_\mu\Phi\partial_\nu\Phi-\frac{1}{2}\mu^2\Phi^2\right)\,,\label{action_scalar_dynamics_sourceless}
\ee
This is the same action of Eq.~\eqref{eq_chapter1_scalar}, but we now specify the potential $V$. Varying the action with respect to the metric and to the scalar field, we find the following equations of motion,
\beq
\nabla_\mu\nabla^\mu\Phi&=&\mu^2\Phi\,,\\
R_{\mu\nu}-\frac{1}{2}g_{\mu\nu}R&=&k\left(\frac{1}{2}\Phi_{,\mu}\Phi_{,\nu}-\frac{1}{4}g_{\mu\nu}\left(\Phi_{,\alpha}\Phi^{,\alpha}+\mu^2\Phi^2\right)\right)\,.\label{Einstein_scalar_eoms}
\eeq
These equations are easy to write down, but they're challenging to solve. In these lectures, we will use a powerful approach that is appropriate for studying many phenomena that take place in the cosmos.

In particular, since we know the solution to the vacuum ($\Phi_0=0$ is a background solution) equations, we consider a perturbative expansion of the equations. The dominant terms in an expansion in powers of $\Phi$ and its derivatives, yields
\beq
R_{\mu\nu}-\frac{1}{2}g_{\mu\nu}R&=&0\nonumber\\
\frac{1}{\sqrt{-g}}\partial_\mu\left(\sqrt{-g}g^{\mu\nu}\partial_\nu\Phi\right)&=&\mu^2\Phi\nonumber
\eeq
We know the unique {\it stationary, asymptotically flat} solution to the first set of equations, it is the Kerr geometry. In this setting, at leading order in the scalar field amplitude and its derivatives, we therefore bypass the enormous task of solving for the geometry.

We are left with only the Klein-Gordon equation to solve. It is still a challenging dynamical equation, since the field depends on three spatial coordinates and time. We can use the symmetries of the background geometry to proceed forward.
\section[Exploiting symmetries]{Exploiting symmetries: expansion in harmonics and master functions}
We will focus on non-spinning black holes for now, described by the geometry~\eqref{eq:Schwarzschild_geometry}. We parametrize the sphere by the angular and azimuthal variables $\theta, \phi$. Hence, given the spherical symmetry of the background we can expand the scalar field as
\be
\Phi=\sum_{\ell m}\frac{\Psi}{r}Y_{\ell m}(\theta, \phi)\,,\label{expansion_harmonics_scalars}
\ee
where $Y_{\ell m}$ is a scalar spherical harmonic function of degree $\ell$ and order $m$. Notice that the spherical harmonics are a complete set, any function $X(\theta,\phi)$ can be expanded in a set of spherical harmonics $Y_{\ell m}$~\cite{ArfkenWeber,Gwaiz}. Hence (c.f. notebook \coderef{code:Perturbations.nb}), we find,
\be
\frac{\partial^2\Psi}{\partial r_*^2}-\frac{\partial^2\Psi}{\partial t^2}-V_{s=0}\Psi=0\,, \label{eqn:wave equation}
\ee
with
\beq
\frac{dr_*}{dr}=\frac{1}{f}=\frac{1}{1-2M/r}\,,\qquad V_{s=0}=f\left(\frac{\ell(\ell+1)}{r^2}+\frac{2M}{r^3}+\mu^2\right)\,,\nonumber
\eeq
Thus, an amazingly complex set of partial differential equations describing the evolution of the geometry and matter (under the form of a scalar field) are reduced to a very simple second-order partial differential equation for one variable, which depends only on a radial and on a time coordinate. We labeled the potential with a subscript $s=0$ standing for the fact that scalars are spin-0 particles.

For vectors and tensors, we need a slightly different procedure. Take Maxwell theory
\be
S=\int d^4x \sqrt{-g}\left(\frac{R}{k}-\frac{1}{4}F_{\mu\nu}F^{\mu\nu}\right)\,,\quad F_{\mu\nu}=\nabla_\mu A_\nu-\nabla_\nu A_\mu\nonumber
\ee
As before, the equations of motion can be expanded in powers of the electromagnetic field, and to leading order, we only need to solve Maxwell's equations $\nabla_\mu F^{\mu\nu}=0$ in a vacuum black hole background. Now, $A_\mu$ is a vector and separation of variables needs care, since components must transform like vector (try expanding each component in {\it scalar} harmonics!). One elegant way out is to construct three vectors with help of scalar harmonics (for instance using the gradient, the curl and projecting onto the radial direction)~\cite{Edmonds:1955fi,mathewsharmonics,DeWitt:1973uma},
\beq
\nabla Y_{\ell m}&=&\left(0,\partial_\theta Y_{\ell m},\partial_\phi Y_{\ell m}\right)\,,\\
{\bm L} Y_{\ell m}&=&\left(0,\frac{i}{\sin\theta}\partial_\phi Y_{\ell m},-i\sin\theta\partial_\theta Y_{\ell m}\right)\,,\\
\vec{r}_r Y_{\ell m}&=&\left(Y_{\ell m},0,0\right)\,.\label{expansion_vector_harmonics}
\eeq
To expand a four-vector using this three vector spherical harmonics we need to include a time component also, which is achieved by adding $\vec{e}_t Y_{\ell m}$. Then it yields
    {\small
        \begin{eqnarray}
            A_{\mu}=\sum_{\ell m}\left(
            \begin{array}{cc}\left[
                    \begin{array}{c} 0                                                        \\ 0 \\
                            \frac{a^{\ell m}(t,r)}{\sin\theta}\,\partial_\phi Y_{\ell m} \\
                            -a^{\ell m}(t,r)\sin\theta\,\partial_\theta Y_{\ell m}
                        \end{array}\right] &
                    +\left[
                        \begin{array}{c}f^{\ell m}(t,r)\,Y_{\ell m}               \\h^{\ell m}(t,r)\,Y_{\ell m} \\
                            k^{\ell m}(t,r) \,\partial_\theta Y_{\ell m} \\ k^{\ell m}(t,r)\, \partial_\phi
                            Y_{\ell m}
                        \end{array}\right]
                \end{array}\right)\,.\label{eq_Maxwell_separation}
        \end{eqnarray}}

We thus have expanded any vector as a superposition of vector spherical harmonics. Note that they split into two distinct components, depending on how they behave according to parity transformation, the simultaneous inversion of all Cartesian axes, corresponding to
$(\theta,\phi)\to (\pi-\theta,\pi+\phi)$.
The first term has parity $(-1)^{\ell+1}$, second term parity $(-1)^{\ell}$. In practice, this distinction is observed when the full decomposition is inserted in Maxwell equations. One will find two distinct, orthogonal terms, that need to vanish separately. We call the first term on the right-hand side of Eq.~\eqref{eq_Maxwell_separation} the axial or odd or magnetic type, and the second term the polar or even or electric type. Inserting decomposition (cf. \coderef{code:Perturbations.nb}) one finds that {\it both sectors} obey the same master equation
\beq
&&\frac{\partial^2\Psi}{\partial r_*^2}-\frac{\partial^2\Psi}{\partial t^2}-V_{s=1}\Psi=0\,,\\
&&V_{s=1}\equiv\left(1-\frac{2M}{r}\right)\frac{\ell(\ell+1)}{r^2}\,.\label{eq:RW_s1}\eeq
The master variable $\Psi$ was chosen to be
\be\label{inversion_em_perturbations}
\Psi= \left\{
\begin{array}{ll}
    a^{\ell m}(t,r)                                                         & {\rm axial} \\
    \frac{r^2}{\ell (\ell+1)} (\partial_t h^{\ell m}-\partial_r f^{\ell m}) & {\rm polar} \\
\end{array}
\right.
\ee
Due to gauge invariance of the Maxwell's theory, one can always impose the Lorentz gauge, namely $\nabla_{\mu}A^{\mu}=0$. Using the decomposition of $A^{\mu}$ into axial and polar parts, one can show that the axial part trivially satisfies the Lorentz condition. So, among the two degrees of freedom of the electromagnetic field the axial part contains only one.
The Lorentz gauge condition imposes a constraint on the three mode functions of the polar sector. Similarly, it can be shown that the time component of the Maxwell's equation can be constructed out of the radial and theta components. This shows that among the three mode functions only one is independent. Therefore, overall one finds two degrees of freedom (two polarizations) for the electromagnetic field.

Note that the potential \eqref{eq:RW_s1} is very similar to the one that describes scalar perturbations, $V_{s=0}$. In particular, they both vanish at the horizon and at infinity.

We can also express the above in terms of the electric and magnetic
fields as seen in an orthonormal frame of static observers (remember we are focusing on non-spinning black holes only). This frame has components
\be
e^\mu_{(t)}=\frac{\delta_\mu^t}{\sqrt{1-2M/r}}\,,\,e^\mu_{(r)}=\sqrt{1-2M/r}\delta_\mu^r\,,\,e^\mu_{(\theta)}=\frac{\delta_\mu^\theta}{r}\,,\,e^\mu_{(\phi)}=\frac{\delta_\mu^\phi}{r\sin\theta}\,.
\ee
Then, the components of the electric and magnetic fields are
\be
{\cal E}_{(a)}=F_{\mu\nu}e^\mu_{(a)}e^{\nu}_{(t)}\,,\quad
{\cal B}_{(a)}=\frac{1}{2}\epsilon_{\mu\nu\lambda\tau}F^{\lambda\tau}e^\mu_{(a)}e^{\nu}_{(t)}\,,
\ee
with $\epsilon$ the covariant Levi-Civita tensor. As such one finds
\beq
{\cal E}_{(\theta)}&=&\frac{ie^{im\phi}m\dot{a}^{\ell m}}{r\sin\theta\sqrt{1-2M/r}}Y_{\ell m}\,,\, {\cal E}_{(\phi)}=-\frac{e^{im\phi}\dot{a}^{\ell m}}{r\sqrt{1-2M/r}}\frac{dY_{\ell m}}{d\theta}\,,\\
{\cal B}_{(\theta)}&=&\frac{ie^{im\phi}\sqrt{r-2M}\,a'^{\ell m}}{r^{3/2}}\frac{dY_{\ell m}}{d\theta}\,,\,{\cal B}_{(\phi)}=-\frac{me^{im\phi}\sqrt{r-2M}\,a'^{\ell m}}{\sin\theta r^{3/2}}Y_{\ell m}\,,\label{electric_magnetic_fields}
\eeq
where dots and primes stand for time and radial derivative respectively. Note that at large distance the fields are radiative indeed, they fall off like $1/r$.

The procedure can be repeated for tensors. But what does it mean to have a tensor in a fixed background? Well, what it means is that we are now looking at vacuum disturbances. Instead of considering a small scalar or electromagnetic field, we take instead a small gravitational field $h_{\mu\nu}$, such that the full geometry is described by
\be
g_{\mu\nu}=g_{\mu\nu}^{(0)}+h_{\mu\nu}\,, \label{perturbative_scheme}
\ee
and $g_{\mu\nu}^{(0)}$ is the geometry of a black hole. In essence, we are studying how vacuum reacts to any fluctuation. For example, how does the waving your hand disturb Minkowski space or a black hole spacetime. To do this, we need to disturb the Einstein equations. Gravitational perturbations $h_{\mu\nu}$ would also need to be considered at the next order in the above setup of scalar or electromagnetic perturbations.

We can repeat the procedure to find the appropriate expansion of tensors in tensor spherical harmonics~\cite{Regge:1957rw,Zerilli:1970wzz,mathewsharmonics,Thorne:1980ru,Nagar:2005ea}. We find that the
perturbations fall into two distinct classes as before: odd (Regge-Wheeler or
vector-type) and even (Zerilli or scalar-type), with parities equal to
$(-1)^{\ell+1}$ and $(-1)^\ell$, respectively
\cite{Nagar:2005ea,Thorne:1980ru,Zerilli:1970wzz,mathewsharmonics}.
In the Regge-Wheeler gauge
\cite{Kokkotas:1999bd,Nollert:1999ji,Nagar:2005ea,Regge:1957rw,Vishveshwara:1970cc},
the perturbations are written as
\begin{eqnarray}
    h_{\mu \nu}= \left[
        \begin{array}{cccc}
            0 & 0 & 0 & h_0(t,r)
            \\ 0 & 0 &0 & h_1(t,r)
            \\ 0 & 0 &0 & 0
            \\ h_0(t,r) & h_1(t,r) &0 &0
        \end{array}\right]
    \left(\sin\theta\frac{\partial}{\partial\theta}\right)
    Y_{\ell m}(\theta,\phi)\,, \label{oddgauge}
\end{eqnarray}
for odd parity, whereas for even parity
\begin{eqnarray}
    h_{\mu \nu}= \left[
        \begin{array}{cccc}
            H_0(t,r) f & H_1(t,r) & 0 & 0
            \\ H_1(t,r) & H_2(t,r)/f  &0 & 0
            \\ 0 & 0 &r^2K(t,r) & 0
            \\ 0 & 0 &0 & r^2K(t,r)\sin^2\theta
        \end{array}\right] \, Y_{\ell m}(\theta,\phi)\,,\label{evengauge}
\end{eqnarray}
with $f=1-2M/r$.
The angular dependence of the perturbations is dictated by the structure of
tensorial spherical harmonics
\cite{Nagar:2005ea,Thorne:1980ru,Zerilli:1970wzz,mathewsharmonics} (and it is understood that each function carries a $\ell m$ dependence, which we omit for clarity). Inserting
this decomposition into Einstein's equations one gets ten coupled second-order
differential equations that fully describe the perturbations: three equations
for the odd radial variables, and seven for the even variables. The odd
perturbations can be combined in a single Regge-Wheeler or vector-type
gravitational variable $\Psi^-_{s=2}$, and the even perturbations can likewise
be combined in a single Zerilli or scalar-type gravitational wavefunction
$\Psi^+_{s=2}$. The Regge-Wheeler and Zerilli functions ($\Psi^-_{s=2}$ and
$\Psi^+_{s=2}$, respectively) satisfy equation \eqref{eq:RW_s1} but now with potentials
\begin{equation}
    V_{s=2}^{-}= f
    \left\lbrack\frac{\ell(\ell+1)}{r^2}-\frac{6M}{r^3}\right\rbrack\,
    \label{vodd}
\end{equation}
and
\begin{equation}
    V_{s=2}^{+}= \frac{2f}{r^3}
    \frac{9M^3+3\lambda^2Mr^2+\lambda^2\left(1+\lambda\right)r^3+9M^2\lambda
        r} {\left(3M+\lambda
        r\right)^2} \,. \label{veven}
\end{equation}
where $\lambda\equiv (\ell-1)(\ell+2)/2$.

We will not deal with the full gymnastics of the equations, but we point that for odd perturbations the metric functions $h_0$ and $h_1$ are related to the master function
$\Psi^-_{s=2}$ as (cf. \coderef{code:Perturbations.nb})
\begin{equation}
    \Psi^-_{s=2}= \frac{f}{r} h_1(r)\,,\qquad
    \frac{\partial h_0}{\partial t}=\frac{\partial}{\partial r_*}\left(r\Psi^-_{s=2}\right)\,.
    \label{qodd}
\end{equation}
For the scalar-type gravitational perturbation,
\begin{eqnarray}
    K&=& \frac{6M^2+\lambda \left(1+\lambda\right)r^2+3M \lambda
        r} {r^2\left(3M+\lambda r\right)}
    \Psi^+_{s=2}
    +\frac{\partial\Psi^+_{s=2}}{\partial r_*} \,, \\
    H_1&=& -\frac{\left(3M^2+3\lambda Mr-\lambda r^2\right)} {r \left(3M+\lambda r\right)
        f(r)}\frac{\partial \Psi^+_{s=2}}{\partial t} +
    \frac{r}{f(r)}\frac{\partial^2 \Psi^+_{s=2}}{\partial t\partial r_*} \,. \label{evenrelations}
\end{eqnarray}
As seen from the above, all the metric fluctuations can be expressed in terms of two master variables $\Psi^\pm_{s=2}$. Thus, there are two polarizations for gravitational waves.

A complete discussion of Regge-Wheeler or vector-type gravitational perturbations of the four-dimensional Schwarzschild geometry can be found in
the original work~\cite{Regge:1957rw} as well as in
Ref.~\cite{Edelstein:1970sk}, where some typos in the original work were corrected. For Zerilli or scalar-type gravitational perturbations, the
fundamental reference is Zerilli's work~\cite{Zerilli:1970se,Zerilli:1970wzz};
typos were corrected in Appendix A of Ref.~\cite{Sago:2002fe}. An elegant,
gauge-invariant decomposition of gravitational perturbations of the
Schwarzschild geometry was described by Moncrief~\cite{Moncrief:1974am} (see
also Refs. \cite{Gerlach:1979rw,Gerlach:1980tx,Sarbach:2001qq,Martel:2005ir}). The extensions to the Schwarzschild-anti de Sitter$_4$ (SAdS$_4$) geometry can be found in Ref.~\cite{Cardoso:2001bb}, while the general higher dimensional treatment was uncovered in a series of elegant works by Kodama and
Ishibashi~\cite{Kodama:2003jz,Ishibashi:2003ap,Kodama:2003kk}.

One can also express the perturbed spacetime in a radiation gauge to calculate fluxes and the usual TT components of the wave. The procedure can be found in Refs.~\cite{Maggiore:2007ulw,JHU}.
\section[The effective potential]{The effective potential: relation to null geodesics, superpartner relations}
The result of the calculations above is that perturbations of non-rotating, neutral black holes in General Relativity can all be reduced to the study of the master equation
\be
\frac{\partial^2\Psi}{\partial r_*^2}-\frac{\partial^2\Psi}{\partial t^2}-V_s\Psi=0\,,\label{wave_equation_spins}
\ee
with the effective potential
\be
V_s\equiv\left(1-\frac{2M}{r}\right)\left(\frac{\ell(\ell+1)}{r^2}+(1-s^2)\frac{2M}{r^3}\right)\,.\label{effective_potential_spins}
\ee
For $s=2$, $V_s\equiv V_s^-$. The description above leaves out the even part of the gravitational sector, but it turns out that there's a remarkable relation between the even and odd potentials~\cite{Chandrasekhar:1985kt,Berti:2009kk,Berti:2025hly}:
\beq
fV^{\pm}_{s=2}&=&W_0^2\mp f\frac{dW_0}{dr}-\frac{\lambda^2(\lambda+1)^2}{9M^2}\,,\\
W_0&=&\frac{6M(2M-r)}{r^2(6M+2\lambda r)}-\frac{\lambda(\lambda+1)}{3M}\,.\label{superpotential}
\eeq
These potentials are sometimes referred to as ``superpartners,'' since they can both be obtained from a ``superpotential.'' This property allows us to infer a number of relations between the two wavefunctions, like isospectrality and reflection coefficients~\cite{Chandrasekhar:1985kt,Berti:2009kk,Berti:2025hly}.

\begin{figure}[!htbp]
    \centering
    \begin{minipage}[b]{0.45\textwidth}
        \centering
        \includegraphics[height=7.5cm]{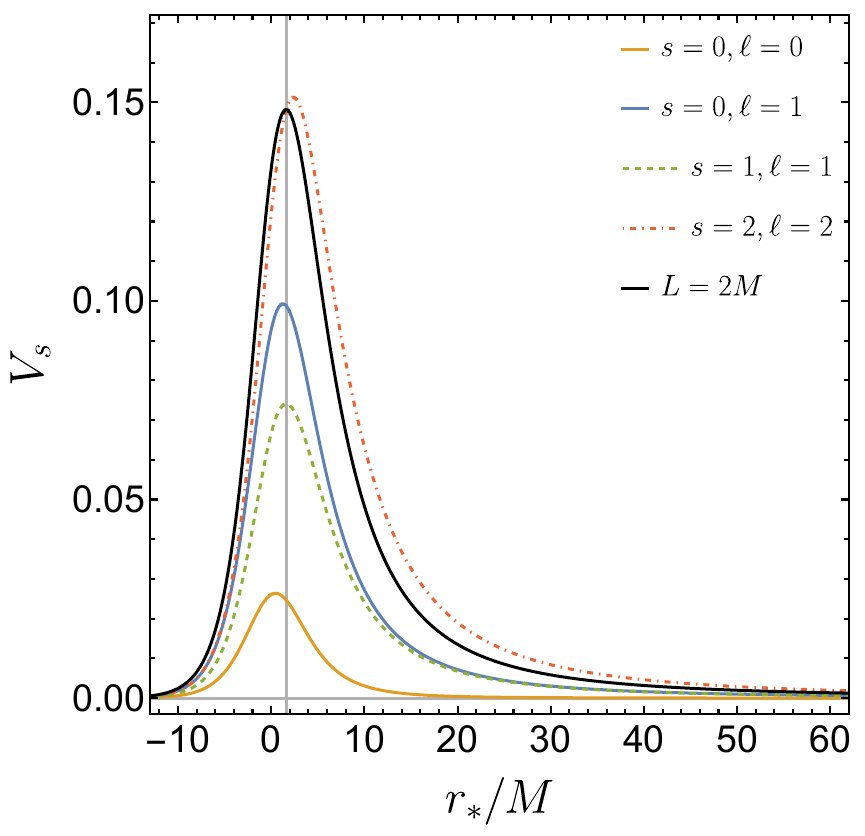}
    \end{minipage}
    \begin{minipage}[b]{0.45\textwidth}
        \centering
        \includegraphics[height=7.5cm]{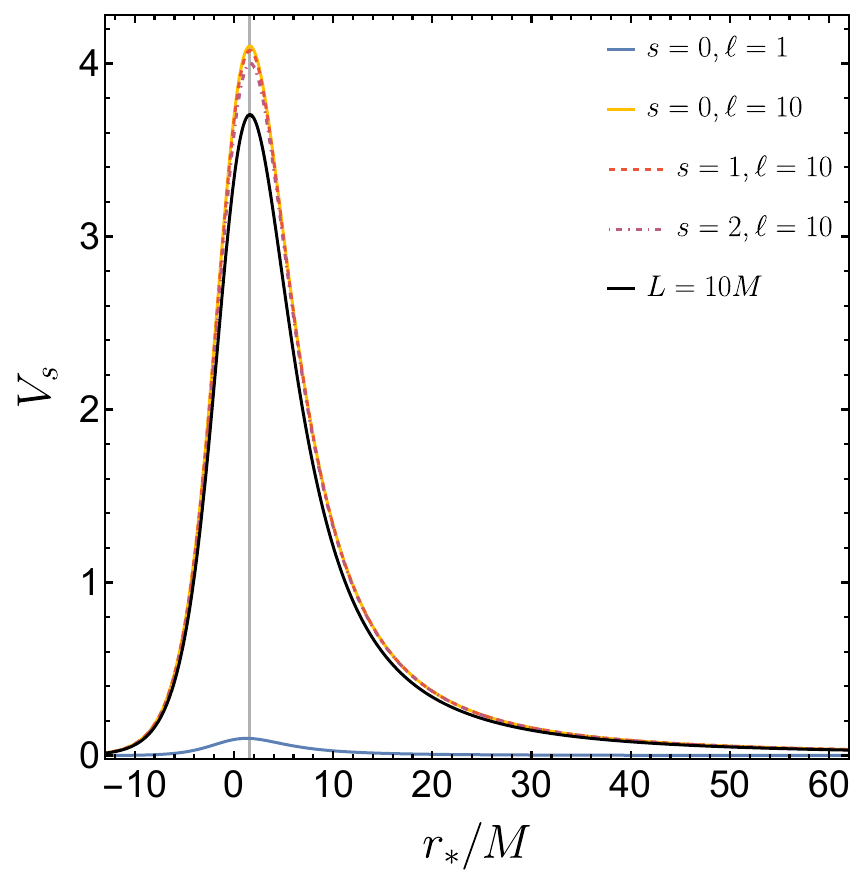}
    \end{minipage}
    \caption{The effective potential $V_s$ as a function of $r_*$ for different values of spin $s$ and multipole indices $\ell$. The solid black curve shows the effective potential $V_{\mathrm{null}}$ encountered by null geodesics around the Schwarzschild black hole, $L$ being the angular momentum per unit mass of the particle. The gray vertical line shows the location of the light ring at $r=3 M$ which corresponds to the position of the maxima of $V_{\mathrm{null}}$.\textbf{ Left Panel:} The shape of $V_s$ for low values of $\ell$. The colored solid curves correspond to scalar perturbations ($s = 0$), while dashed and dot-dashed lines represent electromagnetic ($s = 1$) and gravitational ($s = 2 $) perturbations, respectively. Notice that for $s=1$, $V_s$ resembles $V_{\mathrm{null}}$. \textbf{Right Panel:} In the so-called eikonal limit, that is, when $\ell$ becomes very large, the shape of the potential becomes quantitatively very similar to $V_{\mathrm{null}}$ for all $s$.
        \label{fig:potential_different_s}}
\end{figure}

While separation of the angular variables is a big advantage, the massaging of the equations to a single master equation is not really a requisite to solving the coupled partial differential equations. However, building on knowledge and intuition from quantum mechanics, it certainly helps in understanding the physics. The shape of the effective potential $V_s$ is shown in Fig. \ref{fig:potential_different_s}.
The potential vanishes at both boundaries, spatial infinity and the black hole horizon. Hence, we get essentially a free field in what looks like Minkowski space in the $r_*$ coordinate, at the boundaries. The effective potential peaks close to the light ring, at
\be
r_{\rm peak}= 3M + M\frac{1-s^2}{3\ell^2}\left(1+\frac{1}{\ell}-
\frac{(1+8s^2)}{9\ell^2}\right)+{\cal O}(\ell^{-5})\,,
\ee
for which
\be
V_{\rm peak}= \frac{\ell(\ell+1)}{27M^2} +\frac{(1-s^2)}{81M^2}\left(2+\frac{(1-s^2)(\ell-1)}{9\ell^3}+\frac{(1-s^2)(7+20s^2)}{243\ell^4}\right)+{\cal O}(\ell^{-5})\,,\label{eq:Vpeak}
\ee

Contrast this behavior with that of null geodesics, Eq.~\eqref{eq_Eff_potential_geodesics},
\be
\dot{r}^2=E^2-f\frac{L^2}{r^2}\,,
\ee
The effective potential $V_{\rm null}\equiv f\frac{L^2}{r^2}$ has a local maximum at $r=3M$ exactly and $V_{\rm null}(r=3M)=L^2/(27M^2)$. This is part of a result that allows one to connect null particles to wave propagation in the high frequency regime~\cite{Perlick:2004tq,Dolan:2017zgu}. In particular, as we shall see, in the high $\ell,\omega $ limit one recovers the results we discussed for null particles, including absorption cross-section, and the decay of waves trapped at the light ring.

\section{No hair \label{sec:no_hair}}
Now that we have the master equation \eqref{wave_equation_spins} governing the dynamics of massless fields, we can do physics. The first important question regards the existence of nontrivial static solutions to Eq.~\eqref{wave_equation_spins}, that would then describe a black hole surrounded by some distribution of scalar, electromagnetic or gravitational field. Is this possible?

In staticity, $\frac{\partial \Psi}{\partial t}=\frac{\partial^{2} \Psi}{\partial t^{2}}=0$, and the equation can be re-written as
\be
f^2\frac{d^2\Psi}{dr^2}+ff'\frac{d\Psi}{dr}-V_s\Psi=0\,,\label{wave_equation_spins2}
\ee
with the effective potential~\eqref{effective_potential_spins}. It is straightforward to see that both $r=2M$ and $r=\infty$
are regular singular points of the differential equation above\footnote{Remember that an ordinary differential equation of the form $y''+p_1y'+p_0y=0$ can be expanded as a Taylor series around any ordinary point, i.e., any point at which $p_1\,,p_0$ are analytic. Points $x_0$ at which either $p_1$ or $p_0$ are singular, but both $(x-x_0)p_1$ and $(x-x_0)^2p_0$ are analytic, are called regular singular points. At these points the equation admits independent solutions with a Frobenius expansion of the form $(x-x_0)^\alpha A(x)+(x-x_0)^\beta B(x)$ or $(x-x_0)^\alpha A(x)+(x-x_0)^\alpha\log(x-x_0) A(x)+(x-x_0)^\beta C(x)$, with $A,B,C$ analytic around $x_0$.}, which therefore admits a Frobenius expansion, which can be seen to have the form
\beq
\Psi&=&A_\infty r^{-\ell} + B_\infty r^{\ell+1}\,,\qquad \ell \neq 0,\, r\to \infty\\
\Psi&=& A_+ + B_+ \log{(r-2M)}\,,\quad r\to 2M \label{behavior_static}
\eeq
For a regular, reasonable spacetime, one should enforce $B_+=B_\infty=0$. We are free, of course, to choose the behavior at the horizon to be regular and just integrate towards infinity. But there is no freedom in the problem, no extra knob, that allows us to get a regular solution there, unless by accident. For example, take a monopolar scalar ($s=\ell=0$
), for which a simple, exact solution can be found,
\be
\Psi=k_1 r+k_2 r\log\left(1-2M/r\right)\,.
\ee
Regularity at the horizon requires $k_2=0$, and we are left with a solution that blows up at large distances. In other words, there is no well-behaved solution other than $k_1=0$, which means $\Psi=0$ everywhere.

We can make this statement more rigorous and more general.
Re-write the static equation as\footnote{The argument is easily generalized to any static, spherically symmetric and asymptotically flat geometry},
\be
\left(f\Psi'\right)'-\frac{V_s}{f}\Psi=0\,,
\ee
then multiply by $\Psi^*$ and integrate in the exterior,
\be
\int_{2M}^\infty\left(f\Psi'\right)'\Psi^*-\frac{V_s}{f}|\Psi|^2\, dr=0\,.
\ee
An integration by parts yields,
\be
\left[f\Psi'\Psi^*\right]_{2M}^{\infty}
-\int_{2M}^\infty dr \,f|\Psi'|^2-\int_{2M}^\infty dr\,\frac{V_s}{f}|\Psi|^2=0\,.\label{hair_equation}
\ee
But the regular behavior at the boundaries, Eq.~\eqref{behavior_static}, implies that the boundary term vanishes. For scalar fields, $V_s>0$ anywhere in the exterior, hence the only possible solution to Eq.~\eqref{hair_equation} is $\Psi=0$. This property is also known referred to as {\it black holes have no hair}, no structure protrudes from them, no nontrivial static field is possible in the exterior. Please note that uniqueness or no-hair properties always assume staticity or stationarity. One is always free to wave one's hand outside a black hole, or to shine a lamp or to let some pulse of scalar waves propagate.

There are interesting exceptions to the ``no-hair'' results above. For electromagnetic fluctuations with $\ell=0$, $V_1=0$ and hence one non-trivial solution to Eq.~\eqref{hair_equation} is $\Psi'=0$. But from \eqref{inversion_em_perturbations} and \eqref{expansion_vector_harmonics}, the axial sector is identically zero, thus there is also no axial electromagnetic hair. But for the polar electromagnetic sector, we find the nontrivial relation $(r^2(f^{00})')'=0$, or equivalently $f^{00}=Q/r$. In other words, a nontrivial solution to the perturbed Maxwell equations corresponds to adding a small charge to the black hole. There must then exist (they do!) charged black hole solutions, a consequence of this simple result.

Likewise, $\ell=0,1$ makes the potential for axial gravitational fluctuations negative, evading the no-hair results. The $\ell=0$ is trivial since the harmonics vanish. For $\ell=1$, one finds the general solution
\be
\Psi_1=\frac{k_1+k_2\left(12M^2r+3Mr^2+r^3+24M^3\log(r-2M)\right)}{r}\,,
\ee
and the regular solution is then ($k_2=0$)
\be
\Psi^-_2=\frac{k}{r}\,,\qquad h_1=\frac{k}{f}\,,\qquad h_0=0\,.
\ee
It is not too hard to verify that this corresponds to adding angular momentum to the black hole. Likewise, $\ell=0$ polar perturbations correspond to a small mass change in the spacetime, and $\ell=1$ to an addition of linear momentum~\cite{Zerilli:1970wzz}.

All the above discusses the existence of ``hair'' in a perturbative manner, that is, we are looking for static field configurations on top of a background geometry. Thus, when we do find hairy solutions, for example, an $\ell=0$ electromagnetic component, we can claim that there must be a full nonlinear solution of the field equations that asymptotes the perturbative solution we found. Indeed, in this particular case, it is the Reissner-Nordstr\\"{o}m family of charged black holes.

On the other hand, our perturbative no-hair result cannot exclude other fully nonlinear solutions. It only excludes nonlinear solutions that connect to the background solution we deal with (the Schwarzschild geometry).


\section{No polarizability}
The results above also allow us to work out how black holes polarize when in the presence of an external field. As a warm-up, let's revisit the classical calculation of the polarization of a conducting sphere of radius $R$ in electromagnetism, placed in a uniform-at-infinity electric field $E$. To find the distribution of the electric potential everywhere, note that it satisfies Laplace equation $\nabla^2U=0$ in the exterior of the sphere (the analog of our Eq.\eqref{wave_equation_spins2}, to which it reduces in flat space and spherical coordinates). In spherical coordinates, and after expanding in harmonics $U=\sum_{\ell m} U_r^{\ell m} Y_{\ell m}$ one finds (superscript omitted here only, we choose axis to be aligned with symmetry axis such that only $m=0$ modes survive)
\be
U_r''+\frac{2}{r}U_r'-\frac{\ell (\ell+1)}{r^2}U_r=0\,.
\ee
This equation has a regular singular point at $r=0$, and it is easy to see that it is exactly solvable,
\be
U_r^\ell=A_\ell\,r^\ell+B_\ell\, r^{-\ell-1}\,.
\ee
Now, the electric field $\vec{E}=-\nabla U$ and the asymptotic behavior at large distances that the electric field is constant imposes
\be
A_\ell=0\,,\qquad  {\rm for}\, \ell >1\,.\label{pol_sphere}
\ee
The condition that the sphere is conducting is that $U={\rm constant}$ anywhere on the surface of the sphere, of radius $R$. For this to be possible then, given \eqref{pol_sphere}, we must also have $B_\ell=0$ for $\ell>1$. We are therefore left with
\be
U=\left(A_1 r+\frac{B_1}{r^2}\right)Y_{10}+\left(A_0+\frac{B_0}{r}\right)Y_{00}\,.
\ee
But $Y_{00}=1/\sqrt{4\pi}$ is a constant, hence the last term effectively represents a charged sphere, which we forbid by construction. The $A_0$ term is irrelevant since the potential is defined up to a constant. Requiring a constant potential on the surface for any $\theta$, we find
\be
U=-E r\left(1-\frac{R^3}{r^3}\right)\cos\theta\,.
\ee
The first term is the contribution of the external field. The second is the reaction of the sphere. The sphere is polarized by the external field, behaving like a dipole with moment $p=4\pi \epsilon_0 R^3E$. It is now clear that the polarizability is encoded in the $r^{-\ell-1}$ terms, whereas the external fields are encoded in the $r^\ell$ terms\footnote{An interesting challenge for the reader is to devise a dielectric sphere such that its polarizability vanishes.}.

The example above is also instructive, as it shows that generally any object should polarize, since the external diverging piece of the potential comes with a subleading piece indicating polarization, and it's naturally present. The only way to avoid such terms would be through boundary conditions that would forbid them, while allowing for an external ``tidal'' field. This, it seems, is precisely what happens to black holes. Consider first a scalar field. Equation \eqref{wave_equation_spins2} can still be solved exactly~\cite{Abramowitz:1970as},
\be
\Psi=C_1 r P_\ell (r/M-1) + r C_2 Q_\ell(r/M-1)\,,
\ee
where $P,Q$ are Legendre polynomials. The second diverges logarithmically at the horizon, in accordance with the general behavior \eqref{behavior_static}. Therefore, the only regular solution is
\be
\Psi=C_1 r P_\ell (r/M-1)\,.
\ee
But these polynomials have no decaying terms. For example, for a monopolar and dipolar field, the regular solution at the horizon
\beq
\Psi&=&k_1 r\,,\qquad s=0,\ell=0\,,\\
\Psi&=&k_1 r(r/M-1)\,,\qquad s=0,\ell=1\,.
\eeq

One can repeat the calculation for other fields. A black hole in an external dipolar electric field, for example, behaves as
\be
\Psi=k_1 r^2\,,\qquad s=1,\ell=1\,.
\ee
(thus the electric potential $U\sim r\cos\theta$). Black holes don't polarize under external scalar, electric or gravitational fields, an amazing result~\cite{Binnington:2009bb,Damour:2009vw,Hinderer:2007mb,Cardoso:2017cfl}.
Of course, we have just calculated a number, which represent sub-leading coefficients in a large-r expansion, and in General Relativity we should worry about the gauge invariance of these results~\cite{Gralla:2017djj,Katagiri:2024fpn}. A numerical evaluation of the static, scalar tidal Love number of black holes is done in the \coderef{code:Perturbations.nb} notebook.

One might be tempted to conclude that the vanishing of polarizability is a consequence of the extraordinary properties of black holes. To some extent, this is true. However, note that black holes in higher dimensions do polarize, as well as in spacetime with different asymptotics~\cite{Kol:2011vg,Cardoso:2019vof,Emparan:2017qxd,Franzin:2024cah}.

Incidentally, one can immediately show very general no-polarizability results for scalars on a monopolar tidal field $\ell=0=s$, in an arbitrary geometry (be it a black hole or a star)
\be
ds^2=-fdt^2+\frac{dr^2}{g}+r^2d\Omega^2\,.
\ee
A static scalar then $\Phi(r)=\Psi(r)/r$ obeys the equation
\be
\left(r^2\sqrt{fg}(\Psi/r)'\right)'=0\,,
\ee
where primes stand for radial derivatives. We then find
\be
(\Psi/r)'=\frac{k_1}{r^2\sqrt{fg}}\,,
\ee
with $k$ a constant. But spacetime regularity force $f,g$ to be constants at the origin of coordinates, so the term integrates to a singular field, and one is left with $\Psi=k_2r$ as the only acceptable solution\footnote{This result, as far as we are aware, has not been reported before.}.
\section[Intermezzo]{Intermezzo: Fourier transforms, Laplace transforms and solution of ordinary differential equations\label{sec:Laplace}}
Our purpose now is to learn as much as possible about the content of Eq.~\eqref{wave_equation_spins}. There are two problems of great generality that can be studied. One concerns scattering problems, where a wave of some given frequency interacts with the black hole. The other class of problems concerns initial-value problems, where we specify the field $\Psi$ and its time derivative at some instant and use the partial differential equation to evolve the field in time. There are problems which are formally outside the scope of the homogeneous equation Eq.~\eqref{wave_equation_spins}, and include source terms. For example, the gravitational waves generated when you wave your hand can be studied using the perturbative framework which we have just discussed, but it would require the inclusion of matter source terms in Einstein equations. Of course, one can also look at the waving of your hand as an initial value problem. However, in that case, several more fields (like those making up your hand, the nervous system and all the microscopic details of your brain) would have to be included in the description.

So, for now, let us focus on the homogeneous equation~\eqref{wave_equation_spins}. We multiply the master variable by $e^{-st}$ and define its Laplace transform
\be
\psi\equiv\int_0^\infty e^{-st}\Psi dt\,,\label{def_Laplace_transform}
\ee
(we define $s=-i\omega$ such that the Laplace transform looks formally like a Fourier transform with the usual frequency parameters $\omega$, but note that convergence of the integral requires that $\Re[s]>0$ which demands a small imaginary component of the frequency $\omega$. These are convergence requirements~\cite{ArfkenWeber,Maggiore:2007ulw,Maggiore:2018sht}, and we usually brush over them when doing actual calculations). Integration by parts yields,
\be
\frac{d^2 \psi}{dr_*^2}+\left(\omega^2-V_s\right) \psi=-\frac{\partial \Psi(t=0)}{\partial t}+i\omega \Psi(t=0)\equiv I(r\omega)\,.\label{Laplace_wave_eq}
\ee
Therefore, we have reduced a partial differential equation to a simple ordinary differential equation with a source term. The source is dictated by the field structure at $t=0$ (the starting time is, of course, arbitrary). It is also clear that the more general problem involving a source term can be addressed in the same way as initial value problems.

Given appropriate convergence properties, it is also customary to use a Fourier transform instead
\be
\psi\equiv\int_{-\infty}^{+\infty} e^{i\omega t}\Psi dt\,.\label{def_Fourier_transform}
\ee
Assuming that the field decays sufficiently fast in the past and the future, the transform exists, and one ends up with Eq.~\eqref{Laplace_wave_eq} but without a source term. The initial data was pushed back into the past, where it vanishes. Fourier transforms are appropriate when looking for stationary properties, like a scattering experiment, where we assume we are bombarding a target with a constant energy flux, and measuring the scattered wave. Laplace transforms are more useful for real-world phenomena, like an explosion occurring close to a black hole, or some wavepacket of radiation falling into it.
\subsection{Solution by variation of parameters} \label{subsec:variation-of-param}
There are various ways to solve inhomogeneous ordinary differential equations. One way is to use the method of variation of parameters, where one writes the solution as a linear combination of homogeneous solutions with variable coefficients. Therefore, to solve Eq.~\eqref{Laplace_wave_eq}, let us choose two independent solutions $\psi_{H}$ and $\psi_{\infty}$ to the homogeneous equation which satisfy the following boundary conditions
\be\label{asymptotic_behaviorH}
\psi_{H}\sim  \left\{
\begin{array}{ll}
    e^{-i\omega r_*}                                        & r\to 2M     \\
    A_{\rm{out}}e^{i\omega r_*}+A_{\rm{in}}e^{-i\omega r_*} & r\to \infty \\
\end{array}
\right.
\ee
\be\label{asymptotic_behaviorinfinity}
\psi_{\infty}\sim  \left\{
\begin{array}{ll}
    B_{\rm{out}}e^{i\omega r_*}+B_{\rm{in}} e^{-i \omega r_*} & r\to 2M     \\
    e^{i\omega r_*}                                           & r\to \infty \\
\end{array}
\right.
\ee

This asymptotic behavior comes about because the potential vanishes at the boundaries (and we are left with the equation describing a simple harmonic oscillator). In fact, it turns out that it is useful to seek subleading terms in the above expansion (see \coderef{code:Scattering-Scalar.nb}), and finding, say, for $\psi_H$  (note that $r=\infty$ is an irregular singular point of the wave equation for $\omega \neq 0$, hence we have a rather different asymptotic expansion compared to Eq.~\eqref{behavior_static})
\be\label{asymptotic_behaviorH_accurate}
\psi_{H}\sim  \left\{
\begin{array}{ll}
    e^{-i\omega r_*}\!\left(1+a_H(r-2M)+\cdots\right),                                                          & r\to 2M     \\[6pt]
    A_{\rm out}\,e^{i\omega r_*}\!\left(1+\dfrac{a_{\rm out}}{r}+\dfrac{b_{\rm out}}{r^2}+\cdots\right)         &             \\[2pt]
    \quad +\,A_{\rm in}\,e^{-i\omega r_*}\!\left(1+\dfrac{a_{\rm in}}{r}+\dfrac{b_{\rm in}}{r^2}+\cdots\right), & r\to \infty
\end{array}
\right.\!
\ee
where $a_H, a_{\rm out}, b_{\rm out}, a_{\rm in}, b_{\rm in}$ are constants which can be determined from the master equation (see the first exercise at end of this chapter for their explicit expressions and a guide to calculate other higher-order terms).

Define the Wronskian of the two linearly independent solutions
\be
W\equiv\psi_{\infty}\frac{d\psi_{H}}{dr_*}-\psi_{H}\frac{d\psi_\infty}{dr_*}=-2i\omega A_{\rm in}.
\ee
where in writing the last equality, we used the asymptotic behavior of the solutions at infinity since the Wronskian is constant for all $r_*$, thanks to Abel's identity \cite{Bender:1999box,ArfkenWeber, hassani2013mathematical}\footnote{Recall Abel's result, which you can easily verify: for a differential equation of the form $y''+p_1y'+p_0y=0$, the Wronskian of two independent solutions $y_1,y_2$ satisfies $$W(x)=W(x_0)e^{-\int_{x_0}^xp_1(t)dt},$$ and in terms of the $r_*$ coordinate, the $p_1$ term is absent from Eq.~\eqref{Laplace_wave_eq}, hence the Wronskian is position-independent.}. With the above, Eq.~\eqref{Laplace_wave_eq} can be solved as \cite{Bender:1999box,ArfkenWeber, hassani2013mathematical} \footnote{Given two independent homogeneous solutions $y_1,y_2$ to the equation $y''+p_1y'+p_0y=I$, {\it variation of parameters} (also known as variation of constants) yields the following solution $$y=y_1\int^x Iy_2dx/W+y_2\int_x Iy_1 dx/W+a_1y_1+a_2y_2,$$ where $a_1,a_2$ are constants~\cite{Bender:1999box,ArfkenWeber, hassani2013mathematical}. You should readily check that this solves the inhomogeneous equation.}
\be
\psi(r_*)=\psi_{\infty}(r_*)\int_{-\infty}^{r_*}dr_*'\frac{I(r_*')\psi_{H}(r_*')}{W}+\psi_{H}(r_*)\int_{r_*}^{\infty}dr_*'\frac{I(r_*')\psi_{\infty}(r_*')}{W}\,\label{eq:variation_of_param_integration}.
\ee
Thus, we reduced our problem to the integration of the homogeneous equation, which is relatively easy to do, and an integral over the source term describing the initial conditions. Note that this expression is formally identical to the one we get on solving the inhomogeneous equation using Green's functions, as it should be.

It turns out that our detectors are usually far away from the source, so we are mostly (but not always) interested in calculating the signal $\psi$ at very large distance. For very large values of $r$ the above equation has the following asymptotic form (we replaced $dr_*=dr/f$ for a Schwarzschild geometry)
\be
\psi(r\to\infty)=\frac{e^{i\omega r_*}}{2i\omega A_{\rm in}}
\int_{2M}^{\infty} \frac{I(r')\psi_{H}(r')}{f}dr'\equiv\psi^{\rm out}_{\infty}e^{i\omega r_*}\label{rlm}
\ee
Likewise, at the horizon we get
\be
\psi(r\to 2M)=\frac{e^{-i\omega r_*}}{2i\omega A_{{\rm{in}}}} \int_{2M}^{\infty}\frac{I(r')\psi_{\infty}(r')}{f}dr'\equiv\psi^{\rm in}_{H}e^{-i\omega r_*}\,.\label{rhor}
\ee

The original time dependent field is then, naturally,
\be
\Psi(t,r)=\int \frac{d\omega}{2\pi}e^{-i\omega t}\psi(\omega,r) \,.\label{eq_Laplace_inverse}
\ee
In the above equation, we can substitute, $\psi(\omega, r)$ with either $\psi^{\rm out}_{\infty}e^{i\omega r_*}$ or $\psi^{\rm in}_{H}e^{-i\omega r_*}$ depending on whether we want to extract the signal at infinity or at the horizon.

\subsection{Fluxes at horizon and infinity\label{sec:fluxes}}
Corresponding to timelike Killing vector $\xi^{\mu}_{(t)}$, one can form the current density four-vector $J^{\mu}$ as,
\begin{align}
    J^{\mu}=T^{\mu}{}_{\nu}\xi^{\nu}_{(t)}~,
\end{align}
where $T^{\mu}{}_{\nu}$ is the stress-energy tensor. This leads to the covariant conservation law (that is, the general relativistic version of the continuity equation), $\nabla_{\mu}J^{\mu}=0$. One can interpret the components of $J^{\mu}$ as energy density  and the momentum current densities. Then from the covariant version of the conservation law we can show that,
\begin{align}
    \partial_{t}\int_{\Sigma} d^{3}x \sqrt{\gamma} Q=-\int_{\partial \Sigma} d^{2}x \sqrt{\sigma} F~.
\end{align}
Here, $\Sigma$ is a spacelike hypersurface with  normal $n_{\mu}$ and $\partial\Sigma$ denotes the 2D hypersurface which is the boundary of $\Sigma$, $\gamma$ and $\sigma$ denotes the induced metric of $\Sigma$ and $\partial\Sigma$ respectively. Here $Q$ plays the role of charge density and is given by,
\begin{align}
    Q=-n_{\mu}J^{\mu}~,
\end{align}
and the flux out of $\partial\Sigma$ is given by,
\begin{align}
    F=\alpha N_{i}J^{i}~,
\end{align}
where $\alpha$ is the lapse function and $N_{i}$ is the normal to $\partial\Sigma$.

Accordingly, the energy fluxes to infinity and through the horizon can be calculated as, for non-spinning black holes~\cite{Teukolsky:1974yv,Dolan:2007mj,Vicente:2022ivh}
\beq
E^{\infty, H} =-\lim_{r \to +\infty, 2M} r^2
\int d\Omega f T_{t r}\,,
\eeq
where $d\Omega$ is the area element of the two sphere.

For a scalar field, the stress tensor is,
\be
T_{\mu\nu}=\frac{k}{8\pi}\left(\frac{1}{2}\nabla_\mu\Phi\nabla_\nu\Phi-\frac{1}{4}g_{\mu\nu}\nabla_{\alpha}\Phi \nabla^{\alpha}\Phi\right)
\ee
With the expansion in spherical harmonics \eqref{expansion_harmonics_scalars} and with the asymptotic behavior $\psi=\psi^{\rm out}_\infty e^{i\omega r-i\omega t}$ at large distances, we find
\be
\frac{d^2E}{dtd\Omega}=\frac{k}{16\pi}\int d\omega d\omega' \omega\omega' \,Y_{\ell m}Y_{\ell'm'}e^{i(\omega+\omega')(r-t)}\psi^{\rm out}_\infty \psi^{\rm out'}_\infty\,,
\ee
where $\psi^{\rm out'}_\infty$ means that it is associated with angular numbers $\ell'm'$. The normalization conditions for the spherical harmonics reduce this to
\be
\frac{dE}{dt}=\lim_{r\to \infty}=\frac{k}{16\pi}\int d\omega d\omega' \omega\omega' e^{i(\omega+\omega')(r-t)}\psi^{\rm out}_\infty \psi^{\rm out'}_\infty\,,
\ee
which we can manipulate to write as
\be
\frac{dE}{d\omega}=\lim_{r\to \infty}=\frac{k}{16\pi}\int dt d\omega' \omega\omega' e^{i(\omega+\omega')(r-t)}Z_\infty Z'_\infty\,.
\ee
Using
\be
\delta(x-x_0)=\frac{1}{2\pi}\int e^{it(x-x_0)}dt\,,\label{eq:delta}
\ee
we can get rid of the time integration, to find
\be
\frac{dE}{d\omega}=\frac{k}{8}\omega^2 \left|\psi^{\rm out}_\infty\right|^{2}\,.
\ee
One can repeat the above steps to find the flux at the horizon,
\be
\frac{dE}{d\omega}=\frac{k}{8}\omega^2 \left|\psi^{\rm in}_H\right|^{2}\,.
\ee

One can proceed in the same way for electromagnetic perturbations. Details can be found in \coderef{code:Perturbations.nb}. The electromagnetic stress tensor is (alternative versions include a multiplicative factor of $k/16\pi$, the results below are accordingly modified trivially)
\be
T^{\mu\nu}=F^{\mu}_{\alpha}F^{\nu\alpha}-\frac{1}{4}g^{\mu\nu}F^{\alpha\beta}F_{\alpha\beta}\,.
\ee
With the expansion \eqref{eq_Maxwell_separation} one gets
\beq
\frac{dE}{dt}&=&\lim_{r\to \infty}\int d\vartheta d\varphi\sqrt{-g}T^r_t\,,\nonumber\\
&=&\int d\vartheta d\varphi \sin\vartheta \sum_{\ell\ell'm m'} e^{i(m+m')\varphi}\left(\frac{mm'}{\sin^2\vartheta}Y_{\ell m}Y_{\ell'm'}-\frac{dY_{\ell m}}{d\vartheta}\frac{dY_{\ell' m'}}{d\vartheta}\right)\frac{da_{\ell m}}{dr}\dot{a}_{\ell' m'}\nonumber\\
&=&2\pi\int d\vartheta \sin\vartheta \sum_{\ell\ell'm} \left(-\frac{m^2}{\sin^2\vartheta}Y_{\ell m}Y_{\ell'-m}-\frac{dY_{\ell m}}{d\vartheta}\frac{dY_{\ell' -m}}{d\vartheta}\right)\frac{da_{\ell m}}{dr}\dot{a}_{\ell' m}\nonumber\\
&=&\sum_{\ell m} \ell(\ell+1)\frac{da_{\ell m}}{dr}\dot{a}_{\ell m}\,,
\eeq

Alternatively, one can simply compute the radial component of the Poynting vector at infinity $\vec{E} \times \vec{B}$, using \eqref{electric_magnetic_fields}. We find the same expression. The procedure can easily be generalized to the even sector. Results for the gravitational perturbations can be found in Refs.~\cite{Zerilli:1970wzz,Martel:2005ir,Nagar:2005ea}. Beware of possible different normalizations in harmonics or master variables.

\section{Scattering of waves off black holes\label{sec:scattering}}

We are now ready to start looking at dynamical processes. Let us start with the scattering of waves off black holes, but looking for monochromatic solutions, $\Psi\sim e^{-i\omega t}\psi(r_*)$. Then, the wave equation \eqref{wave_equation_spins} or equivalently \eqref{Laplace_wave_eq} are simply
\be
\frac{d^2 \psi}{dr_*^2}+\left(\omega^2- V_s\right)\psi=0\,.\label{EQ_master_frequencydomain}
\ee
The behavior of the regular solution at the horizon is given by Eq.~\eqref{asymptotic_behaviorH}\footnote{The equation we are solving is linear, hence we could also have chosen instead to work with a solution which differs by a multiplicative factor, hence with the behavior
    \[
        \psi_H= \left\{
        \begin{array}{ll}
            A_{\rm T} e^{-i\omega r_*}                                & r_*\to -\infty    \\
            A_{\rm in} e^{-i\omega r_*} + A_{\rm out} e^{i\omega r_*} & r_*\to +\infty\,.
        \end{array}
        \right.
    \].
}.

\begin{figure}
    \centering
    \includegraphics[width=0.9\textwidth]{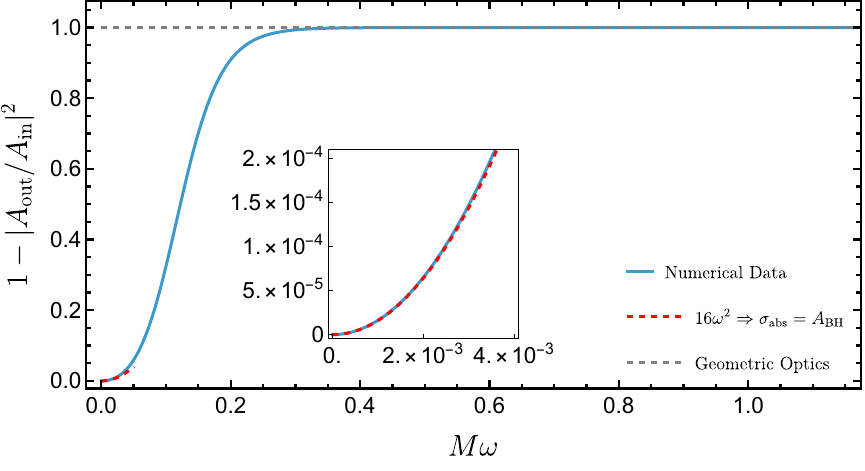}
    \caption{
        {\small Absorption amplitudes of a scalar field for a Schwarzschild black hole. At low frequencies, the $\ell=0$ mode dominates and the absorption cross-section can be calculated to be equal to the horizon area $\sigma_{\rm abs}=16\pi M^2$. Note that at low frequencies most of the wave is reflected, as long wavelength waves find it hard to tunnel through the potential. At high frequencies, most of the signal tunnels through into the horizon, and $A_{\rm out}\sim 0$. In this regime, the cross-section gets contributions from all multipolar numbers and evaluates to the null geodesic calculation $\sigma_{\rm abs}=27\pi M^2$, Eq.~\eqref{cross_section_light}. See also the exercises at the end of this  chapter and the accompanying notebook.}
    }
    \label{fig:absorption}
\end{figure}

The potential $V_s$ is real, so $\psi^*$ is also solution of the differential equation, satisfying correspondingly complex conjugate boundary conditions. We can then consider $\psi_H$ and $\psi_H^*$ as the two independent solutions. We can then evaluate the Wronskian at the horizon and at infinity. We find
\beq
W(r_*\to -\infty)&=&2i\omega\,,\\
W(r_*\to+\infty)&=&2i\omega \left(|A_{\rm in}|^2-|A_{\rm out}|^2\right)\,.
\eeq
Since the Wronskian is independent of position and a constant (due to Abel's identity), we can equate it at the two boundaries to get,
\begin{equation}
    |A_{\rm in}|^2-|A_{\rm out}|^2=1 \,,
\end{equation}
expressing conservation of energy\footnote{As we showed in Section~\ref{sec:fluxes}, at fixed frequency, the energy is indeed proportional to the square of the wavefunction.}.

We still haven't calculated the ingoing or outgoing amplitudes themselves, but established only a relation between them. To compute amplitudes, one needs to solve Eq.~\eqref{EQ_master_frequencydomain} numerically (as shown in \coderef{code:Scattering-Scalar.nb}, or analytically in some regime, as we shall do in Section~\eqref{sec:low_frequency}). However, just by inspecting the effective potential and the structure of the differential equation, we can immediately infer a couple of important aspects.
Energies, or frequencies, smaller than the potential $V_s$ will find it hard to tunnel through the potential down to the black hole horizon. From Eq.~\eqref{eq:Vpeak}, we can predict that frequencies $M\omega \lesssim \ell/(3\sqrt{3})$ will simply be reflected off, whereas those larger than this will be transmitted through the potential and absorbed by the black hole.

The overall picture is confirmed numerically as shown in Fig.~\ref{fig:absorption} for the monopolar $\ell=0$ mode of a scalar field (you can check it for yourself using the notebook \coderef{code:Scattering-Scalar.nb}). The results are qualitatively similar for other modes and fields.


Let us now use the more physical setup of a {\it plane} wave $\Phi_{\rm plane}$ of amplitude $A_{\rm plane}$ coming from $z=-\infty$, scattering off the black hole and continuing on its way. Now,
with the normalization $\int d\theta \sin\theta Y_{\ell m}Y^*_{\ell'm'}=\delta_{\ell \ell'}\delta mm'$, the spherical harmonics are related to Legendre polynomials as
\be
Y_{\ell 0}=\sqrt{\frac{2\ell+1}{4\pi}} P_\ell (\cos\theta)\,.
\ee
The handling of the scattering problem is then based on the identity (which can be easily derived, see Sec. 16.3 of \cite{ArfkenWeber})
\beq
e^{i\omega r\cos\theta}&=&\sum_{\ell=0}^{\infty}i^\ell (2\ell+1)j_\ell(\omega r)P_\ell(\cos\theta)\,,\\
%
%
&\approx&\sum_{\ell=0}^{\infty}i^\ell (2\ell+1)\frac{\sin (\omega r-\ell\pi/2)}{\omega r}P_\ell(\cos\theta)\,,\\
&=&\sum_{\ell=0}^{\infty}i^\ell \sqrt{4\pi (2\ell+1)}\frac{\sin (\omega r-\ell\pi/2)}{\omega r}Y_{\ell 0}(\theta)
\eeq
with $j_\ell$ a \emph{spherical} Bessel function and $j_\ell \approx \sin (\omega r-\ell\pi/2)/\omega r$ for $\omega r \gg l(l+1)/2$~\cite{ArfkenWeber}.

But the scalar was decomposed as
\be
\Phi=\sum_\ell \frac{\psi}{r} Y_{\ell 0}e^{-i\omega t}\,,\label{Phi_spherical}
\ee
with $\psi=\psi_{\rm H}$ governed by Eq.~\eqref{EQ_master_frequencydomain}, and obeying the boundary behavior \eqref{asymptotic_behaviorH}. Based on the above, a plane scalar wave looks like, at large distances,
\be
\Phi_{\rm plane}=A_{\rm plane}e^{-i\omega t}\sum_{\ell=0}^{\infty}i^\ell \sqrt{4\pi (2\ell+1)}\frac{\sin (\omega r_*-\ell\pi/2)}{\omega r}Y_{\ell 0}(\theta)\,,r\to \infty\,.
\ee
After the interaction takes place (note we are in the monochromatic, Fourier, stationary regime) then the signal takes the form of the original incoming wave plus a scattered signal
\be
\Phi=\Phi_{\rm plane}+\frac{f(\theta)}{r}e^{i\omega r-i\omega t}\,,r\to \infty\,. \label{eq:scattered_wave}
\ee
To extract $f(\theta)$ we re-write the plane wave as
\begin{align}
    e^{i\omega t}\Phi_{\rm plane} & =A_{\rm plane}\sum_{\ell=0}^{\infty}i^\ell \sqrt{4\pi (2\ell+1)}\frac{e^{i(\omega r_*-\ell\pi/2)}-e^{-i(\omega r_*-\ell\pi/2)}}{2i\omega r}Y_{\ell 0}(\theta)\,, \\
                                & =A_{\rm plane}\sum_{\ell=0}^{\infty}\sqrt{4\pi (2\ell+1)}\frac{e^{i\omega r_*}-(-1)^\ell e^{-i\omega r_*}}{2i\omega r}Y_{\ell 0}(\theta)
\end{align}
From \eqref{Phi_spherical} we find
\beq
e^{i\omega t}\left(\Phi-\Phi_{\rm plane}\right)=\sum_\ell Y_{\ell 0}\left(\frac{\psi}{r} -A_{\rm plane}\sqrt{4\pi (2\ell+1)}\frac{e^{i\omega r_*}-(-1)^\ell e^{-i\omega r_*}}{2i\omega r}\right)\,.\label{eq:_phi_phiplane}
\eeq
We now need to make sure that no incoming wave at infinity remains in $\Phi-\Phi_{\rm plane}$. If we use \eqref{asymptotic_behaviorH} and insert the expression for $\psi_H$ at infinity into the above equation we see that we need to have
\beq
&&A_{\rm in}=A_{\rm plane}\frac{(-1)^{\ell+1}\sqrt{4\pi (2\ell+1)}}{2i\omega}\,.\label{eq:Ain}
\eeq
From Eq.~\eqref{eq:scattered_wave}, we get,
\beq
re^{i\omega t-i\omega r}\left(\Phi-\Phi_{\rm plane}\right)=f(\theta)
\eeq
Using Eq.~\eqref{eq:_phi_phiplane} and Eq.~\eqref{eq:Ain}, we can write (note, we are using the asymptotic expansion of $r_*$, hence $r_*\sim r$)
\beq
f(\theta)&=&\sum_{\ell=0}^\infty Y_{\ell 0} A_{\rm plane}\left(\frac{A_{\rm out}}{A_{\rm plane}}-\frac{\sqrt{4\pi (2\ell+1)}}{2i\omega}\right)\\
&=&\sum_{\ell=0}^\infty Y_{\ell 0} \frac{A_{\rm plane}\sqrt{\pi (2\ell+1)}}{i\omega}\left(\frac{(-1)^{\ell+1}A_{\rm out}}{A_{\rm in}}-1\right)\,.
\eeq

Now, the plane wave has a flux (energy per time per area) $\omega^2\left| A_{\rm plane}\right|^2$. The total energy (per time) going {\it into} the black hole, is simply $\sum_\ell \omega^2\left(\left|A_{\rm out}\right|^2-\left|A_{\rm in}\right|^2\right)$. Thus, we can define an effective area, or {\it absorption cross-section} $\sigma_{\rm abs}$ (using \eqref{eq:Ain}),
\be
\sigma_{\rm abs}\equiv \frac{\rm Energy}{\rm Flux}=\sum_\ell \frac{\pi(2\ell+1)}{\omega^2}\left(1-\frac{\left|A_{\rm out}\right|^2}{\left|A_{\rm in}\right|^2}\right)\,,\label{cross_section_definition}
\ee
where $A_{\rm in},\,A_{\rm out}$ carry $\ell$ dependence.

At low frequencies, the monopolar $\ell=0$ mode dominates the contribution to the absorption cross-section, which can be analytically calculated to be equal to the horizon area $\sigma_{\rm abs}=16\pi M^2$~\cite{Unruh:1976fm,Das:1996we,Cardoso:2005mh}, see below for a low-frequency solution to the problem. Note that at low frequencies most of the wave is reflected, as long wavelength waves find it hard to tunnel through the potential. At high frequencies, most of the signal tunnels through into the horizon, and $A_{\rm out}\sim 0$. In this regime, the cross-section gets contributions from all multipolar numbers and evaluates to the null geodesic calculation $\sigma_{\rm abs}=27\pi M^2$, Eq.~\eqref{cross_section_light} (see also Fig.~\ref{fig:absorption}). High-frequency approximations can also reproduce this behavior~\cite{Sanchez:1977si,Futterman:1988ni,Decanini:2011xi,Benone:2018rtj}, but we will not dwell on this any further.

\section{Low-frequency solution\label{sec:low_frequency}}
We now show how to obtain simple solutions to Eq.~\eqref{EQ_master_frequencydomain} when the frequency is small, $M\omega\ll 1$. The calculation will be done for a scalar field, but it can easily be generalized for any other massless field, and spin~\cite{Unruh:1976fm,Starobinskil:1974nkd,Cardoso:2005mh,Pani:2012bp,Cardoso:2019nis,Santos:2024tlt,Brito:2015oca}.

The low-energy approximation uses a matching procedure to find a solution on the whole spacetime. Define the near-horizon region $r-2M\ll 1/\omega$ (this is the region within less than a wavelength away from the black hole) and the asymptotic region $r\gg 2M$. Notice that there is an overlap region $2M \ll r \ll 1/\omega$.
To solve the wave equation in the near-horizon region, we set
\be
\Psi_{s=0}=r~R\,,
\ee
in Eq. (\ref{EQ_master_frequencydomain}). With this substitution, the scalar equation is cast in the form (see \coderef{code:Perturbations.nb})
\be \frac{f}{r^{2}}\partial_r \left (fr^{2}\partial_r
R\right )+ {\Biggl [} \omega^2 -\frac{f}{r^2}\ell(\ell+1){\Biggl
    ]}R=0 \,.
\ee
If we approximate $r\sim 2M$ and $f\sim (r-2M)/2M$, we get for the ingoing solution
\be
R={\cal T}(r-2M)^{-2iM\omega}\,,
\ee
where ${\cal T}$ is a constant. This solution is valid close to the black hole. If we expand this same, near-horizon solution at infinity\footnote{Observe that $(r-2M)^{-2iM\omega} = \exp{[-2 i M \omega (\log r + \log(1-2M/r))]}$. You can Taylor expand the second logarithm to get $r^{-2 i M \omega}\exp[4 i M^2 \omega/r]$, which gives you the desired expression if you retain the leading order term.}, we get
\be
\Psi \sim{\cal T} r^{-2iM\omega}(r+4iM^2\omega)\,,
\ee
which can be written as
\be
\Psi \sim{\cal T} (r - 2 i M \omega r \log r + 4 i M^2 \omega + 8 i M^3 \omega^2 \log r) \,. \label{eqn:near_hor_at_infinity}
\ee
In the asymptotic region at large $r$, where we imagine setting up a scattering experiment, the equation simplifies to the spherical Bessel equation of order zero~\cite{ArfkenWeber}
\be \partial^2_r R +\frac{2}{r}\partial_r R+ \omega^2 R=0 \,.
\ee
The solution of this equation is
\be
\Psi = C_1r^{1/2}J_{1/2}(\omega r)+C_2r^{1/2}Y_{1/2}(\omega r)\,,
\ee
where $J_{1/2}(\omega r)$ and $Y_{1/2}(\omega r)$ are Bessel functions. Expanding for small $\omega r$, we obtain
\be
\Psi \sim \sqrt{2/\pi}\left(C_1\sqrt{\omega}r-C_2/\sqrt{\omega}\right)\,.
\ee
Matching the above expression with the near horizon solution Eq.~\eqref{eqn:near_hor_at_infinity} (because, remember, there is a region of overlap) we find
\be
-\frac{C_1}{C_2}=\frac{1}{4iM^2\omega^2}
\ee
On the other hand, the asymptotic behavior of the Bessel functions is~\cite{ArfkenWeber}
\beq
J_n(x)&\sim& \sqrt{\frac{2}{\pi\,x}}\cos \left(x-(2n+1)\pi/4\right)\,,\\
Y_n(x)&\sim& \sqrt{\frac{2}{\pi\,x}}\sin \left(x-(2n+1)\pi/4\right)\,,
\eeq
and the reflection coefficient is therefore
\be
\left|\frac{A_{\rm out}}{A_{\rm in}}\right|^2 =\left|\frac{C_1-iC_2}{C_1+iC_2}\right|^2\sim \left|1-2i\frac{C_2}{C_1}\right|^2=1-16(M\omega)^2\,. \label{reflection_low_frequency}
\ee
We can find the absorption cross-section, for instance, from Eq.~\eqref{cross_section_definition},
\be
\sigma_0=16\pi M^2\,,
\ee
i.e, the area of the black hole, as we had stated previously. The calculation kept only the dominant terms, and as such described well $\ell=0$ modes, but higher $\ell$ modes can easily be captured within the same framework.

\section{Superradiance\label{sec:superradiance}}
\subsection{The wave equation in the presence of rotation}
Most of what is discussed in these lectures are features universal to all spacetimes, and hence we focus particularly on non-spinning geometries. However, as we also saw when discussing the Penrose process, there are qualitatively new phenomena when spin is included.
It turns out that the most appropriate formalism known so far to handle perturbations around rotating black holes is different from the approach we took~\cite{Teukolsky:1973ha}. We will not deal with it, except to say that the vast majority of the methods we use here to understand the physics can be carried over in a straightforward way.

However, we can understand the new features of rotation by focusing on a scalar field. In a Kerr geometry, the spherical harmonics will not separate the Klein-Gordon equation, but we can generalize such harmonics. In particular, letting (the geometry is still axisymmetric)
\be
\Phi=R(r)S(\theta)e^{-i\omega t+im\phi}\,,
\ee
then the angular and radial dependence separate and one finds, in the Kerr background \eqref{kerr_geometry},
\beq
&&\left(\Delta R(r)'\right)'+\left(\frac{K^2}{\Delta}-A-a^2\omega^2+2am\omega\right)R(r)=0\,,\label{kerr_radial}
\eeq
\beq
&&\frac{1}{\sin\theta}\left(\sin\theta S(\theta)'\right)'+\left(a^2\omega^2\cos\theta-\frac{m^2}{\sin^2\theta}+A\right)S(\theta)=0\,.\label{kerr_angular}
\eeq
Here, $K=(r^2+a^2)\omega-am$, $A$ is a separation constant and primes stand for derivatives with respect to the argument of the respective function. It is clear that for $a\omega=0$ we recover the usual spherical harmonic equation for $S$ and therefore $A=\ell(\ell+1)$ in this limit.

The re-scaling
\be
R=\frac{\psi}{\sqrt{r^2+a^2}}\,,
\ee
turns the radial equation onto
\begin{align}
    \frac{d\psi}{dr_*^2}+\left((\omega-m\varpi)^2-\frac{(A+a^2\omega^2-2am\omega)\Delta}{(r^2+a^2)^2} - \frac{r^2\Delta^2}{(r^2+a^2)^4} \right.  \nonumber \\ \left.-\frac{\Delta}{r^2+a^2}\frac{d}{dr}\frac{r\Delta}{(r^2+a^2)^2}\right)\psi=0\,,
\end{align}

with $\varpi=a/(r^2+a^2)$. This equation can be re-written as
\be
\frac{d^2 \psi}{dr_*^2}+\left(\left(\omega-m\varpi\right)^2-V\right)\psi=0\label{eq_rotation},
\ee
with $V$ vanishing at the boundaries. Indeed the perturbation equations for electromagnetic and gravitational fields on a Kerr geometry can be reduced to this form~\cite{Teukolsky:1973ha}. We can therefore use all of the material above to learn that the solution with the correct boundary condition at the horizon is~\cite{Bardeen:1972fi,Brito:2015oca}
\[
    \Psi= \left\{
    \begin{array}{ll}
        A_{\rm T} e^{-i(\omega-m\Omega_{\rm H}) r_*}              & r_*\to -\infty \\
        A_{\rm in} e^{-i\omega r_*} + A_{\rm out} e^{i\omega r_*} & r_*\to +\infty \\
    \end{array}
    \right.
\]
with $\Omega_{\rm H}=\varpi (r=r_+)$ the horizon angular velocity.
Since the potential $V$ is real, $Z^*$ also solution. Abel's result then yields that the Wronskian $W=ZdZ^*/dr_*-Z^*dZ/dr_*$ is constant, thus
\beq
W(-\infty)&=&2i(\omega-m\Omega_{\rm H})  |A_{\rm T}|^2\nonumber\\
W(+\infty)&=&2i\omega \left(|A_{\rm in}|^2-|A_{\rm out}|^2\right)\nonumber\\
&\Rightarrow& |A_{\rm in}|^2-|A_{\rm out}|^2=\frac{\omega-m\Omega_{\rm H}}{\omega}|A_{\rm T}|^2 \nonumber
\eeq
For $\omega<m\Omega_{\rm H}$ something new happens: we get out more than we put in. This is called superradiance and occurs because there is an ergoregion in the spacetime.

\subsection{A thermodynamics argument for superradiance}
In fact the phenomenon of energy extraction via wave scattering is very general~\cite{Brito:2015oca}.
Objects with free microstates (internal degrees of freedom) to absorb radiation, will amplify low-frequency waves at the cost of their rotational energy. The thermodynamic argument takes any {\it axisymmetric} macroscopic body rotating rigidly with constant angular velocity about its symmetry axis, and with a well-defined entropy $S$, rest mass $M$ and temperature $T$. Suppose now that a wavepacket with frequency $(\omega,\omega+d\omega)$ and azimuthal number $m$ is incident upon this body, with a power $P_m(\omega)d\omega$. Radiation with a specific frequency and azimuthal number carries angular momentum at a rate $(m/\omega)P_m(\omega)d\omega$ (see Appendix C in Ref.~\cite{Brito:2015oca}).
Neglecting spontaneous emission by the body (of thermal or any other origin), the latter will absorb a fraction $Z_m$ of the incident energy and angular momentum (dot stands for time derivative),
\begin{equation}
    \dot{E}= Z_m P_md\omega ,\, \dot{J}= Z_m \frac{m}{\omega} P_m d\omega
\end{equation}
Notice that the assumption of axisymmetry and stationarity implies that no precession nor Doppler shifts occur during the
interaction. Both the frequency and
multipolarity of the incident and scattered wave are the same. Now, in the frame co-rotating with the body, the change in energy of the spinning body is simply
\begin{equation}
    dE_0 = E-\Omega dJ = dE\left(1-\frac{m\Omega}{\omega}\right)\,,
\end{equation}
and thus the absorption process is followed by an increase in entropy, $dS = dE_0/T$, of
\begin{equation}
    \dot{S} = \frac{\omega-m\Omega}{\omega T} Z_m P_m(\omega)d\omega\,.\label{eq:entropy}
\end{equation}
Finally, the second law of thermodynamics demands that $(\omega- m\Omega)Z_m>0$ and superradiance ($Z_m<0$) follows in the superradiant regime $\omega-m\Omega< 0$.

\subsection{An analytic example}
To provide a simple example where energy extraction can be solved in closed form, consider a generic action involving one complex, charged scalar $\Psi$ and a massless vector
field $A_{\mu}$ minimally coupled to gravity,
\beq
S&=&\int d^4x \sqrt{-g} \Bigl( \frac{R}{\kappa} - \frac{1}{4}F^{\mu\nu}F_{\mu\nu}
-\frac{1}{2}g^{\mu\nu}\Psi^{\ast}_{,\mu}\Psi^{}_{,\nu} \nonumber\\
&+&i\frac{q}{2}A_{\mu}\left(\Psi\nabla^{\mu}\Psi^{\ast}-\Psi^{\ast}\nabla^{\mu}\Psi\right)-\frac{q^2}{2}A_{\mu}A^{\mu}\Psi\Psi^{\ast}
\Bigr) \,,\label{eq:MFaction}
\eeq
where $\kappa=16\pi$, $F_{\mu\nu} \equiv
    \nabla_{\mu}A_{\nu} - \nabla_{\nu} A_{\mu}$ is the Maxwell tensor and $q$ a scalar charge.
The resulting equations of motion are
\begin{subequations}
    \label{eq:MFEoMgen}
    \begin{eqnarray}
        \label{eq:MFEoMScalar}
        &&\left(\nabla_{\mu}-iqA_{\mu}\right)\left(\nabla^{\mu}-iqA^{\mu}\right)\Psi =0
        \,,\\
        \label{eq:MFEoMVector}
        &&\nabla_{\mu} F^{\mu\nu} =
        -i\frac{q}{2}\left(\Psi\nabla^{\nu}\Psi^{\ast}-\Psi^{\ast}\nabla^{\nu}\Psi\right) \,.
        %
    \end{eqnarray}
\end{subequations}
Now consider small linearized and Fourier-transformed fluctuations of the scalar around a static, spherically symmetric charged black hole background with trivial background scalar field. This is similar in spirit to the perturbative approach we have been considering.
In this case, to leading order the Maxwell equation for the background vector potential reduces to $\nabla_{\mu} F^{\mu\nu}=0$.
Once the Maxwell equations are solved, the solution for the background vector $A_0$ can then be used to solve the Klein-Gordon equation for the scalar.
Fluctuations in this background are generically of the form \eqref{eq_rotation},
\be
\frac{d^2X}{dx^2}+X\left[\Upsilon^2-V\right]=0\,,\label{waveeq}
\ee
where $x$ is a tortoise coordinate whose domain is the entire real axis, $V$ is a coupling-independent function
and $\Upsilon=\omega-qA_0$. The potential $V$ depends on the spacetime geometry and on the angular momentum
of the field. For example, for D-dimensional Reissner-Nordstr\\{o}m BHs, and for spherically symmetric fluctuations $\Psi=r^{-(D-2)/2} X(r) e^{-i\omega t}$, the potential reads
\be
V=f\left(\frac{n(n-2)f}{4r^2}+\frac{f'n}{2r}\right)\,,
\ee
with $f=1-2M/r^{D-3}+Q^2/r^{2(D-3)}$ and $dx/dr=1/f$.
In the large-coupling limit, $q\to \infty$, the particular form of the potential $V$ is irrelevant.

Equation \eqref{waveeq} also describes rotating acoustic holes for~\cite{Berti:2004ju}
\be
%
V=f\left(\frac{m^2-1/4}{r^2}+\frac{5}{4r^4}\right)\,,
\ee
with $f=1-1/r^2$ and $\Upsilon=\omega-Bm/r^2$. Here $B$ describes how fast the hole is spinning and $m$ is an integer, azimuthal number.

The equation is to be solved imposing regular boundary conditions,
\beq
X&\to& {\cal R} e^{i\omega x}+{\cal I}e^{-i\omega x}\,\quad x\to \infty\,,\\
X&\to& {\cal T}e^{-i\Upsilon x}\,\quad x\to -\infty
\eeq

An instructive example concerns the wave equation
\be
X''+\left(\omega-\frac{V_0}{1+e^x}\right)^2X=0\,,
\ee
which is of the form \eqref{waveeq}. The change of variable $z=-e^x$ brings this to the form
\be
z^2X''+zX'+\left(\omega-\frac{V_0}{1-z}\right)^2X=0\,,
\ee
where a prime now stands for $z-$derivative.
This equation can be brought into an hypergeometric form by the substitution
\be
X=z^\alpha(1-z)^\beta F(z)\,,
\ee
with $\alpha=-i(V_0-\omega),\,2\beta=1+\sqrt{1-4V_0^2}$. Then, the general solution is
\begin{align}
    F= & C_1F(\beta-iV_0,\beta-iV_0+2i\omega,1+2\alpha;z)\nonumber                                \\
      & +(-1)^{-2\alpha}z^{-2\alpha}C_2F(\beta+iV_0,\beta+iV_0-2i\omega,1-2\alpha;z)\,.\nonumber
\end{align}
The boundary conditions require that $C_1=0$.
We can use the transformation properties of hypergeometric functions
%
to get finally the amplification factor
\be
{\cal A}=\frac{\cosh(2\pi\,V_0)-\cosh(2\pi(V_0-2\omega))}{\cos(\pi\sqrt{1-4V_0^2})+\cosh(2\pi(V_0-2\omega))}\,.
\ee
The amplification factor asymptotes to $100\%$ {\it from above} when $V_0 \to \infty$, unless $\omega=0$.
However, notice that the amplification factor at small $\omega$ is not necessarily small: at $\omega=V_0/2$,
${\cal A}\sim 1/V_0^2$ for small $V_0$.

\subsection{Large coupling limit and a WKB analysis}
In the large coupling regime\footnote{This analysis was done jointly by V.C. with Vishal Baibhav and Roberto Emparan, but never published before.}, $q\to\infty$ then $\Upsilon$ in Eq.~\eqref{waveeq}
is almost everywhere much larger than the potential $V$. In this regime, $\Upsilon$
has one zero on the real line outside the BH horizon. For finite but large $q$ there will be two closely-spaced real zeroes.
Thus, WKB methods can be used by expanding around the (second-order) turning point of $\Upsilon$. To the left and right of the turning point, one can use standard WKB expansions,
\be
X\approx \Upsilon^{-1/2}\exp\left(\pm i\int_{x_0}^x \Upsilon(t)dt\right)\,,\nonumber
\ee
Close to the turning point, $\Upsilon$ can be approximated by a straight line, and the solutions will become parabolic cylinder functions~\cite{Bender:1999box}.
Defining $k=(\Upsilon')^2$ and $t=\sqrt{2\Upsilon'}e^{i\pi/4}(x-x_0)$, the wave equation \eqref{waveeq} is transformed to
\be
X''(t)-t^2X(t)/4=0\,,
\ee
whose solutions are parabolic cylinder functions,
\be
X=A (4k)^{1/8}D_{-1/2}(t)+B(4k)^{1/8}D_{-1/2}(it)\,.
\ee
At large values of the argument, these solutions have the asymptotic expansion
\beq
X&\approx& Be^{-3i\pi/8}(x-x_0)^{-1/2}e^{i\sqrt{k}(x-x_0)^2/2}\nonumber\\
&+&\left(A+B\sqrt{2}e^{i\pi/4}\right)e^{-i\pi/8}(x-x_0)^{-1/2}e^{-i\sqrt{k}(x-x_0)^2/2}\,,x\gg x_0\nonumber \\
X&\approx& Ae^{3i\pi/8}(x-x_0)^{-1/2}e^{-i\sqrt{k}(x-x_0)^2/2}\nonumber\\
&+&\left(B-iA\sqrt{2}e^{i\pi/4}\right)e^{i\pi/8}(x-x_0)^{-1/2}e^{i\sqrt{k}(x-x_0)^2/2}\,,x\ll x_0\,.\nonumber
\eeq
These expansions can be directly matched onto the WKB solutions (and could also simply be read off from Ref.~\cite{Schutz:1985km}).
The boundary condition at the horizon forces $B=iA\sqrt{2}e^{i\pi/4}$. Finally, the amplification factor can be read from the expansion at $x\gg x_0$ and is precisely
\be
{\cal A}=100\%\,.
\ee
This is one of our main results. The amplification factor saturates, for very large couplings $q$, at precisely 100\%.
This prediction is consistent with all numerical results we are aware of, and includes other, more generic geometries than D-dimensional Reissner-Nordstr\\{o}m.
This result by itself does not guarantee that the amplification factors are bounded by 100\%, but do lend strong support to this conjecture.

\section{Exercises}

{\bf $\blacksquare$\ Q.1.a.} In Sec.~\ref{subsec:variation-of-param} we mentioned that we can expand the master variable as
\be
\psi_{H}\sim  \left\{
\begin{array}{ll}
    e^{-i\omega r_*}\!\left(1+a_H(r-2M)+b_H(r-2M)^2+\dots\right),                                                                                              & r\to 2M     \\[6pt]
    A_{\rm out}\,e^{i\omega r_*}\!\left(1+\dfrac{a_{\rm out}}{r}+\dfrac{b_{\rm out}}{r^2}+\dfrac{c_{\rm out}}{r^3}+\dfrac{d_{\rm out}}{r^4}+\dots\right)       &             \\[2pt]
    \quad +\,A_{\rm in}\,e^{-i\omega r_*}\!\left(1+\dfrac{a_{\rm in}}{r}+\dfrac{b_{\rm in}}{r^2}+\dfrac{c_{\rm in}}{r^3}+\dfrac{d_{\rm in}}{r^4}+\dots\right), & r\to \infty
\end{array}
\right.\! \nonumber
\ee
Determine the constants $a_H, b_H$ for a generic perturbations of spin $s$ around the Schwarzschild background.
Then determine $a_{\rm in/out}, \cdots, d_{\rm in/out}$. Check that $a_{\rm in}=a^*_{\rm out}, \cdots, d_{\rm in}=d^*_{\rm out}$, where $*$ denotes complex conjugation.

\vspace{0.25cm}

\noindent\textbf{{$\square$}\ Solution:} This calculation has been carried out in the \coderef{code:Scattering-Scalar.nb} notebook and here we sketch the main idea. We shall first determine $a_H$ and $b_H$. Let us start by taking the power series ansatz $\psi_{H}\sim e^{-i\omega r_*}(1 + a_H(r-2M)+b_H(r-2M)^2)$, where we have truncated the series expansion near the horizon to include two subleading terms. We shall plug this into the homogeneous wave equation,
\begin{equation}
    f^{2}\,\frac{d^{2}\psi}{dr^{2}}+f f'\frac{d\psi}{dr}+ \bigl(\omega^{2} - V_s\bigr)\psi(r) =0\,,\nonumber
\end{equation}
with the potential \eqref{effective_potential_spins}.
This would give us,
\begin{equation}
    \begin{split}
        & \frac{1}{r^{4}}\Bigl[(r-2M)\bigl(-2M - \ell(1+\ell)r + 2Ms^{2}\bigr)                                \\
        & +\,a_H(r-2M)\bigl(-\ell(1+\ell)r^{2}-4M^{2}(-1+s^{2})+2Mr(\ell+\ell^{2}+s^{2})-2i r^{3}\omega\bigr) \\
        & +\,b_H(r-2M)\bigl(2r^{3}-\ell(1+\ell)r^{3}+8M^{3}(-1+s^{2})-4M^{2}r(\ell+\ell^{2}+2s^{2})           \\
        & -\,4i r^{4}\omega+2Mr^{2}(-1+2\ell(1+\ell)+s^{2}+4ir\omega)\bigr)\Bigr] = 0 \nonumber
    \end{split}
\end{equation}
The above expression is not very illuminating. But it tells us that after substituting the series ansatz into the wave equation, the coefficients of $(r-2M)$ still involve functions such as $1/r$, $r^2$ and $f(r)=1-2M/r$. So the equation does not immediately produce a \emph{clean} power series in $(r-2M)$. To convert it into a genuine power series in $(r-2M)$, we Taylor-expand the resulting expression about $r=2M$ and get,
\begin{equation}
    -\frac{1}{8 M^{3}}
    \Bigl(1 + \ell(\ell + 1)-s^2 + 2 a_H M \bigl(-1 + 4 i M \omega \bigr)\Bigr)\,(r-2M)
    \;+\;\mathcal{O}\!\left((r-2M)^{2}\right) = 0 \,.\nonumber
\end{equation}
Here we have kept only the first order term. It is important to note that after substituting the power series into the wave equation and performing the Taylor expansion up to some arbitrary order, the coefficients of each power of $(r-2M)$ generally involve several of the unknowns $a_H, b_H, \ldots$; thus the expansion determines these coefficients in a sequential manner, with $a_H$ fixed first, then $b_H$ depending on $a_H$, and so on. Moreover, the series solution holds for each power of $(r-2M)$, so the coefficients of each term in the series must vanish individually.

Therefore we can now equate the coefficient of $(r-2M)$ to zero to determine $a_H$ and get,
\begin{equation}
    a_H= i\dfrac{(1+\ell(l+\ell)-s^2)}{2M(i+4M\omega)}\,. \nonumber
\end{equation}

Now that we have the first subleading term, we can now determine the next subsubleading term. Using the expression for $a_H$, we can substitute the series expansion of $\psi_H$ back into the wave equation and expand it about $2M$ but now up to order $\mathcal{O}((r-2M)^{3})$. Since we substituted the value of $a_H$, the leading order term now is proportional to $(r-2M)^2$ and contains only one unknown $b_H$. We can then equate the coefficient of $(r-2M)^2$ to $0$ to determine $b_H$. We can therefore build the series expansion of $\psi_H$ around $2M$ by determining the series coefficients in a sequential manner.

We repeat this process for the series expansion of $\psi_H$ near infinity as well. However, the series solution now contains an ingoing part and an outgoing part. But since the wave equation is linear, we can invoke the principle of superposition to determine the coefficients of the series expansion of the ingoing and outgoing parts individually. Naturally, the Taylor expansion of the wave equation is performed around $1/r$. But note that the leading order term is proportional to $1/r^2$ and hence the Taylor expansion must be done accordingly. In the \coderef{code:Scattering-Scalar.nb} notebook, we have calculated the constants $a_{\mathrm{in/out}}, \cdots, d_{\mathrm{in/out}}$. The first two terms, for example are,
\beq
a_{\rm out}&=&i\frac{\ell(\ell+1)}{2\omega}\,,\nonumber\\
b_{\rm out}&=&-\frac{(\ell-1)\ell(\ell+1)(\ell+2)+4iM\omega (s^2-1)}{8\omega^2}\,.\nonumber
\eeq
Note that $a_{\rm in}=a^*_{\rm out},\,\quad b_{\rm in}=b^*_{\rm out}\,$ etc. This is verified in straightforward manner, for real values of $\omega$. This is not surprising since the ingoing and outgoing parts are complex conjugates of each other.

\vspace{0.5cm}

\noindent{\bf $\blacksquare$\ Q.1.b.} In this exercise we shall attempt to solve the (Laplace transformed) wave equation numerically for $\ell=s=2$. Physically, we want to study the scattering of gravitational waves by a Schwarzschild black hole. Model the initial perturbation as a Gaussian pulse and calculate the radiated flux. Show that the time-domain signal at late times ``rings down,'' and looks like a decaying sinusoid. Determine the frequency of oscillation of the wave along with the decay rate.

\vspace{0.25cm}

\noindent\textbf{{$\square$}\ Solution:} We want to solve Eq.~\eqref{Laplace_wave_eq}, numerically. The idea is to solve for $\psi(\omega)$ for a grid of frequencies and use those to invert the transform and find the time domain response. We first fix the value of $\omega$ and integrate the homogeneous wave equation to find $\psi_H$. To integrate the wave equation, we adopt a direct integration strategy to determine $\psi_H$. We use Eq.~\eqref{asymptotic_behaviorH_accurate} near the horizon to start the integration, and we integrate up to (a numerical) infinity. Note that we are dealing with a second-order equation, so we require also a start value for the radial derivative. We get this by taking the (radial) derivative of the series solution~\eqref{asymptotic_behaviorH_accurate} at the horizon.

We know the behavior of the solution at infinity analytically, up to two undetermined constants, $A_\mathrm{in}$ and $A_\mathrm{out}$ in Eq.~\eqref{asymptotic_behaviorH_accurate}. We therefore match the numerical wavefunction and analytical solution, that is, we solve them algebraically to determine the values of $A_\mathrm{in}$ and $A_\mathrm{out}$. This process is then repeated for various values of $\omega$. Notice that we have two unknowns, so we need two equations to solve them. The second equation is obtained by equating the radial derivatives of the numerical wavefunction and the analytical series solution at infinity.

\begin{figure}[htbp]
    \centering
    \begin{minipage}[b]{0.45\textwidth}
        \centering
        \includegraphics[height=5.5cm]{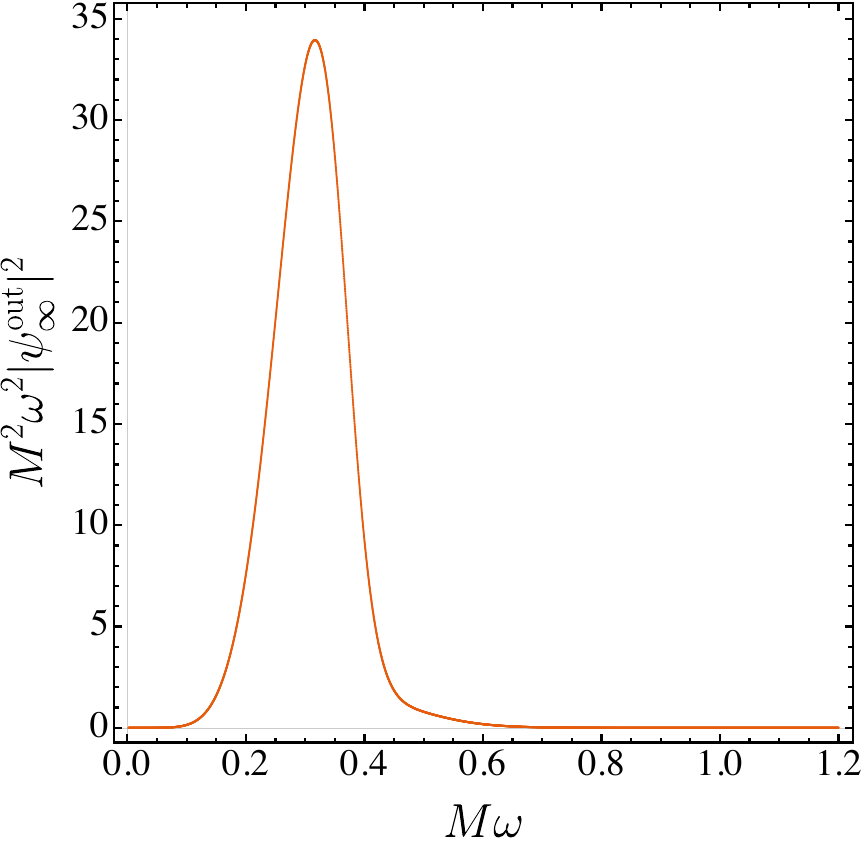}
    \end{minipage}
    \hfill
    \begin{minipage}[b]{0.45\textwidth}
        \centering
        \includegraphics[height=5.5cm]{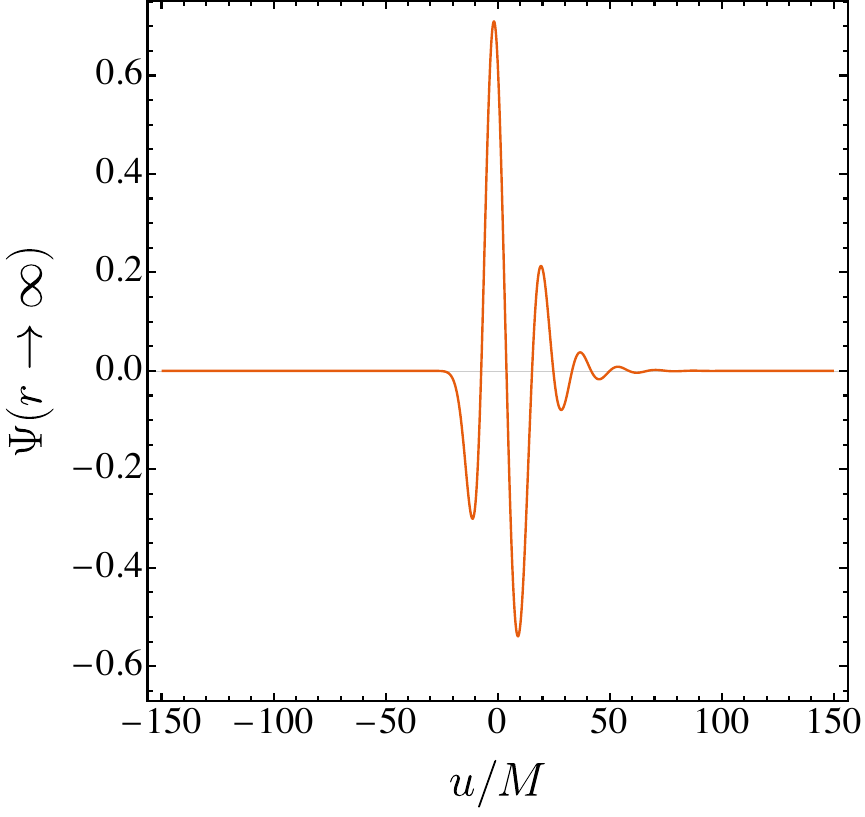}
    \end{minipage}
    \caption{Scattering of gravitational waves by a Schwarzschild black hole. \textbf{Left}: Emitted flux $M^2 \omega^2 |\psi|^2$ as a function of the dimensionless frequency $M\omega$, showing a clear peak corresponding to roughly the real part of the black hole's resonant frequency. \textbf{Right}: Time-domain signal $\Psi(t, r \to \infty)$ obtained via inverse Laplace transform, exhibiting a damped sinusoidal behavior.}
    \label{fig:scattering}
\end{figure}

For each $\omega$, then perform an integration with respect to the initial data (c.f. Eq.~\eqref{eq:variation_of_param_integration}) to find the solution to the inhomogeneous equation. To determine the evolution of the wave in the time domain, we perform an inverse Laplace transform, Eq.~\eqref{eq_Laplace_inverse}. This calculation has been carried out in \coderef{code:Scattering-Scalar.nb}. We first plot the emitted flux as a function of $\omega$ and the time domain signal in left and right panel of Fig.~\ref{fig:scattering} respectively. We note that the radiated flux has a peak at $M \omega \sim 0.32$. At late times, the signal looks like a damped sinusoid. To determine the (what we will soon show to be characteristic) frequency of oscillation and the decay rate, we fit a damped sinusoid, $\psi_t = a_0 e^{i \omega_I t}\cos{(\omega_R t + a_1)}+a_2,$ to the resultant late-time domain profile. The time domain signal has a frequency of oscillation $|M\omega_R| \sim 0.372$ and a decay rate of $\omega_I \sim 0.0897$.

We mention the following salient points that we must keep in mind while performing the simulation.

\begin{enumerate}
    \item At the horizon and at infinity, some terms in the equation to be integrated diverge: indeed the boundaries (the horizon and spatial infinity) are at infinity, inaccessible by any numerical means. Thus, in practice we start the integration from some $r=2M (1+\epsilon)$ with $\epsilon$ a very small number. Our final result should be weakly dependent on the value of $\epsilon$ (as long as it is very small). The fluctuation on the final result as $\epsilon$ varies gives us a quantifiable source of error.

    \item We integrate up to $r=r_{\mathrm{N}}/\omega$ where $r_{\mathrm{N}}$ is a large number. This is our numerical infinity. Notice that we divide $r_{\mathrm{N}}$ by $\omega$. This ensures that our numerical infinity is large enough to fit in (more than a few) complete wave-cycles for all values of $\omega$.

    \item We also start from a very small value of $\omega$, because for $\omega = 0$ the factors $e^{\pm i \omega r_*} = 1$ and the ingoing/outgoing boundary conditions become indistinguishable. Moreover, at $\omega = 0$ the wave equation reduces to $\psi''(r_*) - V\psi(r_*) = 0$, so it no longer describes scattering or wave propagation. We can only talk about $\omega \to 0$ or the zero frequency limit.

    \item The initial data is a Gaussian pulse, and is practically zero beyond four standard deviations about its mean. Therefore, while performing the integration with respect to the source term (c.f. Eq.~\eqref{eq:variation_of_param_integration}), we can keep the upper-limit of the integration as $r_0+4 \sigma$ where $r_0$ and $\sigma$ are the mean and the standard deviation of the Gaussian pulse. In fact, since we are interested in observing the signal perhaps at a gravitational wave observatory located very far away from the black hole, we use Eq.~\eqref{rlm} to determine $\psi^{\mathrm{out}}_\infty$.

\end{enumerate}

Note that we have performed the scattering experiment to determine $A_\mathrm{in}$ and $A_\mathrm{out}$ for {\it real} values of $\omega$. The resulting time-domain signal is well approximated by a decaying sinusoid, which can be described as $e^{-i\omega t}$ but with a {\it complex} frequency $\omega$. As we discuss in the next chapter, the time domain waveform shows evidence for characteristic relaxation of black holes.  We can understand this using the analogy of striking a bell with a hammer: the hammer delivers a short, sharp push that contains a range of frequency components. The bell does not vibrate equally at all these frequencies; instead, it responds most strongly at its natural resonant frequency, which dominates the ringing and gradually dies down as energy is dissipated. Similarly, when a black hole is “struck” by an incoming wave, it excites its characteristic frequencies. This is reflected in the peak of the radiated flux as seen in the left panel of Fig.~\ref{fig:scattering}. Black holes are inherently dissipative systems: waves fall into the horizon, carrying energy away, and additional dissipation occurs through waves leaving the domain at infinity. As a result, these oscillations decay over time as seen the right panel of Fig.~\ref{fig:scattering}. We will see that there are indeed characteristic
frequencies that correspond to such an exponentially damped ringing. Thus, even though the scattering calculation uses real $\omega$, the late-time evolution of the signal is governed by these complex frequencies which characterize the black hole's response to the initial perturbations.

The complex frequencies, called \emph{quasinormal modes}, which show that the initial wave plus decays sinusoidally with time has far-reaching consequences which we shall discuss in subsequent chapters. But for now we note that we have studied the scattering of gravitational waves by a black hole and reproduced the classic result discovered by Vishveshwara~\cite{Vishveshwara:1970zz} that (in retrospect) led to the birth of black hole spectroscopy.

\vspace{0.5cm}

\noindent{\bf $\blacksquare$\ Q.2.} Consider a photon falling into a non-spinning black hole. Provide a precise estimate of the radiated gravitational energy by solving the Zerilli equation numerically. Calculate the waveform of the gravitational waves emitted by the process.
\vspace{0.25cm}

\noindent\textbf{{$\square$}\ Solution:} The problem of a photon colliding with a black hole can be studied with the theoretical and numerical tools we have been discussing. We shall start by considering a massless particle falling into a black hole from infinity along a radial geodesic. To do this, we take the high-energy limit of the motion and stress-tensor of a material, timelike particle, described by Eqs.~\eqref{eqn:dt/dtau}--~\eqref{eqn:tlight} and~\eqref{eqn:particle_stress_tensor}.
Therefore, we can use the machinery developed in the text to write down the following equation for the polar perturbations. Top avoid confusion with the Regge-Wheeler wavefunction, we use here $Z(t,r)$ to denote the Zerilli master function and $z(\omega,r)$ its Fourier transform.
\begin{equation}
    f^{2}\,\frac{d^{2}z}{dr^{2}}+f f'\frac{dz}{dr}+ \bigl(\omega^{2} - V_{s=2}^+\bigr)z(r) =f S\,,
\end{equation}
with the potential given explicitly in Eq.~\eqref{veven}.
This is the Zerilli equation that was introduced in the text but now with a source term $S$, given by
\begin{equation}
    S = i\dfrac{4 Ep_0 \lambda e^{-i \omega r_*}  (4\ell+2)^{1/2}}{\omega(3M+\lambda r)^2}\,,
\end{equation}
where $p_0$ is the momentum along the worldline parameterized by an affine parameter $\lambda$.

We derive the source term \cite{Zerilli:1974ai} in the notebook \coderef{code:Zerilli-Infalling-Photon.nb}. The source term for the Regge-Wheeler equation vanishes for radial infall (see Problem 12.2 in Ref.~\cite{Maggiore:2018sht} for a detailed discussion). Hence to understand how a radially infalling photon affects a black hole, we have to study the Zerilli equation alone. Formally, we say that the axial modes are not excited by a radially infalling test particle, only the polar modes survive.

Due to the presence of the source term, we use direct integration to find the solution to this inhomogeneous equation. The computations have been performed in the notebook \coderef{code:Zerilli-Infalling-Photon.nb}. We have to, of course, repeat the series expansion of $\psi_H$ -- that we did for the homogeneous Regge-Wheeler in \textbf{Problem Q.1.a} -- to the solution of the homogeneous Zerilli equation. Symbolically, the series expansion resembles that of $\psi_H$ when written in terms of the constants $a_H, b_H, a_{\mathrm{in/out}},\cdots,d_{\mathrm{in/out}}$ but the exact expressions of these constant are different.

Here we integrate the inhomogeneous wave equation with the source term from $r=2M(1+\epsilon)$ to $r=r_N/\omega$ for different values of $\omega$. The caveats discussed earlier apply here as well. Recall that while using the method of variation of parameters, we had integrated the homogeneous equation and performed another integration with respect to the source to obtain the response of the black hole. Here, however, we take a different approach: we shall integrate the wave equation with the source term from the start, ``shooting'' to the correct value.

The first task is to compute the frequency domain response of the black hole to the radially infalling photon at various real values of the real frequency $\omega$. We use a low-order expansion to start the integration at $r=2M(1+\epsilon)$, $z_H \sim A_H e^{-i \omega r_*}, z'_H\sim -i\omega f A_H e^{-i \omega r_*}$, with primes standing for radial derivatives. Notice that the amplitude of the ingoing wave is not unity. This is the key difference compared to the previous procedure. We start with an initial guess for $A_H$ and perform the integration to find $A_{\rm in}$ and $A_{\rm out}$. This is done by equating the numerical solution and its derivative at infinity with the analytical series solution and its derivative at infinity. Now, once we get $A_{in}$, we try to see if $A_{in}=0$. If it is not, we repeat the calculation with a different guess value of $A_H$ but for the same $\omega$. Actually, we just have to supply the initial guess, Mathematica's \texttt{FindRoot} routine does the searching for us.

\begin{figure}[ht!]
    \centering
    \centering
    \includegraphics[width=0.8\textwidth]{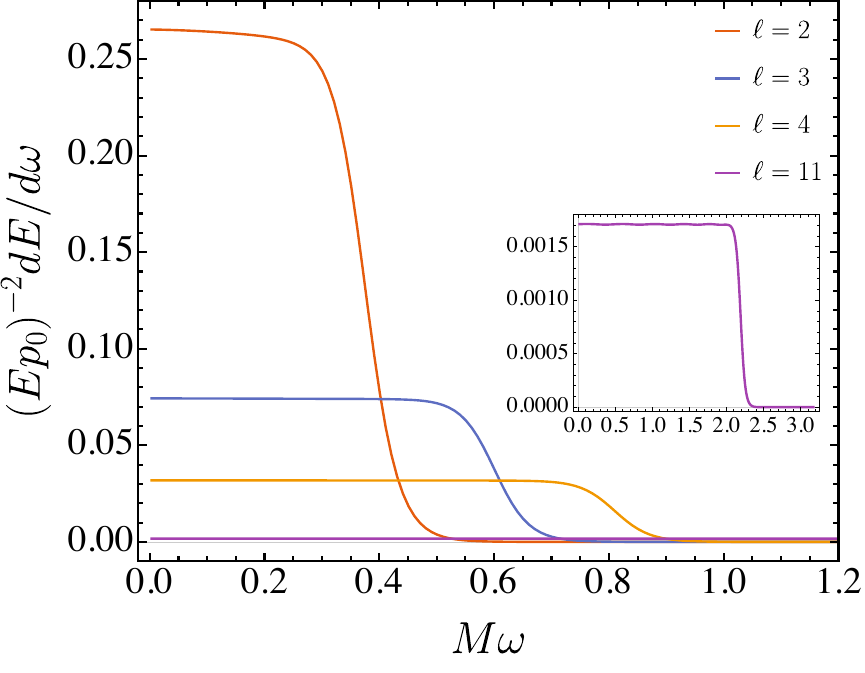}
    \caption{Energy Spectrum corresponding to various values of $\ell$. The spectrum is flat at low frequencies, and well described by a flat space calculation, see Refs.~\cite{Cardoso:2002ay,Berti:2010ce} for further details. Note that the axes labels of the inset are same as that of the main figure.} \label{fig:energy-spectra-Zerilli}
\end{figure}

\begin{figure}[h!]
    \centering
    \begin{minipage}{0.45\textwidth}
        \centering
        \includegraphics[width=\textwidth]{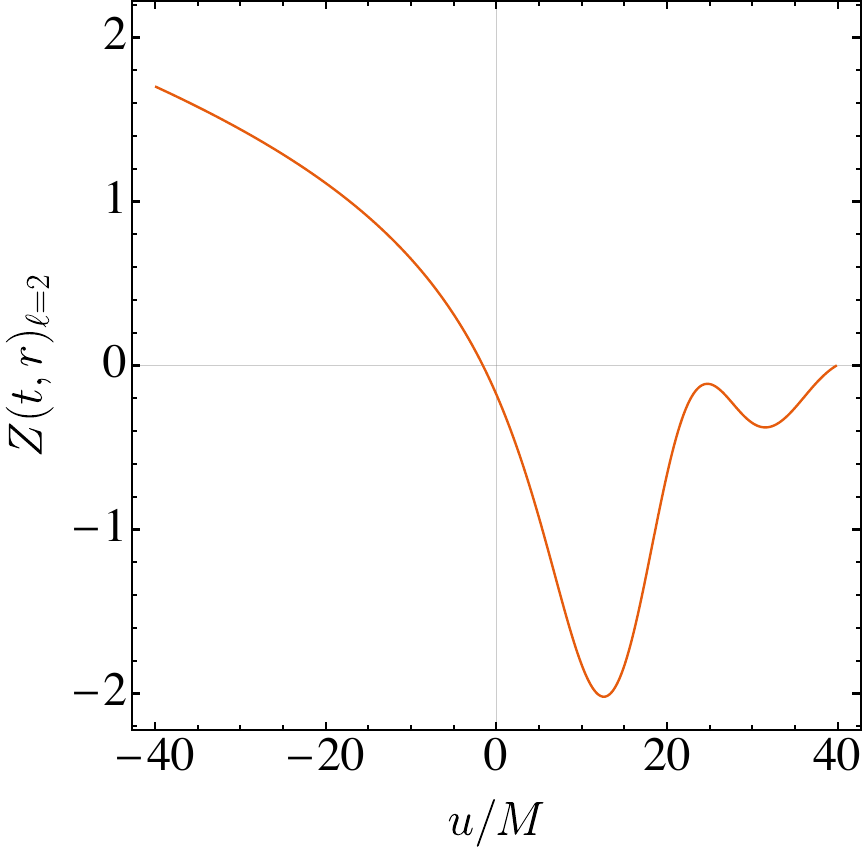}
    \end{minipage}
    \hfill
    \begin{minipage}{0.45\textwidth}
        \centering
        \includegraphics[width=\textwidth]{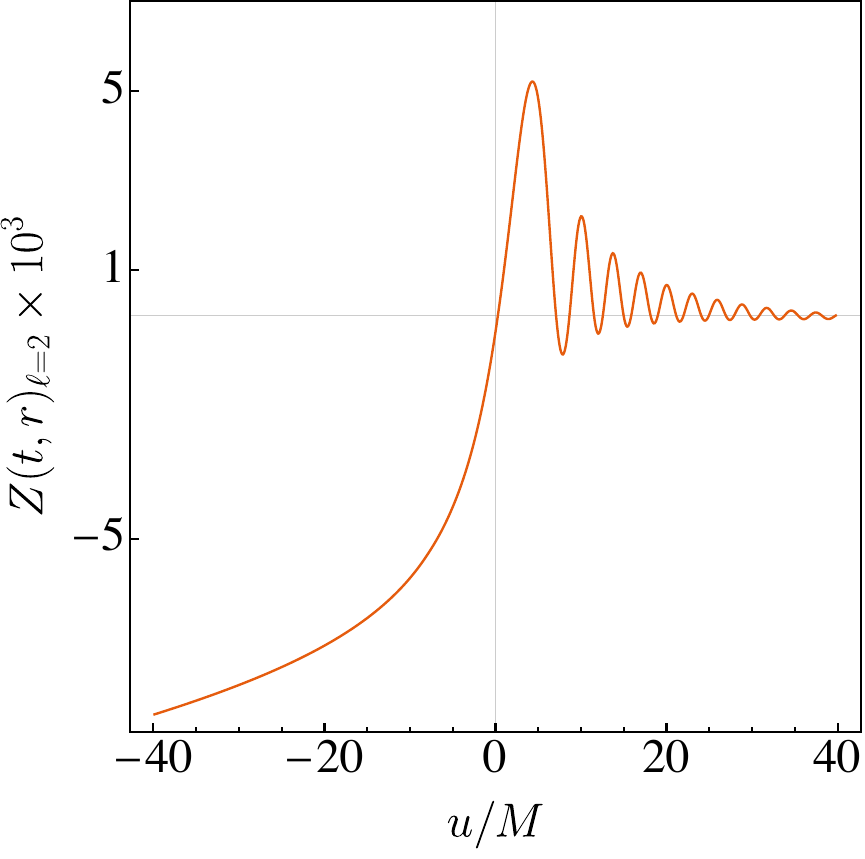}
    \end{minipage}
    \caption{Gravitational waveforms for the multipole indices $\ell=2,3,11$ from left to right, for a massless particle falling from infinity into a Schwarzschild black hole. Here $z(t,r)$ is measured in units of $E p_0$. \label{waveform_photon}}
\end{figure}

Finally, to compute the energy spectra, we use a result which we have not derived but which can readily be obtained using the tools we developed~\cite{Berti:PhD,Maggiore:2007ulw}
\begin{equation}
    \dfrac{dE}{d\omega} = \dfrac{1}{32 \pi} \dfrac{(\ell+2)!}{(\ell-2)!}\omega^2 | z(\omega,r)|^2 \,.
\end{equation}
We carry out this process for $\ell=2,3$ and $\ell=11$ and show the result in Fig.~\ref{fig:energy-spectra-Zerilli}.
We can now reconstruct the wave function $z(t,r)$ by performing an inverse Fourier transform. Results are summarized in Fig.~\ref{waveform_photon}. There are two interesting things to note about our results. First, note that for $\ell=2$, $dE/d\omega \sim 0$ when $M\omega \sim 0.31$. The energy spectrum is flat up to a certain critical value of the frequency, after which it rapidly decreases to zero. Recall that when we performed the Gaussian scattering experiment in Problem \textbf{Q.1.b}, the radiated flux had a peak at a similar value. Moreover, the waveform corresponding to $\ell=2$ again looks like a damped sinusoid, of frequency $\sim 0.35$.

The problem we have just solved is extremely interesting because it shows that a black hole responds in a rather similar manner to distinct physical phenomena (Gaussian scattering vs relativistic collision). Moreover, this setup can be used as a prototype for studying the collision of binary black holes, and in fact shows that linear perturbation theory is somewhat surprisingly capable of describing phenomena which are highly nonlinear (and in this case, falls within the domain of numerical relativity)  \cite{Cardoso:2002ay,Sperhake:2008ga}.
\vspace{0.5cm}

\noindent{\bf $\blacksquare$\ Q.3.} A spinning black hole of mass $M$ and spin $a$ is enclosed in a cavity of constant (Boyer-Lindquist) radius $R$. Calculate the massless scalar modes of this system numerically.

\vspace{0.25cm}

\noindent\textbf{{$\square$}\ Solution:} In this problem, we consider a Kerr black hole enclosed in a cavity. We assume that the walls of the cavity located at $R$ are reflecting surfaces. In other words, the scalar field will vanish at $R$. The frequencies which satisfy the reflecting or Dirichlet boundary condition at the walls cavity and are ingoing at the horizon are called \emph{boxed quasinormal modes} (BQNMs). The Kerr black hole and mirror configuration is also called the black hole bomb \cite{Press:1972zz,Cardoso:2004nk}. To determine the modes of this system, we have to solve the Klein Gordon equation in the Kerr background which separates out into angular and radial equations (c.f. Eqs.~\eqref{kerr_angular} -- Eq.~\eqref{kerr_angular}). Note that by introducing a new variable, $u=\cos{\theta}$ where $-1\leq u \leq 1$, the angular equation (Eq.~\eqref{kerr_angular}) can be transformed to the following form,
\begin{align}
    \left(1-u^2 \right)\dfrac{d^{{2}} S(u)}{du^{{2}}} - 2 u \dfrac{d S(u)}{d u}  \nonumber + \left( \bar{\Lambda} + \gamma^2 - \gamma^2 u^2 - \dfrac{m^2}{1-u^2}\right)S(u) =0, \label{angular_equation_trigfree}
\end{align}
where we have introduced the constants $\gamma$ and $\bar{\Lambda}$ such that $\gamma = i a {\omega}$ and   $A_{\ell m} = \gamma^2 + \bar{\Lambda}$. To find the separation constant $\bar{\Lambda}$ for a specific value of $a, \omega,$ and $m$, we can now easily use the \texttt{SpheroidalEigenValue} function provided by \texttt{Mathematica}. This works for $s=0$ (for general $s$, one may use \texttt{Black Hole Perturbation Toolkit} or compute it from scratch using continued fractions~\cite{Berti:2005gp}).

The radial equation~\eqref{kerr_radial} can be written as,
\begin{equation}
    \frac{d^2}{dr_*^2}\psi + V \psi=0\,,
\end{equation}
where $\psi = (r^2+a^2)^{1/2}R$ and
\begin{equation}
    V = \frac{K^2 - \lambda\Delta}{(r^2 + a^2)^2} - G^2 - \frac{d}{dr_*}G,
\end{equation}
with $K = (r^2 + a^2)\omega - am$, $G = r\Delta(r^2 + a^2)^{-2}$, $\lambda = A_{\ell m} + a^2\omega^2 - 2am\omega$, and $\Delta = (r-r_+)(r-r_-)$.

To find the BQNMs, we have to solve these two equations simultaneously since they are coupled through the separation constant $A_{\ell m}$ and the frequency $\omega$. We focus on the case $\ell =m=1$ and use the method of direct integration to find the modes.

We attempt to integrate (or, shoot) from the horizon to the location of wall of the cavity. In this exercise, we simply use the following boundary condition at the horizon
\begin{equation}
    \psi \sim e^{-i(\omega - m \Omega )r_*} = (r-r_+)^{-i\sigma}.
\end{equation}
Here, we have kept just the leading term in the series expansion. To figure of the expression for $\sigma$, note that near the horizon we can write,
\begin{equation}
    \Delta(r) = \Delta(r_+) + \Delta'(r_+)(r-r_+) + \mathcal{O}[(r-r_+)^2]\,,
\end{equation}
which lets us approximate the tortoise coordinate as,
\begin{equation}
    \dfrac{d r_*}{d r} \sim \dfrac{r_+^2 +a^2}{\Delta'(r_+)(r-r_+)}.
\end{equation}
We can then do the integration and get
\begin{equation}
    \sigma = \dfrac{r_+^2 +a^2}{\Delta'(r_+)}(\omega - m \Omega_\mathrm{H})\,,
\end{equation}
where $\Omega_{\mathrm{H}}=a/(r_+^2+a^2)$.

\begin{figure}[htbp]
    \centering
    \includegraphics[width=0.8\textwidth]{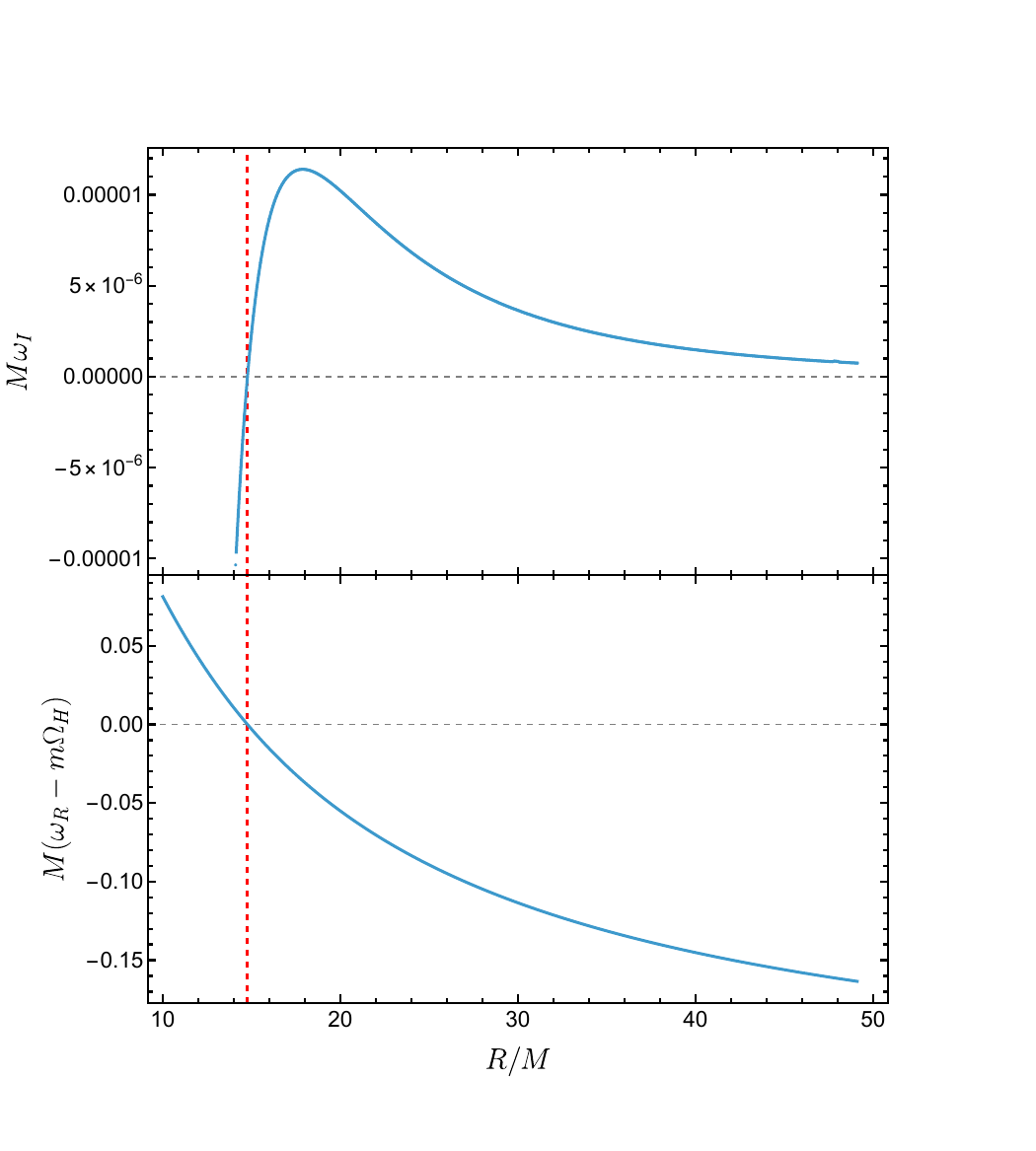}
    \caption{The imaginary frequency $M\omega_I$ (top panel) and the superradiance condition $M(\omega_R - m\Omega_H)$ (bottom panel) are shown as a function of the normalized mirror radius $R/M$ for a black hole-mirror system with spin $a=0.8M$. The instability, corresponding to $\omega_I > 0$, occurs only when the superradiance condition $\omega_R - m\Omega_H < 0$ is met. The vertical dashed line marks the critical radius $R_0$. For $R < R_0$, the system is stable, as the superradiance condition is violated.}
    \label{fig:bh_bomb}
\end{figure}
At the cavity wall, we match the numerical solution with the boundary condition to find the desired mode. Since we impose Dirichlet boundary condition, we simply set the numerical solution to zero. This last stage makes use of a root-finding algorithm to pin down the value of $\omega$ that satisfies the aforementioned condition. The numerical computation has been carried out in \coderef{code:BH-Bomb.nb}. In Fig~\ref{fig:bh_bomb} we show the results for $a=0.8 M$.

Our numerical calculation shows that the black hole mirror system is unstable when the wall of the cavity is located beyond some critical radius $R_0$. This is a consequence of superradiance: the modes extract energy from the ergosphere and when they strike the mirror, they are reflected back to the black hole, whereupon they extract even more energy. This feedback mechanism gives rise to a superradiant instability.
\begin{figure}[htbp]
    \centering
    \includegraphics[width=0.8\textwidth]{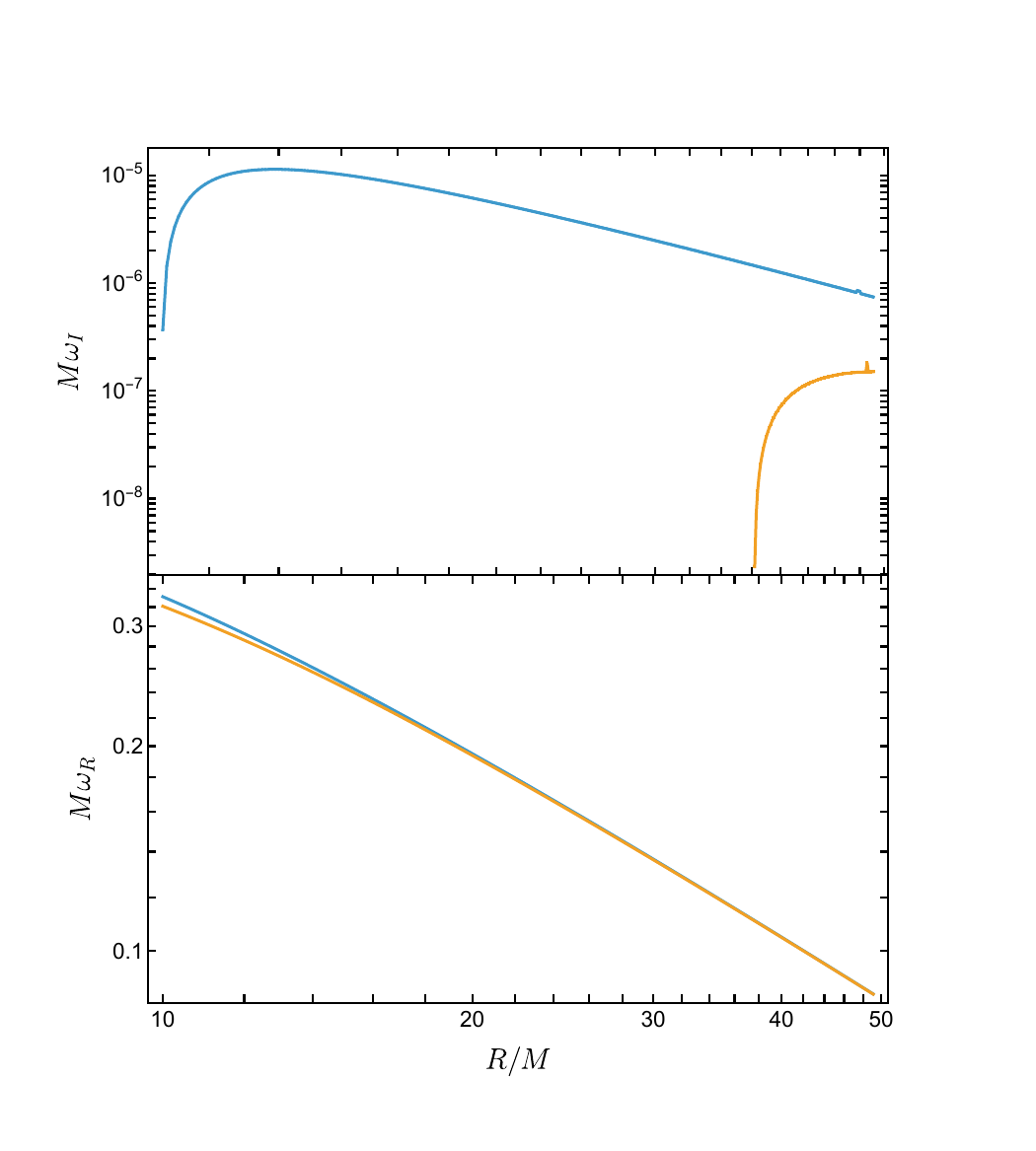}
    \caption{This figure compares the black hole-mirror system for a black hole with spin $a = 0.8 M$ (blue curve) and $a = 0.4 M$ (orange curve). The top panel shows the instability growth rate ($M\omega_I$), while the bottom panel shows the real oscillation frequency ($M\omega_R$), both as a function of mirror radius $R/M$, in log-log scale. The instability for the lower-spin case ($a = 0.4 M$) is triggered at a significantly larger radius than for the higher-spin case ($a = 0.8 M$). The real frequencies for both scenarios, however, are nearly identical.}
    \label{fig:bh_bomb2}
\end{figure}
As we bring the mirror closer to the black hole from infinity, unsurprisingly, the superradiant instability increases (as measured by the timescale $\tau=1/\omega_I$) and reaches a maximum. For very small cavities, the instability rate decreases and the system becomes stable below the critical radius $R_0$. This may be explained by the fact that the superradiance condition $\omega_R-m\Omega_H<0$ is violated below a critical $R$ (since $\omega_R \propto 1/R$)~\cite{Cardoso:2004nk}. Finally, in Fig.~\ref{fig:bh_bomb2}, we compare the black hole mirror system for two different values of spin: to build a bomb with a slowly spinning black hole, we need to place the mirror relatively further (that is, the imaginary part of the BQNM flips sign at large $R$). The oscillation frequency (the real part of the BQNM) however does not strongly depend upon $a$. We refer the reader to \cite{Cardoso:2004nk} for a detailed discussion.


\chapter{Part III}
\label{ch:part3}
\allowdisplaybreaks
\minitoc
\section{The response of black holes to external perturbations\label{sec:response}}
%
\begin{figure}
    \centering
    \includegraphics[width=0.6\textwidth]{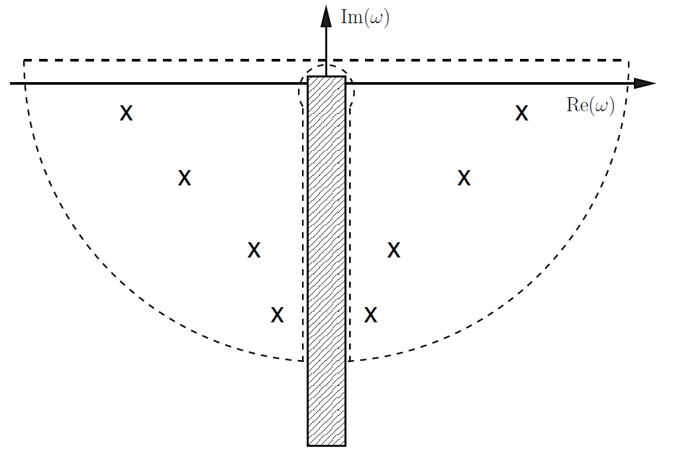}\\
    \includegraphics[width=0.7\textwidth]{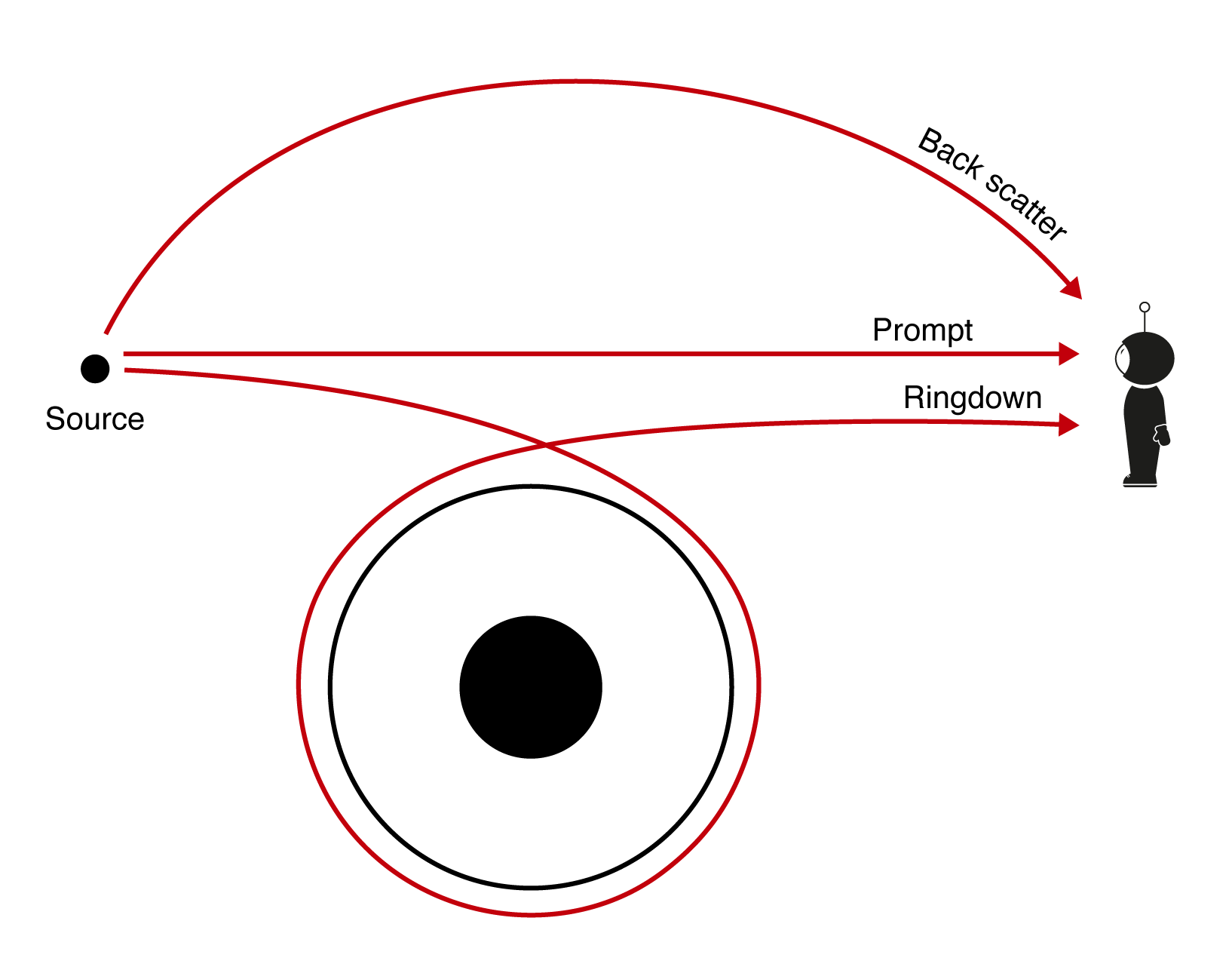}
    \caption{
        {\small {\bf Top:} Contour integration (dashed lines) in complex frequency plane for the inverse~\eqref{eq_Laplace_inverse}. Crosses are poles of Green function, known as the quasinormal frequencies. The vertical bar denotes a branch cut at $\omega=0$. {\bf Bottom:} Cartoon of wave propagation on black hole background. A source (black dot) emits waves which eventually reach an observer. These waves can reach the observer in different ways, either direct propagation (the contribution from the zero-frequency portion of the diagram above), or circling the black hole a number of times, close to the critical circular null orbit, before finally escaping to measurement device. This latter picture gives us a physical understanding of black hole {\it ringdown} in terms of its quasinormal modes, therefore localized close to the light ring}. This stage is associated to the poles (crosses) in diagram above. Finally, waves can be back-scattered by spacetime curvature, resulting in power-law tails, which originate mathematically in the branch cut in top diagram.
    }
    \label{fig:contour}
\end{figure}
%
\subsection{Prompt response, ringdown, late-time tails}
We have dealt with the main issues in scattering of monochromatic waves, so let us now try to understand the dynamics with a nontrivial time dependence. For that, instead of a Fourier transform, we use a Laplace transform defined in Section~\ref{sec:Laplace}. In essence, we will be concerned with the solutions \eqref{rlm}-\eqref{rhor} at asymptotically large distances and close to the horizon.

The time-domain signal is obtained by inverting the Laplace transform, Eq.\eqref{eq_Laplace_inverse} (please note that we are using the Fourier $\omega$ variable, and not the usual Laplace $s$, see discussion around Eqs.~\eqref{def_Laplace_transform}--\eqref{Laplace_wave_eq}). One can follow the usual contour integration techniques to perform the inverse~\cite{ArfkenWeber,DeAmicis:2025xuh}. The contour is drawn in Fig.~\ref{fig:contour}, top panel. It goes around the branch cut at $\omega=0$ and encircles a number of poles of the integrand (i.e., zeroes of the Wronskian, or equivalently of $A_{\rm in}$ as defined by \eqref{asymptotic_behaviorH}). This structure of the Green's function of our problem determines characteristics of the time domain signal.

An observer receiving the signal from a source, as in the lower panel of Fig.~\ref{fig:contour} will receive signals that come from different spacetime regions and can be roughly associated with the structure of the top panel. First, a prompt signal hits the observer, the analog of flat space, on-light-cone propagation. The prompt signal is a consequence of the $\omega=0$ behavior (in general, a pole; it is not static but slowly varying since it probes the vicinity of the $\omega=0$ pole).

\begin{figure}[t!]
    \centering
    \includegraphics[width=0.8\textwidth]{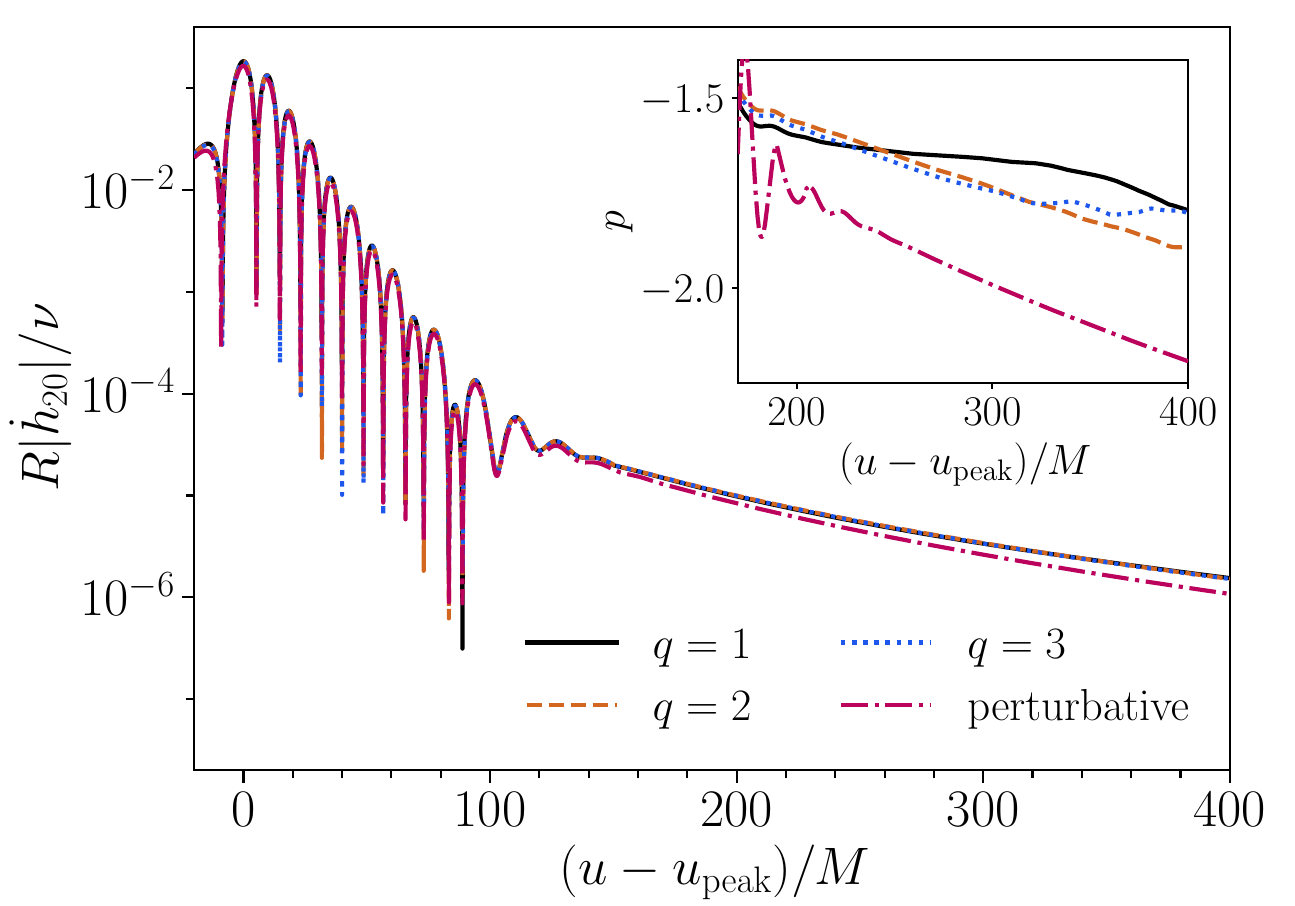}
    \caption{Mass-rescaled GW, as a function of retarded time from the peak, for the head-on collision of two black holes. Inset shows tail exponent, as defined by $h\sim u^p$ at late times.
        Thick lines represent nonlinear evolution of head-on comparable-mass collisions, dot-dashed line represents a perturbative evolution of a radial infall with compatible initial data. From Ref.~\cite{DeAmicis:2024eoy}.
    }
    \label{fig:SXS_RWZ}
\end{figure}
As we have discussed, the spacetime describing very compact objects, in particular black holes, has peculiar features. Radiation that travels with a near critical impact parameter\footnote{See Eq.~\eqref{critical_b_light} and discussion around it. Note that we are establishing a rough physical picture, waves don't travel on null geodesics.} will orbit around the black hole a number of times, and can still reach the observer. Thus, a detector close to a black hole will see radiation from the source and then from the black hole, exponentially damped pulses following roughly the behavior~\eqref{null_geoedesic_decay} if this picture is correct (it is indeed, as we will see shortly). This ringdown or quasinormal mode signal is directly associated to the poles in the Green's function, the crosses in the top panel of Fig.~\ref{fig:contour}, which are called quasinormal mode frequencies~\cite{Berti:2009kk,Baibhav:2023clw,Berti:2025hly}.

Finally, some of the radiation emitted to large distances backscatters to the observer. This radiation is associated to the branch-cut in the top panel of Fig.~\ref{fig:contour}, and gives rise to a late-time power law tail, where the field decays generically as $t^{-2\ell-3}$~\cite{Price:1971fb,Gundlach:1993tp,Ching:1994bd,DeAmicis:2024eoy,Cardoso:2024jme,Ling:2025wfv}.

The above picture, the result of decades of research on partial differential equations, black hole physics, and numerical simulations, provides a very compelling description of how a black hole relaxes to its final quiescent state. A given perturbation, say a star orbiting a black hole in its immediate neighborhood, triggers the emission of gravitational radiation.

The emission causes the orbit of the star to approach the black hole, and observers far away see a mix of all of the stages mentioned above: radiation that mostly travels from source to observer, radiation that is backscattered from large distances, and radiation that gets close to the black hole and is trapped at the light ring. But the radiation is sourced by the star continuously. Once the star approaches the black hole horizon and gets deep into the gravitational potential of the black hole, the observer now sees a sourceless region, and the emission process is dominated by recycling of waves at the light ring, it observes radiation that was emitted towards the light ring a few light crossing times ago. This corresponds to the breathing or relaxation of the black hole. At very late times, radiation scatter by spacetime curvature, typically of lower amplitude, dominates the signal, until the spacetime asymptotes to a vacuum, quiescent black hole state. These different stages are borne out of state-of-the-art nonlinear simulations, summarized in Fig.~\ref{fig:SXS_RWZ}.

\subsection{Example 1: A lossy string}
A string attached at both ends is an example of a conservative system, with well-known {\it normal} modes. If one end is allowed to be dissipative, then its modes become quasinormal, even if the equation describing the system is still the wave equation. Consider, therefore, $\psi(x)$ governed by
\beq
&&\frac{d^2\psi}{dx^2}+\omega^2\psi=0\,.\\
&&\psi(0)=0\,,\quad \psi'(L)=i\omega\delta \psi(L)\,,
\eeq
where primes are derivatives with respect to argument and $\delta$ is a constant. This represents a string fixed at $x=0$ and dissipating at $x=L$.
(Note that a purely right-moving wave behaves as $\psi\sim e^{-i\omega (t-x)}$. Hence, $\psi'=i\omega \psi$ for such a wave. The parameter $\delta$ is a measure of how much reflection one gets, see below).
The asymptotic behavior at $\pi$ is of the form \eqref{asymptotic_behaviorH}. In fact, it is an exact solution of the equation, $\psi=A_{\rm in}e^{-i\omega x}+A_{\rm out}e^{i\omega x}$. Imposing the boundary condition at $x=L$, we find the reflection coefficient
\be
R\equiv \frac{A_{\rm in}}{A_{\rm out}}=\frac{1-\delta}{1+\delta}e^{2i\omega L}\,.
\ee
Finally, requiring the boundary condition at $x=0$,
\be
-1=\frac{1-\delta}{1+\delta}e^{2i\omega L}\,,
\ee
or~\cite{DRISCOLL1996125}
\be
L\omega=\pi\left(n+\frac{1}{2}\right)-\frac{i}{2}\log\frac{1+\delta}{1-\delta}\,.
\ee
We see that dissipation introduces an imaginary part to the modes. Indeed, in the small $\delta$ limit, $L\omega=\pi(n+1/2)-i\delta$.
\subsection{Example 2: A delta-function potential \label{sec:deltaqnms}}
We now show another simple example illustrating well what has been discussed. This example concerns the scattering off a delta potential (which is to be considered a first very rough approximation to the Regge-Wheeler-Zerilli potential in Fig.~\ref{fig:potential_different_s}) in $1+1$ dimensions, i.e., the problem
\be\label{master equation}
\frac{d^2\psi}{dx^2}+\left(\omega^2-2V_0\delta(x-x_0)\right)\psi=I~.
\ee
There are two linearly independent solutions for the homogeneous equation, one behaves as
\begin{align}\label{psi-L}
    \psi_{L}(x,\omega)=
    \begin{cases}
        e^{-i\omega x}~\text{for}~x<0
        \\\\
        A_{\rm in}(\omega) e^{-i\omega x}+A_{\rm out}(\omega) e^{i\omega x}~\text{for}~x>0~.
    \end{cases}
\end{align}
That is, $\psi_{L}(x)$ is a purely left-moving wave at $x=-\infty$. Similarly, one can have purely right moving waves at spatial infinity as, $\psi_{R}(x\rightarrow\infty)\rightarrow e^{i\omega x}$. As we have seen, the Wronskian $W[\psi_{L},\psi_{R}]=\psi_{L}\frac{d\psi_{R}}{dx}-\psi_{R}\frac{d\psi_{L}}{dx}$ is a constant, $W=2i\omega A_{\rm in}(\omega)$. To proceed further, we recall that if $\psi_{L}$ and $\psi_{R}$ are the two linearly independent solutions satisfying the QNM boundary conditions on left and right then, following (\ref{eq:variation_of_param_integration}) we can write the solution to the above inhomogeneous problem as,
\be
\psi(x,\omega)=\psi_{R}(x,\omega)\int_{-\infty}^{x}dx'\frac{I(x',\omega)\psi_{L}(x',\omega)}{W(\omega)}+\psi_{L}(x,\omega)\int_{x}^{\infty}dx'\frac{I(x',\omega)\psi_{R}(x',\omega)}{W(\omega)}\,
\ee\label{Sol-Inhom}
In the limit, $x\rightarrow\infty$, only the first term will contribute.
Here $I(x,\omega)$ is a source term that appears in the right-hand side of (\ref{master equation}). Such a term may appear due to initial conditions or due to the presence of some source generating the gravitational waves \cite{Oshita:2024wgt,Berti:2006wq}.
Demanding that the wave function $\psi_{L}(x,\omega)$ is continuous across the delta function barrier and exploiting the jump discontinuity of ${d\psi_{L}}/{dx}$ across the barrier, one can find that
\begin{align}\label{A's for delta-barrier}
    & A_{\rm in}=\left(1+\frac{iV_{0}}{\omega}\right), \\
    & A_{\rm out}=-\frac{iV_{0}}{\omega}~.
\end{align}
At this point, it is important to note that when $A_{\rm in}$ vanishes then $\psi_{L}$ satisfies the quasinormal mode boundary conditions. Therefore, from this simple example, we find that\cite{Correia:2018apm}
\begin{align}\label{toy-QNM}
    \omega_{\rm QNM}=-iV_{0}~.
\end{align}
Interestingly enough, the delta function barrier supports only one quasinormal mode and it is purely imaginary. In physical scenarios, unlike the delta function case, the potential will have non-compact support outside the horizon and the quasinormal modes will have both real and imaginary parts.
Now, let us move on to solving the inhomogeneous problem. We consider only sources due to the initial conditions. Mathematically, one can show that such a source term is (see the discussion around \eqref{Laplace_wave_eq})
\begin{align}
    I(\omega,x)=-\{-i\omega\psi(t,x)+\partial_{t}\psi\}|_{t=0}~.
\end{align}
For simplicity, here we take the source term to be,

$$S(x,\omega)=i\omega\alpha\delta(x-x_{0})+\gamma\delta(x-x_{0}).$$
Inserting this source term in (\ref{Sol-Inhom}) and assuming that $x\gg -\infty$, we get,
\begin{align}
    \psi(x,\omega)=(i\omega\alpha+\gamma)\frac{\psi_{L}(x_{0})e^{i\omega x}}{2i\omega A_{\rm in}}
\end{align}
Further, assuming that $x_{0}>0$ we can write,
\begin{align}
    \psi(x,\omega)=\left(\frac{\alpha}{2}+\frac{\gamma}{2i\omega}\right)\bigg[e^{i\omega(x-x_{0})}+\frac{A_{\text{out}}}{A_{\text{in}}}e^{i\omega(x+x_{0})}\bigg]~.
\end{align}
The solution in the time domain can be obtained by performing a Fourier transform of the above result in the following way
\begin{align}
    \Psi(x,t)=\frac{1}{2\pi}\int^{\infty}_{-\infty}e^{-i\omega t}\psi(x,\omega)d\omega.
\end{align}
Now analytically continuing the integral in the complex plane and deforming the contours in such a way that we can avoid the singularities on the real axis\footnote{Note that such integrals can also be done by using the $i\epsilon$ prescription, see for example \cite{Dennery1996MathematicsFP}.}, we have\footnote{In obtaining the above, we have used the following results,
    \beq
    \int^{\infty}_{-\infty}\frac{e^{i\omega(x+x_{0}-t)}}{(2\omega+V_{0})}d\omega&=&i\pi e^{\frac{V_{0}}{2}(x+x_{0}-t)}\theta(t-x_{0}-x)\,,\label{QNM-Contribution}\\
    \int^{\infty}_{-\infty}\frac{e^{i\omega(x+x_{0}-t)}}{\omega}d\omega&=&i\pi~\text{Sign}(x+x_{0}-t)\,. \label{pole-w}
    \eeq
}
%
\begin{align}\label{Green-Final}
    \tilde{\psi}(x,t)=\frac{\alpha}{2}\delta(x-x_{0}-t)+\theta(t-x-x_{0})\left(\frac{V_{0}\alpha +\gamma}{2}e^{V_{0}(x+x_{0}-t)}+\frac{\gamma}{2}\right)~.
\end{align}

In the above expression, each term has a physical interpretation. The first term contributes only if $x-x_{0}-t=0$. This is nothing but the time taken by signal to reach the observer at $x$ from the source point at $x_{0}$ directly. Thus, it represents a direct signal from source to observer. For ``early times'', when $t-(x+x_{0})<0$, the second and the third term vanish.
The second term starts playing its role at ``late times'' when $t>(x+x_{0})$. But this is the time that a signal propagating at the speed of light takes to hit the barrier at $x=0$ and then reach the observer. Thus, the second term comes from the QNM contribution to the integral \eqref{QNM-Contribution} at $\omega_{\rm QNM}=-i V_{0}$.
On the other hand, the last term comes from the contribution of the pole to the real axis $\omega=0$ which can be checked from \eqref{pole-w}. That is, the pole on the real axis will contribute at a late time.

\section{Quasinormal modes and black hole spectroscopy}
%
\begin{figure}
    \centering
    \includegraphics[width=0.7\textwidth]{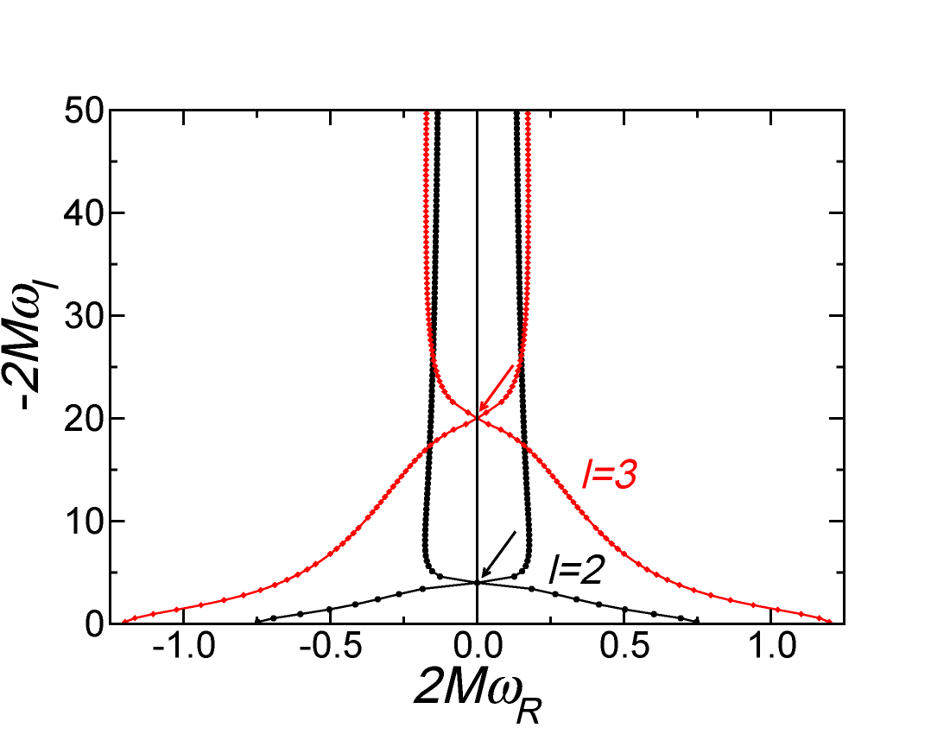}
    \caption{
        {\small Gravitational quasinormal mode frequencies of a Schwarzschild black hole for multipoles $\ell=2,3$ and the first $\sim 100$ overtones~\cite{Berti:2005ys,Berti:2009kk,Berti:2025hly,CoG,GRIT}. In both cases the arrow marks a mode that hits the imaginary axis.}
    }
    \label{fig:qnms}
\end{figure}
The examples above show that the calculation of the characteristic modes is conceptually clear. In asymptotically flat spacetimes, one imposes that waves go into the black hole horizon or to asymptotically large distances. Unfortunately, for physically relevant objects like black holes, simple closed-form solutions cannot be found and one needs to resort to numerical calculations. We are looking for solutions that have $A_{\rm in}=0$ in \eqref{asymptotic_behaviorH}. However, as we have seen in the examples above, this is the coefficient of $e^{-i\omega r_*}$ which is exponentially suppressed for frequencies with a negative imaginary part\footnote{In fact, if the spacetime is relevant at all, it will have all modes with a negative imaginary part, otherwise it will blow up in time, it would be unstable.}. The solution to this problem is to perform accurate analytic asymptotic expansions of the homogeneous equation \eqref{Laplace_wave_eq} and to factor out the exponentially divergent behavior. There are other approaches similar in spirit~\cite{Leaver:1985ax,Lo:2025njp}, but in the notebook \coderef{code:QNMs-DI.nb}, we demonstrate how a simple-minded direct integration can work, at least for some modes~\cite{Chandrasekhar:1975zza}.

\begin{table}[ht!]
    \centering
    \setlength{\tabcolsep}{9pt} 
    \renewcommand{\arraystretch}{0.7} 
    \begin{tabular}{|c||cc|}
        \hline
        Overtone & $\omega_{\rm QNM}=\Re{\omega}+i\Im{\omega}$ & $e^{\Im{\omega}\times 2\pi/T_0}$ \\
        \hline\hline
        $n=0$    & $0.373672 - 0.088962i$                      & $0.224$                          \\\hline
        $n=1$    & $0.346711 - 0.273915i$                      & $0.00999$                        \\ \hline
        $n=2$    & $0.301053 - 0.478277i$                      & $3.22\times 10^{-4}$             \\ \hline
        $n=3$    & $0.251505 - 0.705148i$                      & $7.09\times 10^{-6}$             \\\hline
        $n=4$    & $0.207515 - 0.946845i$                      & $1.22\times 10^{-7}$             \\ \hline
        $n=5$    & $0.169298 - 1.195606i$                      & $1.86\times 10^{-9}$             \\ \hline
        \hline
    \end{tabular}
    \caption{First six lowest lying modes of a Schwarzschild black hole for gravitational $\ell=2$ perturbations. A measure of how fast these modes decay is given in the last column, where we show how much a mode decreases its amplitude after one period of the fundamental $n=0$ mode $T_0=2\pi/0.373672$.}
    \label{tab:BH_modes}
\end{table}
A complete characterization of the modes can be found elsewhere~\cite{Berti:2009kk,Berti:2025hly,CoG,GRIT,Lo:2025njp}. At fixed $\ell$, we find an infinite number of frequencies that satisfy the boundary conditions, we label each with an overtone index $n$. We order them by their increasing imaginary part, $n=0$ for the fundamental, longest lived mode.

The first few modes are shown in Table~\ref{tab:BH_modes}. The Table shows the most troublesome aspect of black hole quasinormal modes: even the longest lived mode decreases by a factor of five, in amplitude, after one period. The first hundred modes or so are shown in Fig.~\ref{fig:qnms}.
At large overtone $n$ the quasinormal mode frequencies asymptote to~\cite{Motl:2003cd}
\be
M\omega=\frac{\log 3}{8\pi}-i\frac{(2n+1)}{8}\,,
\ee
and their physical significance is, therefore, suppressed, since they decay to insignificant amplitudes after a short timescale set by the mass of the black hole. The asymptotic structure above was the subject of a vibrant discussion in the context of black hole area quantization~\cite{Hod:1998vk}. It is unclear at the moment if there are any observational consequences of having a spectrum accumulating at a fixed point.
Note that in the large overtone $n$ limit, the frequencies are always small, we are never in a regime where an eikonal or ray approximation can be used or trusted.

On the other hand, at large $\ell$ one finds~\cite{Iyer:1986np,Barreto:1997,Berti:2009kk,Berti:2025hly}
\be
M\omega\approx \frac{(\ell+1/2)}{3\sqrt{3}}-i\frac{(n+1/2)}{3\sqrt{3}}=(\ell+1/2) M\Omega_{\rm LR}-i (n+1/2) M\lambda\nonumber
\ee
which is precisely the behavior of null geodesics on the light ring, Eq.~\eqref{null_geoedesic_decay}. This agreement extends to rotating black holes~\cite{Cardoso:2008bp,Yang:2012he}.

\begin{figure}
    \centering
    \includegraphics[width=0.8\textwidth]{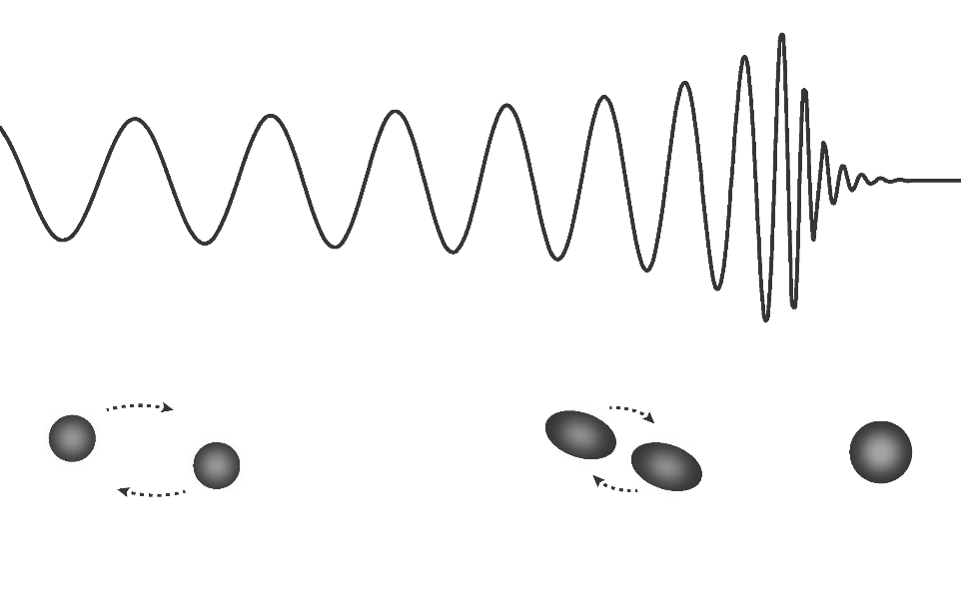}
    \caption{{\small Gravitational-wave driven stages in the life of a binary of compact objects, such as black holes. When separations are large the system emits nearly monochromatic, low frequency radiation. Energy loss drives the system to smaller separations and larger velocities. When they approach each other, a common horizon and light ring forms (roughly corresponding to the peak of the signal), which is filled with radiation that leaks out to the observer as ringdown or quasinormal modes.}
    }
    \label{fig:ringdown}
\end{figure}
The agreement between a local, null geodesic calculation, and a full wave analysis with boundary conditions at the boundary of spacetime is remarkable. These are the modes that seem to be present in waveforms generated in linear perturbation theory of in full nonlinear General Relativity, a signal such as that in Fig.~\ref{fig:ringdown}. Why? Because the boundary conditions imposed on quasinormal modes are radiative, they let the fields propagate freely, exactly as they do in our local null geodesic analysis around \eqref{null_geoedesic_decay}. As we discuss shortly in Section~\ref{sec:spectral_stability}, once we change boundary conditions or introduce extra structure in the effective potential, the mode structure changes and may longer agree with a local geodesic calculation. Nevertheless, transient behavior corresponding to the quasinormal modes in Table~\ref{tab:BH_modes} and Fig.~\ref{fig:qnms} is always expected to be there as long as a vacuum light ring exists in the spacetime, sufficiently far from matter (we will be quantifying this statement). In summary, we know the characteristic frequencies of black holes to high precision, and they are related to properties of spacetime in the vicinity of the light ring.

\subsection{Black hole spectroscopy}
In the pictorial description of quasinormal modes, they are excited when a source starts ``filling'' the light ring with radiation. The reason that quasinormal modes play a key role in gravitational wave astronomy is that when a source generating the waves gets very close to the light ring, then they are excited to a very large amplitude (see discussion around Eqs.~\eqref{null_geoedesic_decay} and \eqref{eq:Vpeak}). Indeed, Fig.~\ref{fig:ringdown} represents the gravitational wave amplitude measured at large distances when two black hole collide. The pulse of radiation emitted after the black holes coalesce is intense and mostly a superposition of quasinormal modes. We can use this spectacular emission of energy to infer black hole properties and to test the theory. This {\it black hole spectroscopy} program aims to understand how black holes relax, via a superposition of modes. Since each mode is in essence {\it two quantities}, $\Re{\omega}$ and $\Im{\omega}$, one can use it to infer the black hole mass and spin, the two quantities characterizing {\it any} neutral black hole in the universe, according to General Relativity in vacuum (see also Section~\ref{sec:no_hair}). In other words, we fit the observational data to
\be
h=A e^{-\frac{(t-t_0)}{\tau}}\cos\left(2\pi f (t-t_0)+\phi_0\right)\,,\label{ringdown_waveform}
\ee
but since $\tau=\tau(M,a),\,f=f(M,a)$, one can invert these relations to find $(M,a)$.

\begin{figure}
    \centering
    \includegraphics[width=0.8\textwidth]{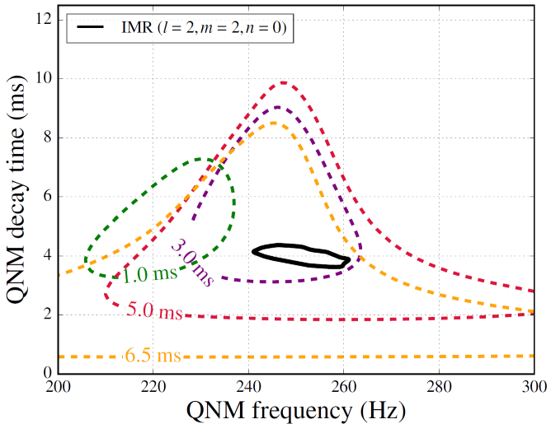}
    \caption{
        {\small Black hole spectroscopy with LIGO. Shown 90\% posterior distributions.
                Black solid is 90\% posterior of fundamental $\ell=2$ quasinormal mode frequency as derived from the posterior mass and spin of remnant. From Ref.~\cite{LIGOScientific:2016lio}.}
    }
    \label{fig:ringdown_ligo}
\end{figure}
The measurements of two modes would give back two estimates for mass and spin and some associated errors. If General Relativity is correct, these must agree within uncertainties, providing a powerful test of General Relativity in the strong field, highly dynamical, regime. If additional structure is present in the spacetime (say, matter around the black holes or if there is no horizon) then the relaxation must be different from that of a black hole in vacuum. It is useful to plug in numbers. The characteristic vibration of a (non-spinning) black hole for the fundamental quadrupolar mode, which usually\footnote{The dominance of a given mode depends on the amplitude to which it is excited, and on its lifetime. Numerical studies have shown that coalescences from quasi-circular inspirals of two black holes, the first overtone is excited to a comparable amplitude to that of the fundamental mode. Nevertheless, because it is short-lived, the fundamental mode ends up dominating the late-time portion of the signal.} dominates gravitational-wave signals, has frequency and timescale
\beq
f&=&\frac{\Re{\omega}}{2\pi}=1.207\left(\frac{10M_\odot}{M}\right)\, {\rm kHz}\\
\tau&=&\frac{1}{|\Im(\omega)|}=0.5537\left(\frac{M}{10M_\odot}\right)\, {\rm ms}\label{fundamental_qnm_numbers}
\eeq

Figure~\ref{fig:ringdown_ligo} summarizes the first observation of a black hole ringing down, and also the first test of General Relativity using black hole spectroscopy~\cite{LIGOScientific:2016lio,LIGOScientific:2020tif,Cotesta:2022pci,Carullo:2023gtf}. The underlying idea is very simple: one picks observational data, that look like a noisy version of Fig.~\ref{fig:ringdown}, and one fits the late-time signal to~\eqref{ringdown_waveform}. The start time $t_0$ is such that one expects to be within the validity of the linearized approximation (the start time is indicated in the figure; earlier start times tend to ``push'' the frequencies to lower values, possibly so as to pick up signal from the lower-frequency, inspiral component~\cite{Cotesta:2022pci}). The dashed lines in the plot represent the resulting uncertainties in the estimates of the decay time $\tau$ and of the frequency $f$. When the start time $t_0$ is large and away from the peak of the waveform, the remaining signal is very weak, hence uncertainties very large. Notice the agreement with the predictions from the full waveform, represented as black solid line in the figure.

\begin{figure}
    \centering
    \includegraphics[width=0.7\textwidth]{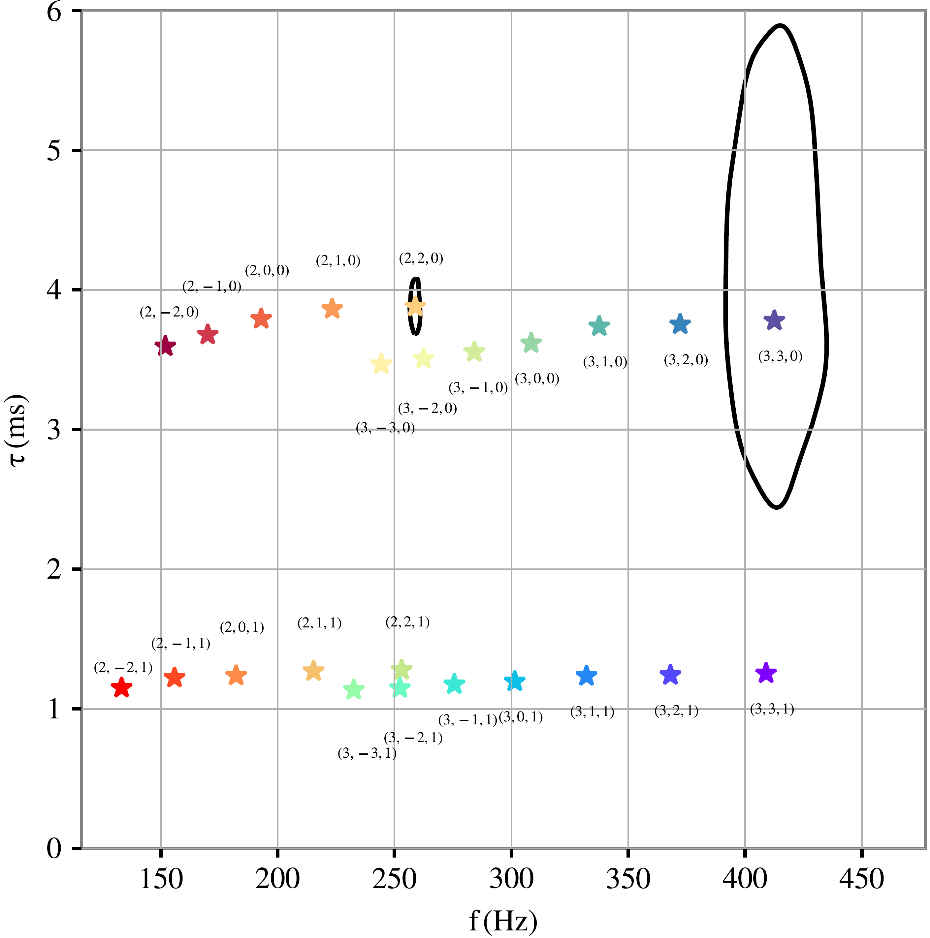}
    \caption{
        {\small Black hole spectroscopy in the near-future, assuming an event like GW150914 but with a signal-to-noise ratio in ringdown of 40. Now, a number of modes are visible and purely spectroscopic tests of General Relativity are possible. Courtesy of Gregorio Carullo.
            }}
    \label{fig:ringdown_ligo2}
\end{figure}
The first gravitational wave observations provided truly revolutionary ways of testing the black hole paradigm, but this is just the beginning. The upgrade of existing instruments or new detectors will considerably decrease the errors in Fig.~\ref{fig:ringdown_ligo}. A simulation of a detection with a signal-to-noise ratio in ringdown of 40 (for reference, the ringdown signal-to-noise ratio in the GW150914 event was of order 7) achievable within a decade, is shown in Fig.~\ref{fig:ringdown_ligo2}. Now, the dominant and subdominant modes will be apparent, providing robust measurements of mass and spin and of tests of the nature of the object. Indeed, Nature is so kind that while these lectures were being finalized, we were graced with the event GW250114, similar in all aspects to GW150914, but much clearer since the instruments have improved~\cite{KAGRA:2025oiz}. A two-mode detection has been robustly identified, and this work provides a glimpse of what black hole spectroscopy will look like in the future.

\subsection{Overtones, nonlinearities and the complexity of General Relativity}
Figure~\ref{fig:ringdown_ligo} shows that the uncertainties become larger when we start fitting later. The reason is very simple: the signal decays exponentially, thus the signal-to-noise ratio decreases and the errors associated with the measurement increase. On the other hand, it is at late times that one trusts the linearized expansion \eqref{perturbative_scheme}. As one approaches the instant at which the black hole was newly formed, linearized expansions are bound to break down. In fact, Fig.~\ref{fig:ringdown_ligo} shows that  fitting too early to a single exponentially damped sinusoid is not appropriate, even if done after formation of a common horizon\footnote{This statement refers of course to nonlinear simulations of black hole mergers.}. There can be several reasons that would explain this feature.

Overtones may play an important role at early times. Indeed, there is evidence that overtones are excited to comparable amplitude upon coalescence~\cite{Zhang:2013ksa,Giesler:2024hcr}. However, given the mode structure, with higher overtones having {\it lower} frequency and larger damping times (see Fig.~\ref{fig:qnms} and Table~\ref{tab:BH_modes}), they are convenient to use when fitting to any lower frequency feature, but difficult to find robust evidence that they {\it are} there. Ultimately, the lack of a notion of completeness or proper projection operator to find quasinormal modes in a time series is a serious challenge to claim the presence of very high order overtones, to which one needs to add the current low signal-to-noise ratio of events~\cite{Cotesta:2022pci,Baibhav:2023clw,Giesler:2024hcr}.

At very late times in the life of a process disturbing a black hole, one expects to find a stationary Kerr solution. Thus, linearized perturbation theory is more suitable the later we observe the system. However, if one wishes to describe the dynamics at early times, one should consider nonlinear corrections to the picture we have developed.
An evolving spacetime must be affected in the way it rings down since its mass is changing, giving rise to secular variations of the ringdown frequencies~\cite{Redondo-Yuste:2023ipg}. In addition, $h_{\mu\nu}=h_{\mu\nu}^{(1)}$ in Eq.~\eqref{perturbative_scheme} is just the first term in a perturbation ladder. The perturbative scheme written should be expanded to higher orders,
\be
g_{\mu\nu}=g_{\mu\nu}^{(0)}+h_{\mu\nu}^{(1)}+h_{\mu\nu}^{(2)}+\,...\,, \label{perturbative_scheme_complete}
\ee
It turns out that one can still write master equations having the same structure as Eq.~\eqref{Laplace_wave_eq} at any order in perturbation theory~\cite{Gleiser:1995gx,Nakano:2007cj,Brizuela:2009qd}. In particular, for second order quantities one finds
\be
\frac{d^2 \psi^{(2)}}{dr_*^2}+\left(\omega^2-V_s\right) \psi^{(2)}={\cal S}\propto \left(\psi^{(1)}\right)^2\,.\label{Laplace_wave_eq_secondorder}
\ee
A glance at this equation, where we suppressed angular indices (but they dictate which combinations of first order quantities are allowed as source) indicates that in the ringdown stage extra frequencies $\propto 2\omega_{\rm QNM}$ should be present in the spectrum. These have recently been reported, together with the relative amplitude of the second order modes~\cite{Cheung:2022rbm,Mitman:2022qdl,Redondo-Yuste:2023seq,Bourg:2024jme,Bucciotti:2024zyp,Bucciotti:2025rxa}.

Finally, during the relaxation stage, backscattering off spacetime curvature is already taking place (cf. Figs.~\ref{fig:contour}-\ref{fig:SXS_RWZ}), contaminating what should otherwise be a pure sum of exponentially damped sinusoids.

In other words, at each stage, the gravitational wave signal from a black hole system is rich. We can, of course, try to perform nonlinear simulations and take the numerical output as a bona fide response of the black hole. While we can do this, and certainly should do it too, dissecting the response and finding evidence for a rich structure in it, is certainly worthwhile doing. It will strengthen our confidence in the predictions of General Relativity, while providing robust tools to test alternatives.
\subsection{Resonant excitation of modes}
Thinking back to Fig.~\ref{fig:contour}, one might consider whether there are sources that could excite resonantly a black hole characteristic frequency. This would be similar in spirit to the excitation of the resonant modes of a glass when a wet finger circles its rim. For particles without a structure on circular motion this is impossible, since the modes of an isolated black hole are localized at the light ring, as we saw ample evidence for. Then to excite the light ring one needs matter orbiting at the speed of light, which together with stability requirements seem to preclude resonant excitation in a simple fashion.

Infalling particles thrown into a black hole at just the right time intervals can resonantly excite the modes of a black hole~\cite{Berti:2006hb}. However, realist accretion scenarios will be stochastically accreting, and the high damping rate of black hole modes suppresses resonances~\cite{Yuan:2025fde}.

A promising mechanism to excite resonantly the modes of a black hole consists on stellar-mass binaries orbiting supermassive black holes (so-called b-EMRI systems). The evolution of the stellar-mass binary via gravitational wave emission or dynamical friction provides a natural tuning mechanism: as it sweeps upwards in frequency it will naturally resonate with one or more of the black hole modes~\cite{Cardoso:2021vjq,Yin:2024nyz,Santos:2025ass}.

\section{Spectral stability\label{sec:spectral_stability}}
\begin{figure}[!ht]
    \centering
    \includegraphics[width=0.9\textwidth]{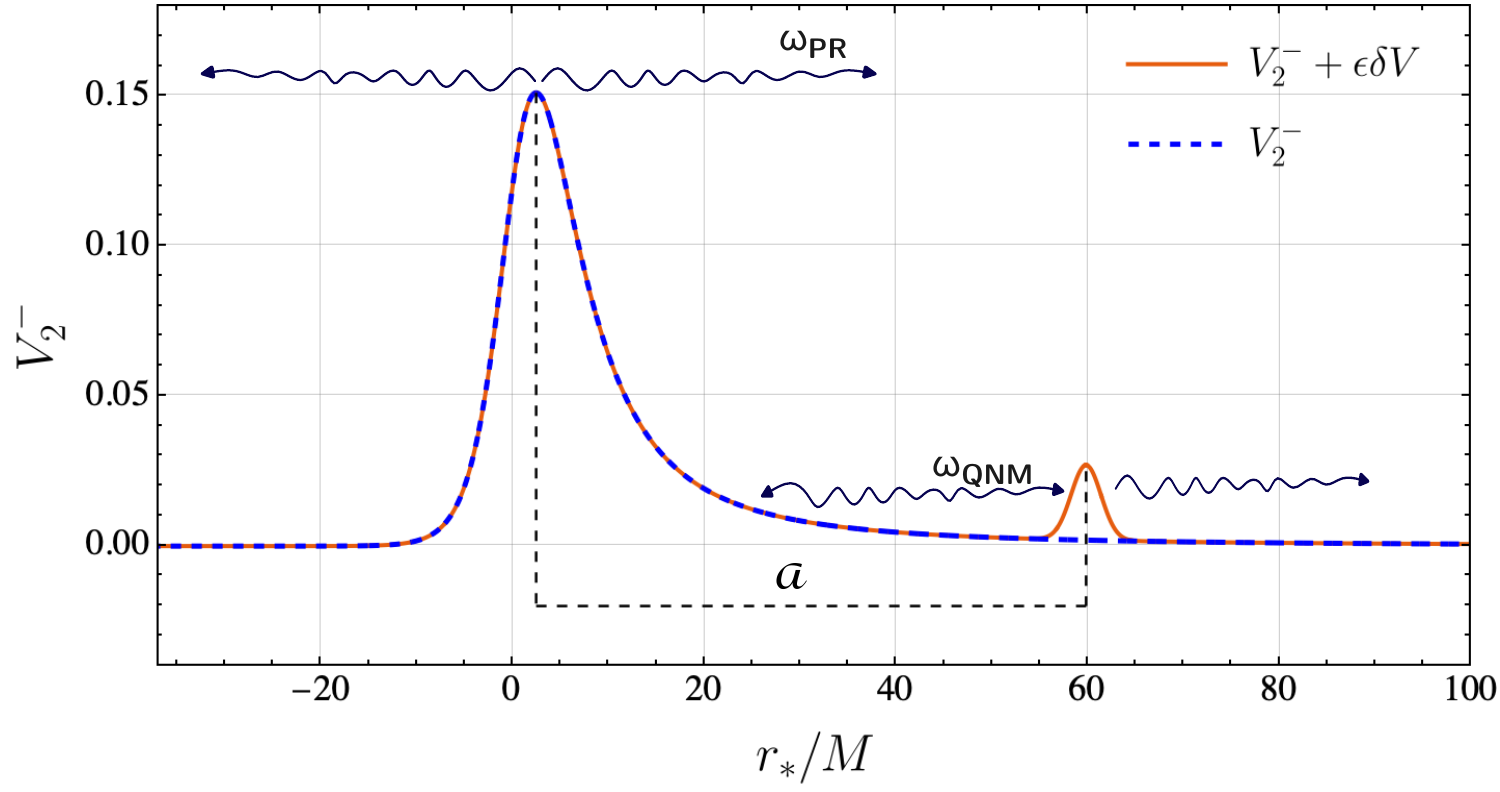}
    \caption{
        {\small Spectral instability of black holes: the unperturbed Regge-Wheeler potential, $V_2^-$, is shown as an orange solid line, while the superimposed blue dashed line shows the potential with an added perturbation, $V_2^- + \epsilon\delta V$. This perturbation is a Gaussian bump whose location relative to the main potential's maximum is indicated by the horizontal dashed line of length $a$ in units of $M$. This feature mimics the presence of external matter (stars, gas or dust, etc.) far from the black hole, and creates a potential cavity that traps low frequency waves as demonstrated in Ref.~\cite{Cheung:2021bol}. The modes of the vacuum potential admit resonances which give rise to prompt ringdown ($\omega_{\rm PR}$ in the sketch) in the signal, and the cavity gives rise to the longest lived modes of the full system, the quasinormal modes ($\omega_{\rm QNM}$ in the sketch).}
        \label{fig:spectral_instabilitya}
    }
\end{figure}

We have been discussing exclusively vacuum spacetimes, perhaps perturbed by some massless field. As we have showed, there are no stationary black hole configurations supporting stationary fundamental fields, but our universe is dynamic, and full of matter. Are the results concerning the modes of black holes stable against small environmental perturbations, or those from an improved description of the gravitational interaction?

A primitive but possible way to think about the robustness of a QNM result against environmental effects, is by... well, by adding matter. Modeling general-relativistic spacetimes containing black holes and matter is a notoriously difficult problem that we consider below in Section~\ref{exact_environment_BH_cluster}. We will here use a dirty and foggy shortcut. Since the governing master equation -- particularly the effective potential $V_s$ in Eq.~\eqref{wave_equation_spins} is sensitive to the spacetime geometry -- and hence to the matter distribution, one can simply deform it as in Fig.~\ref{fig:spectral_instabilitya}, by promoting the potential to (we focus for simplicity on odd gravitational perturbations but the conclusions are very general)
\be
V_\epsilon=V_2^-+\epsilon V_{\rm bump}\,,
\ee
with $\epsilon$ a small number and $V_{\rm bump}$ some arbitrary localized function that decays sufficiently fast. In this way, matter external to the black hole (like a star, a planet or dark matter) is assumed to cause a tiny, extra bump in the potential.

\begin{figure}[!ht]
    \centering
    \includegraphics[width=0.8\textwidth]{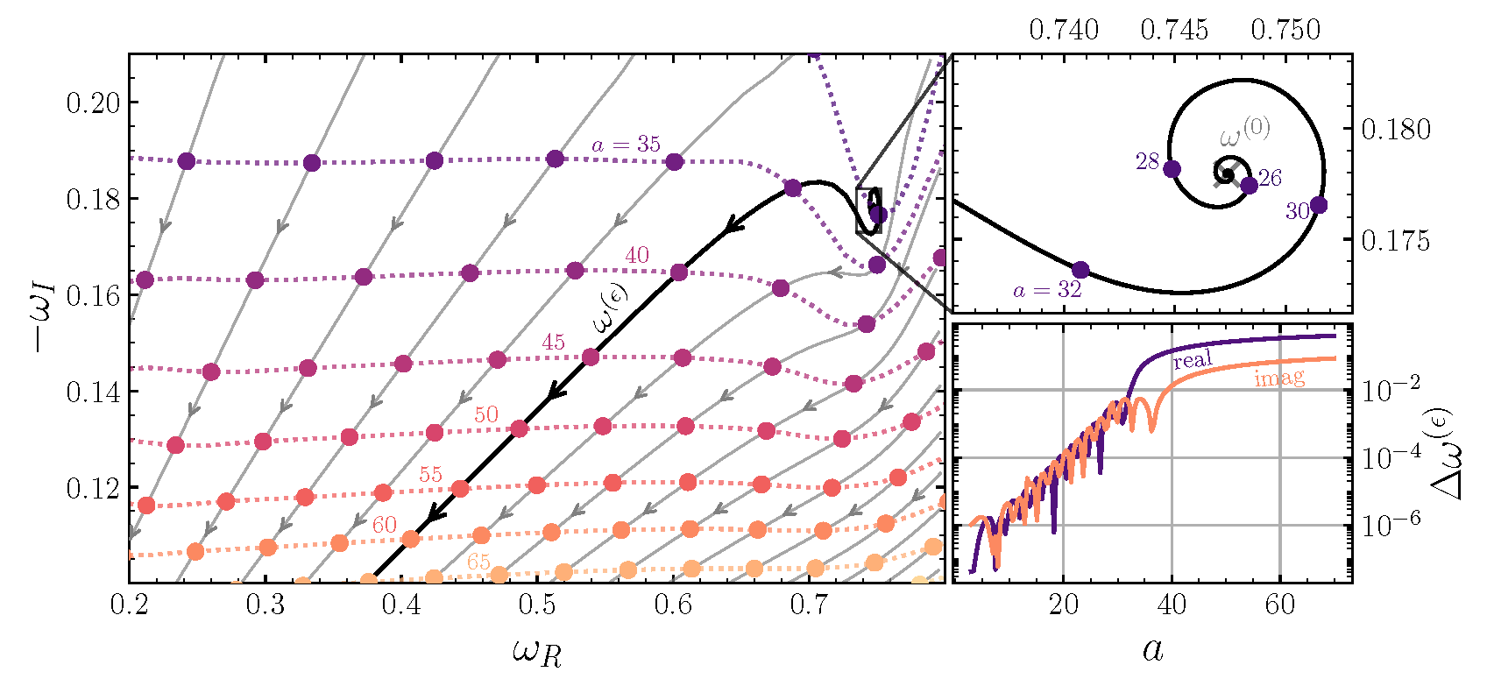}
    \caption{
    {\small The change in the dominant, lowest-lying mode as the distance $a$ is varied, for a Gaussian of height $\epsilon=10^{-6}$, see Fig.~\ref{fig:spectral_instabilitya}. Modes with the same value of a have the same color, and are connected with dotted lines. As one can see even for very small disturbances the spectrum can change sizeably, in what we term ``spectral instability''. From Ref.~\cite{Cheung:2021bol}.}
    }
    \label{fig:spectral_instabilityb}
\end{figure}
The {\it fundamental} mode of the effective potential $V_\epsilon$ is shown in Fig.~\ref{fig:spectral_instabilityb}. They show a surprising -- if disturbing! -- feature: the fundamental mode changes drastically when the distance $a$ between the bump and the light ring is increased to large values~\cite{Barausse:2014tra,Cheung:2021bol}. Please note that the height of the potential is extremely small, $\epsilon=10^{-6}$ and yet for these separations the fundamental mode changes by factors of $\sim 100\%$. Instabilities in higher overtones were also observed with other types of small modifications to the effective potential~\cite{Jaramillo:2020tuu,Cardoso:2024mrw,Siqueira:2025lww}.
In fact, if instead of a ``bump'' the perturbed potential looks like a ``well,'' then the system may even become unstable~\cite{Mai:2025cva}.
Does this mean that we should discard all of the beautiful and hard-won results from black hole perturbation theory and quasinormal mode calculations in vacuum? No!

\subsection{Temporal stability: a guitar entering a concert hall\label{sec:guitar}}
%
\begin{figure}[!ht]
    \centering
    \includegraphics[width=0.9\textwidth]{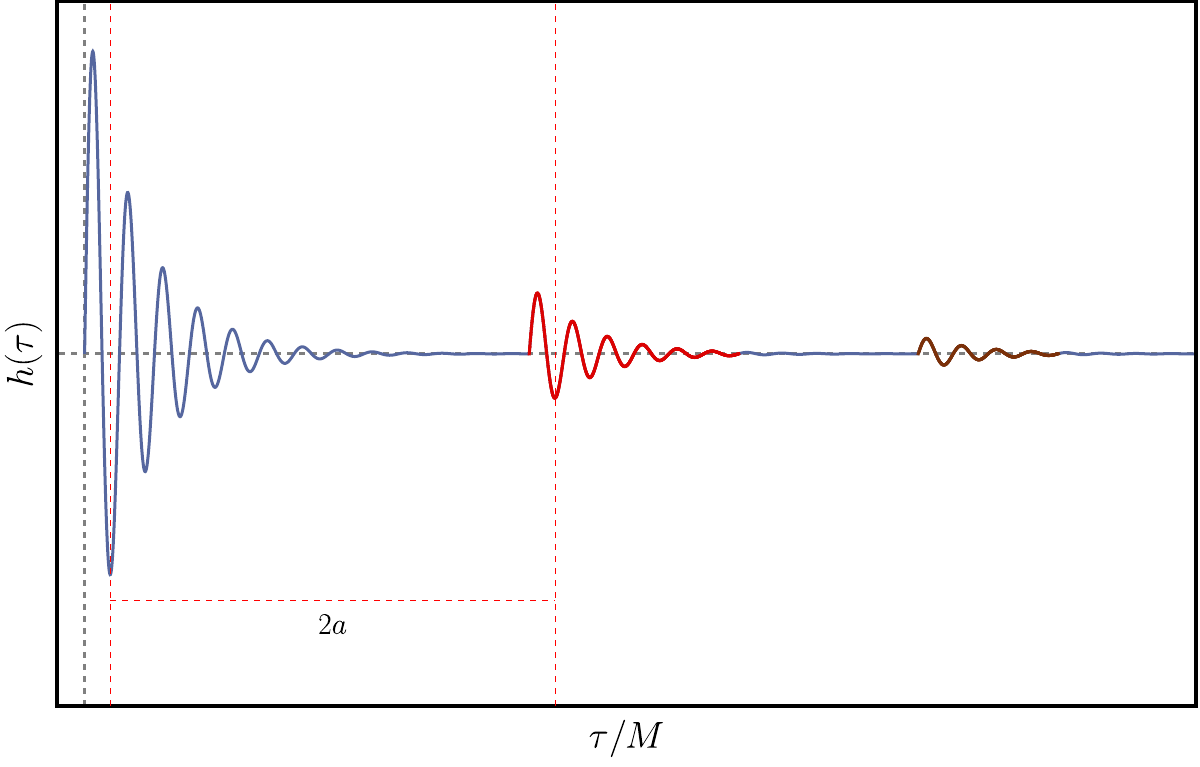}
    \caption{
        {\small A schematic depiction of the time-domain signal $h(\tau)$ generated when a source radiates in the setup of Fig.~\ref{fig:spectral_instabilitya}. As the source crosses the light ring it excites prompt ringdown (the early, blue portion of the signal) localized at the light ring. The (relatively large frequency) wave traveling outwards tunnels through the small height ``bump'', except for a low frequency component which gets trapped inside the cavity formed by the light ring and the bump. This low-frequency component gives rise to successive echos (indicated in red and brown) after a light crossing time inside the cavity, which eventually becomes simply a low-frequency, exponentially signal dominated by the lowest quasinormal mode of the system, as shown in an accurate numerical evolution in Fig.~\ref{fig:spectral_instability3}. Note, incidentally, that if the disturbance (the source) is instead localized far away and never approaches the black hole, or if it only emits low frequency radiation, prompt light ring ringdown is strongly suppressed.}}
    \label{fig:spectral_instability3}
\end{figure}
\begin{figure}[!ht]
    \centering
    \includegraphics[width=0.8\textwidth]{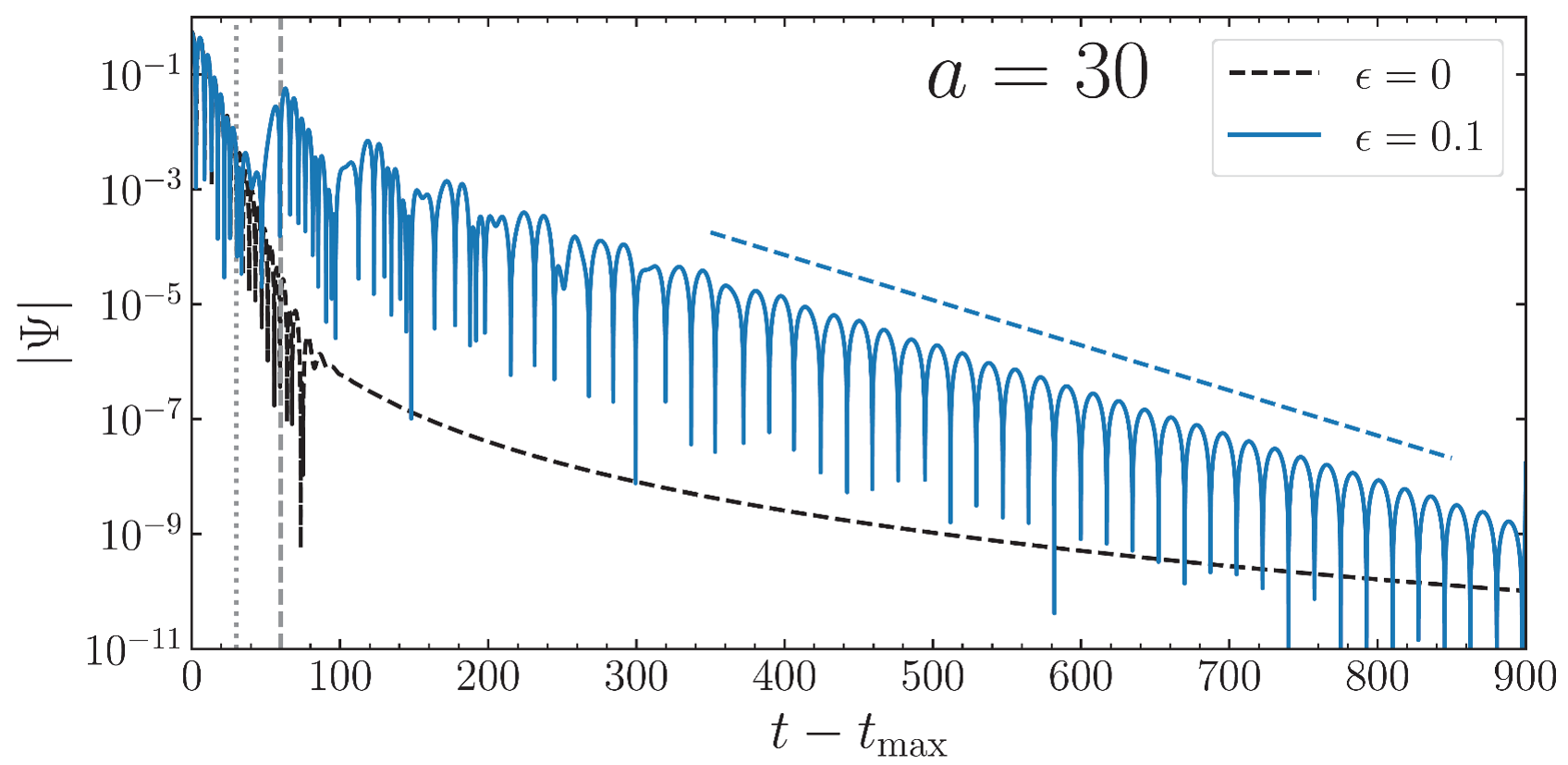}
    \caption{
        {\small The time evolution of a wavepacket released closed to the light ring (i.e. close to the global maximum of the potential in Fig.~\ref{fig:spectral_instabilitya}) excites prompt ringdown, which subsequently give rise to echoes and to a slow late-time decay. The decay is described not by the vacuum black hole quasinormal modes, but by the new fundamental mode of the system, as in Fig.~\ref{fig:spectral_instabilityb}. From Ref.~\cite{Berti:2022xfj}.
            }
        \label{fig:spectral_instability4}}
\end{figure}
We have been discussing the spectral content of equation~\eqref{Laplace_wave_eq}, but any finite duration observation is more concerned with the time-domain signal instead, the solution of \eqref{wave_equation_spins}. A time domain waveform has been sketched in Fig.~\ref{fig:spectral_instability3} and the results of a numerical calculation has been shown in Fig.~\ref{fig:spectral_instability4}.

If we're interested in the physics related to the coalescence of compact objects, we need to consider radiation emission triggered by processes close to the horizon. For example, a star or blob of matter being accreted by a black hole. What emerges is the following clear picture: as the source crosses the light ring, black hole ringdown is excited to large amplitudes. It is dominated by the vacuum black hole quasinormal mode, even if it plays no obvious role in the quasinormal mode spectrum of the full system (cf. Fig.~\ref{fig:spectral_instabilityb}). As we argued, this is a necessity, since the source emits gravitons to the light ring and this is a local relaxation which cannot possibly yet have known about the existence of a bump or special feature far away. We can call this a prompt ringdown, or light-ring ringdown.

The burst of radiation that ensues then travels outwards and interacts with the potential bump. Since light-ring ringdown has a high frequency ($\Omega_{\rm LR} \sim 1/M$), larger than the other frequency scale in the problem ($1/a$), this burst tunnels out very easily. It leaves behind a low-frequency component that is reflected by the bump and travels back to the black hole, where it encounters the light ring potential peak, sending it back to the bump, where a fraction tunnels out. We get echoes of the original signal in our detectors as seen in Figs.~\ref{fig:spectral_instability3}--\ref{fig:spectral_instability4}. The quasinormal modes of the full system, shown in Fig.~\ref{fig:spectral_instabilityb}, only show up clearly at very late times\footnote{Fourier-domain calculations, done in frequency domain, have the time information folded in through the corresponding integral. It turns out, as we are discussing, that light ring relaxation is not always clearly visible in the spectrum, when other scales are present.}.
Then, the low frequency signal that survives is continuously leaking out, and one finds an exponentially decaying mode, the real part of which $\Re \propto 1/a$ and the imaginary part is lower than the vacuum black hole quasinormal frequency. For further details, see Refs.~\cite{Berti:2022xfj,Cardoso:2024mrw,Ianniccari:2024ysv}. This separation of phases requires a separation of scales. In particular, the existence of clear echoes requires that the light ring had time to relax before the first echo arrives. But the light ring relaxation timescale is of order $3/\lambda$ (see Eq.~\eqref{null_geoedesic_decay}, we are taking three e-fold times), whereas the echoes delay is $2a$. Thus, for
\be
a\gtrsim 9\sqrt{3}/2 M\,,
\ee
one finds distinct phases: prompt ringdown and echoes.

What we have been discussing is in essence also what happens when a guitar string is struck inside a room. Taken in isolation, we look at the modes of the guitar string, accounting for losses due to coupling to the air. These modes would describe well the sound of a guitar. But the room itself has its own modes, which have lower frequency. Whether we consider the string modes -- which will tell us about prompt sound -- or the concert hall modes -- which will inform us on the late time vibrations -- is to some extent a question of the physics we are interested in~\cite{Oshita:2025ibu}.
\subsection{Black holes in cavities \& echoes}

The previous discussion generalizes in two important ways:

\vskip 2mm
\noindent {\bf i. Black holes within cavities.} The setup of Fig.~\ref{fig:spectral_instabilitya} can be used to understand black holes inside cavities, such as anti-de Sitter black holes~\cite{Cardoso:2015fga}.
Pump up the potential such that nothing tunnels out. In that case, an observer sitting between the light ring and the bump (i.e., inside the cavity) will see a prompt black hole ringdown\footnote{A process, which, we insist, is due to light ring relaxation, a local process, following the quasinormal modes of Schwarzschild in vacuum.}. It is easy to understand that, even in the absence of a true branch cut, a decay $t^{-2\ell+3}$ will also be observed at intermediate times. Then, the observer will see ``echoes'' caused by reflection at the boundary of the cavity~\cite{Cardoso:2015fga}.

One can estimate the lowest quasinormal mode of a black hole in a cavity (our ``guitar in a concert hall''). Consider the $\ell=0$ mode for simplicity, but the calculation generalizes easily~\cite{Brito:2015oca}. The normal  modes $\Re \omega\sim 1/a$ for a cavity of size $a$ are. For small black holes, $M/a \ll 1$, this means we are in the low frequency regime and can use result \eqref{reflection_low_frequency}.
Thus, a wave impinging on the black hole with amplitude $A_{0}$ is scattered with amplitude $\left|A_{\rm out}\right|^2 = \left|A_{0}\right|^2(1-16(M\omega)^2)$. But the wave is trapped, and will undergo reflections at the cavity boundary (the bump) and scatters at the black hole. After a time t the wave interacted $N = t/a$ times, and its amplitude changed to $\left|A_{\rm out}\right|^2 = \left|A_{0}\right|^2(1-16(M\omega)^2)^N$, which can be approximated to $\sim \left|A_{0}\right|^2(1-16N(M\omega)^2)$. We then get
\be
\left|A_{\rm out}\right|^2 =\left|A_{0}\right|^2\left(1-\frac{16(M\omega)^2t}{a}\right) \,.
\ee
This small absorption at the horizon adds a small imaginary part to the frequency, which we can estimate by promoting the above to the Taylor expansion of an exponential, $A_{\rm out} \sim A_0 e^{\Im \omega t} \sim A_0(1 -\Im\omega t)$. We find immediately an estimate for the quasinormal frequency $\omega_{\rm QNM}\sim 1/a-i8M/a^3$.

\vskip 2mm
\begin{figure}
    \centering
    \includegraphics[width=0.7\textwidth]{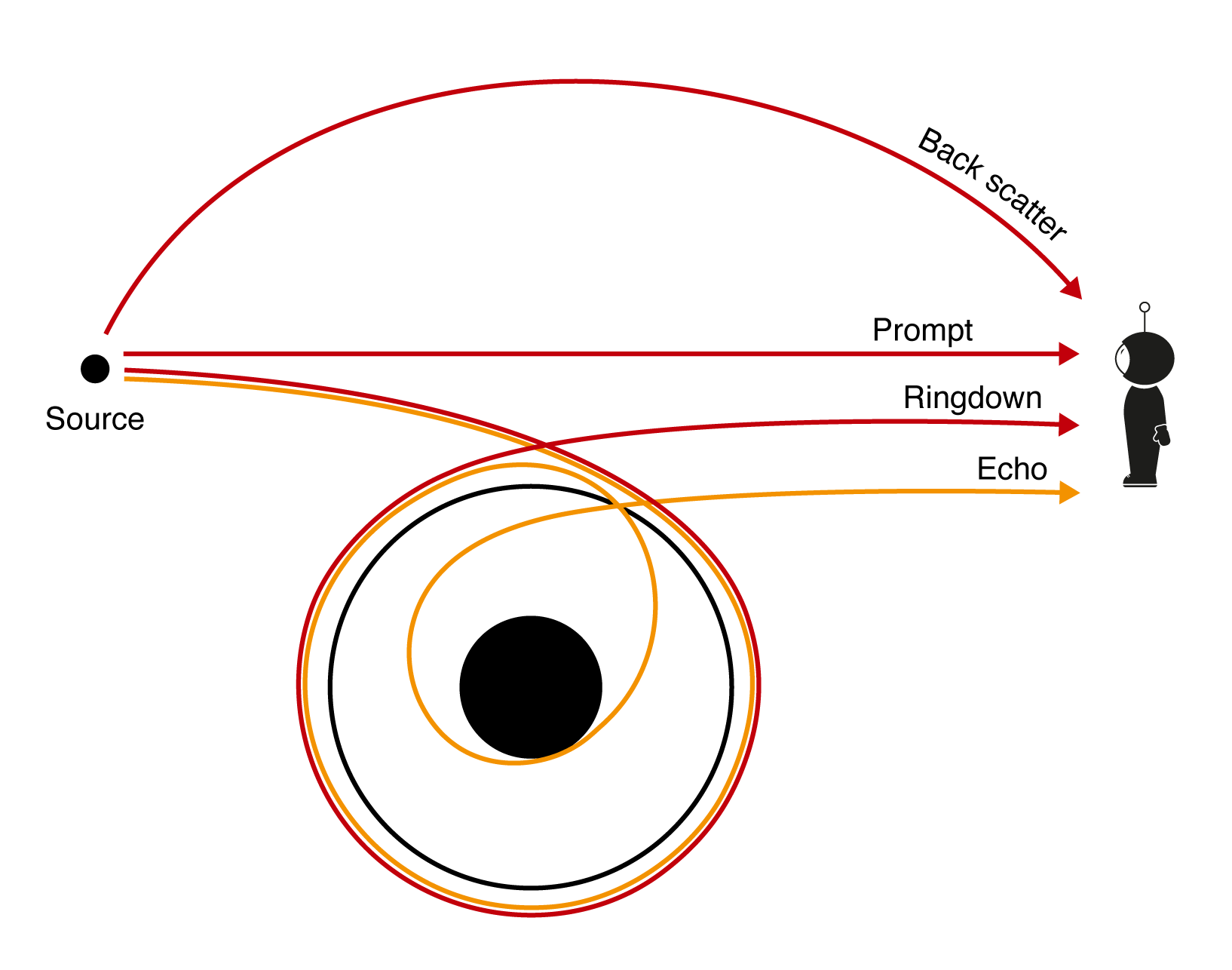}
    \caption{
    {\small Cartoon of wave propagation on the background of a very compact but horizonless object. See caption of Fig.~\ref{fig:contour}. A source (black dot) emits waves which eventually reach an observer. These waves can reach the observer in different ways, either direct propagation (the contribution from the zero-frequency portion of the diagram above), or circling the black hole a number of times, close to the critical circular null orbit, before finally escaping to measurement device. This latter picture gives us a physical understanding of {\it prompt ringdown}, localized close to the light ring}. Some of the waves close to the light ring will scatter from the object, giving rise to late-time {\it echoes}. Finally, waves can be back-scattered by spacetime curvature, resulting in power-law tails.
    }
    \label{fig:contour2}
\end{figure}
\noindent {\bf ii. Near-horizon structure}
What we just discussed concerning ``bumpy'' potentials, with structure added at large distances far from the black hole, applies equally to bumps or features {\it close} to the Schwarzschild radii. This is a specially relevant setup, as it allows us to understand what would happen if horizons simply don't exist, and are instead replaced by something parametrically close to where they'd develop\footnote{If this were true, then we would not have to worry about singularities nor about issues with loss of information, making it an appealing possibility.}. Take for example a bump at a location
\be
r_{\rm bump}=2M(2+\epsilon)\,.
\ee
For $\epsilon\lesssim 0.019$ the roundtrip time between the light ring and the bump is larger than the light ring relaxation time (also called the Lyapunov timescale), and echoes are again expected in the relaxation of this object~\cite{Cardoso:2016rao,Cardoso:2016oxy,Siemonsen:2024snb,Cardoso:2019rvt}. A cartoon depicting the extra feature that an observer would register is shown in Fig.~\ref{fig:contour2}.

\begin{figure}[!ht]
    \centering
    \includegraphics[width=0.6\textwidth]{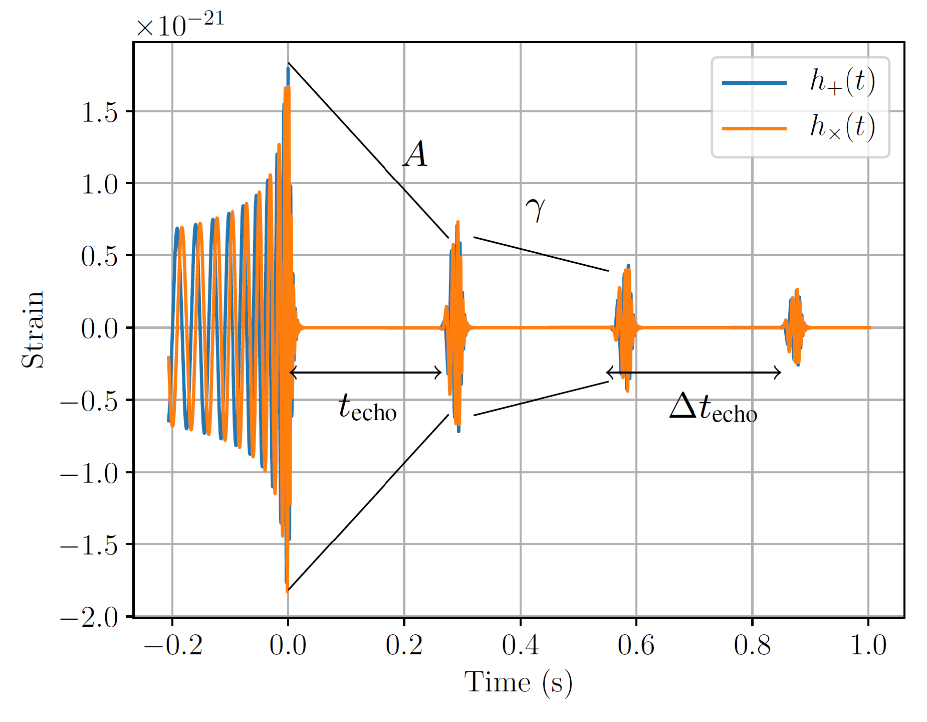}\\
    \includegraphics[width=0.7\textwidth]{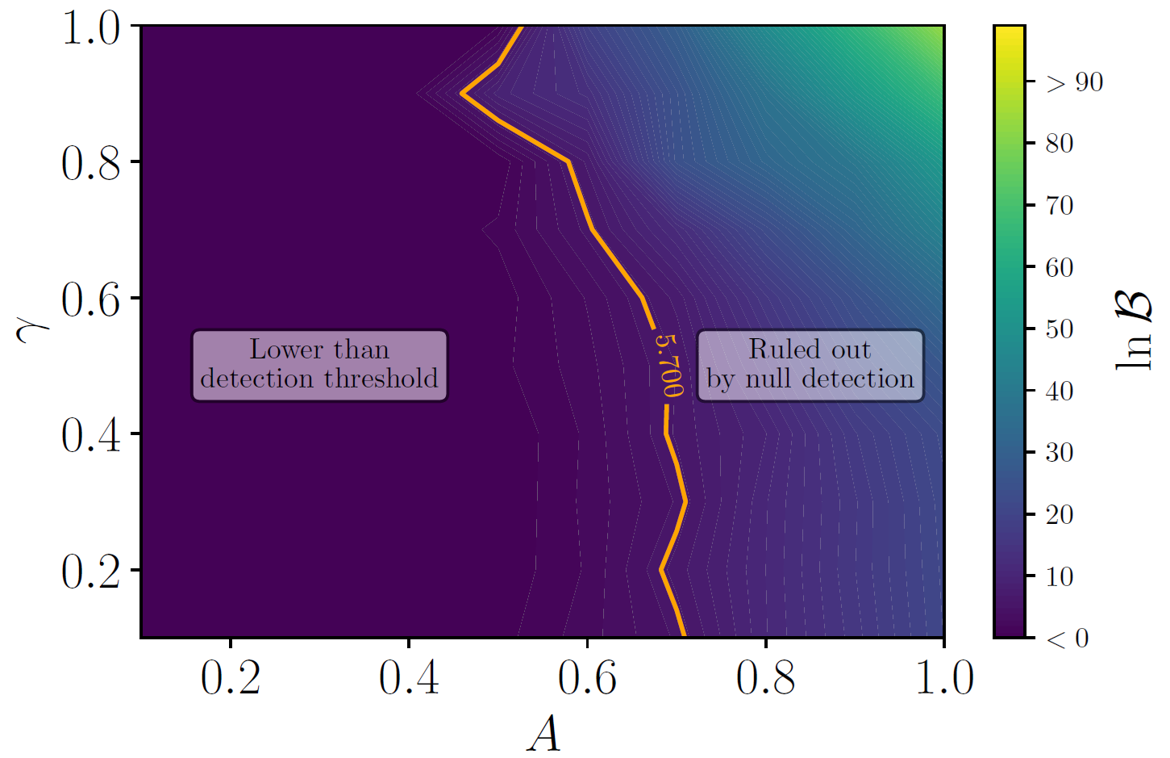}
    \caption{
        {\small {\bf Top panel:} A possible parametrization of postmerger echoes from inspiralling compact objects, in terms of a damping $\gamma$, time interval $\Delta t$ and a first revival of amplitude drop $A$ separated by $t_{\rm echo}$ from the main burst. {\bf Bottom Panel:} Exclusion diagram that can be obtained by LIGO-Virgo-KAGRA-like instruments on the relevant parameters describing the postmerger signal. From Ref.~\cite{Lo:2023cto}.
            }
        \label{fig:Echoes_Rico}}
\end{figure}
As should be clear by now, echoes in gravitational-wave signals are robust predictions of near horizon physics, or of large distance structure~\cite{Cardoso:2019rvt}. Observations are now starting to place the first constraints on the existence of post-merger echoes~\cite{Abedi:2016hgu,Westerweck:2017hus,Lo:2018sep,Uchikata:2019frs,LIGOScientific:2020tif,LIGOScientific:2021sio,Uchikata:2023zcu}. For that, one possible search methods through matched filtering, uses a parametrization of the waveform, represented in Fig.~\ref{fig:Echoes_Rico} (top panel). The parametrization assumes that a coalescence takes place and that the main burst is followed by a sequence of echoes, which have a damping rate $\gamma$ and are equidistant in time\footnote{The first pulse after the main burst occupies a special role in the modelling, since they have not experienced reflection by the potential barrier.}.

Typical exclusion diagrams on the main parameter look like Fig.~\ref{fig:Echoes_Rico}. The non observation of any postmerger features rules out physical models that do away with horizons. Tighter and tighter constraints (i.e., upper bounds on the echo amplitude and lower bounds on the separation between echoes) will quantify and strengthen enormously the black hole paradigm. For the state-of-the-art see the overview in Ref.~\cite{Berti:2025hly}.

\section{Radiation from pointlike particles}
We haven't dealt yet with sources to the perturbation equations. So far, dynamics are triggered by nontrivial initial data, as we explained in Section~\ref{sec:Laplace}. Under the assumption that they follow geodesics, the handling of sources in the formalism is quite straightforward. Take for example the simple theory of a minimally coupled scalar field $\Phi$ \eqref{action_scalar_dynamics_sourceless}, but now include a source for the scalar, a (dimensionless) scalar charge $\gamma$\footnote{You will notice that there is an overall factor difference from Eq.~\eqref{action_scalar_dynamics_sourceless}, we here use this convention to make it easier to compare against the results in Ref.~\cite{Cardoso:2019nis}.},
\be
S=\int d^4x \sqrt{-g}\left(\frac{R}{16\pi}-g^{\mu\nu}\partial_\mu\Phi\partial_\nu\Phi-2\gamma \Phi T\right)\,,\label{action_scalar_dynamics_source}
\ee
with $T$ the trace of the stress tensor of the particle. If we assume, for example, that the particle is pointlike and follows a circular geodesic $z^{\mu}$ of radius $r=r_p$ and orbital frequency $\Omega_p=\sqrt{M/r_p^3}$,
\beq
T^{\mu\nu}&=&m_p\int_{-\infty}^{+\infty}\delta^{(4)}(x-z(\tau))\frac{dz^\mu}{d\tau}\frac{dz^\nu}{d\tau}d\tau\,,\\
&=&m_p\frac{dt}{d\tau}\frac{dz^\mu}{d\tau}\frac{dz^\nu}{d\tau}\frac{\delta(r-r_p(\tau))}{r^2}\delta^{(2)}(\Omega-\Omega(\tau))\,,
\eeq
with $m_p$ the mass of the pointlike object. The equation of motion for the scalar takes the form
\be
\frac{1}{\sqrt{-g}}\partial_\mu\left(\sqrt{-g}g^{\mu\nu}\partial_\nu\Phi\right)=\gamma T\Phi\,.\label{scalar_motion_source}
\ee
Fourier decomposing the field and the source, as in Section~\ref{sec:Laplace}, we find
\be
\frac{d^2\psi}{dr_*^2}+\left(\omega^2-V_0\right)\psi=f\frac{m_p\gamma}{r}\frac{Y_{\ell m}(\pi/2)}{U^t}\delta(r-r_p)\delta(\omega-m\Omega_p)\,,
\ee
with $U^t=(1-3M/r)^{-1/2}$. We can now use our previous result \eqref{rlm} to find
\be
\psi=e^{i\omega r_*}\frac{m_p\gamma}{r_p}\frac{Y_{\ell m}(\pi/2)}{2i\omega A_{\rm in}U^t}\psi_{\rm H}(r_p)\delta(\omega-m\Omega_p)\,.
\ee
In other words, we need only evaluate the homogeneous solution at the location of the particle to solve our problem. This was done in the scattering calculation of the previous chapter, see notebook \coderef{code:Scattering-Scalar.nb}.
We can use the techniques of Section~\ref{sec:low_frequency}, to evaluate the solutions in the slow-motion, low-frequency regime to find~\cite{Poisson:1993vp,Cardoso:2019nis,Santos:2024tlt},
\beq
\dot{E}_{s=0}&=&\frac{1}{12\pi}\frac{\gamma^2m_p^2M^2}{r_p^4}\,,\label{dipolar_scalar_flux_PP}\\
\dot{E}_{s=1}&=&\frac{2}{3}\frac{q^2M^2}{r_p^4}\,,\\
\dot{E}_{s=2}&=&\frac{32}{5}\frac{m_p^2M^3}{r_p^5}\,,\label{Quadrupolar_flux_PP}
\eeq
where $q$ is the electromagnetic charge of the orbiting particle. Note that for $s=0,1$ fluxes scale with separation in a different way from the gravitational $s=2$ case; in this regime scalar and electromagnetic particles generate mostly dipole radiation, which is forbidden for gravitational waves. It is satisfying to recover the Larmor result for electromagnetic radiation and the quadrupole formula for gravitational waves~\cite{Poisson:1993vp,Cardoso:2019nis,Santos:2024tlt}.

We can also evaluate the gravitational-wave amplitudes at infinity from the field $\psi$~\cite{Maggiore:2007ulw,JHU}, and find that for low frequencies and for circular motion,
\beq
h_+&=&-\frac{2Gm_p}{c^2r}\left(\frac{GM\Omega}{c^3}\right)^{2/3}(1+\cos^2\theta)\cos2\psi\,,\label{wave_aplitude_PP}\\
h_\times&=&-\frac{4Gm_p}{c^2r}\left(\frac{GM\Omega}{c^3}\right)^{2/3}\cos\theta\sin2\psi\,,\\
\psi&=&\Omega(t-r)-\phi\,,
\eeq
which agrees with the quadrupole result in the point particle limit~\cite{Blanchet:1996pi}.
We will not go through the (complex) details of radiation in arbitrary orbits, nor on how one can go beyond this leading order approximation. However, we highlight noteworthy aspects of the above results:

\noindent {\bf a.} One can expect to be able to extrapolate the above to the equal mass case and still get meaningful, correct order-of-magnitude estimates. If we promote $m_p\to M$ and maximize energy flux by making $r_p\to 6M$ one finds $\dot{E}\lesssim {\cal O}(\gamma^2,q^2/M^2, 1)$ for $s=0,1,2$ respectively. But for particles to get that close to a black hole, we expect that they are black holes themselves, hence subject to the bound $(\gamma,q/M)<1$. Thus, the luminosity (which is a dimensionless quantity in geometric units) is always less than unity. There seem to be deep connections between this bound and black hole formation, in what is known as the Thorne-Dyson conjecture, $\dot{E}\lesssim c^5/G$, momentarily restoring units~\cite{Cardoso:2018nkg}. The extrapolation to equal-mass results is formally not allowed. However, nonlinear calculations have shown that there is an ``unreasonable'' agreement with results extrapolated from perturbation theory, for reasons not yet fully understood~\cite{Berti:2007fi,Sperhake:2008ga,DeAmicis:2024eoy}.

\noindent {\bf b.} We have been treating emitting systems in a classical way. If a periodic system emits waves of typical frequency $\sim \Omega$, then if $\dot{E}/\Omega\lesssim \hbar \Omega$ the system is behaving quantum mechanically, in the sense that in one period less than one quantum of energy is emitted. One can easily check (see Exercise at end of Chapter \ref{ch:part4}) that astrophysically relevant systems behave classically indeed.

\noindent {\bf c.} Energy release as radiation of massless waves will force the system to evolve. As we will discuss later, in vacuum orbits tend to circularize and to become tighter (of smaller radius) leading to plunge and coalescence, and therefore to the formation of a single black hole as end product of the collision of two compact objects. A beautiful ringdown and relaxation via tail ensues, and therefore all that has been described above will arise as a consequence of gravitational-wave driven systems.

\section{Exercises}

\noindent{\bf $\blacksquare$\ Q.1.} In Section \eqref{sec:deltaqnms} we calculated the QNM spectrum of a delta-function potential. Show that the spectrum can be destabilized via the addition of a second ``weak'' delta-like ``bump.''

\vspace{0.25cm}

\noindent \textbf{{$\square$}\ Solution:}  Now let us replace the single delta potential of the previous problem by a double delta potential, that is, consider (see \ref{QNM-Double-Bump})
\begin{align}
    V(x)=V_{0}\delta(x)+V_{1}\delta(x-a)~.
\end{align}
The two Dirac delta potential divides the x-range into three regions. In this case we have to solve the master equation in the three regions respectively, see also Fig.~\ref{fig:spectral_instabilitya}

\noindent\textbf{Region-I:} In this region, located to the left of the global maximum in Fig.~\ref{fig:spectral_instabilitya}, the solution to the master equation is:
\begin{align}\label{region-1}
    \psi_{I}=C_{1}e^{-i\omega x}+C_{2}e^{i\omega x}~.
\end{align}
\textbf{Region-II:}
This is the cavity region, located within the two barriers in Fig.~\ref{fig:spectral_instabilitya}. Here, the solution to the master equation is:
\begin{align}\label{region-2}
    \psi_{II}=C_{3}e^{-i\omega x}+C_{4}e^{i\omega x}~.
\end{align}
\textbf{Region-III:}
In this region, to the right of the potential barriers, the solution to the master equation is:
\begin{align}\label{region-3}
    \psi_{III}=C_{5}e^{-i\omega x}+C_{6}e^{i\omega x}~.
\end{align}

To proceed further, let us demand that the solution is continuous across $x=0$ and $x=a$. This yields,
\begin{align}
    & C_{1}-C_{3}=(C_{4}-C_{2})~~\text{and}      \\
    & C_{4}-C_{6}=(C_{5}-C_{3})e^{-2i\omega a}~.
\end{align}
Similarly, exploiting the jump discontinuity of the derivative of the solution of the master equation at $x=0$ and at $x=a$ we get
\begin{align}
    & \left[(C_{4}-C_{2})+(C_{1}-C_{3})\right]=\frac{V_{0}}{i\omega}(C_{1}+C_{2})~~\text{and}                                             \\
    & \left[(C_{6}-C_{4})e^{i\omega a}+(C_{3}-C_{5})e^{-i\omega a}\right]=\frac{V_{1}}{i\omega}(C_{3}e^{-i\omega a}+C_{4}e^{i\omega a})~.
\end{align}
For the quasinormal mode problem we impose the following boundary condition,
\begin{align}
    & C_{1}=1,~C_{2}=0,                                         \\
    & C_{5}=\text{A}_{\text{in}},~C_{6}=\text{A}_{\text{out}}~.
\end{align}
Using these boundary conditions, we can solve for $C_{i}$'s (for $i=3,4,5,6$). The relevant guys are,
\begin{align}
    C_{3}=\frac{1}{2}\left(2+\frac{iV_{0}}{\omega}\right)~,
    C_{4}=-\frac{iV_{0}}{2\omega}~, \\
    C_{5}=C_{3}+\frac{iV_{1}}{2\omega}\left(C_{3}+C_{4}e^{2i\omega a}\right)~.
\end{align}
Now the quasinormal modes are characterized by the boundary conditions that modes will be purely outgoing at spatial infinity (or null infinity), that is by demanding $C_{5}=0$. This leads to,
\begin{align}\label{QNM-Double-Bump}
    (2\omega+iV_{0})(2\omega+iV_{1})+V_{0}V_{1}e^{2i\omega a}=0~.
\end{align}
The key physical message is that unlike the single delta potential, in this case we have a spectrum of quasinormal frequencies. One can solve the equation in the perturbatively in the regime $\frac{V_{1}}{V_{0}}e^{aV_{0}}\ll 1$~\cite{Barausse:2014tra}, the solution is,
\begin{align}
    \omega=-\frac{iV_{0}}{2}\left(1+\epsilon e^{aV_{0}}\right)~,\text{with}~\epsilon=\frac{V_{1}}{V_{0}}.
\end{align}
Thus, this represents a slight change in the original spectrum, slightly modified by the presence of a second barrier. Notice that the mode remains purely imaginary.

However, there is another branch of solutions for large $a$. To obtain it, let us segregate the real and imaginary parts of Eq.~\eqref{QNM-Double-Bump}. This gives,
\begin{align}\label{Re-Im-QNM-Double-Bump}
    4\omega_{R}^{2}-(2\omega_{I}+V_{1})(2\omega_{I}+V_{0})=-V_{1}V_{0}\cos(2\omega_{R}a)e^{-2a\omega_{I}}~, \\
    2\omega_{R}(4\omega_{I}+V_{0}+V_{1})=-\sin(2\omega_{R}a)e^{-2a\omega_{I}}V_{0}V_{1}~.
\end{align}
These two equations can be combined to yield,
\begin{align}
    \cot(2\omega_{R}a)=\frac{2\omega_{R}-(\frac{V_{1}}{2\omega_{R}}+\frac{\omega_{I}}{\omega_{R}})(1+\frac{\omega_{I}}{V_{0}})}{1+\frac{4\omega_{I}}{V_{0}}+\frac{V_{1}}{V_{0}}}
\end{align}
Now assuming $0<V_{1}\ll V_{0}\ll \omega_{R}$, $|\omega_{I}|\ll V_{1}$ and $a|\omega_{I}|\gg 1$, we find,
\begin{align}
    \omega_{R}=\frac{n\pi}{2a}~\text{with}~n=\text{odd~integers}~.
\end{align}
This solution corresponds, as we described earlier, to a low-frequency mode trapped in the two-barrier cavity. See also Fig.~\ref{fig:spectral_instability3}. Note that, given the approximations we used, solutions for even integers do not exist.
Finally, plugging the expression of $\omega_{R}$ in (\ref{Re-Im-QNM-Double-Bump}) and using the approximations, we find
\begin{align}
    \omega_{I}=-\frac{1}{2a} \ln\frac{n^{2}\pi^{2}}{a^{2}V_{0}V_{1}}~.
\end{align}
The simple example shows that adding a bump at large distance destabilizes -- in the sense that the longest-lived mode can be arbitrarily different from the original fundamental quasinormal mode of the system. The new family corresponds to a trapped mode oscillating with the cavity, slowly leaking out via tunneling.

\vspace{0.5cm}
\noindent{\bf $\blacksquare$\ Q.2.} Calculate the quasinormal frequencies of a Schwarzschild black hole.

\vspace{0.25cm}

\noindent{\textbf{{$\square$}\ Solution:}} We shall now carry out a numerical computation that forms the cornerstone of these lectures: the quasinormal modes of a black hole. We limit ourselves to spherical symmetry and for concreteness, we take $\ell=2$ and compute the QNM frequency for gravitational perturbations ($s=\pm2$). Building up on the numerical recipe introduced in the previous chapter, we adopt a very straightforward direct integration method. Our strategy is as follows:
\begin{itemize}
    \item We rewrite the wave equation Eq.~\eqref{EQ_master_frequencydomain} as
        \begin{equation}
            (f(r))^2 \frac{d^2 \psi}{dr^2} + f(r) f'(r) \frac{d \psi}{dr} + (\omega^2 - V_s)\psi = 0 \label{eq:wave_expanded}.
        \end{equation}
    \item We start by constructing the following series solution about $r=2 M$ 
        \begin{equation}
            \psi = (r - 2M)^\beta \sum_{j=0}^{N} h^H_{j} (r - 2M)^j
        \end{equation}

    \item Inserting the above ansatz into Eq.~\eqref{eq:wave_expanded}, we get
        \begin{align}
            & 4M^2(4M^2\omega^2 + \beta^2)h^H_0 + (2M(3 - \beta + 2\beta^2 - \ell(\ell + 1) + 16M^2\omega^2)h^H_0 \nonumber \\&+ 4M^2(1 + 4M^2\omega^2 + 2\beta + \beta^2)h^H_1 (r - 2M) + (\cdots)(r - 2M)^2 + \dots = 0
        \end{align}
    \item The above near-horizon expansion tell us that $\beta = \pm 2 i M \omega$. To impose ingoing boundary conditions at the horizon we set $\beta = -2i M \omega$ since $e^{- i \omega r_*} \approx (r - 2M)^{-2i M \omega}$ near the horizon. This directly follows from the definition of the tortoise coordinate. We can use the above series expansion to determine the various $h^H_{i>0}$'s in terms of $h^H_0$, $\ell$, $\omega$ and $M$. This task can be automated by equating the coefficients of various powers of $(r-2M)$ to zero and solving simultaneously for the various $h_i$'s. Using this analytical solution and its derivative as the two initial conditions, we can integrate the wave equation up to a point $r_m>2M$.
    \item We then carry out a similar step near $r \to \infty$. But here we take the ansatz
        \begin{equation}
            \psi = e^{\kappa r} r^n \sum_{j=0}^{N} \frac{h^{\infty}_j}{r^j}\,.
        \end{equation}
    \item Putting this ansatz into Eq.~\eqref{eq:wave_expanded}, we get,
        \begin{equation}
            (\kappa^2 + \omega^2)h^{\infty}_0 + \dfrac{(-4\kappa^2 M + 2\kappa n)h^{\infty}_0 + (\kappa^2 + \omega^2)h^{\infty}_1}{r} + \dfrac{(\dots)}{r^2} + \dots = 0
        \end{equation}
        and we can conclude that
        \begin{equation}
            \kappa^2 = -\omega^2, \qquad n = -\dfrac{2M\omega^2}{\kappa}.
        \end{equation}
        Since we are interested in the outgoing solution at infinity, we choose $\kappa = i \omega$. Notice that this again lines up with what we expect from the definition of the tortoise coordinate, since near infinity $r_* \approx r + 2M \ln r$ and $e^{i \omega r_*} \approx e^{i \omega r}r^{2Mi\omega}$. Now, we use this analytical solution and its derivative as initial conditions to integrate the wave equation from infinity to $r_m$.
    \item We therefore end up with two numerical solutions ($\psi^N_H$ and $\psi^N_\infty$) approaching $r_m$ from either sides (we also set $h^{H/\infty}_0=1$) and these two solutions must be linearly dependent. In other words, the Wronskian must vanish.

    \item We therefore pick those values of $\omega$ which makes the Wronskian vanish, and we do this in the following way: first we demand continuity at the matching point $r_m$ by defining the scaling factor $b = \psi^N_H(r_m) / \psi^N_\infty(r_m)$. We then enforce smoothness by finding those values of $\omega$ that makes the derivatives continuous by solving $\Delta\psi'(\omega) \equiv \psi^N_H{'}(r_m) - b \cdot \psi^N_\infty{'}(r_m) = 0$. It should be evident that $\Delta\psi'(\omega)$ is nothing but the good old Wronskian. For numerical stability, we may normalize this quantity by $\psi^N_H{'}(r_m)$.
\end{itemize}

This algorithm has been implemented in the notebook \coderef{code:QNMs-DI.nb}. We find that the fundamental quasinormal mode frequency of the Schwarzschild black hole for gravitational perturbation for the $\ell=2$ mode is
$$\omega M=0.373672 - i 0.0889623.$$
We have also demonstrated that the result does not depend on the matching point $r_m$ and given error estimates. One may also calculate the first overtone for the $\ell=s=2$ mode (but this requires a much higher order series expansion about infinity to get a stable solution).

This approach closely follows the pioneering work of Chandrasekhar and Detweiler \cite{Chandrasekhar:1975zza} (although they had chosen to integrate the Zerilli equation after transforming it into the Riccati equation for numerical convenience). We close this exercise with a few more comments: one can tweak the prescription we have described here to, say, match the solution at the boundaries after shooting (for example, see \cite{Aimer2023}). In fact, there exist a plethora of methods and strategies to find quasinormal modes \cite{Berti:2009kk}, each with their own set of advantages and quirks. These methods often complement each other, and one's choice depends on the demands of the physics and the system one is looking to investigate~\cite{Berti:2009kk,Berti:2025hly}.

\vspace{0.5cm}
\noindent \textbf{$\blacksquare$\ Q.3.} Show that the quasinormal mode (QNM) eigenfunctions become \emph{ill-defined}
in the limits $r_* \to \pm \infty$ when the standard quasinormal boundary conditions are imposed.
Explain the physical and geometric reason for this divergence,
and show how adopting a \emph{hyperboloidal foliation} of spacetime
renders the eigenfunctions regular at the event horizon and at future null infinity.

\vspace{0.25cm}

\begin{figure}[htbp]
    \centering
    \begin{minipage}[t]{0.32\textwidth}
        \centering
        \includegraphics[width=\linewidth]{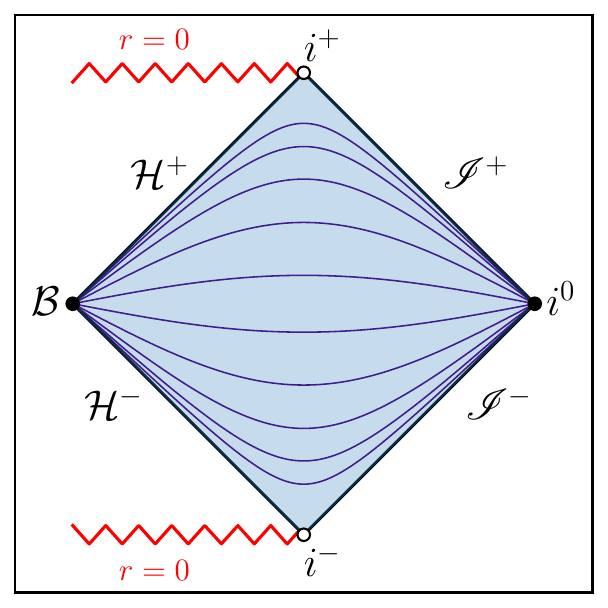}
    \end{minipage}
    \hfill
    \begin{minipage}[t]{0.32\textwidth}
        \centering
        \includegraphics[width=\linewidth]{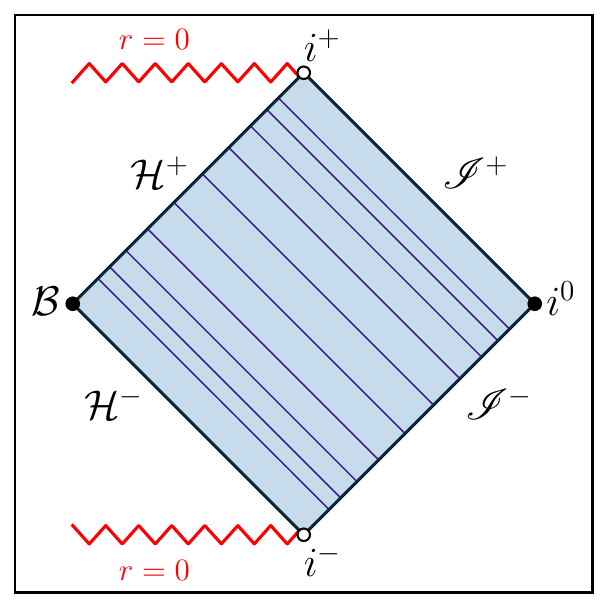}
    \end{minipage}
    \hfill
    \begin{minipage}[t]{0.32\textwidth}
        \centering
        \includegraphics[width=\linewidth]{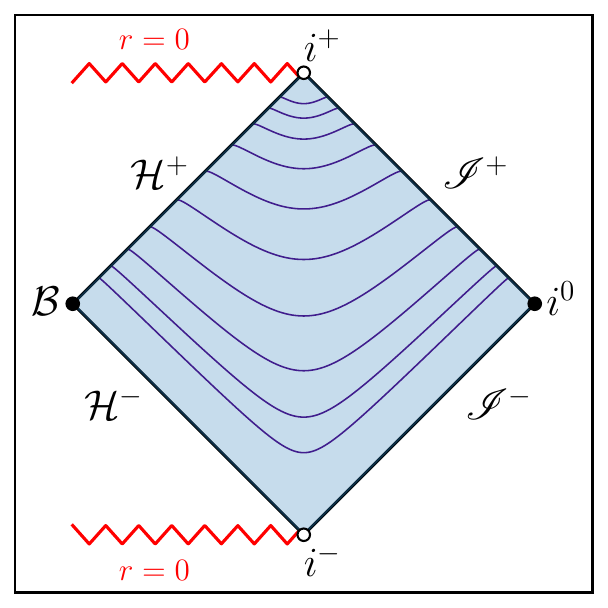}
    \end{minipage}
    \caption{Conformal (Penrose) diagrams showing constant $t$ (right panel), constant $v$ (middle panel), and constant $\tau$ foliations of the Schwarzschild spacetime with $M = 0.5$. In these figures, $\mathcal{H}^+$ and $\mathcal{H}^-$ are the future and past event horizons, $\mathcal{B}$ is the bifurcation sphere at the intersection of the two event horizons, $i^+$ and $i^-$ are the future and past timelike infinity, $\mathscr{I}^+$ and $\mathscr{I}^-$ are future and past null infinity, respectively, $i^0$ is the spatial infinity, and $r=0$ represents the curvature singularity. The figure on the extreme right shows the hyperboloidal slicing of a Schwarzschild black hole which is emerging as a natural arena for studying perturbations in various black hole spacetimes. The constant time slices were generated using the transformation equations involved in obtaining the conformal compactification of the Schwarzschild metric.}
    \label{fig:penrose_all}
\end{figure}

\noindent\textbf{{$\square$}\ Solution:}We have seen that solutions to the wave equation when subjected to the quasinormal mode boundary conditions have the form,
$$\Psi_n(t, r_*) = e^{-i \omega_n t}\psi_n(r_*)\,,$$
and that these perturbations must decay over time if we want the black hole spacetime to be stable, that is,
\begin{equation}
    \Im \omega_n \equiv \omega_I <0.
\end{equation}
The above feature seems physically meaningful, the wave resembles a damped sinusoid. However, it results in a rather strange behavior at the spatial infinities. By definition, the QNMs satisfy the following boundary conditions,
\begin{equation}
    \psi(\omega,r_*) \simeq e^{\pm i \omega r_*} = e^{\pm i \omega_R r_*}e^{\mp \omega_I r_*}= e^{\pm i \omega_R r_*}e^{\pm \varpi_I r_*} , ~~r_* \to \pm \infty, \label{blowup}
\end{equation}
where (suppressing the index $n$) $\omega_R$ and $\omega_I$ correspond to the real and imaginary parts of the quasinormal mode frequency, and we have defined $\varpi \equiv -\omega_I>0$. Clearly, since $\varpi$ is positive, the modes diverge exponentially both at spatial infinity and at the bifurcation sphere\footnote{In the Penrose diagram, the limit $r_* \to \infty$ with $t$ fixed corresponds to the point $i^0$ which is called spatial infinity. It is the meeting point of the future and past null infinities. The bifurcation sphere $\mathcal{B}$ corresponds to the point we can reach by imposing the limit $r_* \to -\infty$ with $t$ fixed, and it corresponds to the point where the future and past event horizons meet. We refer the reader to \cite{Hawking:1973uf} for a more detailed introduction to Penrose diagrams.} [c.f. extreme left panel of Fig.~\ref{fig:penrose_all}].

The divergence of the modes at the bifurcation sphere and at spatial infinity can be treated in a simple manner. The issue can be traced back to the fact that we carried out our computations on $t=$ constant hypersurfaces, that is, we kept $t$ fixed while taking $r^*\to \pm \infty$ [c.f. \eqref{blowup}].
It can be shown that if we consider an alternative foliation of the spacetime such that the constant time hypersurfaces are made to penetrate the event horizon $\mathcal{H^+}$ and future null infinity $\mathscr{I}^+$, then we can avoid the pathological behavior at $i^0$ and $\mathcal{B}$. Such spacelike surfaces are called hyperboloidal foliations. Although very popular in numerical relativity, the idea that we could use such foliations in the study of black hole perturbations was first proposed in Ref.~\cite{Zenginoglu:2011jz} which we shall follow closely below.

The reason why such hyperboloidal coordinates do the trick is rather easy to understand. The pathology in the QNMs stems from the boundary conditions [c.f. \eqref{blowup}] which, loosely speaking, were put in by hand at $i^0$ and $\mathcal{B}$. The bad behavior of the eigenmodes at $\mathcal{B}$ can be cured if we instead use a coordinate system that is ingoing at the event horizon $\mathcal{H}^+$. Such a  horizon penetrating foliation would then naturally respect the ingoing boundary conditions at $\mathcal{H}^+$. For example, the ingoing Eddington-Finkelstein coordinates shown in the middle panel of Fig. \ref{fig:penrose_all} are indeed horizon penetrating. However, these coordinates do not penetrate $\mathscr{I}^+$. Hence, if we construct a set of coordinates that are ingoing at $\mathcal{H}^+$ and outgoing at $\mathscr{I}^+$, then the resultant hyperboloidal foliation will implement the QNM boundary conditions geometrically.
The foliation by construction points outside the domain of interest and have been shown in the extreme right panel of Fig. \ref{fig:penrose_all}.

To construct our hyperboloidal coordinates, let us introduce a new time coordinate $\tau$, such that
\begin{equation}
    \tau(t,r_*) = t - H(r_*), \label{newtau}
\end{equation}
where $H(r_*)$ is called the height function. Under this coordinate transformation, the Schwarzschild metric \eqref{eq:Schwarzschild_geometry} becomes,
\begin{equation}
    ds^2 = -f(r)d \tau^2 - f(r)\left(2 \mathcal{H} d \tau d r_* -\left[1 - \mathcal{H}^2\right] d r_*^2\right) + r^2 d \Omega^2_2,
\end{equation}
where we have introduced $\mathcal{H}:= dH/dr_*$. The purpose of making this coordinate transformation was to make the above metric ingoing at the event horizon $\mathcal{H}^+$ and outgoing at $\mathscr{I}^+$. In other words, the coordinate $\tau$ should asymptote to the null coordinates $v$ and $u$ at $\mathcal{H^+}$ and $\mathscr{I^+}$ respectively. Therefore, this metric should resemble the ingoing and outgoing Eddington-Finkelstein metrics at $\mathcal{H}^+$ and $\mathscr{I}^+$ respectively.  We, therefore, impose the following conditions on $\mathcal{H}$,
\begin{equation}
    |\mathcal{H}|\leq 1, ~~ \lim_{r_* \to \pm \infty} \mathcal{H}= \pm 1, ~~   \lim_{r_* \to \pm \infty} \dfrac{d \mathcal{H}}{d r_*}=0\,. \label{assymp}
\end{equation}

It is now straightforward to show that under the transformation given by \eqref{newtau}, the wave equation \eqref{eqn:wave equation} becomes\cite{Jaramillo:2020tuu},
\begin{equation}
    \left(\dfrac{\partial^2}{\partial r_*^2}  -   \left[ 2  \mathcal{H}\dfrac{\partial^2}{\partial_\tau \partial r_*}       + \left(1- \mathcal{H}^2\right) \dfrac{\partial^2}{\partial \tau_*^2}      + \dfrac{d \mathcal{H}}{d r_*}\dfrac{\partial}{\partial_\tau}   \right]  -V_\ell(r_*) \right)\Psi(\tau,r_*)=0.
\end{equation}
We can perform a Fourier transform,
\begin{equation}
    \Psi(\tau,r ) = \dfrac{1}{2 \pi}\int_{-\infty}^{\infty}e^{-i \omega \tau} \psi_H(\omega,r) d \omega. \label{ft2}
\end{equation}
and recast the above equation into,
\begin{equation}
    \left(\dfrac{d^2}{d r_*^2}  +    2 i \omega \mathcal{H}\dfrac{d}{d r_*}     + \omega^2 \left(1- \mathcal{H}^2\right)       + i \omega \dfrac{d \mathcal{H}}{d r_*}     -V_\ell(r_*) \right)\psi_H(\omega,r_*)=0.
\end{equation}
Note that from Eqs.~\eqref{def_Fourier_transform}, \eqref{newtau} and \eqref{ft2}, we can see that changing to the hyperboloidal foliation is equivalent to a rescaling of the modes, that is $\psi_H(\omega,r_*)=e^{-i\omega H}\psi(\omega,r_*)$, in the frequency domain. Note that before performing numerics in the usual $(t, r)$, we often factor out the bad behavior of the eigenmodes at the boundary by invoking the boundary conditions and choosing an appropriate ansatz. The choice of such ansatz can be directly related to the aforementioned rescaling and hence the hyperboloidal foliation provides a nice geometrical picture to what may seem as a purely algebraic manipulation \cite{PanossoMacedo:2018hab}.

Now, by invoking \eqref{assymp}, we can then write down the asymptotic form of the above equation, viz.,
\begin{equation}
    \left(\dfrac{d^2}{d r_*^2}  \pm  2 i \omega \dfrac{d}{d r_*}  \right)\psi_H(\omega,r_*)=0 ~~ \mathrm{as} ~~ r_* \to \pm \infty,
\end{equation}
which has the {general} solution,
\begin{equation}
    \psi_H(\omega, r_*) \simeq C_0 \pm C_1 e^{\mp 2 i \omega r_*} ~~ \mathrm{as} ~~ r_* \to \pm \infty.
\end{equation}
The above solution is regular at the boundaries. So the boundary conditions that we must  now apply reduces to saying that the solution should be of the order unity at $H^+$ and $\mathscr{I^+}$. As a result, the rescaled  quasinormal modes are now well-behaved in the exterior of the black hole. However, notice that we have quite a bit of gauge freedom in choosing how these hyperboloidal slices intersect $\mathscr{I}^+$ and there are several choices one can make. In Fig.~\ref{fig:penrose_all}, while plotting the $\tau$ constant hypersurfaces, we chose $H(r)=\sqrt{1+r^2}$ \cite{Zenginoglu:2006rj, Zenginoglu:2007jw}. Moreover, the hyperboloidal approach can be used in conjunction with a class of numerical techniques called \emph{spectral methods} to convert the problem of finding quasinormal mode frequencies into a linear algebra problem that involves finding the eigenvalues of (somewhat large) matrices. These advances have led to the development of new tools (imported from disciplines like fluid dynamics) that have increased our understanding issues such as spectral instabilities. We refer the reader to Refs.~\cite{Jaramillo:2020tuu, PanossoMacedo:2023qzp, Destounis:2023ruj, PanossoMacedo:2024nkw, Zenginoglu:2025sft,Berti:2025hly} for more details.



\chapter{Part IV}
\label{ch:part4}
\allowdisplaybreaks
\minitoc

\section[Tracking the motion and dynamics of compact objects]{Tracking the motion and dynamics of compact objects: inspiral and ringdown}

We went through a possible strategy to infer black hole parameters or to test General Relativity from ringdown data. The approach was relatively straightforward since the ringdown was modeled in a fairly simple, but robust way\footnote{Based on, as we saw, a solid physical and mathematical foundation.}: a series of exponentially damped sinusoids. In general, any dynamical process radiates gravitational waves, with a pattern that encodes (and in general, mimics) the motion of the source but which can be very rich and complex. For example, the simplest of gravitational-wave sources is a binary system on a quasi-circular orbital motion, losing energy to gravitational radiation. As time progresses, the binary components get closer, their relative velocity increase, and eventually they coalesce. If the end product is a black hole, it will relax by emitting radiation in quasinormal modes of the final remnant black hole, as we have discussed in the previous chapter. The pre-merger phase is {\it long} and the approach to capture the entire signal needs a higher level of sophistication. Therefore, let us focus on the pre-merger, large distance stage.

In vacuum General Relativity, the dominant (in a velocity or orbital frequency expansion) fluxes are quadrupolar. For points particles they are expressed by Eq.~\eqref{Quadrupolar_flux_PP} and read~\cite{Maggiore:2007ulw}\footnote{Here, for later purposes we consider the energy loss or gain rate of the bound system, the minus sign translates the fact that it is loosing energy.},
\be
\dot{E}_{\rm quad}=-\frac{32}{5}\mu^2a^4\Omega^6\,.\label{E_quad}
\ee
The reduced mass $\mu=M_1M_2/(M_1+M_2)$ can be expressed in terms of the chirp mass ${\cal M}$ as
\be
\mu=\left(\frac{{\cal M}^5}{M^2}\right)^{1/3}\,,
\ee
where the chirp mass is defined by
\be
{\cal M}=\frac{(M_{1}M_{2})^{3/5}}{(M_{1}+M_{2})^{1/5}}\,.
\ee

For circular motion, the semi-major axis $a$ can be expressed in terms of the orbital frequency ($\Omega$) via Kepler's law
\beq
a=\left(\frac{M}{\Omega^2}\right)^{1/3}\,,\label{Kepler_law}
\eeq
and the orbital frequency can in turn be expressed in terms of the GW frequency
\be
\Omega=\pi f_{\rm gw}\,.\label{orbital_gw_frequency}
\ee
The energy lost by the system needs to come from the only available place, which is the binding energy. Since the total energy of a binary is
\be
E_{\rm orb}=-\frac{M\mu}{2a}\,,
\ee
we find the following relation for the evolution of the frequency (and therefore separation),
\beq
&&\frac{dE_{\rm orb}}{dt}+\dot{E}_{\rm quad}=0\,\\
&&f_{\rm gw}=\frac{5^{3/8}}{8\pi{\cal M}^{5/8}(t_c-t)^{3/8}}\,,\label{fgw_pn_leading}
\eeq
where $t_c$ is some coalescence time at which the approximation\footnote{The approximation being that the system is evolving adiabatically through a series of quasi-circular orbits, or equivalently, $\dot{\Omega}\ll \Omega^{2}$.} breaks down. The result for the time dependence of the gravitational-wave frequency together with Kepler's law \eqref{Kepler_law} -- \eqref{orbital_gw_frequency} then implies that the separation {\it decreases} with time, the flux output and wave amplitude \eqref{wave_aplitude_PP} and \eqref{E_quad} {\it increase} with time. This gives rise to the now famous ``chirp'' of the gravitational-wave signal as shown in Fig.~\ref{fig:ringdown} indicating that the frequency and amplitude increase with time.

To be more general, we can consider an orbit with some eccentricity $e$, and write its energy and angular momentum as
\beq
E&=&-\frac{M_1M_2}{2L}\,,\nonumber\\
L^2&=&\frac{M_1^2M_2^2a(1-e^2)}{M}\nonumber
\eeq
On the other hand, a generic result for radiation is
\be
\frac{\dot{L}^{\rm rad}}{\dot{E}^{\rm rad}}=\frac{m}{\omega}=\frac{1}{\Omega}	\nonumber
\ee
Equating $\dot{E}^{\rm rad}=-\dot{E}$ and $\dot{L}^{\rm rad}=-\dot{L}$, we find
\beq
\dot{a}&=&-\frac{2a^2\dot{E}^{\rm rad}}{M_1M_2}\leq0 \label{semimajor_axis_evol_GR}\\
\dot{e}&=&\sqrt{\frac{M}{a}}\frac{\sqrt{1-e^2}}{e}\frac{\dot{E}^{\rm rad}}{M_1M_2}\left(\frac{\dot{L}^{\rm rad}}{\dot{E}^{\rm rad}}-\frac{\sqrt{1-e^2}}{\Omega}\right)\,.
\eeq
Thus, eccentricity evolution is a delicate balance~\cite{Cardoso:2020iji}.

Gravitational-wave detectors are sensitive to a combination $h$, of the two $h_+,\,h_\times$ polarizations, weighed by the detector form factors~\cite{Cutler:1994ys}:
\be
h=\left(\frac{384}{5}\right)^{1/2}\frac{\pi^{2/3}Q\mu M}{Da}\cos\int 2\pi f_{\rm gw}dt\,,
\ee
where $D$ is the distance from detector to the binary. The factor $Q$ depends on the position and orientation of the source, but we will not be concerned with it any further.

The results \eqref{dipolar_scalar_flux_PP} --\eqref{Quadrupolar_flux_PP}, or result~\eqref{E_quad} are in fact only the leading order term of an expansion in powers of the orbital angular frequency $\Omega$. In other words, these expressions provide the dominant contribution to the energy fluxes at large separations, within vacuum General Relativity. The corrections become more and more important at higher velocities, hence at higher frequencies, and hence expression \eqref{fgw_pn_leading} for the phase of the gravitational wave will also receive corrections.

Now we know how to evolve a low-frequency binary and a low frequency gravitational-wave signal. Ideally, one would be able to ``stitch'' this evolving signal to a ringdown signal, hence obtaining the full waveform. However, signal from black hole binaries is so complex and rich that this procedure requires a lot of fiddling and careful comparisons of higher order slow motion expansions and nonlinear, numerical simulations~\cite{Buonanno:2006ui,Berti:2007fi,Aylott:2009tn,Hinder:2013oqa}.

\section{The post-Newtonian expansion of the phase}
The tool of excellence in gravitational-wave astronomy to search for signals is matched filtering~\cite{Cutler:1994ys,Maggiore:2007ulw,Maggiore:2018sht}. It requires a precise knowledge of the gravitational wave that is expected to impinge on the detector. A matched filter is the one that maximizes the signal-to-noise ratio of an event in the presence of Gaussian white noise. It is matched because it coincides with the signal itself. Although we will not deal with data analysis issues in these lectures, understanding how searches are done is important knowledge. But for our immediate purpose in these lectures, it is relevant to note that it provides a framework to deal with mis-modeled events which might be missed altogether, due to ignorance or non-inclusion of the physics at work.

For many applications, specially those concerning data analysis, it is best to work in Fourier domain and to define the transform
\be
\tilde{h}(f)=\int_{-\infty}^{+\infty}e^{2\pi i f t}h(t)dt \,.
\ee
Note that the Fourier transform $\tilde{h}$ is nothing but the transform defined in \eqref{def_Fourier_transform} with the frequency $2\pi f\equiv \omega$.

Highly oscillatory integrands are cumbersome to evaluate, and Fourier transforms fall in that category. However, one can find a good approximation in certain conditions. In particular, if $h(t)=A(t)\cos\phi(t)$ and $A(t)$ is a slowly varying function of time, in the sense that $d\log A/dt\ll d\phi/dt$ and $d^2\phi/dt^2\ll (d\phi/dt)^2$ then a stationary phase approximation yields~\cite{Cutler:1994ys},
\be
\tilde{h}(f)=\frac{A}{2}\left(\frac{df}{dt}\right)^{-1/2}e^{i(2\pi f t-\phi(f)-\pi/4)}\,,
\ee
where $t$ here is defined as the instant at which $d\phi/dt=2\pi f$, and $\phi(f)=\phi(t(f))$.

With these definitions, from Eq.~\eqref{fgw_pn_leading} we find
\be
t=t_c-5\left(8\pi f_{\rm gw}\right)^{-8/3}{\cal M}^{-5/3}\,.
\ee
The phase defined above is then
\beq
\phi(f)&=&\int 2\pi f_{\rm gw}dt=\phi_c-2\frac{5^{-5/8}(t_c-t)^{5/8}}{{\cal M}^{5/8}}\nonumber\\
&=&\phi_c-2\left(8\pi {\cal M}f_{\rm gw}\right)^{-5/3}\,.
\eeq
Therefore, finally,
\beq
\tilde{h}&=&\frac{Q}{D}{\cal M}^{5/6}{f_{\rm gw}}^{-7/6}e^{i\Psi_{\rm gw}}\,,\\
\Psi_{\rm gw}&=&2\pi f_{\rm gw} t_c-\phi_c-\frac{\pi}{4}+\frac{3}{128}(\pi {\cal M}f_{\rm gw})^{-5/3}\,.
\eeq
Interestingly, in the frequency domain the amplitude decreases with frequency.

These results are a good description of the signal and its phase at low orbital frequencies. However, the phase of the signal has a very important bearing on matched filtering, and hence on maximizing the signal-to-noise ratio. Thus, getting an accurate tracking of the phase of gravitational-waves is important. In vacuum General Relativity, several tools can be used to calculate next-to-leading order corrections to the above result. One of them, consists of expanding the Einstein equations in powers of $v/c$, with $v$ being a typical velocity of the interacting objects and $c$ the speed of light. One calls it a ``post-Newtonian'' expansion, of which we have seen the dominant term in fluxes and phase. One can calculate in General Relativity higher order corrections,
which will affect both the amplitude and phase $\Psi$ of the waveform. Since the phase of the signal is much more important for matched filtering, we usually parametrize only the phase, and write:
\beq
\Psi&=&\frac{3}{128}(\pi {\cal M}f_{\rm gw})^{-5/3}\nonumber \\
&\times&\left(...+\alpha_{-4{\rm pN}}x^{-4}+...+\alpha_{-1{\rm pN}}x^{-1}+1+\alpha_{1{\rm pN}}x+...\right)\,,\label{pN_expansion_phase}
\eeq
where $x=(\pi M f_{\rm gw})^{2/3}$ is a velocity parameter. In vacuum GR all pN coefficients of negative order vanish and, up to 1.5 pN,
\be
\Psi_{\rm gw}=\frac{3}{128}(\pi {\cal M}f_{\rm gw})^{-5/3}\left(1+\frac{20}{9}\left(\frac{743}{336}+\frac{11\mu}{4M}\right)x-16\pi x^{3/2}\right)\,.\label{1p5pN}
\ee

It is therefore customary and instructive to quantify the impact of a non-vacuum setup or of a modified theory of gravity on the gravitational-wave phase $\Psi_{\rm gw}$, via the above post-Newtonian form. Constraints on some of the coefficients are readily available~\cite{Seoane:2021kkk,Perkins:2020tra}. Now we have a framework in place to understand how to calculate the gravitational wave phase in vacuum General Relativity, given knowledge of fluxes and energy balance. Our purpose in the following is to outline what needs to be changed once one starts considering compact binaries evolving within astrophysical environments, and how they might affect the gravitational-wave phase.

\section{Environmental effects}
%
\begin{figure}
    \centering
    \includegraphics[width=1\textwidth]{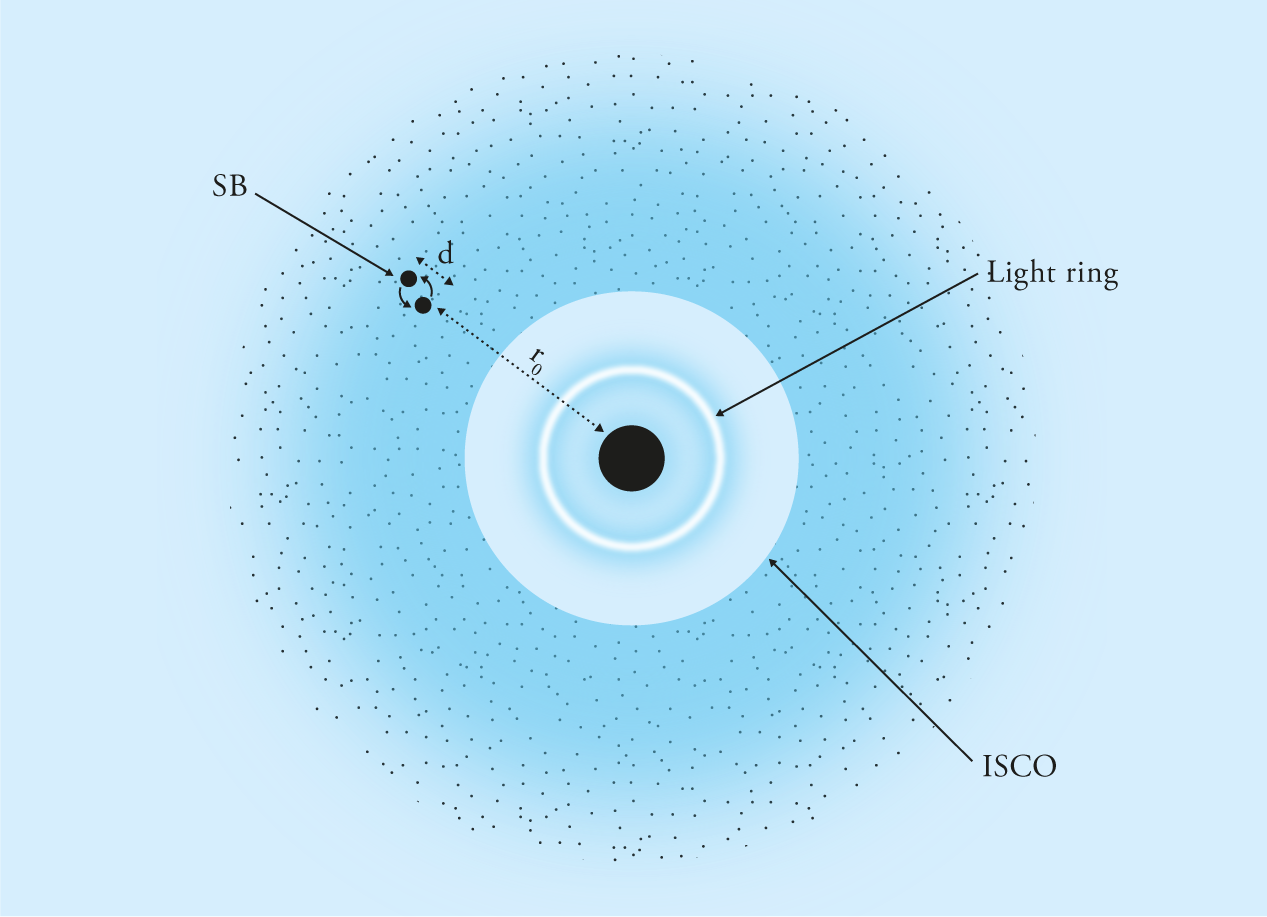}
    \caption{
        {\small A supermassive black hole, surrounded by an environment (plasma, interstellar dust, satellites, stellar-mass black holes or stellar-mass binaries (SB) etc.). The environment changes the light ring properties and the way the black hole relaxes. A stellar-mass binary evolving in its vicinity will be affects by accretion and drag of the environment or by tidal effects from the supermassive central object. Gravitational or kinematic Doppler effects, etc. will all contribute to change the signal with respect to the vacuum, isolated scenario, and they should all be taken into account, both for parameter estimation and for tests of General Relativity.}
    }
    \label{fig:BEMRI}
\end{figure}

In General Relativity and especially in black hole physics, it is standard practice to assume vacuum or that the presence of matter causes negligible back-reaction\footnote{Then, the equations to solve are relatively simple, with a simple equation of state for black holes. This has built a frame of mind in the community that all problems can be precisely posed and must have a quantitative precise answer, with no regards for the extreme uncertainty and variation in astrophysical environments.}. These arguments typically involve arguing that, for example, the density of matter
$G^3\rho M^2/c^6=1.6\times 10^{-18} \rho/\rho_{\rm water} M^2/M_{\odot}^2$, or that the gravitational potential of a galaxy is of order $10^{-5}$. The numbers are correct, but it is important to quantify how much a small effect piles up as a binary performs thousands or millions of cycles.

We would now like to overview environmental effects that may have an impact on the phase \eqref{pN_expansion_phase}, potentially affecting our ability to detect some events or significantly bias parameter estimation, i.e., leading us to believe that the waves are coming from a system (mass, spin, distance, etc.) which is different from what it actually is. We envision {\it any} possible effect coming from the real universe. A possible example concerning the immediate vicinity of a supermassive black hole is depicted in Fig.~\eqref{fig:BEMRI}. A supermassive black hole is surrounded by gas, stars and planets and other black holes, some of them bound to other black holes (for example, a stellar-mass black hole binary might be orbiting the supermassive black hole). The phase of the gravitational-wave signal from the stellar-mass binary will be affected by the supermassive black hole and by the environment, and this will in turn influence the relaxation of the supermassive black hole itself.

How extreme can environments be? The Sagittarius A$^*$ (Sgr A$^{*}$) source at the center of our galaxy is a supermassive black hole candidate with mass $M\sim 4\times 10^6M_\odot$. It is surrounded by a galaxy, the Milky Way, of total mass $M_{\rm H} \sim 5 \times  10^{11}M_\odot$ within a radius
of $a_{\rm H} \sim 20 {\rm kpc}$ (corresponding to $a_{\rm H} \sim 10^4 M_{\rm H}$ in geometrical units)~\cite{2019A&A...621A..56P,Jiao:2023aci}. Thus, one should question how accurate is a Kerr description of the spacetime in these circumstances.

The merging of two neutron stars to form a black hole, or even hyperluminous accretion disks can produce very high accretion rates around black holes. In these circumstances, the stationary assumption about the geometry also may break down in immediate postmerger events. Including environments is also crucial to test our ideas about fundamental issues such as spectral instability, see for example the ``guitar in a room'' discussion of Section~\ref{sec:guitar}. Finally, we want to test General Relativity. Any meaningful attempt to test the underlying description of gravity must ensure that all the astrophysical systematics are under control.

How do we model the full complexity of astrophysical systems out there in a fully General Relativistic approach? Moreover, how do we solve for the dynamics of compact objects in such environments, when even dynamics in vacuum is itself such a challenging problem? To analyze data, we need proper astrophysical models, which include the impact of environments on gravitational radiation from compact objects.
Including environments in relativistic, gravitational-wave emitting systems is currently a very active field of research, with only partial answers. We will make headway on two fronts, by first modeling environments in a fashion that is as simple as possible. These will be constant-density environments and the language will be, mostly, Newtonian. We will then increase the level of complexity by adding General Relativistic effects, at the level of the background solution, motion and gravitational-wave emission.
\section{A black hole moving in a constant density medium}

\subsection{Accretion\label{sec:accretion}}
When a black hole or compact object moves in an environment,
it interacts electromagnetically and gravitationally with it. For electrically neutral environments, contact forces will cause friction proportional to the object area. We consider only black holes here, in which case there are only long-range gravitational forces at play. If the interaction happens too close to the black hole, the material will be accreted. Accretion depends on what exactly composes the environment, and the forces (pressure) at play. For dust-like particles, we can use geodesics to understand accretion\footnote{This is a very reasonable approach, as the near-horizon region is to a very good extent matter-depleted and backreaction is extremely small.}. More broadly, when the mean-free path of the particles composing the medium is larger than the size of the perturber, we are in a collisionless regime of accretion\footnote{V.C.~was reminded during these lectures of the distinction between collisionless and fluid-like environments is very much similar to the different flow of traffic in India and in Europe.}.
But we studied timelike geodesics. For low-energy particles, the absorption cross-section was found to be Eq.~\eqref{cross_section_lowE}, $\sigma_{\rm abs}=16\pi\frac{m_p^2}{v^2}$\footnote{Here and in what follows, the black hole accreting or being dragged will have mass $m_p$. We want to eventually make it orbit around a supermassive black hole of mass $M$, but this will be done later on.}.

Given a black hole moving with velocity $v$ in an environment of density $\rho$, accretion in the collisionless regime is then described by
\be
\dot{m}_p=\sigma_{\rm abs} \rho v\,.
\ee

On the other hand, if the radius $R_p$ of the object is comparable to or larger than the mean free path,
$R_p\gg \lambda$, then accretion becomes a macroscopic process and cohesion forces and matter compressibility must be taken into account. When the small compact
object is a black hole, this type of accretion is described by the
Bondi-Hoyle formula~\cite{1941MNRAS.101..227H,Bondi:1944rnk,Bondi:1952ni,Shapiro:1983du,Giddings:2008gr}
\be
\dot{m}_p=4\pi \kappa \frac{\rho m_p^2}{(v^2+c_s^2)^{3/2}}\,,\nonumber
\ee
where $\kappa$ is a constant of order unity and $c_s$ the local sound speed.

All of the above assumes that the Compton wavelength of the fields making up the environment are small, the constituents behave as particles. For accretion of fields with a large wavelength, one can use the formalism and some of the results from the scattering of waves in Section~\ref{sec:scattering}. A complete analysis of massive scalar fields accreting onto rotating black holes can be found in Refs.~\cite{Cardoso:2022nzc,Vicente:2022ivh,Traykova:2023qyv}.

\subsection{Dynamical friction}
Black holes and compact objects must be able to transfer energy to the medium they evolve in, even without accretion. This is a classical and important calculation in electrodynamics, important for detection of charged particles and radiation damaging etc~\cite{Bohr:1948}. In gravitational physics, the calculation proceeds in the same way, since electromagnetism and gravity both fall like the inverse square of the distance.

Consider two pointlike objects of mass $m_p,\,m_2$ with object 2 at rest. An exact calculation gives that the energy transferred from particle 1 to particle 2 as~\cite{Bohr:1948}
\be
T=\frac{2m_p^2m_2}{v^2}\frac{1}{b^2+(m_p+m_2)^2/v^4 }\,,
\ee
with $\mu$ the reduced mass and $b$ the impact parameter. Let's try to understand what is happening.
\begin{figure}
    \centering
    \includegraphics[width=1.0\textwidth]{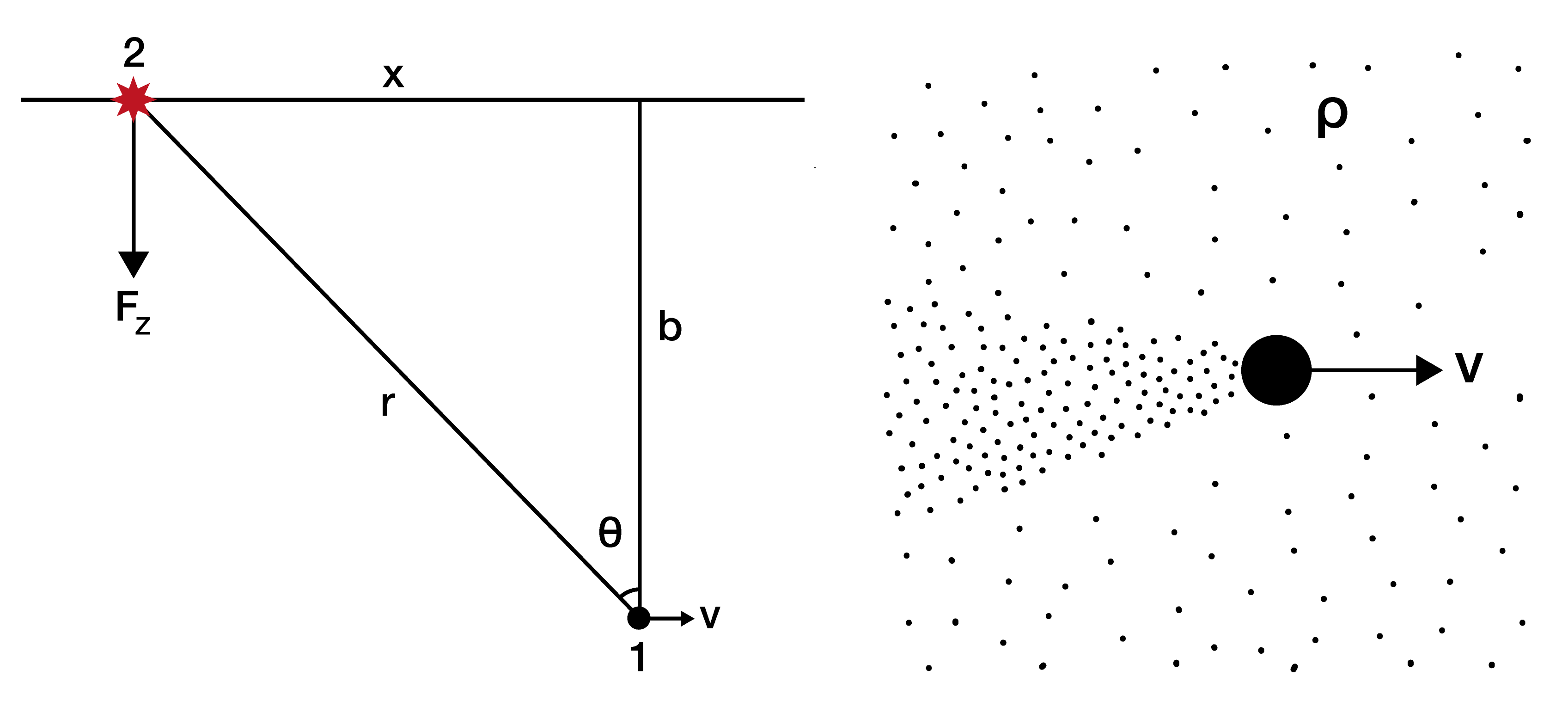}
    \caption{
        {\small {\bf Left:} A black hole (Object 1) is moving in an environment composed of individual stars, and passes close to one (Object 2) with impact parameter $b$ and velocity $v$. Part of the Object 1 kinetic energy will be transferred to Object 2. {\bf Right:} The stars move, creating an overdensity wake along the path of the black hole, that pulls on it, slowing it down, in what we term dynamical friction or gravitational drag.}
        \label{fig:passingstar}}
\end{figure}
The setup is illustrated in Fig.~\ref{fig:passingstar}. A black hole (or any massive perturber within an environment) of mass $m_p$ is traveling with a velocity $v$ relative to a medium, of density $\rho$. Let us focus on the immediate neighbor, a star of mass $m_2$ at rest in this frame of the medium. Before and after the passage of the black hole, the component of the gravitational force along the velocity averages out, but the orthogonal component $F_z$ does not,
\beq
F_z&=&\frac{m_pm_2}{b^2+x^2}\cos\theta=\frac{m_pm_2b}{(b^2+x^2)^{3/2}}\nonumber\\
&=&\frac{m_pm_2}{b^2}\left(1+\left(\frac{vt}{b}\right)^2\right)^{-3/2}\,,
\eeq
With this, we can calculate the momentum transfer and hence the change in velocity $\delta v$ and kinetic energy $\delta E$ acquired by the Object 2,
\beq
\delta v&=&\frac{1}{m_2}\int_{-\infty}^{+\infty}dt F_z=\frac{2m_p}{bv} \,,\\
\delta E&=&\frac{1}{2}m_2\delta v^2=\frac{2m_p^2m_2}{b^2v^2}\,.
\eeq

Now imagine the black hole to be moving through a constant density environment, as in the right panel of Fig.~\ref{fig:passingstar}. Stars, planets and dust and plasma will move upon passage, creating an overdensity wake trailing the black hole. The gravitational attraction from the wake on the moving black hole will cause it to slow down. This kind of frictional force is different from that of normal viscous forces which originate due to velocity gradients between various layers of fluids. In this scenario, one can infer from the above, taking an annulus of radius $b$ and width $db$ that
\beq
\frac{dE}{dt}&=&-\int_{R_{\rm min}}^{R_{\rm max}}\frac{2m_p^2\rho\times 2\pi bdb v}{b^2v^2}\nonumber\\
&=&-\frac{4\pi m_p^2\rho}{v} \log \left(\frac{R_{\rm max}}{R_{\rm min}}\right)\,.
\eeq
The result can also be framed in terms of a friction force $F_{\rm df} v=dE/dt$. In the above, $R_{\rm max}$ is the extent of the environment and $R_{\rm min}$ is the size of the moving object; in the context of black holes, we can take $R_{\rm min}\sim 3\sqrt{3}m_p$ as the critical impact parameter, the distance below which particles are simply accreted.

The $1/r$ nature of the gravitational potential causes the
total transferred energy to diverge; this phenomenon is
akin to the divergence of the scattering cross-section of
the Coulomb potential~\cite{Merzbacher:1998fi}. Within our approach the transferred energy always needs to be smaller than the energy of the moving black hole. Within General Relativity, a natural cutoff radius arises, since for radii $r^2\rho\gtrsim 1$ a black hole would form.

The general situation, where the (collisionless) medium has a velocity distribution was handled by Chandrasekhar~\cite{Chandrasekhar:1943ys}. If the environment is isotropic and made of constituents with a Maxwellian velocity distribution with dispersion $\sigma$, then the drag is expressed as a force directed along the motion
\be
F=-\frac{4\pi m_p^2\rho \log \left(\frac{R_{\rm max}}{R_{\rm min}}\right)}{v^2}\left({\rm erf}{X}-\frac{2Xe^{-X^2}}{\sqrt{\pi}}\right)\,,\label{chandra_drag}
\ee
with $X=v/(\sqrt{2}\,\sigma)$. One will notice the appearance of a similar ``cutoff'' logarithm as above. Shortcomings and specially the suitability of this result for collisionless systems are discussed by Binney and Tremaine~\cite{2008gady.book.....B}.

The above was a Newtonian calculation. In fact, it is interesting that one can also do a similar computation within
General Relativity~\cite{Cardoso:2019dte}. We will not repeat the calculation here, but merely point out that friction {\it has} to occur even in a fully relativistic setting. Consider a moving black hole and throw a photon, of energy $E_i$ in the same frame where the black hole moves, such that it bounces off close to the light ring and is reflected at an angle of $180\degree$ by the strong-field region (such orbits do exist). Then, a trivial change of frames and consequent blueshift yields, for the reflected photon the energy $E_f$~\cite{Cardoso:2019dte}
\be
E_f=\frac{1+v}{1-v}E_i\,.
\ee
Here, $v$ is the velocity of the black hole.
If the black hole is approaching the photon, $v$ is positive and the energy of the reflected photon is larger than that of the incoming photon. The above formula is exactly the energy gain of a photon reflecting off a moving mirror. It's all kinematics.

The treatment of fluid-like environments was done by Ostriker~\cite{Ostriker:1998fa}. The result reads as follows
\be
F_{\rm df}=-\frac{4\pi m_p^2\rho}{v^2}I\,,
\ee
with
\be
I=\frac{1}{2}\log\left(\frac{1+v/c_s}{1-v/c_s}\right)-v/c_s\,,
\ee
for subsonic motion $v<c_s$, and
\be
I=\frac{1}{2}\log\left(1-c_s^2/v^2\right)+\log\left(\frac{vt}{R_{\rm min}}\right)\,,
\ee
Here, $v$ is the relative velocity between the moving object and the medium, which is assumed homogeneous. Note that the scalings with black hole mass and velocity are the same as the collisionless behavior \eqref{chandra_drag}.

All the above refers to linear motion in a medium where the composing fundamental particles have a very small Compton wavelength. For scalars, generic formulae assuming a fixed background are available and supported by fully relativistic numerical simulations~\cite{Vicente:2022ivh,Dyson:2024qrq,Traykova:2023qyv}. These results predict a number of interesting features when a black hole crosses a medium, such as a Magnus force, lifting, etc.

\section{Isotropic spherically symmetric environments}

Now place a supermassive black hole of mass $M$ at the center of a collisionless distribution of stars or gas or dark matter. As we saw, both accretion and dynamical friction depend on relative velocities, hence the exact nature of the environment plays a crucial role in understanding the impact of the dynamics of compact binaries.

We start with perhaps the simplest possible scenario: an environment where average velocities of the particles composing the environment are zero, they are random and isotropic, and the environmental density is spherically symmetric around the center where a supermassive black hole of mass $M$ sits. Take a stellar or intermediate mass black hole of mass $m_p$ to be orbiting, on a circular motion of radius $r_p$, the supermassive black hole. Consider a general matter distribution with density~\cite{Eda:2013gg,Macedo:2013qea}
\be
\rho =\frac{\alpha M_{\rm H}}{4\pi {\cal L}^\alpha}r^{\alpha-3}\,,
\ee
with $\alpha$ a constant, and ${\cal L}$ a typical length-scale, corresponding to an enclosed environment mass
\be
m(r) =M+M_{\rm H}\left(\frac{r}{{\cal L}}\right)^\alpha\,,
\ee
%

\subsection{Accretion}
Neglecting dynamical friction for the time being, the motion of the stellar-mass object is governed by the gravitational force
\be
F_g=\frac{m(r)m_p}{r^2}\,.
\ee
Without radiation reaction, evolution of this system is determined by the fact that the intrinsic angular momentum of the environment vanishes, and thus
\be
L^2=r_p^2m_p^2v^2=m(r)m_p^2r_p\,,
\ee
is a conserved quantity. Taking time derivatives we find immediately,
\be
\dot{r}_p=-2\frac{r_p\dot{m}_p}{m_p(1+r_pm'/m)}\,,
\ee
with the accretion rate described in Section~\ref{sec:accretion}. Notice that, in absence of central supermassive black hole, $1+rm'/m\equiv \beta=1+\alpha$~\cite{Macedo:2013qea}. Also note that for most environments with a central supermassive black hole and a perturber sufficiently close to it, then $1+rm'/m\sim 1$.

For collisionless matter, we then find
\be
\dot{r}_p=-32\pi\frac{r_p m_p \rho}{v \beta}\,,\label{accretion_constant_density}
\ee
where the Keplerian velocity $v^2=m(r_p)/r_p$. We stop here since knowledge of $\dot{r}$ is sufficient to calculate the leading order dephasing (with respect to vacuum) in the gravitational wave.
\subsection{Dynamical friction}
In this setup, dynamical friction will change the orbital angular momentum by
\be
\frac{dL}{dt}=-rF_{\rm df}\sim -\kappa m_p^2 r_p \frac{\rho(r_p)}{v^2}\,,
\ee
where $\kappa$ is a dimensionless constant of order unity, which depends on whether the environment is collisionless or fluid-like.

To find the orbital evolution, we use Kepler's law $v^2=m(r_p)/r_p$ to find $L=m_p\sqrt{m(r_p) r_p}$ and hence
\be
2\frac{dL}{dt}=m_p\sqrt{\frac{m(r_p)}{r_p}}\beta\dot{r}_p\,,
\ee
where, as before $\beta=(1+\alpha, 1)$ depending on whether there's a supermassive black hole at the center of the configuration or not. Equating one gets
\be
\dot{r}_p=-2\frac{\kappa}{\beta}\frac{m_p  r_p^{5/2}}{m^{3/2}}\,\rho\,.\label{DF_constant_density}
\ee
A number of interesting results can be worked out from here~\cite{2008gady.book.....B}, but this is sufficient for our purposes.
\section[A black hole orbiting a disk]{A black hole orbiting a disk: migration, tidal torques and gaps}
The calculations above show that bodies are indeed affected by environments, but we assumed linear motion in a constant density environment. Extension to circular motion is known, but is limited to constant density environments~\cite{Kim:2007zb,Buehler:2022tmr}.
Let us use our procedure above to derive a more general description of friction in the presence of circular motion, both for the perturber and for the medium. We have in mind an environment like an accretion disk, orbiting around a supermassive black hole, within which a stellar-mass object is immersed. Since a significant fraction of the drag is caused by the closest particles, we can use an impulse approximation similar to the above. However, in addition to energy transfer, energy momentum transfer is key to determine the fate of the orbiting object, and that requires to keep one order higher of the approximations~\cite{2008gady.book.....B,1979MNRAS.186..799L,1979MNRAS.188..191L,1993prpl.conf..749L,1986ApJ...309..846L,1980ApJ...241..425G,Armitage:2002uu,Shapiro:2013qsa}.

Take a stellar-mass object of mass $m_p$ within a disk, moving with some Keplerian velocity on a circular trajectory of radius $r_p$ around some supermassive black hole of mass $M$. Consider a small lump of disk of mass $m_2$. Then, if the disk matter $m_2$ is at a radius larger than that of the mass $m_p$, it will drag it: it is moving slower and hence, like the star in Fig.~\ref{fig:passingstar} will remove energy and angular momentum from $m_p$. The reaction of the on this lump is to transfer angular momentum and energy to it, driving it to larger distances from the central body.

However, disk matter in an inner orbit, will produce the opposite effect, it will accelerate the mass $m_p$ and transfer energy to it. The reaction on this part of the disk is to extract angular momentum and energy from it, driving it to {\it smaller} orbital distances from the central supermassive black hole. Since the effect is more pronounced in regions close to $m_p$, the stellar-mass object will open an {\it annular gap} around it. Or in other words, an orbiting satellite within a disk empties the disk in its immediate vicinity. For collisionless fluids, this gap will remain and widen as the stellar-mass object evolves. For baryonic matter, competition between viscous forces and pressure may close the gap, depending on the timescales involved. In the following we neglect tidal truncation.

The transfer of angular momentum during a close encounter of the sort discussed in the context of Fig.~\ref{fig:passingstar} is~\cite{1979MNRAS.186..799L,1979MNRAS.188..191L,1993prpl.conf..749L,1986ApJ...309..846L,1995MNRAS.277..758S,Armitage:2002uu}
\be
\Delta L=\frac{m_p^2m_2r_p}{b^2v^3}\,.
\ee

We can proceed in a similar fashion by integrating over impact parameter. This time however, it is convenient to define an accretion disk surface density $\Sigma(t,r)$ such that the disk is thin along the $z$ direction and $d\Sigma=\rho dz$. It is also convenient to calibrate the final result to numerical simulations~\cite{Armitage:2002uu}. One therefore finds
\be
\frac{dL}{dt}= \int 2\pi\Sigma \Lambda r dr\,,
\ee
with
\be
\label{tidal_torque}
\Lambda= \left\{
\begin{array}{ll}
    -\frac{fq^2M}{2r}\left(\frac{r}{\Delta_p}\right)^4  & r<r_p \\
    \frac{fq^2M}{2r}\left(\frac{r_p}{\Delta_p}\right)^4 & r>r_p \\
\end{array}
\right.
\ee
Here $f\sim 0.01$ is a dimensionless constant and $\Delta_p={\rm max}(h, |r-r_p|)$. Assuming Kepler's law $L^2=Mm_p^2r_p$ 
(hence $2L\dot{L}=M m_p^2\dot{r}_p$), the orbital evolution of the stellar-mass object is governed by,
\be
\dot{r}_p=\frac{\sqrt{r_p}}{m_p\sqrt{M}}\int_{r_{\rm ISCO}}^\infty 4\pi \Sigma \Lambda r dr\,.\label{satellite_migration}
\ee
%

\section{Gravitational-wave dephasing due to accretion and drag}
We can now make contact between environmental effects and gravitational wave physics, via the contribution of the former to the dephasing \eqref{pN_expansion_phase}. Let's express the rate of change of separation of a binary in circular motion \eqref{semimajor_axis_evol_GR} with the help of the quadrupole formula \eqref{E_quad} but allowing for other effects,
\be
\dot{r}_p=-\frac{64}{5}m_pM^2r_p^{-3} - \epsilon K \frac{r_p^{\gamma}}{M^\gamma}\,,
\ee
where $\epsilon$ is a small quantity (proportional to the environmental density for dynamical friction or accretion), $\gamma$ depends on the physics under consideration and $K$ the overall proportionality constant.

Expanding to first order in $\epsilon$ and integrating one finds
\be
t=\frac{5}{4096 M^4 m_p^2}\left(
-\frac{16m_p M^{10/3}}{(\pi f_{\rm gw})^{8/3}}
+5K\epsilon\frac{M^{(7-2\gamma)/3}(\pi f_{\rm gw})^{-2(\gamma+7)/3}}{(7+\gamma)}
\right)\,,
\ee
where we used Kepler's law and relation \eqref{orbital_gw_frequency}. We can thus find the corrections to the gravitational-wave phase as
\beq
\Psi_{\rm gw}&=&\frac{3}{128}\left(\pi m_p^{3/5} M^{2/5}\,f_{\rm gw}\right)^{-5/3}\left(1-K_{\Psi_{\rm gw}}x^{-3-\gamma}\right)\,,\\
K_{\Psi_{\rm gw}}&=&\frac{25\epsilon KM^{-1-2\gamma/3}\pi^{-2(3+\gamma)/3}}{16(7+\gamma)(11+2\gamma)m_p}
\eeq

With this we can see for example that if the black holes were charged they would emit dipolar radiation, for which $\gamma=-2$~\cite{Cardoso:2020iji} affects the phase at -1PN order. For constant density environments, Eqs.~\eqref{accretion_constant_density} -- \eqref{DF_constant_density} give\footnote{The -7.5 for accretion disks assumes that $\Sigma$ is a constant and that the evolution of the perturber depleted the disk for $r<r_p$. We then use \eqref{satellite_migration}.}
%
%

\begin{equation}
    -3-\gamma = 
    \begin{cases}
        -4.5 & \text{collisionless accretion} \\
        -5.5 & \text{isotropic collisionless friction} \\
        -7.5 & \text{depleted accretion disk}
    \end{cases}
    \label{PN_dephasing_correction_environments}
\end{equation}

\begin{figure}
    \centering
    \includegraphics[width=0.9\textwidth]{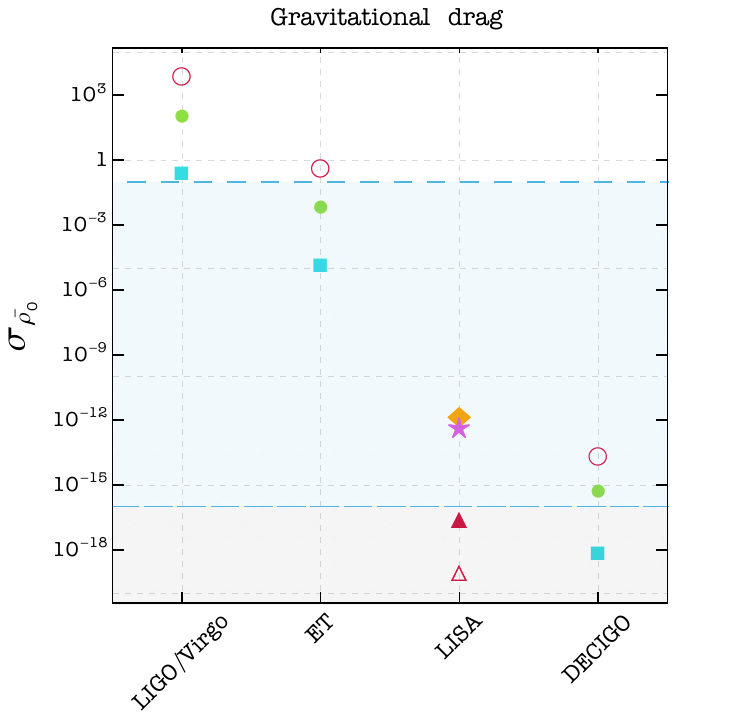}
    \caption{
    {\small $1-\sigma$ uncertainties on the density parameter $\rho$ normalized to the average density of water $\rho_{H_2O} \approx 10^3{\rm kg}/{\rm m}^3$ for different sources and detector configurations, obtained via (lack of) dephasing from dynamical friction. Different point markers
    identify distinct sources. For ground-based detectors and for DECIGO we consider sources with the same parameters of GW150914, GW170608 and GW170817~\cite{LIGOScientific:2018mvr}. For LISA we consider a massive and an intermediate-massive systems, as well as IMRI and EMRI, all located
    at $d = 1$ Gpc from the detector.
    The shaded blue area denotes densities typical or smaller than that of accretion
    disks, while the gray shaded region denotes densities typical of dark matter. From Ref.~\cite{Cardoso:2019rou}.}
    \label{fig:drag_maselli}
    }
\end{figure}
The leading order corrections to the gravitational-wave phase for constant density, isotropic environments coincide with those derived with other methods~\cite{Barausse:2014tra,Cardoso:2019rou} and can be used to set observational constraints on the type of astrophysical environments that compact binaries evolve in~\cite{Cardoso:2019rou,CanevaSantoro:2023aol}. One example is shown in Fig.~\ref{fig:drag_maselli}: future detectors will be able to impose impressive constraints on the density of astrophysical environments.

It should be clear that the approach described above is a primitive way to consider environmental effects. Accretion and drag were mostly calculated within a setup where densities are constant or very simple, and gravitational drag force is a Newtonian concept, which was calculated only for slowly moving objects. The background where the objects move was also taken to be a vacuum background upon which ad hoc matter distributions are placed, without any regard for backreaction. It is, nevertheless, a robust starting point, and for most cases the best we have. A few fully semi-relativistic results exist. A self-consistent approach would build a consistent relativistic solution in General Relativity and then perform systematic studies of its dynamics. As we mentioned, this is a vibrant field, we now briefly overview a couple of selected topics.

\section{Environmental effects and strong-field gravity}

\subsection{A tidal perturber}
As the simplest case, take a companion of mass $M_c$ some distance $R$ away from a supermassive black hole of mass $M$, inducing a small perturbation in the spacetime close to the black hole such that
\be
g_{\mu\nu}=g_{\mu\nu}^{\rm Sch}+\epsilon h_{\mu\nu}\,.
\ee
As we described, small fluctuations can be handled by expanding in tensorial harmonics. If we take the companion far away and nearly static, then
    {\small
        \begin{eqnarray}
            h_{\mu\nu}^{\rm polar}=\sum_{\ell m}\left(\begin{array}{cccc}
                    f H_0^{\ell m} & H_1^{\ell m}        & 0             & 0                          \\
                    H_1^{\ell m}   & f^{-1} H_2^{\ell m} & 0             & 0                          \\
                    0              & 0                   & r^2K^{\ell m} & 0                          \\
                    0              & 0                   & 0             & r^2\sin^2\theta K^{\ell m}
                \end{array}\right) Y_{\ell m}\nonumber
        \end{eqnarray}}
We get $H_1=0,\, H_2=H_0$ and the regular solution at the horizon

\be
H_2^{\ell=2}=-{\cal E}_2r^2(1-2M/r)\,,\label{solH0_tides}
\ee
with ${\cal E}_2$ the tidal moment. To find this moment, we can expand the Newtonian potential of a monopole of mass $M_c$ located at a distance $R\gg r$, as\footnote{Notice that at large distances the Newtonian potential grows without bound. We are not allowed to use the full expression at very large $r$, since it is an expansion in powers of $r/R$. A similar comment can be made with respect to the General Relativistic solution~\eqref{solH0_tides}. A proper handling of tides would involve calculating the spacetime describing two masses, a highly challenging task. One usually ascribes the growing piece to the far companion.}
\beq
U&=&\frac{M}{r}+\frac{M_c}{|{\vec{r}}-\vec{R}|}\nonumber\\
&=&\frac{M}{r}+M_c\sum_{\ell m}\frac{4\pi}{2\ell+1}\frac{r^\ell}{R^{\ell+1}}Y^*_{\ell m}(\theta_c,\phi_c)Y_{\ell m}(\theta_\phi)\,.
\eeq
Therefore for the dominant $\ell=2$ component,
\be
g_{tt}^{PN}=-1+\frac{2M}{r}+M_c\sum_{m}\frac{8\pi}{5}\frac{r^2}{R^{3}}Y^*_{2 m}(\theta_c,\phi_c)Y_{2 m}(\theta_\phi)\,.
\ee

We thus found the tidal moment and metric perturbation parameter
\beq
{\cal E}_2&=&-\frac{8\pi M_c}{5R^3}\sum_{m} Y^*_{2 m}(\theta_c,\phi_c)\,,\\
\epsilon&=&\frac{M^2M_c}{R^3}\,.
\eeq
In other words,
\beq
H_0&=&H_2=-\frac{8\pi \epsilon}{5M^2}r^2(1-2M/r)\sum_{m} Y^*_{2 m}(\theta_c,\phi_c)\,,\\
K&=&\frac{8\pi \epsilon}{5M^2}(r^2-2M^2)\sum_{m} Y^*_{2 m}(\theta_c,\phi_c)\,.
\eeq
Now, it is straightforward to find corrections to geodesics. For instance timelike motion is now described by Kepler's modified law
\be
\Omega^2=\frac{M(1-2\epsilon)}{r^3} + \frac{\epsilon}{M^2}\,,
\ee
which will also affect the gravitational-wave phase at some order.

The quasinormal modes of tidally deformed black hole geometries are unknown\footnote{Note that, as we saw previously, the tidal Love numbers of black holes vanish in General Relativity. Nevertheless, their geometry {\it is} affected by companions, and therefore their dynamics and motion in their vicinity too.}. However, as we argued, there's very strong evidence suggesting that prompt ringdown is locally associated with the properties of null geodesics, which can easily be computed. Again for a companion on the pole of a massive black hole, the location and frequency of the light ring reads~\cite{Cardoso:2021qqu}
\be
r_{\rm LR}=3M(1+5\epsilon)\,,\qquad M\Omega_{\rm LR}=\frac{1+5\epsilon}{3\sqrt{3}}\,.
\ee
Unfortunately, we know close to nothing about the wave dynamics of black holes close to compact companions. In particular numerical simulations capturing the full geometry of tidally deformed black holes are urgently needed.
\subsection{Exact solutions for black hole geometries in environments\label{exact_environment_BH_cluster}}
We have dealt at some length with a non-rotating black hole in vacuum, and we know how to study its reaction under ``small'' external perturbations. But as we already stated, all black holes are surrounded by matter, some of them by {\it a lot} of matter. To test our ideas about {\it real} black holes in the universe, we now aim to construct a stationary solution to the Einstein equations that does not use approximations, and that represents a black hole surrounded by an arbitrarily large and heavy distribution of matter.

Solving the Einstein equations {\it for realistic} matter states is not a trivial task, and black hole solutions might not even exist if one demands, in addition, stationarity. Nothing in the cosmos is stationary, but on some timescales stationarity might be a good approximation to a certain problem. The absence of solutions is more of a problem of our ignorance of how to find interesting slow-evolving solutions, rather than a property of the physics at play.

Nevertheless, let us continue with the assumption of stationarity, since it allows a valuable comparison to the results obtained in vacuum. In spherical symmetry, one can think of taking matter on circular orbits across all of space, and averaging over all directions. One would end up with a spherically symmetric, stationary configuration with some effective stress tensor. This approach was devised by Einstein~\cite{Einstein:1939ms,Geralico:2012jt}, and is sometimes known as the Einstein Cluster. The Einstein construction assumes a stress tensor $\langle T_{\mu\nu}\rangle =
    \frac{n}{m_p} \langle P_\mu P_\nu \rangle$, with $n$ the number density of particles with mass
$m_p$ and $P$ the four-momentum satisfying the geodesic equations.

Alternatively, it is easily shown that this construction
is equivalent to assuming an anisotropic material with only
tangential pressure $P_t$, and vanishing radial pressure,
$T_{\mu\nu} ={\rm diag}\left(-\rho,0,P_t,P_t \right)$. We can find a black hole geometry, taking the ansatz~\cite{Cardoso:2021wlq}
\be
ds^2=-f\, dt^2+\frac{dr^2}{1-2m(r)/r}+r^2d\Omega^2\,. \label{eq:galactic_ansatz}
\ee
The field equations yield
\beq
\frac{rf'}{2f}&=&\frac{m(r)}{r-2m(r)}\,,\\
2P_t&=&\frac{m(r)\rho}{r-2m(r)}\,,\\
m'&=&4\pi r^2\rho\,.
\eeq
Therefore, given a density profile (which, admittedly, varies from galaxy to galaxy, if the goal is to describe the geometry of supermassive black holes sitting at the center of galaxies), hence a mass profile, the problem is closed. Imposing $f\to 1$ at large distances one can solve for all quantities, given a density profile, inspired by observations. For Hernquist-type profiles~\cite{Cardoso:2021wlq},
\beq
m(r)&=&M_{\rm BH}+\frac{Mr^2}{(a_0+r)^2}\left(1-\frac{2M_{\rm BH}}{r}\right)^2\,. \label{Eq:Hernquist-type-profile}
\eeq
This profile describes a black hole of mass $M_{\rm BH}$ at the center of ``halo'' of mass $M$ and typical length-scale $a_0$. Note that when $M_{\rm BH}=0$ one recovers the true Hernquist profile describing some galaxies. The factor $\left(1-\frac{2M_{\rm BH}}{r}\right)^2$ was inserted by hand, to have a simple closed for solution to the field equations. In particular, we find,
\beq
f(r)&=&\left(1-\frac{2M_{\rm BH}}{r}\right)e^{\Upsilon}\,,\\
\Upsilon&=&-\pi\sqrt{\frac{M}{\xi}}+2\sqrt{\frac{M}{\xi}}\arctan{\frac{r+a_0-M}{M\xi}}\,,\\
\xi&=&2a_0-M+4M_{\rm BH}\,.
\eeq

We derive this solution in an exercise at the end of this chapter using the notebook \coderef{code:Galactic-BH.nb}. The procedure can be generalized to any density distribution in a straightforward manner. Note that we can also express
\be
f=\left(1-\frac{2M_{\rm BH}}{r}\right)e^{2U}\,,
\ee
where $U$ is the Newtonian potential of matter distribution, indeed a general result when the external matter is not strongly gravitating~\cite{Duque:2023seg}.
Notice that for a mass $M$ distributed on a scale $a_0\gg M_{\rm BH}$ then $U_{\rm LR} \sim U(0)\sim -M/a_0$, and indeed this coincides with the large $a_0$ limit of the exact solution. The potential $U_{\rm LR}$ gives the redshift of a photon as it crosses the galaxy, due to its matter content.

We will not explore the above geometry in any detail, except to point out that at large distances the geometry is dictated by the external matter, whereas close to the horizon at $r=2M_{\rm BH}$ the geometry is dictated by the black hole, and there is a redshift, an imprint of the external mass. We learned that strong-field phenomena of interest depend on the light ring. For low matter compactness $a_0/M$, we find
\beq
r_{\rm LR}&=&3M_{\rm BH}\left(1+\frac{M M_{\rm BH}}{a_0^2}+{\cal O}(1/a_0^3)\right)\,,\\
M_{\rm BH}\Omega_{\rm LR}&=&\frac{1}{3\sqrt{3}}\left(1-\frac{M}{a_0}+\frac{M(M+18M_{\rm BH})}{6a_0^2}\right)\,,\\
M_{\rm BH}\lambda&=&\frac{1}{3\sqrt{3}}\left(1-\frac{M}{a_0}+\frac{M^2}{6a_0^2}\right)\,.
\eeq
To the dominant order in the Newtonian potential, $r_{\rm LR}=3M_{\rm BH}(1+MU_{\rm LR}')$ with $U_{\rm LR}'=U'(r=3M_{\rm BH})$, leading to the orbital frequency and Lyapunov scale (as measured by distant observers),
\be
M_{\rm BH}\Omega_{\rm LR}=\frac{1+U_{\rm LR}}{3\sqrt{3}}\,,\qquad
M_{\rm BH}\lambda=\frac{1+U_{\rm LR}}{3\sqrt{3}}\,.
\ee
The above then indicates that the leading order consequence of matter distributions is a redshift relative to vacuum modes, as has also been observed via time-domain signals~\cite{Spieksma:2024voy,Pezzella:2024tkf}. In other words, for low-compactness distributions, it emerges that ringdown is affected through only a redshift of the corresponding frequencies. Thus, the ringdown of a matter-surrounded black hole, is to a good extent, indistinguishable from that of a heavier vacuum one~\cite{Spieksma:2024voy}.

We will not describe the active field of research that is currently exploring these and other solutions~\cite{Fernandes:2025osu,Destounis:2025tjn}, working out not only the ringdown, but also the inspiralling stage of compact binaries~\cite{Barausse:2014tra,Cardoso:2022whc,Figueiredo:2023gas,Speeney:2024mas,Pezzella:2024tkf,Alnasheet:2025mtr,Dyson:2025dlj,Duque:2023seg,Santos:2025ass,Duque:2025yfm,Vicente:2025gsg,Duque:2024mfw,Vicente:2022ivh}. Suffice it to say that the tools and principles behind this field of work rely on the techniques explored in these notes, and aim to extend the results we described to spacetimes with arbitrary matter distributions.


\section{Exercises}
\noindent{\bf $\blacksquare$\ Q.1.}  I am waving my hand at a friend, producing a time-varying quadrupole moment. Do I emit gravitational waves?
\vspace{0.25cm}

\noindent\textbf{{$\square$}\ Solution}~In this case we model the waving of hands as motion in the x-direction. If one of the hand has mass $M$ then the matter density (in the slow motion approximation this is actually the energy density) $\rho(x^{i})$ is
\begin{align}\label{}
    \rho(x^{i})=M\delta(x-\bar{x}(t))\delta(y)\delta(z)~,
\end{align}
where $\bar{x}(t)$ is the instantaneous position of hand, and mathematically it is modeled as $\bar{x}(t)=A \cos (\omega t)$ with $\omega=\frac{2\pi}{T}$, where $T$ is the period of the oscillation. Using the mass quadrupole formula in flat spacetime, namely
\begin{align}
    Q_{ij}=\int d^3x \rho(x^i)\left(x_ix_j-\frac{1}{3}r^2\delta_{ij}\right)~,
\end{align}
one can show that only the diagonal terms of the quadrupole tensor survives and these are given by
\begin{align}
    Q_{xx}=\frac{2}{3}M(\bar{x}(t))^{2},~Q_{yy}=Q_{zz}=\frac{1}{3}M(\bar{x}(t))^{2}~.
\end{align}
Therefore, the flux of gravitational wave becomes,
\begin{align}
    \dot{E}=\frac{G}{5c^{5}}\left<\left(\frac{d^{3}Q_{xx}}{dt^{3}}\right)^{2}+\left(\frac{d^{3}Q_{yy}}{dt^{3}}\right)^{2}+\left(\frac{d^{3}Q_{zz}}{dt^{3}}\right)^{2}\right>~\,,
\end{align}
where we will take the average $\left<\ldots\right>$ over one period of oscillation. Then it can be shown that
\begin{align}
    & \dddot{Q}_{xx}=\frac{4M}{3}\left(\bar{x}\dddot{\bar{x}}+3\dot{\bar{x}}\ddot{\bar{x}}\right)=\frac{8M}{3}A^{2}\omega^{3}\sin(2\omega t)~\text{and}~      \\
    & \dddot{Q}_{yy}=\dddot{Q}_{zz}=\frac{2M}{3}\left(\bar{x}\dddot{\bar{x}}+3\dot{\bar{x}}\ddot{\bar{x}}\right)=\frac{4M}{3}A^{2}\omega^{3}\sin(2\omega t)~.
\end{align}
Then we get the following expression,
\begin{align}
    \dot{E}=\frac{G}{45c^{5}}M^{2}A^{4}\omega^{6}(64+16+16)\left<\sin^{2}(2\omega t)\right>=\frac{96G}{45c^{5}}M^{2}A^{4}\omega^{6}\times \frac{1}{2}=\frac{3072G}{45 c^{5}}\frac{A^{4}M^{2}}{T^{6}}\pi^{6}
\end{align}
If we consider this as a quantum process, then the number of graviton emitted in one period is,
for a waving hand of mass $M\sim 1\,{\rm Kg}$ and doing motion of amplitude $A=1\,{m}$ and period $T=1\,{\rm sec}$,
\begin{align}
    \frac{\dot{E}T}{\hbar\omega}=   \frac{1536~GA^4M^2\pi^5}{45\hbar c^5T^4}\sim 10^{-15}
\end{align}
Thus, the waving of a hand is, from the point of view of gravitational radiation, a quantum process, no ``graviton'' (if gravitons exist) is emitted in a single period!


\vskip 1cm

\noindent{\bf $\blacksquare$\ Q.2.} Black holes or neutron stars could carry electromagnetic charge. In some theories of gravity, such as scalar-tensor theories, they can also acquire a scalar charge. The net effect of such charges is that the system emits dipolar radiation, as well as gravitational waves. Use Larmor's formula, to estimate the dephasing of gravitational waves caused by emission of dipolar radiation.

\vspace{0.25cm}
\noindent\textbf{{$\square$}\ Solution:}
In this problem we consider that the members of the binary carries electric charges. In particular let us assume that the body with mass $m_{1}$ carries charge $q_{1}$ and the body with mass $m_{2}$ carries charge $q_{2}$. Then the equation of motion of the first body is,
\begin{align}\label{EOM-1}
    m_{1}\Vec{a}_{1}=-\frac{Gm_{1}m_{2}}{r^{3}}\Vec{r}+\frac{1}{4\pi \epsilon_{0}}\frac{q_{1}q_{2}}{r^{3}}\Vec{r}~.
\end{align}
Here $\Vec{r}$ is the separation vector of the first body with respect to the second body. By introducing the effective gravitational constant $G_{\rm eff}$ we can write \eqref{EOM-1} as,
\begin{align}
    m_{1}\Vec{a}_{1}=-\frac{G_{\rm eff}m_{1}m_{2}}{r^{3}}\Vec{r}~,
\end{align}
where $G_{\rm eff}\equiv G(1-\lambda_{1}\lambda_{2})$~with $\lambda_{i}=\frac{q_{i}}{m_{i}}\frac{1}{\sqrt{4\pi \epsilon_{0}G}}$~, $i=1,2$. Similarly, the equation of motion of the second body is,
\begin{align}\label{EOM-2}
    m_{2}\Vec{a}_{2}=\frac{G_{\rm eff}m_{1}m_{2}}{r^{3}}\Vec{r}~.
\end{align}
For simplicity, we consider that the binary evolves in a circular orbit of radius $R$. Then combining \eqref{EOM-1} and \eqref{EOM-2} one can show that the angular velocity of the orbital velocity is given by,
\begin{align}\label{angular-velocity}
    \Omega^{2}=\frac{G_{\rm eff} M }{R^{3}}~.
\end{align}
Here $M=m_{1}+m_{2}$~. The orbital energy of the system (which includes kinetic energy and both gravitational and electrostatic potential energy) is given by,
\begin{align}
    E_{\rm orbit}=-\frac{G_{\rm eff}M\mu}{2 R}~\text{with}~\mu=\frac{m_{1}m_{2}}{M}~.
\end{align}

The accelerated charges will produce dipole radiation which is captured by Larmor's formula~\cite{Jackson:1998nia}. Because of the relative motion of the charges we consider dipole moment($\Vec{d}$) of the whole system, which is given by,
\begin{align}\label{dipole-whole}
    \Vec{d}=\mu\left(\frac{q_{1}}{m_{1}}-\frac{q_{2}}{m_{2}}\right)\Vec{r}~.
\end{align}
Then accordingly to Larmor formula, the rate of loss of energy due to dipole radiation is,
\begin{align}\label{dipole-total}
    \frac{dE_{\text{dip}}}{dt} & =\frac{2}{3}\frac{\kappa}{c^{3}}|\ddot{\Vec{d}}|^{2}~\text{with}~\kappa=\frac{1}{4\pi\epsilon_{0}}                                                             \\
                            & =\frac{2}{3}\frac{\kappa G_{\rm eff}^{2}}{c^{3}R^{4}}m_{1}^{2}m_{2}^{2}Q^{2}~\text{with}~Q^{2}\equiv\left(\frac{q_{1}}{m_{1}}-\frac{q_{2}}{m_{2}}\right)^{2}~.
\end{align}

On the other hand, the emission of gravitational radiation is captured by the quadrupole formula\footnote{The careful reader should note that the $G$ appearing in the first line comes directly from Einstein's equation.},
\begin{align}\label{E-GW}
    \dot{E}_{\rm GW} & =\frac{32 G}{5 c^{5}}\mu^{2}R^{4}\Omega^{6}=\frac{32 G}{5c^{5}}G G_{\rm eff}^{3}\frac{\mu^{2}M^{3}}{R^{5}}
\end{align}
As mentioned in the text we will work under adiabatic approximation, which is manifested as,
\begin{align}\label{evolution}
    \frac{dE_{\rm orbit}}{dt}+\dot{E}_{\text{dip}}+\dot{E}_{\text{GW}}=0~.
\end{align}
Then using \eqref{angular-velocity} in the above condition one gets,
\begin{align}\label{omega-evolution}
    \frac{2}{3}\Omega^{-5/3}(MG_{\rm eff})^{1/3}d\Omega=\left(\frac{\alpha_{\rm GW}\Omega^{2}}{MG_{\rm eff}}+\frac{\alpha_{\rm EM}\Omega^{4/3}} {(MG_{\rm eff})^{2/3}}\right)dt~.
\end{align}
Here we have defined $\alpha_{\rm GW}=\frac{64}{5c^{5}}\mu M^{2}G_{\rm eff}^{2}G$ and $\alpha_{\rm EM}=\frac{4\kappa}{3c^{3}}G_{\rm eff}\frac{m_{1}^{2}m_{2}^{2}}{M\mu}Q^{2}$~.

Now since the post-Newtonian theory is found to be consistent with the observation of binary black hole mergers, we assume that the putative dipole contribution must be small, that is,
\begin{align}\label{PN-approx}
    \dot{E}_{\text{dip}}\ll\dot{E}_{\text{GW}}~.
\end{align}
Under this approximation, \eqref{omega-evolution} can be integrated to yield,
\begin{align}\label{tc}
    t=t_{c}-\frac{(MG_{\rm eff})^{4/3}}{4\alpha_{GW}}\Omega^{-8/3}+\frac{\alpha_{\rm EM}}{5\alpha^{2}_{\rm GW}}(MG_{\rm eff})^{5/3}\Omega^{-10/3}~.
\end{align}
Where the coalescence time is defined as $\Omega(t_{c})=\infty$. Under this approximation $\phi(t)$ becomes,
\begin{align}\label{phi}
    \phi(t) & =\int 2\pi f_{\rm gw} dt=2\int \frac{\Omega}{\dot{\Omega}}d\Omega                                                                                                         \\
            & =\phi_{c}-\frac{4}{5}\frac{(MG_{\rm eff})^{4/3}}{\alpha_{\rm GW}}\Omega^{-5/3}\left[1-\frac{5\alpha_{\rm EM}}{7\alpha_{\rm GW}}(MG_{\rm eff})^{1/3}\Omega^{-2/3}\right]~.
\end{align}
Using \eqref{tc} and \eqref{phi} the phase of the gravitational waves in frequency domain $\Psi_{\text{gw}}=2\pi f_{\rm gw}t-\phi(t)-\frac{\pi}{4}$ becomes,
\begin{align}
    \Psi_{\text{gw}}=\left(2\pi f_{\rm gw}t_{c}-\phi_{c}-\frac{\pi}{4}\right)+\frac{3}{128}\frac{G_{\rm eff}}{G}\left(\frac{M_{c}G_{\rm eff}\pi f_{\rm gw}}{c^{3}}\right)^{-5/3}\left[1-\frac{5}{84}\left(\frac{\mu}{M}\right)^{2/5}\frac{\kappa Q^{2}}{G}\left(\frac{M_{c}G_{\rm eff}\pi f_{\rm gw}}{c^{3}}\right)^{-2/3}\right]~.
\end{align}
This exercise then shows that under dipole radiation GW phase gets $-1\text{PN}$ correction as mentioned in the text, and consistent with the terminology around Eq.~\eqref{pN_expansion_phase}. Results in the literature displaying bounds on dipolar radiation from gravitational-wave measurements include Refs.~\cite{Barausse:2016eii,Cardoso:2016olt,Seoane:2021kkk,Perkins:2020tra}.

\vspace{0.25cm}

\noindent{\bf $\blacksquare$\ Q.3.a.} In the text, we have discussed how we might model a black hole residing at the center of galaxies. In particular, we have encountered a spherically symmetric solution to the Einstein field equations which describes a black hole at small scales but at large scales describes a Hernquist-type matter distribution. Use the strategy described in the text to explicitly derive the galactic black hole solution, first by assuming that the galaxy follows a Hernquist like density distribution. Show that the solution corresponds to a matter density
\begin{equation}
    4 \pi \rho = \dfrac{m'}{r^2}=\dfrac{2 M(a_0+2 M_{\mathrm{BH}})(1-2M_{\mathrm{BH}}/r)}{r(r+a_0)^3}\,,
\end{equation}
where $m=m(r)$ but we omit the argument for simplicity of notation, the primes stand for partial derivatives with respect to $r$. Verify that at large distances, the fall-off is
\begin{equation}
    \rho \sim \dfrac{M(a_0+2M_{\mathrm{BH}})}{r^4} \approx \dfrac{1}{r^4}.
\end{equation}
Calculate the ISCO radius and the corresponding angular frequency, the location of the photon sphere,  light ring frequency, Lyapunov exponent of this black hole solution. Estimate the critical impact parameter for the capture of high frequency photons and gravitational waves.

\vspace{0.25cm}

\noindent\textbf{{$\square$}\ Solution:} The solution has been worked out in the notebook \coderef{code:Galactic-BH.nb}. In the very first exercise of the first lecture, we saw how to derive the Schwarzschild solution, that is, a vacuum solution to the Einstein Field Equations (EFEs) in \textsc{Mathematica}\textsuperscript{\textregistered}. To bring our journey to a full circle, we shall attempt to derive a solution the EFEs in presence of matter, viz.,
\begin{equation}
    T^\mu_\nu = \mathrm{diag}(-\rho, P_r, P_t, P_t). \label{eq:radial_ode}
\end{equation}
Keeping the construction of the Einstein cluster in mind, we shall assume that our stress-tensor describes an anisotropic fluid with only tangential pressure and set $P_r=0$ in the final equations. So we start with the ansatz given by Eq.~\eqref{eq:galactic_ansatz} and compute the Einstein tensor in a manner identical to that followed in \coderef{code:Field-Equations.nb}. We then set $G^r_r = 8 \pi T^r_r$ to get
\begin{equation}
    \dfrac{rf'}{2f}=\dfrac{m}{r-2m}\,. \label{eq:fprimeDM}
\end{equation}
To obtain the expression for the tangential pressure, we substitute the expression for $f'$ obtained above into the Bianchi identity $\nabla_\mu G^{\mu}_\theta=0$ and write,
\begin{equation}
    2P_t=\dfrac{m\rho}{r-2m}\,.
\end{equation}
Finally, from the $G^\theta_\theta = 8 \pi T^\theta_\theta$ equation, we get
\be
\frac{r (r - 2m) \, (f' )^2}{f}
+ 2 f' \big(m + r(m'-1)\big) + \frac{4 f\, ( 8\pi r^3 P_t + r m'-m)}{r}
= 2r (r - 2m ) f''\,.
\ee
We plug in the expressions for $f'(r)$ and $P_t$ in the above equation to get
\begin{equation}
    4 \pi \rho = \dfrac{m'}{r^2} = \dfrac{2 P_t\left(r-2 m\right)}{m}\,. \label{eq:ex_rho}
\end{equation}
Now, consider a matter distribution governed by a Hernquist-type profile~\cite{Cardoso:2021wlq,hernquist1990analytical},
\beq
m&=&M_{\rm BH}+\frac{Mr^2}{(a_0+r)^2}\left(1-\frac{2M_{\rm BH}}{r}\right)^2\,.
\eeq
Using Eq.~\eqref{eq:ex_rho}, we can immediately write,
\begin{equation}
    \rho = \dfrac{ M (a_0 + 2 M_\mathrm{BH}) (r-2 M_\mathrm{BH})}{ 2 \pi r^2 (a_0 + r)^3}\,,
\end{equation}
and
\begin{equation}
    P_t = \dfrac{M  (a_0 + 2 M_\mathrm{BH})  (r - 2 M_\mathrm{BH})  m
    }{4 \pi r^2  (a_0 + r)^3 (r - 2 m)}\,.
\end{equation}
Note that, if we perform a series expansion of $\rho$ about infinity, we get
\begin{equation}
    \rho \approx \frac{M (a_0+2 M_\mathrm{BH})}{2 \pi  r^4}-\frac{M (3 a_0+2 M_\mathrm{BH}) ({a_0}+2 M_\mathrm{BH})}{2 \pi  r^5}+\mathcal{O}\left(\dfrac{1}{r^6}\right)\,,
\end{equation}
which confirms our expectation that the matter density has an $r^{-4}$ fall off.
We now solve Eq.~\eqref{eq:radial_ode} to get the radial function $f(r)$. For symbolic computation, we may use \texttt{DSolve} but it gives the general solution, viz, $$f(r)=C_\infty e^{I(r)}$$ where $C_\infty$ is a constant and $I(r)$ is a very messy function of $r$. We fix $C_\infty$ by imposing asymptotic flatness: we calculate the limit of $f(r)$ at $r \to \infty$ to get,
\begin{equation}
    \lim_{r \to \infty} f(r) = C_\infty \sqrt{\dfrac{M} {\xi}} \equiv K_\infty \,,
\end{equation}
where $\xi = 2 a_0 -M + 4 M_\mathrm{BH}$. To arrive at the final answer, we now simply normalize $f(r)$, such that
\begin{equation}
    f(r) \equiv f(r)/K_\infty = \left(1-\frac{2M_{\rm BH}}{r}\right)e^{\Upsilon}\,,
\end{equation}
where $\Upsilon=-\pi\sqrt{{M}/{\xi}}+2\sqrt{{M}/{\xi}}\arctan({({r+a_0-M})/{M\xi}})$.
This ensures asymptotic flatness, that is, $f(r\to \infty) =1$.

Once we obtain $f(r)$, we can use it, along with $m(r)$, to determine the geodesic properties of massless and massive particles. Note that, like the Schwarzschild spacetime, we can identify two conserved quantities, the energy per unit mass $E = f \dot{t}$ and the specific angular momentum $L = r^2 \dot{\phi}$ at infinity (c.f., Eqs.~\eqref{eqn:tdot_geodesic} and \eqref{eqn:tdot_geodesic}), associated with the line element given by Eq.~\eqref{eq:galactic_ansatz} since it possesses two Killing vectors as well. So the geodesics are planar and we can set $\theta=\pi/2$. Note that in Chapter \ref{ch:part1} we had used $f=g$, this is obviously not true for the present case but the calculation can be generalized easily to the present case. In particular, the normalization condition $u^{\mu}u_\mu=\delta_1$ gives us
\begin{equation}
    \dot{r}^2=\dfrac{1-2m/r}{f}V \equiv V_\mathrm{BH+DM} \label{eqn:hernquist_normalization}
\end{equation}
where $V = E^2-f \dfrac{L^2}{r^2}+f \delta_1$ (c.f. Eq.~\eqref{eq_Eff_potential_geodesics}) with $\delta_1=-1,0$ for massive and massless particles respectively.

For massive particles moving on circular orbits\footnote{We indicate the radius of the circular orbit with $r$ instead of $r_c$ in the following and restore it only if necessary.} ($r=r_c$, and $\dot{r}=\ddot{r}=0$), we can use the radial geodesic equation and put $\delta_1=-1$ in \eqref{eqn:hernquist_normalization} to write,
\begin{align}
    -f'\dot{t}^2+2 r \dot{\phi}^2 =0, \quad 1-f \dot{t}^2 + r^2 \dot{\phi}^2 =0 .
\end{align}
Now, using Eq.~\eqref{eq:fprimeDM}, $E = f \dot{t}$ and $L = r^2 \dot{\phi}$, we arrive at (see \coderef{code:Galactic-BH.nb})
\begin{equation}
    E = \left[ \frac{r - 2m}{\,r - 3m\,} \, f\right]^{1/2}_{r = r_c},
    \qquad
    L = \left[r \left(\frac{m}{\,r - 3m\,}\right)^{1/2}\right]_{r = r_c}.
\end{equation}
The subscript $r = r_c$ indicates that the above relationships hold for a massive particle orbiting the black hole in a circular orbit of radius $r_c$.
Note that from Eq.~\eqref{eqn:hernquist_normalization}, we can write,
\begin{equation}
    E^2=f\left(1+\dfrac{L^2}{r^2}\right) \equiv V_\mathrm{eff}(r,L),
\end{equation}
for circular orbits, where we already have $V'_\mathrm{eff}=0$
at $r_c$ to ensure that the particle remains on the circular orbit. Recall that for marginally stable orbits, we should have $V''_\mathrm{eff}=0$ at $r_c$. Taking the total derivative of the $V_{\mathrm{eff}}' = 0$ condition along the family of circular orbits (where $L$ must be a function of $r$), this then implies
\begin{equation}
    \dfrac{d}{dr} V'_{\mathrm{eff}}(r, L(r))= V''_{\mathrm{eff}}(r, L)+ \dfrac{\partial V'_{\mathrm{eff}}}{\partial L}\, L'(r)= 0.
\end{equation}
At a circular orbit, ${\partial V'_{\mathrm{eff}}}/{\partial L}$ is nonzero, so
\begin{equation}
    V''_{\mathrm{eff}} = 0 \Rightarrow L'(r) = 0,
\end{equation}
which readily gives us the equation,
\begin{equation}
    r^2 m' + r m - 6 m^2 =0. \label{eqn:ISCO_eqn}
\end{equation}
The solution of the above equation gives us the radius of the innermost stable circular orbit, and we find,
\begin{equation}
    r_\mathrm{ISCO} \sim 6 M_\mathrm{BH}\left(1-\dfrac{32 M M_\mathrm{BH}}{a_0^2}\right)\,.
\end{equation}
To obtain the above result, one should impose a hierarchy of scales, that is, $M_\mathrm{BH} \ll M \ll a_0$ (which for the description of galactic haloes is also imposed on us by observations). Specifically, using our choice of $m$, we expand Eq.~\eqref{eqn:ISCO_eqn} around $r_\mathrm{ISCO}=6M_\mathrm{BH}(1+\varepsilon)$, and retain terms that are of the first order in $M/a_0$.

We can solve this equation to determine $\varepsilon$ in terms of $M_\mathrm{BH}$ and $a_0$. This has been implemented in \coderef{code:Galactic-BH.nb} We can obtain higher order corrections in a similar way. We can now easily calculate the corresponding ISCO frequency as follows,
\begin{align}
    \Omega_\mathrm{ISCO} & \equiv \dfrac {\dot{\phi}}{\dot{t}} = \dfrac{L}{r^2}\dfrac{f}{E} = \left[\dfrac{1}{r^2} \dfrac{f m}{r-2m}\right]_{r=r_\mathrm{ISCO}}^{1/2} \\
                        & \sim \dfrac{1}{6\sqrt{6}M_\mathrm{BH}}\left(1-\dfrac{M}{a_0}+\dfrac{M(M+396 M_\mathrm{BH})}{6a_0^2}\right).
\end{align}
Note that the zeroth order term in $r_\mathrm{ISCO}$ and $\Omega_\mathrm{ISCO}$ correspond to those of the Schwarzschild geometry.

Let us now move on to massless particles.
From Eq.~\eqref{eqn:hernquist_normalization}, we can write for null particles moving on circular orbits,
\begin{equation}
    V'=0 \Rightarrow  \left( \dfrac{f}{r^2}\right)'=0 \Rightarrow \dfrac{f'}{2f} = \dfrac{1}{r}\,.\label{eqn:vprime}
\end{equation}
We can now immediately use Eq.~\eqref{eq:fprimeDM} to write down the photon sphere equation,
\begin{equation}
    r - 3 m = 0 \,.
\end{equation}
We can therefore deduce location of the light ring of the black hole immersed in a Hernquist-type halo,
\begin{equation}
    r_\mathrm{LR} \sim 3 M_\mathrm{BH}\left(1+\dfrac{M M_\mathrm{BH}}{a_0^2} -\dfrac{6 M M_\mathrm{BH}}{a_0^3}\right) \,,
\end{equation}
where we use the approximation $M_\mathrm{BH} \ll M \ll a_0$ as before (see \coderef{code:Galactic-BH.nb}). Now, from $u^\mu u_\mu=0$, the light ring frequency turns out to be
\begin{equation}
    \Omega_\mathrm{LR} = \dfrac{\dot{\phi}}{\dot{t}} = \dfrac{f^{1/2}}{r_\mathrm{LR}} \sim \dfrac{1}{3\sqrt{3}}\left(1-\dfrac{M}{a_0} +\dfrac{M(M+18M_\mathrm{BH})}{6 a_0^2}\right)\,. \label{eqn:omegaLR}
\end{equation}
Lastly, we can calculate the Lyapunov exponent (c.f. Eq.~\eqref{null_geoedesic_decay}),
\begin{equation}
    \lambda=\left[\sqrt{\frac{W_\mathrm{BH+DM}''\,f^2}{2}} \right]_{r=r_\mathrm{LR}}
\end{equation}
where
\begin{equation}
    W_\mathrm{BH+DM} =  \dfrac{V_\mathrm{BH+DM}}{E^2}=\left(1-\dfrac{2m}{r}\right)\left[\dfrac{1}{f} -\dfrac{b^2}{r^2}\right]\,,
\end{equation}
with $b=L/E$ is the usual impact parameter. Using $b=r/f^{1/2}$, $2 f=r f'$ and $\delta_1=0$ for null circular orbits (c.f. Eq.~\eqref{eqn:omegaLR} and  Eq.~\eqref{eqn:vprime}), we get
\begin{equation}
    \lambda \sim \dfrac{1}{3 \sqrt{3}M_\mathrm{BH}}\left( 1 - \dfrac{M}{a_0}+\dfrac{M^2}{6 a_0^2}\right).
\end{equation}

Finally, we compute the critical impact parameter for the capture of high frequency photons or gravitational waves in \coderef{code:Galactic-BH.nb},
\begin{equation}
    b_{\mathrm{crit}}=\dfrac{r}{f^{1/2}}\Biggr{|}_{r=r_\mathrm{LR}} = 3 \sqrt{3}{M_\mathrm{BH}}\left(1 +\dfrac{M}{a_0}
    +\dfrac{M (5 M-18 M_\mathrm{BH})}{2 {a_0}^2}\right)
\end{equation}

\vspace{0.25cm}

\begin{figure}[ht]
    \centering

    \begin{minipage}[c]{0.48\textwidth}
        \centering
        \includegraphics[width=\textwidth]{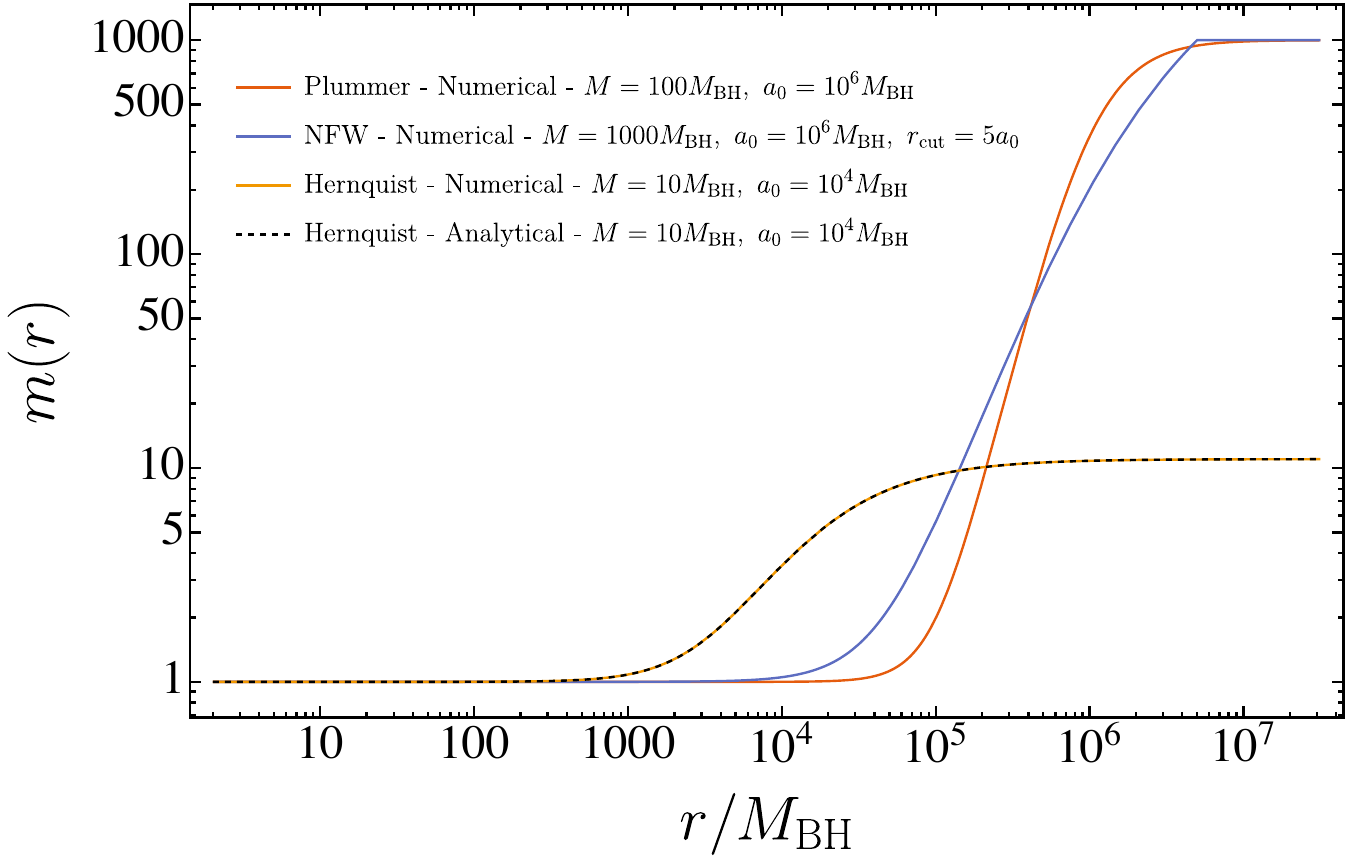}
    \end{minipage}
    \hfill
    \begin{minipage}[c]{0.48\textwidth}
        \centering
        \includegraphics[width=\textwidth]{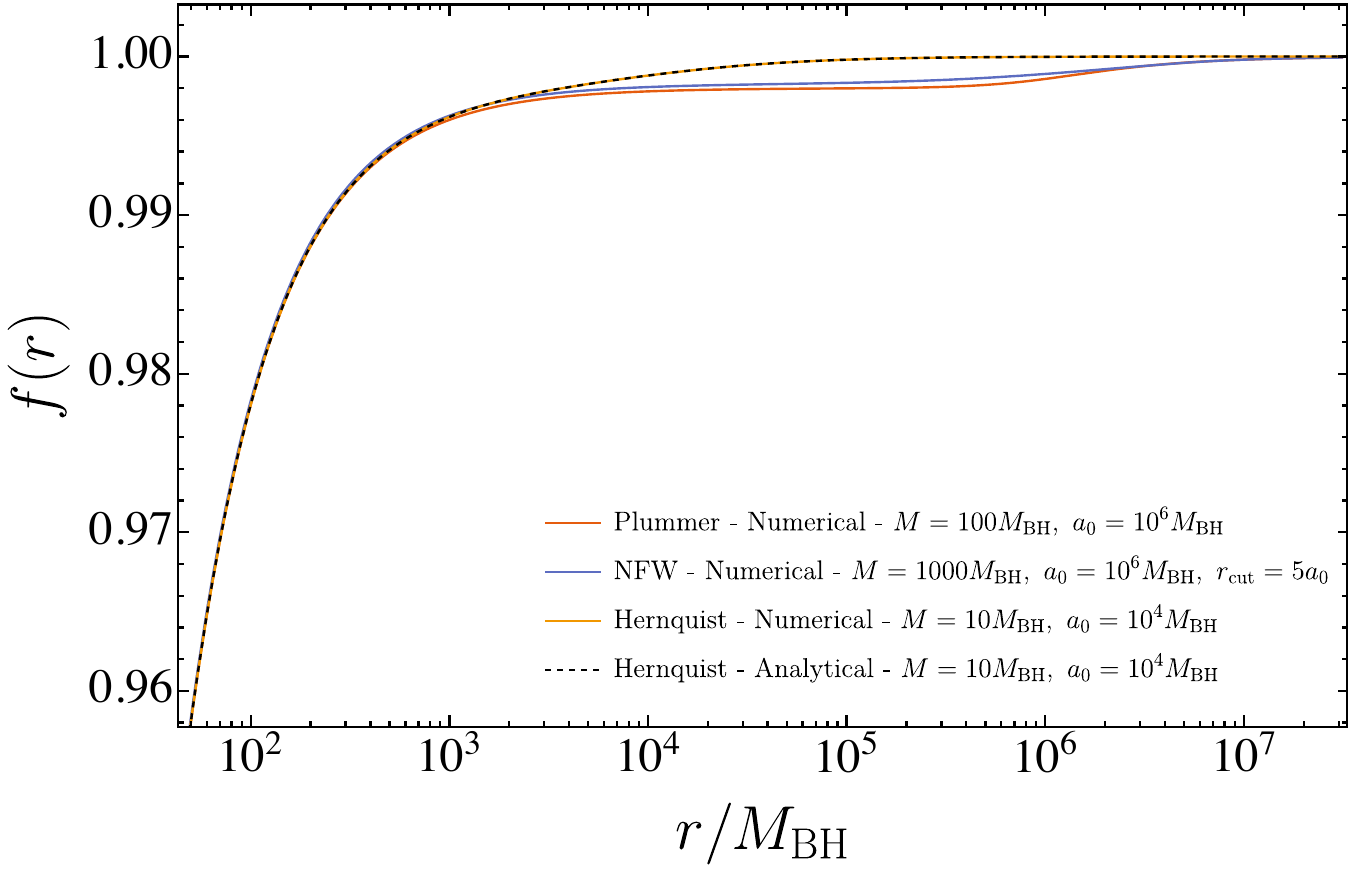}
    \end{minipage}

    \caption{Behavior of the mass function $m(r)$ and the metric function $f(r)$ for a black hole at the center of a galaxy embedded in a dark-matter halo described by Plummer, Navarro–Frenk–White (NFW), and Hernquist profiles. \textbf{Left:} $m(r)$ in log–log scale showing the transition from the black-hole–dominated region at small $r$ to the halo-dominated regime at large $r$, where $m(r)\to M_{\mathrm{BH}}+M$.  \textbf{Right:} $f(r)$ in $\log_{10}$–linear scale showing how the halo slightly modifies the Schwarzschild geometry at large $r$. Differences between the models appear at large radii but eventually $f(r)\to 1$. In both figures, the numerical results for the Hernquist profile show excellent agreement with the analytical solution.}
    \label{fig:generic_dark_matter_halo_m_f}
\end{figure}

\noindent{\bf $\blacksquare$\ Q.3.b.} Let us end with an extremely useful numerical exercise. We shall focus on more generic dark matter profiles \cite{Zhao:1995cp,Graham:2005xx}. Why? Well, for example, while the Hernquist profile is useful for modeling elliptic galaxies and bulges of spiral galaxies \cite{hernquist1990} but if we are looking at globular clusters, it is more appropriate to model the matter density using the Plummer profile \cite{plummer1911},
\begin{equation}
    \rho = \dfrac{3a_0^2M}{4 \pi (r^2+a_0^2)^{5/2}}\,,
\end{equation}
which falls off as $r^{-5}$ at large distances. Now although the Plummer model is a popular toy model used in $N-$body simulations \cite{2008gady.book.....B}, the Navarro–Frenk–White (NFW) profile \cite{Navarro:1995iw, Navarro:1996gj} is more commonly used to describe a more realistic matter distribution with a dark matter halos \cite{Springel:2008cc}. The NFW model,
\begin{equation}
    \rho = \dfrac{\rho_0 a_0}{r(a_0+r)^2}\,,
\end{equation}
has an $r^{-3}$ fall-off at large distances, and the constants $\rho_0$ and $a_0$ vary from halo to halo. Model galactic black holes described by the Plummer and NFW profiles.

\vspace{0.25cm}

\noindent\textbf{{$\square$}\ Solution:} We start by noting that the density distributions that we have talked about can all be obtained from following relation,
\begin{equation}
    \rho = \rho_0 \left(\dfrac{r}{a_0}\right)^{\gamma}\left[1+\left(\dfrac{r}{a_0}\right)^{\alpha}\right]^{(\gamma-\beta)/\alpha}\,.
\end{equation}
Here, the parameters $\beta$ and $\gamma$ fix the behavior of the density distribution at large and small distances, while $\alpha$ determines the sharpness of the transition. The Hernquist profile corresponds to $(\alpha,\beta,\gamma)=(1,4,1)$, while for Plummer and NFW profiles we have $(\alpha,\beta,\gamma)=(2,5,0)$ and $(\alpha,\beta,\gamma)=(1,3,1)$ respectively. To relate $\rho_0$ to the halo mass $M$ we must perform the integration,
\begin{equation}
    \int^{\infty}_0 4 \pi r^2 \rho = M\,.
\end{equation}
We must be a bit careful with the NFW profile. 
A naive attempt at integrating the above equation for the NFW profile will show that it diverges. Therefore, we must prescribe a radial cut-off $r_\mathrm{cut}$ while performing the integration such that $M(r>r_\mathrm{cut})=0$. We demonstrate this in \coderef{code:Galactic-BH.nb} as well. Now, our job is to prescribe the numerical parameters of the problem and choose a particular $\rho$, and solve
\begin{equation}
    m'= 4 \pi r^2 \rho
\end{equation}
to determine the mass function $m$. We start our integration from $r=2 M_\mathrm{BH}$, the black hole horizon, where $m = M_\mathrm{BH}$ provides the initial condition. We end at a sufficiently large radius like $r_\mathrm{out}= 10^{7}a_0$. Our results should not depend on the value of $r_\mathrm{out}$.

Once we have obtained $m$, we can use the data to construct an interpolating function, \texttt{NDSolve} does this automatically. We can now solve for $f(r)$ using Eq.~\eqref{eq:radial_ode}. We integrate from $r_\mathrm{out}$ to $r=2 M_\mathrm{BH}+\epsilon$ where $\epsilon$ is very small. We do this because Eq.~$\eqref{eq:radial_ode}$ diverges at the black hole horizon. We impose the initial condition,
\begin{equation}
    f(r) = 1 - \dfrac{2(M_\mathrm{BH}+M)}{r} + \mathcal{O}(r^{-3})\,.
\end{equation}
This follows from $m(r\to r_\mathrm{out})=M_\mathrm{BH}+M$. We also set $M_\mathrm{BH}=1$ which essentially sets the length scale of the problem. We can then choose different values of $M$ and $a_0$ to model different halos. We first benchmark our method by using the Hernquist profile and comparing the numerical results with the analytical expressions. We find excellent agreement as shown in Fig.~\ref{fig:generic_dark_matter_halo_m_f}. We can then use the method for any generic dark matter distributions. The method has been implemented in \coderef{code:Galactic-BH.nb} and we show the plot of the mass function $m(r)$ and the metric function $f(r)$ for the Plummer and NFW profiles in the right and left panels of Fig~\ref{fig:generic_dark_matter_halo_m_f} respectively. We can then in principle calculate the tangential pressure $P_t$, geodesic quantities, quasinormal modes and so much more. We refer the reader to Refs. \cite{Cardoso:2021wlq,Figueiredo:2023gas,Speeney:2024mas} for details.




\chapter{Denouement}
\label{ch:conclusion}
\allowdisplaybreaks
\minitoc

\section*{Discussion and Summary}




\epigraph{All that is red is beautiful,\\
    And all that is new is bright,\\
    all that is high is lovely, all that is familiar is bitter.\\
    The unknown is honoured, the known is neglected,\\
    until all knowledge is known.}
    {Anonymous, Irish, ninth century,\\ \textit{The Sick-Bed of C\'{u} Chulainn.}}

We have centuries of experience in observations of the cosmos in the electromagnetic spectrum. We have built instruments across a wide range of frequencies from radio waves to gamma rays, and we built a mental picture of the various phenomena and of how to interpret and catalog electromagnetic events. Our brain, language, and civilization dwell on electromagnetic observations on a daily basis.

By contrast, we have barely begun to understand the universe through gravitational waves. The path was laid over the course of the last century: beautiful results concerning gravitational collapse leading to the formation of black holes, the structure of black hole spacetimes and their uniqueness in vacuum. We have powerful mathematical results governing gravitational radiation and the conformal structure of spacetimes. We understand how to write Einstein equations in a variety of forms, some amenable to numerical simulations, and we have discovered what seem to be the dominating aspects on the dynamics of black hole spacetimes. It is also amazing to note how much we can understand about black holes using ``simple'' perturbative techniques. Some twenty years after the publication of Regge and Wheeler's pioneering work \cite{Regge:1957rw} on the perturbative stability of the Schwarzschild ``singularity'', Detweiler remarked  \cite{Detweiler1979aaa} in 1979 that he found it ``remarkable that after all of this time the subject is still yielding interesting information and has ample opportunities for research projects''. We should perhaps consider ourselves doubly fortunate, because the cup is yet to run dry, and we finally have the data to test our predictions. These lecture notes described some of the most relevant results and traced a few of the path and trails of this great expedition. If the path doesn't lead anywhere, let's at least enjoy the walking.


\newpage
\vspace{1cm}
\thispagestyle{plain}
\addcontentsline{toc}{chapter}{\nameref{sec:Acknowledgments}}
\section*{Acknowledgments} \label{sec:Acknowledgments}
We are indebted to Vishal Baibhav and Roberto Emparan for permission to share previously unpublished work with V.C. on superradiant amplification.
We are thankful to the organizers Sumanta Chakraborty and Sudipta Sarkar and participants of the program \emph{Beyond the Horizon: Testing the Black Hole Paradigm} held at the International Centre for Theoretical Sciences (ICTS-TIFR) Bengaluru. The lively and in-depth discussions stimulated us to have the lectures written up and renewed our interest in some of these problems. We also thank the members of the Astrophysical Relativity Group and the staff at ICTS for their warm hospitality during the program.
The Center of Gravity is a Center of Excellence funded by the Danish National Research Foundation under grant No. DNRF184.
We acknowledge support by VILLUM Foundation (grant no.\ VIL37766).
V.C. acknowledges financial support provided under the European Union’s H2020 ERC Advanced Grant “Black holes: gravitational engines of discovery” grant agreement no.\ Gravitas–101052587.
Views and opinions expressed are however those of the author only and do not necessarily reflect those of the European Union or the European Research Council. Neither the European Union nor the granting authority can be held responsible for them.
S.B. thanks IACS, Kolkata for providing his PhD fellowship since a part of this work was done when he was a PhD student there.
S. S. acknowledges funding from the Institutions of Eminence initiative of the Government of India through the Centre for Industrial Consultancy and Sponsored Research (IC\&SR) of the Indian Institute of Technology Madras for the project titled Centre for Strings, Gravitation and Cosmology (Project No.: SB22231259PHE-TWO008479).
This project has received funding from the European Union's Horizon 2020 research and innovation programme under the Marie Sk{\l}odowska-Curie grant agreement No 101007855 and No 101131233.
The Tycho supercomputer hosted at the SCIENCE HPC center at the University of Copenhagen was used for supporting this work.
This work is supported by Simons Foundation International \cite{sfi} and the Simons Foundation \cite{sf} through Simons Foundation grant SFI-MPS-BH-00012593-11.
This work was supported in part by the International Centre for Theoretical Sciences (ICTS-TIFR), Bengaluru for participating in the program \emph{Beyond the Horizon: Testing the Black Hole Paradigm} (code: ICTS/BTH2025/03), held from 24 March 2025 to 04 April 2025.\\ \\
\textbf{Note:} S.B. and S.S. contributed equally to this manuscript and should be considered as joint second authors.


\appendix
\chapter{List of  \texorpdfstring{\textsc{Mathematica}\textsuperscript{\textregistered}}{Mathematica Notebooks} Notebooks}
\label{appendix_codes}
\allowdisplaybreaks
\minitoc

The \textsc{Mathematica}\textsuperscript{\textregistered} notebooks listed here have been developed to accompany the lecture notes and are available online~\cite{CoG,GRIT}. They clarify and expand many of the analytical and numerical results discussed in the main text, particularly those appearing in the exercises. These notebooks will therefore assist the reader in reproducing several important results in the field and may even help in carrying out state-of-the-art research. We also believe that some of them can even be adopted for a course in General Relativity, bridging topics between graduate textbooks and current literature. For convenience, we also provide a brief summary of the contents of each notebook below. If you find the codes in these notebooks useful for your research, please consider citing the lecture notes and the references listed in the corresponding \texttt{.nb} files.

\section{Part I}\label{sec:part1_notebooks}
\begin{itemize}
    \codefile[code:Field-Equations.nb]{\textcolor{red}{Field$_-$Equations.nb}} : Computes Christoffel symbols, Riemann and Ricci tensors, Ricci scalar, and Einstein tensor for a given metric. Solves Einstein’s equations under spherical symmetry (Schwarzschild solution) and verifies that the Kerr metric is a vacuum solution to the Einstein field equations.
    \codefile[code:Kerr-Accretion.nb]{\textcolor{red}{Kerr$_-$Accretion.nb}} : Computes the radial geodesic equation in the Kerr background and studies the process of accretion onto a Kerr black hole from the innermost stable circular orbit (ISCO). Derives the equations in Bardeen's paper \cite{Bardeen:1970zz} and studies the change in the ISCO radius with respect to the black hole spin and mass. Finally relates the final mass to the black hole to the accreted mass.
\end{itemize}

\section{Part II}\label{sec:part2_notebooks}
\begin{itemize}
    \codefile[code:Perturbations.nb]{\textcolor{red}{Perturbations.nb}}: Separates and decouples the equation of motion for scalar, vector, and (axial) gravitational perturbations in the Schwarzschild background.  Includes a numerical evaluation of the static, scalar tidal Love number of the Schwarzschild black hole. Calculates the energy fluxes for the axial and polar electromagnetic perturbations. Introduces \texttt{xAct} \cite{xAct} and shows how to separate the massive scalar wave equation in a spherically symmetric background.
    \codefile[code:Scattering-Scalar.nb]{\textcolor{red}{Scattering$_-$Scalar.nb}}: Computes the coefficients of the series expansion of the wave equation in spherical symmetry near the horizon and infinity up to the desired order. Studies the scattering of a Gaussian pulse off a Schwarzschild black hole and reproduces Vishveshwara's results \cite{Vishveshwara:1970zz}, demonstrating ringdown, and calculates the emitted flux and the frequency of the fundamental mode. Studies the scattering of a monochromatic scalar pulse by a Schwarzschild black hole, and calculates the ingoing or outgoing amplitudes, and hence the absorption amplitudes of a scalar field for the black hole.
    \codefile[code:Zerilli-Infalling-Photon.nb]{\textcolor{red}{Zerilli$_-$Infalling$_-$Photon.nb}}: Computes the gravitational waveform and the flux from the collision of a massless particle falling into a Schwarzschild black hole by solving the Zerilli equation \cite{Cardoso:2002ay}. Briefly illustrates how to perform the series expansion for the Zerilli equation. Calculates the source term for the Zerilli equation algebraically\cite{Zerilli:1970wzz, Zerilli:1974ai}.
    \codefile[code:BH-Bomb.nb]{\textcolor{red}{BH$_-$Bomb.nb}}: Simulates a black hole bomb, that is, solves the Klein Gordon equation in the Kerr background with a mirror placed at a finite radius \cite{Cardoso:2004nk}, demonstrates the occurrence of superradiant instability. Calculates boxed quasinormal mode frequencies. Shows that the instability occurs when the superradiance condition is satisfied. Studies the dependence of the frequencies on the spin of the black hole.

\end{itemize}

\section{Part III}\label{sec:part3_notebooks}
\begin{itemize}
    \codefile[code:QNMs-DI.nb]{\textcolor{red}{QNMs\_DI.nb}}: Uses the method of direct integration to calculate the quasinormal modes of the Schwarzschild black hole, based on the pioneering work of Chandrasekhar and Detweiler \cite{Chandrasekhar:1975zza}. Demonstrates that the result does not depend on the matching point and given error estimates. Also calculates the first overtone for the $l= s = 2$ mode.
\end{itemize}

\section{Part IV}\label{sec:part4_notebooks}
\begin{itemize}
    \codefile[code:Galactic-BH.nb]{\textcolor{red}{Galactic\_BH.nb}}: Solves the Einstein field equations in presence of matter, using the Einstein cluster construction and Hernquist-type matter distribution to find an analytical solution describing a black hole embedded in a dark matter halo \cite{Cardoso:2021wlq}. Derives the ISCO and light ring equations, and computes the leading order corrections to the ISCO radius, ISCO frequency, light ring radius, light ring frequency, Lyapunov exponent, and the critical impact parameter in presence of the dark matter halo. Derive the same solution numerically and compares it with the analytical result. Finally extends the numerical procedure to arbitrary dark matter profiles and obtains the numerical BH solutions for the Plummer and Navarro-Frenk-White profiles \cite{Figueiredo:2023gas}.
\end{itemize}



\providecommand{\href}[2]{#2}\begingroup\raggedright\endgroup

\end{document}